\documentclass{amsart}
\numberwithin{equation}{section}
\usepackage[utf8]{inputenc}
\usepackage[T1]{fontenc} 
\usepackage{lmodern} 
\usepackage{geometry}
\usepackage[usenames, dvipsnames]{xcolor}
\usepackage{graphicx}
\usepackage{subcaption}
\usepackage{mathtools}
\usepackage{amssymb}
\usepackage{amsthm}
\usepackage{xparse}
\usepackage{xspace}
\usepackage{mathrsfs}

\usepackage{setspace}
\usepackage[backref=page]{hyperref}
\usepackage[all]{hypcap}
\usepackage[capitalize, noabbrev]{cleveref}
\usepackage{booktabs}

\usepackage{quiver}

\usepackage{spectralsequences}
\usepackage{adamsmacros}

\newcommand{\C}{\mathbb C} 
\newcommand{\E}{\mathbb E} 
\newcommand{\F}{\mathbb F} 
\AtBeginDocument{ 

}
\newcommand{\N}{\mathbb N} 
\newcommand{\Q}{\mathbb Q} 
\newcommand{\R}{\mathbb R} 
\newcommand{\Sph}{\mathbb S} 
\newcommand{\Z}{\mathbb Z} 
\newcommand{\RP}{\mathbb{RP}} 
\newcommand{\CP}{\mathbb{CP}} 
\newcommand{\cA}{\mathcal A} 
\newcommand{\cE}{\mathcal E}

\AtBeginDocument{ 
 \renewcommand{\O}{\mathrm O} 
}
\newcommand{\SO}{\mathrm{SO}} 
\newcommand{\U}{\mathrm U} 
\newcommand{\SU}{\mathrm{SU}} 
\newcommand{\Sp}{\mathrm{Sp}} 
\newcommand{\Spin}{\mathrm{Spin}} 
\newcommand{\Pin}{\mathrm{Pin}} 
\newcommand{\DPin}{\mathrm{DPin}} 

\newcommand{\Euc}{\mathrm{Euc}}

\DeclareRobustCommand*{\RaiseBoxByDepth}{%
    \raisebox{\depth}%
}
\newcommand{\uQ}{\RaiseBoxByDepth{\protect\rotatebox{180}{$Q$}}}

\newcommand{\Spinc}{\relax\ifmmode{\Spin^c}\else Spin\textsuperscript{$c$}\xspace\fi}
\newcommand{\spinc}{spin\textsuperscript{$c$}\xspace}
\newcommand{\Pinc}{\relax\ifmmode{\Pin^c}\else Pin\textsuperscript{$c$}\xspace\fi}
\newcommand{\pinc}{pin\textsuperscript{$c$}\xspace}
\newcommand{\Pinp}{\relax\ifmmode{\Pin^+}\else Pin\textsuperscript{$+$}\xspace\fi}
\newcommand{\pinp}{pin\textsuperscript{$+$}\xspace}
\newcommand{\Pinm}{\relax\ifmmode{\Pin^-}\else Pin\textsuperscript{$-$}\xspace\fi}
\newcommand{\pinm}{pin\textsuperscript{$-$}\xspace}

\newcommand{\spinh}{spin\textsuperscript{$h$}\xspace}

\makeatletter
\usepackage[mathscr]{eucal}
\def\instring#1#2{TT\fi\begingroup
  \edef\x{\endgroup\noexpand\in@{#1}{#2}}\x\ifin@}
\def\isuppercase#1{%
  \instring{#1}{ABCDEFGHIJKLMNOPQRSTUVWXYZ}%
}%
\newcommand{\C@lIfUpper}[1]{
 \if\isuppercase{#1}\mathscr{#1}%
 \else #1%
 \fi
}
\newcommand{\cat}[1]{\mathit{\@tfor\next:=#1\do{\C@lIfUpper{\next}}}}
\makeatother
\newcommand{\newcat}[2]{\newcommand{#1}{\cat{#2}}}
\AtBeginDocument{
 \newcat{\fC}{C}
 \newcat{\fD}{D}
 \newcat{\Set}{Set} 
 \newcat{\Grp}{Grp} 
 \newcat{\Gpd}{Gpd} 
 \newcat{\Ab}{Ab} 
 \newcat{\Ring}{Ring} 
 \newcat{\Mod}{Mod} 
 \newcat{\Alg}{Alg} 
 \newcat{\Vect}{Vect} 
 \newcat{\sVect}{sVect} 
 \def\Top{\cat{Top}} 
}
\newtheorem{thm}[equation]{Theorem}
\newtheorem{lem}[equation]{Lemma}
\newtheorem{cor}[equation]{Corollary}
\newtheorem{prop}[equation]{Proposition}
\newtheorem{ansatz}[equation]{Ansatz} 

\newtheorem*{longsummand}{\cref{melk_calc,only_one_long_summand}}
\newtheorem*{phithm}{\cref{int_Bott_spiral_1}}
\newtheorem*{psithm}{\cref{MSpin_psi,special_case_psi_ko}}
\newtheorem*{ftoithm}{\cref{only_first_summand_real}}
\newtheorem*{ouransatz}{\cref{our_F2I_ansatz}}
\newtheorem*{IFTclass}{\cref{IFT_class}}
\newtheorem*{Elksmash}{\cref{lem_shearing_Elk}}

\theoremstyle{definition}

\newtheorem{exmx}[equation]{Example}
\newtheorem{defn}[equation]{Definition}
\newtheorem{physres}[equation]{Physical Result} 

\newtheorem{conj}[equation]{Conjecture}

\newtheorem*{ftoidefn}{\cref{Elk_F2I}}
\newtheorem*{absdefn}{\cref{defn_generalized_ABS}}
\newtheorem*{spiraldefn}{\cref{spiral_maps_of_spectra,spiral_maps_of_IFTs}}
\newtheorem*{MElkdefn}{\cref{MElk_defn}}
\newtheorem*{Elkdefn}{\cref{elk_definition}}
\newtheorem*{Gfprimedefn}{\cref{G_f_prime}}

\theoremstyle{remark}
\newtheorem{remx}[equation]{Remark}
\newenvironment{exm}
  {\pushQED{\qed}\exmx}
  {\popQED\endexmx}
\newenvironment{rem}
  {\pushQED{\qed}\remx}
  {\popQED\endremx}

\crefname{thm}{Theorem}{Theorems}
\crefname{lem}{Lemma}{Lemmas}
\crefname{cor}{Corollary}{Corollaries}
\crefname{prop}{Proposition}{Propositions}
\crefname{ex}{Exercise}{Exercises}
\crefname{exmx}{Example}{Examples}
\crefname{defn}{Definition}{Definitions}
\crefname{claim}{Claim}{Claims}
\crefname{remx}{Remark}{Remarks}
\crefname{goal}{Goal}{Goals}
\crefname{ansatz}{Ansatz}{Ansatzes}

\crefformat{section}{§#2#1#3}

\newcommand{\nocontentsline}[3]{}
\newcommand\stoptoc{%
   \let\origcontentsline\addcontentsline
   \let\addcontentsline\nocontentsline
}
\newcommand\resumetoc{%
   \let\addcontentsline\origcontentsline
}

\newcommand{\term}{\emph} 

\newcommand{\RenewMathOperator}[2]{\renewcommand{#1}{\operatorname{#2}}}
\newcommand{\vp}{\varphi}

\newcommand{\inj}{\hookrightarrow}
\newcommand{\surj}{\twoheadrightarrow}
\newcommand{\id}{\mathrm{id}}
\newcommand{\pt}{\mathrm{pt}}

\let\shortmapsto\mapsto
\renewcommand{\mapsto}{\mathchoice{\longmapsto}{\shortmapsto}{\shortmapsto}{\shortmapsto}}
\AtBeginDocument{

}
\DeclarePairedDelimiter\paren{(}{)}
\DeclarePairedDelimiter\ang{\langle}{\rangle}
\DeclarePairedDelimiter\abs{\lvert}{\rvert}

\DeclarePairedDelimiter\bkt{[}{]}
\DeclarePairedDelimiter\set{\{}{\}}

\makeatletter
\let\oldparen\paren
\def\paren{\@ifstar{\oldparen}{\oldparen*}}
\let\oldbkt\bkt
\def\bkt{\@ifstar{\oldbkt}{\oldbkt*}}
\makeatother
\newcommand{\newoperator}[1]{\expandafter\DeclareMathOperator\csname #1\endcsname{\operatorname{#1}}}
\newoperator{Aut}
\newoperator{coker}
\newoperator{End}
\newoperator{Ext}
\newoperator{Hom}
\AtBeginDocument{ 
 \RenewMathOperator{\Im}{Im} 
}
\newcommand{\op}{^{\cat{op}}} 
\newoperator{Pic}
\newoperator{rank}
\newoperator{Sym}
\newoperator{Tor}
\newoperator{tr}
\newcommand{\bl}{\text{--}}

\newcommand{\Cl}{\mathit{C\ell}}
\newcommand{\Cxl}{\C \ell}
\newcommand{\KO}{\mathit{KO}}
\newcommand{\KU}{\mathit{KU}}
\newcommand{\KR}{\mathit{KR}}
\newcommand{\ko}{\mathit{ko}}
\newcommand{\ku}{\mathit{ku}}
\newcommand{\ksp}{\mathit{ksp}}
\newcommand{\Sq}{\mathrm{Sq}}

\newcommand{\NewThomSpectrum}[1]{\expandafter\newcommand\csname M#1\endcsname{M\mathrm{#1}}}
\newcommand{\NewMTSpectrum}[1]{\expandafter\newcommand\csname MT#1\endcsname{MT\mathrm{#1}}}
\newcommand{\BothThomSpectra}[1]{\NewThomSpectrum{#1}\NewMTSpectrum{#1}}
\BothThomSpectra{O}
\BothThomSpectra{SO}
\BothThomSpectra{Spin}
\BothThomSpectra{Pin}
\BothThomSpectra{U} 
\AtBeginDocument{
 \RenewMathOperator{\Re}{Re}
}
\AtBeginDocument{
 \usepackage{microtype}
}
\usepackage{hyperref}
\hypersetup{
 colorlinks,
 linkcolor={red!50!black},
 citecolor={green!50!black},
 urlcolor={blue!80!black}
}

\usepackage{tikz-cd}

\newcommand{\NewCommenter}[2]{\expandafter\newcommand\csname #1\endcsname[1]{\textcolor{#2}{[{\bfseries #1:} ##1]}}}

\newcommand{\ME}{\mathit{ME}}
\newcommand{\ABS}{\mathit{ABS}}
\newcommand{\sm}{\mathrm{sm}}

\newcommand{\ftens}{\mathbin{\hat\times}}

\newcat{\sAlg}{sAlg}
\newcat{\sLine}{sLine} 

\definecolor{softred}{rgb}{0.92, 0, 0.17}

\newtheorem{goal}[equation]{Goal}
\newtheorem{notn}[equation]{Notation}

\newcommand{\CMQ}{\mathrm{CMQ}}

\newcommand{\spint}{\mathrm{sp}^\phi}
\DeclareDocumentCommand{\shortexact}{s O{} O{} mmmm}{
\IfBooleanTF{#1}{ 
\begin{tikzcd}[ampersand replacement=\&]
        {1} \& {#4} \& {#5} \& {#6} \& {1#7}
        \arrow[from=1-1, to=1-2]
        \arrow["#2", from=1-2, to=1-3]
        \arrow["#3", from=1-3, to=1-4]
        \arrow[from=1-4, to=1-5]
\end{tikzcd}
}{ 
\begin{tikzcd}[ampersand replacement=\&]
        {0} \& {#4} \& {#5} \& {#6} \& {0#7}
        \arrow[from=1-1, to=1-2]
        \arrow["#2", from=1-2, to=1-3]
        \arrow["#3", from=1-3, to=1-4]
        \arrow[from=1-4, to=1-5]
\end{tikzcd}
}}

\definecolor{ceruleanblue}{rgb}{0.16, 0.32, 0.75} 

\newcommand{\ldeg}{\mathrm{ldeg}}

\newcommand\bonusspiral{} 
\def\bonusspiral[#1](#2)(#3:#4)(#5:#6)[#7]{
\pgfmathsetmacro{\domain}{#4+#7*360}
\pgfmathsetmacro{\growth}{180*(#6-#5)/(pi*(\domain-#3))}
\draw [#1,
       shift={(#2)},
       domain=#3*pi/180:\domain*pi/180,
       variable=\t,
       smooth,
       samples=int(\domain/5)] plot ({-\t r}: {#5+\growth*\t-\growth*#3*pi/180})
}

\title{Unraveling the Bott spiral}

\author{Arun Debray}
\address{Department of Mathematics, The University of Kentucky, 719 Patterson Office Tower,
Lexington, \indent KY 40506, USA}
\email{\href{mailto:a.debray@uky.edu}{a.debray@uky.edu}}
\urladdr{\href{https://adebray.github.io/}{https://adebray.github.io/}}

\author{Cameron Krulewski}
\address{Department of Mathematics \& Statistics,
Dalhousie University,
6316 Coburg Road,
PO Box 15000,
\indent Halifax, NS, B3H 4R2, Canada}
\email{\href{mailto:ckrulewski@dal.ca}{ckrulewski@dal.ca}}
\urladdr{\href{https://cakrulewski.github.io/}{https://cakrulewski.github.io/}}
\author{Luuk Stehouwer}
\address{
Durham University, Department of Mathematical Sciences, Upper Mountjoy, 
Durham DH1 3LE, \indent United Kingdom}
\email{\href{luukstehouwer@gmail.com}{luukstehouwer@gmail.com}}
\urladdr{\href{https://luukstehouwer.com/}{https://luukstehouwer.com/}}

\date{\today}

\begin{document}
\maketitle

\begin{abstract}
We construct and compute a homotopy-theoretic model for the \textit{Bott spiral} of symmetry-protected topological phases (SPTs) studied by Queiroz--Khalaf--Stern. We model free and interacting fermionic SPTs using $K$-theory and reflection-positive invertible field theories (IFTs), resp., and define a twisted generalization of the Atiyah--Bott--Shapiro orientation to produce a \textit{free-to-interacting map}. We also define and compute \textit{spiral maps} of IFTs to model dimensional reduction in this context, answering a question of Hason--Komargodski--Thorngren.
Our analysis highlights two general aspects of homotopical free-to-interacting maps. 
First,
IFTs are more sensitive than $K$-theory is to the input symmetry data;
in particular, the specification of an Altland--Zirnbauer class is insufficient information to define symmetry type for an IFT.
Second,
the remnant of Bott periodicity on the interacting side relies on an isomorphism of two extraspecial groups of order $32$.
Our computations use a novel $4$-periodic description of a sector of the twisted $\ko$-homology of elementary abelian $2$-groups.

\end{abstract}

\tableofcontents

\section*{Introduction}
    \stoptoc
\subsection*{Background and motivation}

For the last half-century, what have come to be called \textit{symmetry protected topological phases} (SPTs) have inspired a fruitful interplay between algebraic topology and quantum physics, in which topological invariants make physical predictions, while physical desiderata and computations also inspire new mathematics.
SPTs are gapped quantum phases of matter whose stability relies on the presence of a symmetry---on the physics side, often encoded as a group action on a lattice Hamiltonian---and which are invertible under a stacking operation, meaning that the set of SPTs with fixed dimension and \term{symmetry type}\footnote{We will define symmetry types precisely later in the introduction.}---the collection of symmetries acting on the system---forms a group.

\textit{Free fermion} SPTs in particular have received a great deal of interest due to the property that their energy spectrum is exactly solvable, making them convenient models, and due to their applicability to the theory of topological insulators and superconductors, which include experimentally-realized phases like the quantum Hall insulator~\cite{von_klitzing_1980} and quantum spin Hall insulator~\cite{konig_quantum_2007}.
Free fermion models are so-called because their Hamiltonians feature no interactions between particles, and this property ultimately allows the ground states of such models to be described by Clifford modules and for the phases to be classified by $K$-theory \cite{kitaev_periodic_2009, ryu_topological_2010}.
Our preferred mathematical phrasing of this classification will be briefly recalled in \cref{ff_and_k}; specifically, see \cref{neutralluuk_main_thm_citation}.

However, as first discovered by Fidkowski--Kitaev in 2009, free fermion models are in general \textit{not} sufficient for describing interacting systems \cite{fidkowski_effects_2010}.
They found that eight stacked copies of the time-reversal symmetric Majorana chain, which defines a nontrivial phase within free fermion models, becomes trivializable when considered in a
space of interacting gapped Hamiltonian models with unique ground states. %
This indicates that when comparing a moduli space of free fermion 
Hamiltonians with a corresponding space of interacting Hamiltonians, the number of connected components can change. %
There are also SPTs, called \textit{intrinsically interacting}, that are not adiabatically connected to free fermion models.
Since it is often a reasonable physical assumption that fermionic systems do have interactions, it has become a major open question in condensed matter to determine when and how the free and interacting classifications differ.
A special case of this question is the main topic of this paper, and rather than physical, our methods will be homotopy-theoretic.

On the free side, the application of homotopy theory (specifically, $K$-theory) toward classifying 
free fermion SPTs is a well-known and highly general story \cite{kitaev_periodic_2009,ryu_topological_2010,thiang2015topological,freed_twisted_2013,freed_k-theory_2016,gomi_freed-moore_2021,neutralluuk}. 
On the interacting side, the application of homotopy theory is more recent.
Mathematical models for interacting fermionic systems are less developed, as these systems are more subtle primarily because
there is at present no agreed-upon definition of a moduli space of interacting lattice Hamiltonians.
However, it is believed that any such moduli space should deformation retract onto the space of ground states of the theories, meaning that deformation classes of interacting Hamiltonians should be determined by properties of their ground states, which are governed by the low-energy field theory of the SPT \cite{Kit13,Kit15}.
According to an ansatz (see \cref{LEFT_ansatz}) of Freed--Hopkins (see \cite{freed_reflection_2021}, \cite[Section 9.1]{Fre19}), which builds on work of Kapustin--Thorngren--Turzillo--Wang~\cite{kapustin_fermionic_2015}, the low-energy theory of an SPT should be modeled as an \textit{invertible field theory} (IFT); specifically, a reflection-positive fully extended invertible field theory on manifolds with (in the fermionic case) twisted spin structures.
This version of invertible field theory is a mathematically well-defined object, and by results of Freed--Hopkins~\cite[Remark 8.39]{freed_reflection_2021} and Grady~\cite{Gra23}, such theories (up to deformation) admit a classification in terms of \textit{Anderson-dual bordism groups}.
Standard homotopical methods such as the Adams spectral sequence have been used to compute these groups~\cite{freed_reflection_2021, Cam17, BC18, debray_invertible_2021}, finding agreement with physical computations.  %
We recall more details on the invertible field theory model for interacting SPTs in \cref{interacting_IFT}, and state their classification in \cref{IFT_class}.

Formalizing the process of flowing to the low-energy limit is an area of active and exciting research~\cite{Gaiotto_SPT,MR4298021,ogata_classification_2022,kubota2025stable}, but in this paper, due to the success of \cref{LEFT_ansatz} in prior work, we will simply assume it and observe what it predicts computationally in new settings. These computations provide further support to the ansatz.

\subsection*{Symmetry types and the tenfold way}
In any problem motivated by physics, one of the first things to understand is the collection of symmetries of the system. This turned out to be an important but subtle point in our analysis of the Bott spiral, and in free-to-interacting maps in general, so here we review the usual classification of symmetries of free fermion systems in the \textit{tenfold way}.

A free fermion Hamiltonian may have time-reversal and an internal symmetry group $K$ equal to $\O_1$, $\U_1$, or $\Sp_1$.
Kitaev~\cite{kitaev_periodic_2009} showed that these symmetries can combine in ten ways, called \term{Altland--Zirnbauer classes}~\cite{altland_novel_1997}, labeled by a subset of Cartan's labels~\cite{cartan_classe_1927} for symmetric spaces: A, AI, AII, AIII, BDI, C, CI, CII, D, and DIII.\footnote{See~\cite{dyson_threefold_1962, altland_novel_1997, heinzner2005symmetry, ryu_topological_2010, freed_twisted_2013, WS14, kennedy_Bott_2016, agarwala_tenfold_2017, lieu_tenfold_2020, cornfeld_tenfold_2021, freed_reflection_2021, geiko_Dyson_2021, serrano_magnetic_2025, neutralluuk, mussnich_weakly_2026} for related but distinct approaches to this ``tenfold way'' classification, as well as \cite{zirnbauer_symmetry_2011, moore_tenfold_notes, chiu_classification_2016, baez_tenfold_2020} for reviews.} These classes correspond to the eight real and two complex Morita classes of Clifford algebras~\cite{kitaev_periodic_2009,abramovici_clifford_2012}. Moreover, Kitaev showed that the classifications of free fermion phases in these classes are periodic, matching  the real $K$-theory groups $\Z/2$, $\Z/2$, $0$, $\Z$, $0$, $0$, $0$, $\Z$ for the eight real cases. 
Phases in the two remaining classes, A and AIII, are classified by the two shifts of the complex $K$-theory groups. Thus Bott periodicity in $K$-theory appears twice in the classification of free fermion phases: 
first, in the periodicity in spatial dimension for a fixed symmetry class,
and second, more fundamentally, in the classification of the possible symmetries itself.%
\footnote{In the presence of additional symmetries, such as spatial reflection symmetry, the classification can get more complicated. Following Queiroz--Khalaf--Stern~\cite{queiroz_dimensional_2016}, we will sometimes consider one of the ten Altland--Zirnbauer symmetry classes together with a spatial reflection symmetry $\mathcal R$ squaring to fermion parity, acting trivially on the $\U_1$ charge, and commuting with time-reversal symmetry (if present). We will denote this modification with a prime, e.g.\ classes $\mathrm{D'}$, $\mathrm{A'}$, etc. See \cref{CEP} for more information and Song--Schnyder~\cite[\S II.A.1]{song_interaction_2017} for a discussion of more general Altland--Zirnbauer classes incorporating a spatial reflection.} %

Passing to interacting SPT phases, then to IFTs, we would like to describe the symmetries of the system as a group $G$ that generates the Clifford algebra for a given Altland--Zirnbauer class in an appropriate sense. Complicating matters, however, classifications of IFTs lack Bott periodicity in \emph{both} of the above senses: the groups of IFTs in dimensions $d$ and $d+8$ are generally nonisomorphic, and independently, there are ambiguities in choosing the symmetry group corresponding to a given Altland--Zirnbauer class, preventing a clean eight- or two-periodic description.

More specifically, consider the collection $E_{\ell,k}$ of $\ell$ time-reversing symmetries with square $1$ and $k$ time-reversing symmetries with square $-1$, all required to anticommute. 
In the context of free phases, this infinite collection of symmetry groups reduces to eight symmetry classes only dependent on $\ell-k$ modulo $8$.
The underlying mathematical reason is that $E_{\ell,k}$ reproduces a Clifford algebra $\Cl_{\ell,k}$ (Corollary \ref{elk_alg}), which in $\KO$-theory gives nothing but a shift in dimension by $\ell-k$.
Bott periodicity does not persist for interacting phases, the only exception being that $E_{\ell,k+4}$ and $E_{\ell+4,k}$ do still agree (\cref{40_04}), the implications of which we will explain later.

Another important subtlety we have to circumvent regarding symmetry classes is as follows: the pin group $\Pin_{\ell,k}$ also generates the Clifford algebra $\Cl_{\ell,k}$ under addition and multiplication. Freed--Hopkins' choices of symmetry groups representing the ten Altland--Zirnbauer classes~\cite[\S 9.2.1, Tables (9.34), (9.35)]{freed_reflection_2021} include six pin groups with $\ell$ and $k$ small. 
$E_{\ell,k}$ and $\Pin_{\ell,k}$ represent the same Altland--Zirnbauer class, yet in general give rise to different classifications of interacting SPTs, as we show in \cref{discrete_ne_cts}. This is an instance of a general phenomenon also discussed in~\cite{song_interaction_2017,stehouwer_interacting_2022}.

Therefore, to formulate free-to-interacting maps in examples of interest in physics, we have to confront this ambiguity and determine how to match the symmetry data on the free and interacting sides.

\subsection*{The Bott spiral}

Our primary goal in this paper is to compare free and interacting fermionic SPTs in the special case of the \textit{Bott spiral}\footnote{The term ``Bott spiral'' is due to Hason--Komargodski--Thorngren~\cite[\S 5]{hason_anomaly_2020}.} of fermionic SPTs studied by Queiroz--Khalaf--Stern~\cite{queiroz_dimensional_2016}.
The class of SPTs that they study exhibits drastically different behavior depending on whether interactions are allowed: while the free classification is insensitive to certain changes in dimension and symmetry type, the interacting classification depends strongly on these parameters and specifically experiences an exponential growth in size as dimension increases.

\begin{figure}[h!]
\includegraphics{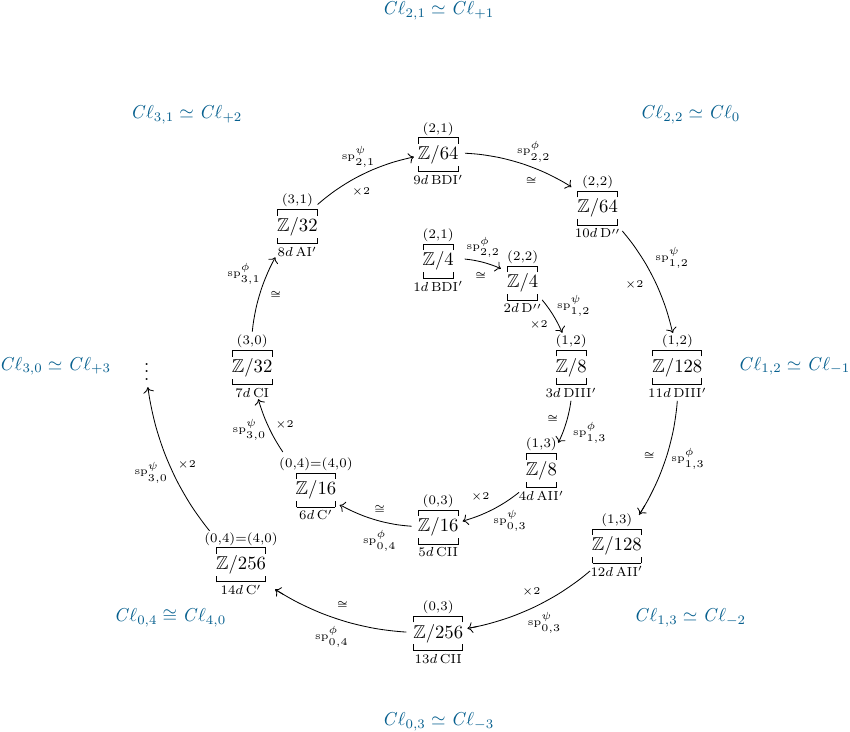}
\caption{The Bott spiral starting with 1d class $\mathrm{BDI}'$: our mathematical model for~\cite[Figure 1(a)]{queiroz_dimensional_2016},
which displays the groups of interacting SPTs admitting free fermion realizations in a specified symmetry class and indicates dimensional reduction maps between them.
Our model for an interacting SPT class is a deformation class of invertible field theory (IFT; see \cref{LEFT_ansatz}), and our models for dimensional reduction are 
the spiral maps $\mathrm{sp}^\phi_{\ell,k}$ and $\mathrm{sp}^\psi_{\ell,k}$ defined in \cref{spiral_maps_of_IFTs} and calculated in \cref{int_Bott_spiral_1,MSpin_psi}.
In each entry of the spiral, $\Z/2^N$ is the group of IFTs 
in the image of a map from $\KO$-theory,
calculated in \cref{melk_calc,only_first_summand_real}.
The bracket below the group denotes the spatial dimension and discrete Altland--Zirnbauer class of this group of IFTs: see \cref{tenfold_table}.
This symmetry class is represented by the fermionic group $E_{\ell,k}$, where the bracket above an entry gives $(\ell,k)$: see \cref{elk_definition}. The identification $(0, 4) = (4, 0)$ in class $\mathrm{C'}$ is \cref{40_04}.
We see that every other step in the spiral, the relevant classification order increases by a factor of two, resulting in a 16-fold increase every eight steps around: the remnant of Bott periodicity for interacting theories.
On the outside, we compare with the Bott clock: the twisted group algebras of the eight symmetry groups we use recover all eight Morita classes of super division algebras; $\simeq$ indicates Morita equivalence.
}
\label{intro_spiral}
\end{figure}

\begin{figure}[h!]
\includegraphics{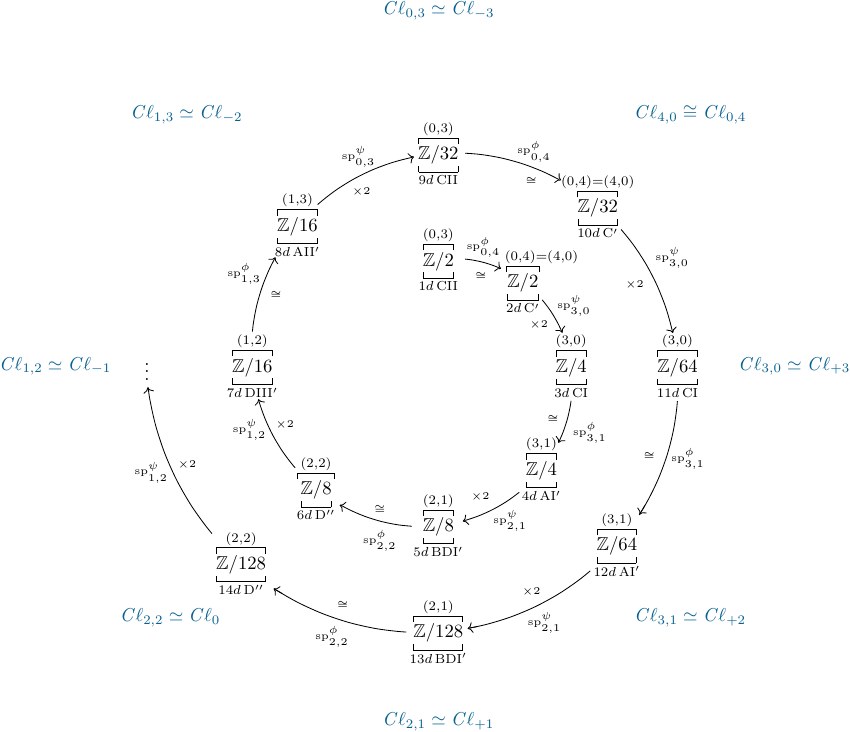}
\caption{The Bott spiral starting with 1d class $\mathrm{CII}$: our mathematical model for~\cite[Figure 1(b)]{queiroz_dimensional_2016}.
All notation is as in \cref{intro_spiral}.
}
\label{intro_spiral_2}
\end{figure}

Queiroz--Khalaf--Stern's paradigm is very close to the tenfold way discussed above.
Indeed, on the free side, the symmetry groups $E_{\ell,k}$ for a fixed $\ell-k\bmod 8$ model a certain Altland--Zirnbauer class.
However, on the interacting side, 
SPTs with symmetry types $E_{\ell,k}$ and $E_{\ell',k'}$ in the same Altland--Zirnbauer class (i.e. with $\ell-k\equiv \ell'-k'\bmod 8$) can be measurably different, as we will discuss.

Queiroz--Khalaf--Stern connect different SPT-phases in  
the spiral using a version of \textit{dimensional reduction}\footnote{See \cref{rem:dim_red_terminology}.} similar to that of e.g.\ Ryu--Schnyder--Furusaki--Ludwig~\cite{ryu_topological_2010},
noting that as the dimension of the system increases, the classification of free fermion SPTs under study repeats every eight (resp.\ two) steps due to real (resp.\ complex) Bott periodicity.
However, they physically derive that the classification of interacting SPTs connected to these free models 
often doubles in size
as the dimension grows,
leading to e.g.\ a 16-fold increase in the number of distinct interacting SPTs every eight steps in the real spirals \cite[(1)]{queiroz_dimensional_2016}.

As we mentioned above, Freed--Hopkins' symmetry types modeling the tenfold way~\cite[Tables (9.34), (9.35)]{freed_reflection_2021} do \emph{not} suffice for describing the Bott spiral: there is a mismatch between their groups of IFTs and what Queiroz--Khalaf--Stern report as the orders of the interacting SPTs in the Bott spiral. See e.g.\ \cref{discrete_ne_cts} for a relevant comparison.

\begin{goal}
\label{goalz1}
    Determine the symmetry types for the free and interacting SPTs in the Bott spiral.
\end{goal}

Developing and computing a model for the interacting SPTs in the Bott spiral, to complement the $K$-theory model for the corresponding free SPTs, is only part of the story.
To describe the relationship between free and interacting SPTs, we not only wish to compute both classifications, but actually produce a map from one to the other.
Generalizing work of Freed--Hopkins~\cite[\S 9, 10]{freed_reflection_2021}, our approach is to define a homotopical map from certain $K$-theory groups---which model the free fermion SPTs in the Bott spiral---to corresponding groups of (deformation classes of) IFTs---modeling the low-energy field theories of the interacting SPTs in the spiral. We call such a map a \textit{free-to-interacting map}.
For each symmetry type of SPT in the Bott spiral, we actually define a map of spectra, whose homotopy in degree $-d$ yields the free-to-interacting map for groups of SPTs in spatial dimension $d$.\footnote{Heuristically, one can think of this map as arising from an index construction for families of Dirac operators. In this paper, however, we do not use such a description. Instead, we construct the map as a homotopical refinement of the 
spin-orientation of $\KO$, suitably twisted by the symmetry data.}

Free-to-interacting maps encode a great deal of information:
we interpret their kernels as consisting of the free phases that are unstable to interactions, while we interpret their cokernels as the intrinsically interacting phases.
Given a free fermion Hamiltonian, if we can compute its class in $K$-theory, the free-to-interacting map also predicts exactly how many copies of this Hamiltonian must be stacked in order to reach a trivializable phase; if the map on groups is a surjection $\Z\to\Z/m$, for example, then we expect that $m$ stacked copies of the input Hamiltonian is adiabatically connected to the trivial phase.
Therefore, once such a map is defined, we may compute it in low dimensions and answer (assuming \cref{LEFT_ansatz}) a large number of physical questions.

To study the phenomenon of the Bott spiral, it suffices to define free-to-interacting maps and compute their \textit{images}, since the interacting SPTs in the Bott spiral that exhibit the 16-fold periodicity are all connected to free fermion models.
In other words, since the intrinsically interacting SPTs of the relevant symmetry types do not play a role in the behavior we wish to study, we will ignore the cokernel of the free-to-interacting maps in this paper.
\begin{goal}\label{goalz2}
    Define and compute a map from $K$-theory to IFTs whose image in the relevant degrees is the group $\Z/2^a$ computed in \cite[(1)]{queiroz_dimensional_2016}.

\end{goal}

A third goal in this paper, related to the previous two, is to use IFTs to model not only the interacting SPTs, but also the dimensional reduction process.
This operation takes a SPT in spatial dimension $d$ to a $d+1$-dimensional SPT with a modified symmetry type. The following goal was raised as a question by Hason--Komargodski--Thorngren~\cite[\S 5]{hason_anomaly_2020}.
\begin{goal}\label{goalz3}
    Define and compute maps of IFTs that 
    model the dimensional reduction procedure of Queiroz--Khalaf--Stern.

\end{goal}
Our inspiration for this goal comes from the success of dualized \textit{Smith homomorphisms} of IFTs toward computing the behavior of SPTs under symmetry defect creation, which has been studied in e.g.\ \cite{hason_anomaly_2020, COSY, kobayashi_anomaly_2021, DNT24, copetti_anomalies_2025, debray_long_2024, jones_charge_2025, manjunath_anomalous_2025}.
Specifically, Hason--Komargodski--Thorngren~\cite{hason_anomaly_2020} suggested a relationship between the growing factors of two in Queiroz--Khalaf--Stern's real classification and a certain 4-periodic sequence of Smith homomorphisms (see \cref{1st_Z2_Smith_exm,spin_sigma_symm_types}).
We find that Smith homomorphisms are not the entire story, but they are a central ingredient in our constructions and aid in our computations.
In fact, we confirm that the periodicity of this family of Smith homomorphisms underlies the 16-fold periodicity of the interacting SPTs in the Bott spiral, as we note in \cref{smith_periodicity_remark}.
We call the maps of IFTs we produce \textit{spiral maps}.

We provide precise definitions of the free-to-interacting maps and spiral maps in the next subsection.

\subsection*{Main results}

The first challenge in 
defining free-to-interacting maps in general is in efficiently encoding the symmetries of the SPTs of interest.
In our chosen framework, we can specify the symmetries of a free fermion system by providing a superalgebra.
On the interacting side, symmetries are encoded by requiring certain tangential structures on the manifolds evaluated by the IFT.
Specifically, for fermionic theories, we will specify twisted spin structures, such as pin structures.

Our approach to fermionic symmetries is to use \textit{fermionic groups} (\cref{defn_ferm_group}), as first introduced by Benson~\cite[\S 7]{Ben88}. These are
triples $(G_f,(-1)^F,\theta)$ of a compact Lie group $G_f$, a central element $(-1)^F$ of order $2$, and a homomorphism $\theta\colon G_f\to \Z/2$ called the \term{grading}, such that $\theta((-1)^F) = 0$.
Physically, the element $(-1)^F$ encodes \textit{fermion parity}, while the grading $\theta$ distinguishes unitary and antiunitary symmetries. We model the symmetry type of an interacting SPT as a fermionic group.

Fermionic groups come with two useful constructions.
\begin{enumerate}
    \item The fermionic group algebra $\R^f[G_f]$ (\cite[\S 7]{Ben88}, \cref{subsec:twistedgroupsuperalgebras}) is defined so that modules over it are those $G_f$-representations that are relevant for physical systems with $G_f$-symmetry. They may therefore be used as an input to the free fermion classification using the $K$-theory of superalgebras, see \cref{sec:Ktheoryoffermgps}.
    \item As we discuss in \cref{ss:xi_to_Gf}, Stolz~\cite[\S 2.6]{Sto98} associates to any fermionic group $G_f$ a topological group $H(G_f)$ with a map $H(G_f)\to \O$, the infinite orthogonal group, generalizing e.g.\ the assignment $\Spin_2\mapsto (\Spin\times_{\set{\pm 1}}\Spin_2)\cong\Spin^c$. Taking classifying spaces, we obtain a stable tangential structure $BH(G_f)\to B\O$ in the sense of Lashof~\cite{Las63}, and can therefore discuss $H(G_f)$-structures on manifolds. This models the topological structure on spacetime in a field theory of fermions with a $G_f$-symmetry: see~\cite[\S 2.1]{stehouwer_interacting_2022}. We thus use IFTs of $H(G_f)$-manifolds to mathematically model interacting fermionic SPTs with $G_f$-symmetry.
\end{enumerate}
In \cref{gpalg_tensor,interacting_symm_mon}, we show these constructions are compatible with products $G_f\ftens K_f$ of fermionic groups (see \cref{ferm_tens_defn}).
This compatibility allows our formalism to use the same input symmetry data for both the free SPTs (classified by $K$-theory) and interacting SPTs (classified by groups of (deformation classes of) IFTs).

With these $K$-theoretic, resp.\ field-theoretic models for the free and interacting classifications of SPTs, we now discuss our mathematical models of free-to-interacting maps.

The tangential structure $H(G_f)$ gives rise to a stable Madsen--Tillmann spectrum $\mathit{MTH}(G_f)$ (see, e.g. \cite[\S 7.1]{freed_reflection_2021}, \cite[\S 2.9]{BC18}), which is a tangential version the Thom spectrum of $H(G_f)$. 
The homotopy groups of $\mathit{MTH}(G_f)$ are bordism groups of manifolds with (tangential) $H(G_f)$ structures.
We will write $\KO$ and $\KU$ for the real and complex $K$-theory spectra, resp.

The main construction involved in a free-to-interacting map is a
\textit{generalized Atiyah--Bott--Shapiro (ABS) map}: an analog of the spin orientation $\mathit{ABS}_0^r\colon \MTSpin\to \KO$ \cite{atiyah_clifford_1963} of $\KO$-theory (resp.\ the \spinc orientation $\mathit{ABS}_0^c$ of $\KU$), but for manifolds with a given notion of twisted spin structure.
Informed by
the approach of Freed--Hopkins \cite[\S 9,10]{freed_reflection_2021}, we define a generalized ABS map for spin manifolds twisted by a pair of bundles.
\begin{absdefn}
Let $G_1$ and $G_2$ be compact Lie groups. Let $V_1\to BG_1$ and $V_2\to BG_2$ be vector bundles of rank $r_1$ and $r_2$, resp. 
    A $(V_1,V_2)$-\textit{twisted real ABS map} is the composition
    \begin{equation}
    \begin{tikzcd}
        MT\Spin\wedge (BG_1)^{V_1-r_1}\wedge (BG_2)^{\,r_2-V_2} \ar[d,"\ABS_0^r\wedge \lambda_{V_1}\wedge e_{V_2}"] \\
        \KO\wedge \Sigma^{-r_1}\KO \wedge \Sigma^{r_2}(BG_2)_+ \ar[d,"\id^{\wedge r_1+1}\wedge \Sigma^{r_2} \underline{1}"] \\
        \KO\wedge \Sigma^{-r_1}\KO\wedge \Sigma^{r_2} \KO \ar[d,"\mu"] \\
        \Sigma^{r_2-r_1} \KO.
    \end{tikzcd}
\end{equation}
\end{absdefn}
Above, $\lambda_{V_1}$ (\cref{lambda_V_defn}) is a $\KO$-class formed from the difference bundle construction applied to $V_1$, while $e_{V_2}$ (\cref{smith_map_of_spectra_specific}) is the map of spectra induced by the zero section map of $V_2$.
On homotopy in degree $n$, this map assigns a class in $\KO^{-n+r_1-r_2}$ to an input twisted spin $n$-manifold $M$; specifically, $M$ is equipped with maps $f_i\colon M\to BG_i$ and a spin structure on the bundle $TM+f_1^*V_1-f_2^*V_2$.

Given such a twisted ABS map, we obtain a free-to-interacting map by taking the \textit{Anderson dual} $I_\Z$~\cite{And69}, as we describe in \cref{ss:F2I,F2I_from_general_ABS_map}.
The Anderson dual of the sphere implements a general form of the universal coefficient theorem, and when applied to an ABS map, it produces a map from dualized $\KO$-theory to dualized spin bordism.
By \cite[Theorem
4.16]{And69} (see \cref{and_KO}), 
$\KO$-theory is Anderson-self dual up to a fourfold shift, meaning that the source of our map is $\KO$-theory.
Meanwhile, Anderson-dual bordism groups, in the target, are exactly what classify deformation classes of our IFTs of interest:

\begin{IFTclass}[{Freed--Hopkins~\cite[Theorem 1.1]{freed_reflection_2021}, Grady~\cite[Theorem 1]{Gra23}}]
Let $G_f$ be a fermionic group.
There is an isomorphism, natural in $G_f$, from the abelian group of deformation classes of reflection-positive IFTs on spacetime manifolds with $H(G_f)$-structure of dimension $n = d+1$ to $[\mathit{MTH}(G_f), \Sigma^{n+1}I_\Z]$.
\end{IFTclass}

Now we can revisit \cref{goalz1,,goalz2,,goalz3}:
\begin{enumerate}
    \item\label{need_elk} Determine fermionic groups $G_f(d)$, depending on the dimension $d$, describing the symmetries of the Bott spiral such that the groups of IFTs of $d$-dimensional $H(G_f(d))$-manifolds match Queiroz--Khalaf--Stern's classifications of interacting SPTs. 
    \item Define a twisted ABS map of spectra whose Anderson dual models the free-to-interacting map for the phases in the Bott spiral.
    \item Define maps between the spectra classifying IFTs for different steps of the Bott spiral, modeling Queiroz--Khalaf--Stern's dimensional reduction maps.
\end{enumerate}
We begin by answering~\eqref{need_elk}. Within the formalism of fermionic groups, the mathematical definition of the symmetry type $E_{\ell,k}$ mentioned earlier in the context of the tenfold way is:
\begin{Elkdefn}
The fermionic group $E_{\ell,k}$ is the pullback of the diagonal subgroup $\set{\pm 1}^{\ell+k}\hookrightarrow\O_{\ell,k}$ under the double cover $\Pin_{\ell,k}\to\O_{\ell,k}$.\footnote{\label{pin_convention}$\Pin_{\ell,k}$ is defined to be the pin group, in the sense of Atiyah--Bott--Shapiro~\cite[Theorem 3.11]{atiyah_clifford_1963}, associated to the quadratic form on $\R^{\ell+k}$ given by $x_1^2 + x_2^2 + \dotsb + x_\ell^2 - y_1^2 - \dotsb - y_k^2$. Thus, for example, $\Pin_{\ell,0} = \Pin_\ell^+$ and $\Pin_{0,k} = \Pin_k^-$.}
\end{Elkdefn}
The fermionic group data comes from the pullback as follows: $(-1)^F$ is the unique nonzero element of the kernel of $E_{\ell,k}\to\set{\pm 1}^{\ell+k}$, and the map $s\colon E_{\ell,k}\to\set{\pm 1}$ is the map to $\O_{\ell,k}$ followed by the determinant. %

Queiroz--Khalaf--Stern use \textit{primed phases} for half of the symmetry types they consider. 
Physically, given an SPT, its primed version consists of two stacked copies with opposite invariants, whose cancellation is prevented by the addition of a mixed time-reversal and reflection symmetry.
See \cref{primed_AZ} for further discussion.
Mathematically, we choose to model the primed symmetry type of a given fermionic group $G_f$ as the fermionic tensor product of itself with a fermionic group $E_{1,1}$ encoding the two additional symmetry generators:
\begin{Gfprimedefn}
Let $G_f$ be a fermionic group. Define $G_f' \coloneqq G_f \ftens E_{1,1}$.
\end{Gfprimedefn}
Here $\ftens$ is the product of fermionic groups (\cref{ferm_tens_defn}).
\begin{ansatz}\label{fg_ansatz}
The Altland--Zirnbauer classes in the Bott spiral are represented by the following fermionic groups
\begin{equation}\label{tenfold_eqn}
\begin{tabular}{ccccccccc}
    \toprule
         \rm{C} & \rm{CII} & \rm{AII} & \rm{DIII} & \rm D & \rm{BDI} & \rm{AI} & \rm{CI} \\
         \midrule
         $Q_8$ & $E_{0,3}$ & $E_{0,2}$ & $E_{0,1}$ & $\Spin_1$ & $E_{1,0}$ & $E_{2,0}$ & $E_{3,0}$ \\
    \bottomrule
\end{tabular}
\end{equation}and Queiroz--Khalaf--Stern's primed symmetry classes are modeled by \cref{G_f_prime}.
\end{ansatz}
For the quaternion group $Q_8$, $\theta$ is trivial, and $(-1)^F = -1$. See \cref{tenfold_table} for the analog in the complex case.

In classes D, DIII, BDI, A, and AIII, the fermionic groups in~\eqref{tenfold_eqn} are isomorphic to Freed--Hopkins' models for these Altland--Zirnbauer types; for classes AI, AII, C, CI, and CII, they are nonisomorphic (see \cref{tenfold_table}). We prove in \cref{discrete_ne_cts} that in all five of these cases, the respective groups of IFTs are also nonisomorphic. This illustrates our point above that, when studying free-to-interacting maps, one must be careful in saying what a given Altland--Zirnbauer class means.

\Cref{fg_ansatz} answers \cref{goalz1}, and almost answers \cref{goalz2}---we just have to express $\mathit{MTH}(E_{\ell,k})$ in the form $\MTSpin\wedge (BG_1)^{V_1-r_1}\wedge (BG_2)^{r_2-V_2}$ from \cref{defn_generalized_ABS}.

\begin{MElkdefn}
Let $\sigma\to B\Z/2$ denote the universal line bundle.
    For $k,\ell\ge 0$, let 
    \begin{equation}
        \ME_{\ell,k}\coloneqq ((B\Z/2)^{\sigma-1})^{\wedge \ell} \wedge ((B\Z/2)^{1-\sigma})^{\wedge k}.
    \end{equation}
\end{MElkdefn}
\begin{Elksmash}
There is an equivalence of $\MTSpin$-module spectra
\begin{equation}
\mathit{MTH}(E_{\ell,k})\simeq \MTSpin\wedge\ME_{\ell,k}.
\end{equation}
\end{Elksmash}
Then, as a corollary of~\eqref{external_internal_products}, if $G_1 = (\Z/2)^\ell$, $G_2 = (\Z/2)^k$, and $V_1$ and $V_2$ the external direct sums of the universal bundles $\sigma\to B\Z/2$ over their respective $\Z/2$ factors, then
\begin{equation}
\mathit{MTH}(E_{\ell,k})\simeq \MTSpin\wedge (BG_1)^{V_1-r_1}\wedge (BG_2)^{r_2-V_2},
\end{equation}
so \cref{defn_generalized_ABS} defines a twisted ABS map for $H(E_{\ell,k})$-manifolds. See \cref{defn:gen_ABS}. This answers \cref{goalz2}.
\begin{notn}
We will call an $H(E_{\ell,k})$-structure a \term{spin-$(\ell,k)$ structure}, and we will let
 \begin{equation}
            \mho^{n}_{\ell,k} \coloneqq [\MTSpin\wedge\ME_{\ell,k}, \Sigma^{n+1} I_\Z],
\end{equation}
the Anderson-dual bordism groups classifying the $E_{\ell,k}$-type IFTs of our interest. 
\end{notn}

The free-to-interacting map is obtained by dualizing \cref{defn_generalized_ABS}.
\begin{ftoidefn}
The free-to-interacting map of type $(\ell,k)$ is
\begin{equation}\label{intro_F2I}
    F2I_{\ell,k} \coloneqq \Sigma^{-k+\ell-2}\KO \simeq \Sigma^2 I_\Z( \Sigma^{k-\ell} \KO) \xrightarrow{I_\Z \ABS_{\ell,k}} \Sigma^2 I_\Z (MT\Spin\wedge \ME_{\ell,k}).
\end{equation}
\end{ftoidefn}
For example, setting $\ell=k=1$ yields a free-to-interacting map for class D$'$ phases. In spatial dimension $d=2$, this map is a surjection $KO^0\cong \Z \twoheadrightarrow \mho^{4}_{1,1}\cong \Z/8$, agreeing with the physical computations of Queiroz--Khalaf--Stern~\cite{queiroz_dimensional_2016}, Qi~\cite{qi_new_2013}, Ryu--Zhang~\cite{ryu_interacting_2012}, and Yao--Ryu~\cite{yao_interaction_2013}. See \cref{ex:Dpin_F2I_im}.

We give the analogous complex free-to-interacting map in \Cref{complex_F2I_map_defn} and the (discrete) class C analog in \cref{Q8_F2I}.
\begin{ouransatz}
\hfill
\begin{enumerate}
    \item Under the identifications in \cref{neutralluuk_main_thm_citation,IFT_class} identifying the groups of $E_{\ell,k}$-symmetric $d$-dimensional free fermion Hamiltonians with $\KO^{d+s-2}$ and reflection-positive IFTs on spin-$(\ell,k)$-manifolds with $\mho_{\ell,k}^{d+2}$, resp., the map $F2I_{\ell,k}$ from \eqref{intro_F2I} is the homomorphism assigning to a free fermion Hamiltonian its low-energy effective field theory.
    \item The same is true mutatis mutandis with $\spinc$-$(\ell,k)$ in place of spin-$(\ell,k)$, $\mho_{\Spin^c\text{-}(\ell,k)}^{d+2}$ in place of $\mho_{\ell,k}^{d+2}$, $\KU$ in place of $\KO$, and the map $F2I_m^c$ from \cref{complex_F2I_map_defn} in place of $F2I_{\ell,k}$, where $m = \ell+k$.
\end{enumerate}
\end{ouransatz}

Before discussing computational results for the free-to-interacting maps, we will introduce the spiral maps of IFTs, answering \cref{goalz3}.
Physically, these maps model the (dual process to) the dimensional reduction used by Queiroz--Khalaf--Stern, which, given an SPT in the Bott spiral, produces an SPT in one dimension lower with a modified symmetry type.

\begin{rem}\label{rem:dim_red_terminology}
    In the context of functorial field theory, \textit{dimensional reduction} is often used synonymously with \textit{compactification}. Let $\cat{Bord}_n^\xi$ denote the bordism (higher) category of $\xi$-manifolds and $\cat C$ denote a symmetric monoidal (higher) category.
    The compactification of a topological field theory $\alpha\colon \cat{Bord}_n^\xi\longrightarrow \cat C$ on a $k$-manifold $M$  with $1\leq k< n$ is the topological field theory $\alpha_M\colon \cat{Bord}_{n-k} \to \cat{C}$ whose value on any manifold or bordism $N$ of dimension at most $n-k$ is given by $\alpha(M\times N)$.\footnote{For more on dimensional reduction of TFTs, see Schommer-Pries~\cite[\S 9]{SP18} and Yamashita--Yonekura~\cite[\S 7.3]{YY23}.}

    However, dimensional reduction in a physics sense can also be modeled by other kinds of maps of topological field theories. For example, the dimensional reduction of SPTs induced by the condensation of a charged order parameter can be modeled by dualized \textit{Smith homomorphisms}, as studied in \cite{hason_anomaly_2020,COSY,debray_long_2024,debray_smith_2024}.   
    It is this second kind of model that we find more applicable to our situation of interest.

    These two constructions are related, but we will not detail that here. Instead, we will discuss ``dimensional reduction'' only in the physics sense, and use the terminology ``spiral map'' for the mathematical maps of IFTs that we construct.
\end{rem}

Queiroz--Khalaf--Stern define their spiral maps of SPTs differently depending on whether space is even- or odd-dimensional. As such, we define two different---but related---families of maps of spectra in order to model their spiral maps.
Each has as its first component $e_\sigma$, the zero-section map for $\sigma\to B\Z/2$, which is the same map underlying the Smith homomorphism $\sm_\sigma$, and which on IFTs induces a process dual to taking a codimension-one submanifold. See \cref{smith_homs_intro}.
The other two ingredients are $c$, the \textit{crush map} (\cref{crush_map_defn}), which enforces forgetting a $\Z/2$-bundle, and $\Delta$ (\cref{diag_map_defn}), which we use to model the formation of a primed phase.
Physically, $c$ forgets the symmetry and $\Delta$ is a `doubling.'
We provide more of the physical motivation in \cref{ss:spiral_model}.

\begin{spiraldefn}
For $\ell \geq 0$ and $k > 0$, let $\phi_{\ell,k}$ be the composition
    \begin{equation}\label{spiral_1_defn_intro}
        \phi_{\ell,k}\colon \ME_{\ell,k} = \ME_{\ell,k-1}\wedge (B\Z/2)^{1-\sigma} 
        \xrightarrow{\id\wedge e_\sigma} 
        \ME_{\ell,k-1} \wedge \Sigma (B\Z/2)_+
        \xrightarrow{\id\wedge c}
        \Sigma\ME_{\ell,k-1},
    \end{equation}
    and let $\psi_{\ell,k}$ be the composition
    \begin{equation}\label{spiral_2_defn_intro}
        \psi_{\ell,k}\colon \ME_{\ell,k} = \ME_{\ell,k-1}\wedge (B\Z/2)^{1-\sigma}
        \xrightarrow{\id\wedge e_\sigma}
        \ME_{\ell,k-1}\wedge \Sigma (B\Z/2)_+
        \xrightarrow{\id\wedge \Delta}
        \ME_{\ell,k-1}\wedge \Sigma \ME_{1,1} = \Sigma\ME_{\ell+1,k}.
    \end{equation}

    Let
    \begin{equation}
        \mathrm{sp}^\phi_{\ell,k} \colon \mho^{d+1}_{\ell,k-1}
        \to
        \mho^{d+2}_{\ell,k}
    \end{equation}
    be the map induced by $\phi_{\ell,k}$,
    and let
    \begin{equation}
        \mathrm{sp}^\psi_{\ell,k} \colon 
        \mho^{d+1}_{\ell+1,k}
        \to
        \mho^{d+2}_{\ell,k}.
    \end{equation}
    be the map induced by $\psi_{\ell,k}$.
    We refer to these maps as \textit{spiral maps} and distinguish one from the other by the superscript.\footnote{We also define and use maps $\mathrm{sp}_{\ell,0}^\psi$, with $\ell\ge 3$. These have to be defined in a different way: see \cref{Q8_psi}.}
\end{spiraldefn}
These maps will be our answer to \cref{goalz3} once we compute what they do on homotopy groups.

For example, the spiral map $\mathrm{sp}^\phi_{1,1}\colon \mho_{1,0}^2\to\mho_{1,1}^3$ is dual to the process of taking the submanifold $\RP^2\subset \RP^3$ and forgetting a line bundle.
It provides an isomorphism between the groups of IFTs generated by (assuming \cref{LEFT_ansatz}) the time-reversal symmetric Majorana chain and the $2+1$d class D$'$ superconductor, resp.
See \cref{d_pin_spiral_exm}.

Next, we discuss the computation of the free-to-interacting and spiral maps.
Before we get into the details,
we need the following definition.
\begin{defn}
\label{ldeg_defn}
Let $\ell,k\in\N$ and $j\coloneqq\lfloor \ell/4\rfloor$. Define
\begin{equation}
    \ldeg(\ell, k)\coloneqq \begin{cases}
        4j+k-1, &\ell\equiv 0\bmod 4\\
        4j+k, &\text{otherwise.}
    \end{cases}
\end{equation}
\end{defn}
We will show in \cref{only_first_summand_real} that $\ldeg(\ell, k)$ is the lowest degree in which the free-to-interacting map $F2I_{\ell,k}$ is nonzero.

The following result describes the ``long summands,'' the subgroups of $E_{\ell,k}$-symmetric IFTs that we focus on. For the most part, we only study the IFTs in this summand, as they have the most interesting behavior with respect to the spiral maps and the free-to-interacting map.
\begin{longsummand}
Suppose at least one of $\ell$ or $k$ is positive, and write $\ell = 4j+i$ with $0\le i\le 3$. Let $\widetilde m(n)\coloneqq \lfloor (n-4j-k)/8\rfloor$.
\begin{enumerate}
    \item If $n\equiv \ell+k-2i-1\bmod 4$ and $n\ge\ldeg(\ell,k)$, $\mho_{\ell,k}^n$ is the direct sum of a summand $L$ of the form
    \begin{equation}
        L\cong \begin{cases}
            \Z/2^{4\widetilde m(n)+4-i}, & n \equiv \ell+k-2i+3\bmod 8\\
            \Z/2^{4\widetilde m(n)+5-i}, & n \equiv \ell+k-2i-1\bmod 8
        \end{cases}
    \end{equation}
    and cyclic groups of orders at most that of $L$, and in fact strictly smaller if $\abs L >2$. 
    \item If $n\not\equiv \ell+k-2i-1\bmod 4$, $\mho_{\ell,k}^n$ is simple $2$-torsion.
\end{enumerate}
\end{longsummand}

We call $L$ the \term{long summand} of $\mho_{\ell,k}^n$ when it exists.\footnote{In general, $\mho_{\ell,k}^n$ has more than one direct-sum splitting satisfying \cref{only_one_long_summand}, so it is a priori ambiguous to say ``the'' long summand. However, by \cref{only_first_summand_real}, there is a canonical choice of long summand: the image of the free-to-interacting map. In what follows, we make this choice.}

Our first computational result is that, in the dimensions where they exist, the long summands are precisely the image of the free-to-interacting map.
\begin{ftoithm}
Let $\ell,k\ge 0$ and $(\ell,k)\ne (0,0)$.
In degrees $d\equiv k-\ell- 2\bmod 4$, $d\ge\ldeg(\ell,k)-1$, the free-to-interacting map $F2I_{\ell,k}\colon \KO^{d+\ell-k-2}\to \mho_{\ell,k}^{d+2}$ has image equal to the long summand of $\mho_{\ell,k}^{d+2}$.
\end{ftoithm}

Next, we calculate the images of the long summands under the two spiral maps.
\begin{phithm}
Let $\ell,k\ge 0$ and $(\ell,k)\ne (0,0)$.
The map
$\mathrm{sp}^\phi_{\ell,k}\colon\mho_{\ell,k-1}^n\to\mho_{\ell,k}^{n+1}$ maps the long summand of $\mho_{\ell,k-1}^n$ isomorphically onto the long summand of $\mho_{\ell,k}^{n+1}$.
\end{phithm}

\begin{psithm}
Suppose $\ell>0$ and $k>0$ or $\ell\ge 3$.
The map
$\mathrm{sp}^\psi_{\ell,k}\colon\mho_{\ell+1,k}^n\to\mho_{\ell,k}^{n+1}$ maps the long summand of $\mho_{\ell+1,k}^n$ injectively into the long summand of $\mho_{\ell,k}^{n+1}$, with cokernel $\Z/2$.
\end{psithm}
Thus we think of $\mathrm{sp}^\psi$ as ``multiplication by $2$'' on long summands.

The analogs of these theorems for the complex Bott spiral are \cref{cplx_F2I_image_thm,,ku_int_Bott,,complex_psi}.

Equipped with these computations, we can build a Bott spiral starting in any dimension and discrete\footnote{Using the computations in \cref{other_real_AZ_computation}, one can also begin the Bott spiral from a continuous Altland--Zirnbauer class. However, as we illustrate in \cref{mixed_DC_spiral}, if the corresponding discrete and continuous fermionic groups are not isomorphic, as happens in classes AI, AII, C, CI, and CII, this variant of the Bott spiral is not periodic: it terminates after finitely many steps. We interpret this as further evidence that it is the discrete fermionic groups, rather than the continuous ones, that correctly model the Bott spiral.}
Altland--Zirnbauer class by alternately applying the maps $\mathrm{sp}^\phi$ and $\mathrm{sp}^\psi$ and using \cref{40_04} where necessary to identify $E_{\ell,k+4}\cong E_{\ell+4,k}$. As mentioned above, this isomorphism is an interacting analog of Bott periodicity. After eight steps, one arrives back at the same Altland--Zirnbauer class, eight dimensions higher. By the above theorems, if one starts in a dimension and class that contains a long summand of order $N$, after eight steps one also has a long summand of order $16N$. Thus the ``Bott clock'' of periodicity for free fermions has unraveled into a ``Bott spiral'' that gets bigger and bigger every period.

We specifically consider the three spirals that start with classes $\mathrm{BDI}'$, CII, and AIII in dimension $d = 1$, as these are the cases Queiroz--Khalaf--Stern~\cite{queiroz_dimensional_2016} consider. Comparing their results (\textit{ibid.}, Figure 1) with ours (\cref{intro_spiral}, \cref{intro_spiral_2} \cref{melk_calc}, and \cref{only_first_summand_real}), we have almost complete agreement. We discuss the discrepancy in \cref{mu2_strikes_back}, where we argue that it arises from a subtlety difficult to detect on the free fermion side of the story, and that our field-theoretic analysis leads to the correct symmetry types in the spiral.

From our models and computations, we gained some insights towards the more general question of studying free-to-interacting maps.
\begin{itemize}
    \item \textbf{Discrete versus continuous Altland--Zirnbauer classes}, or more generally \textbf{how to specify the symmetries of a fermionic system.} In order to define and study free-to-interacting maps at all, one must specify the symmetries acting on the phases of interest. We found that \textbf{this information flows from the interacting to the free setting,} by choosing a fermionic group $G_f$ of symmetries on the interacting side, then obtaining the symmetry algebra $\R^f[G_f]$ on the free side. This is opposite to most prior work on free-to-interacting maps, in which one begins with an Altland--Zirnbauer class, perhaps with some extra data, then attempts to find the best-matching fermionic group realizing this data. Because nonisomorphic fermionic groups may have Morita-equivalent fermionic group superalgebras, extracting the group from the Morita class algebra is in general ambiguous. %

    Though this point was first raised in~\cite{stehouwer_interacting_2022}, it underlies several of our key observations in this paper, and was the most important conceptual step in modeling the Bott spiral. Specifically, our choice to use the fermionic groups $E_{\ell,k}$ to model the ten Altland--Zirnbauer classes (the ``discrete,'' rather than the ``continuous,'' choice) is invisible on the free side, but is essential on the interacting side in several ways.
    \begin{enumerate}
        \item The groups of IFTs in a given dimension and Altland--Zirnbauer class depend on this choice of fermionic group.
        Freed--Hopkins' computations~\cite[\S 9.3]{freed_reflection_2021} imply that the continuous fermionic groups do not in general match Queiroz--Khalaf--Stern's results; see \cref{discrete_ne_cts}. 
        \item It is unclear how to define the spiral maps $\mathrm{sp}^\phi$ and $\mathrm{sp}^\psi$ for the continuous fermionic groups. Even trying to make a ``mixed discrete-continuous'' Bott spiral, by mixing $E_{\ell,k}$ with a continuous fermionic group, is doomed: as we explain in \cref{mixed_DC_spiral}, such a spiral terminates after finitely many steps. By contrast, Bott spirals using $E_{\ell,k}$ fully match the periodicity identified by Queiroz--Khalaf--Stern.
        \item Queiroz--Khalaf--Stern~\cite{queiroz_dimensional_2016} use a parameter $\mu$ to modify their general pattern for the classification of SPTs in Altland--Zirnbauer types BDI, $\mathrm{D}'$, and DIII. In \cref{mu2_strikes_back}, we clarify this phenomenon as arising from the presence of an additional prime on the fermionic group in these three cases (see \cref{G_f_prime}). This behavior is unexpected, and important for correctly modeling the Bott spiral, but cannot be seen solely from the free side.
    \end{enumerate}
    It would be interesting to study other examples of this ``Morita variance'' phenomenon.
    \item \textbf{``Interacting Bott periodicity'' and the Atiyah--Bott--Shapiro orientation.} One major motivation for studying the Bott spiral is to understand what remains of Bott periodicity when one passes from free to interacting phases: see~\cite[Figure 1]{queiroz_dimensional_2016}.
    Since Bott classes in $\KO_8$, resp.\ $\KU_2$, lift to the spin, resp.\ \spinc bordism classes of a Bott manifold, resp.\ $\CP^1$, compactifying on these manifolds implements an analog of Bott periodicity for interacting SPTs, though this process is generally neither injective nor surjective. See~\cite[\S 1.4.4]{ADKPSS24}. Our computations imply that in the cases relevant for the Bott spiral, namely the long summands of $\mho_{\ell,k}^*$, this compactification is surjective, with kernel isomorphic to $\Z/16$ in the real case and $\Z/2$ in the complex case. This agrees with Queiroz--Khalaf--Stern's results~\cite{queiroz_dimensional_2016}.

    However, there is also the question of the periodicity of the Altland--Zirnbauer classes themselves. Freed--Hopkins' choices of fermionic groups representing the eight real Altland--Zirnbauer classes~\cite[(9.25)]{freed_reflection_2021} are not obviously periodic; Freed--Hopkins were not to our knowledge trying to construct a periodic sequence of interacting symmetry types.
    This once again suggests that, in order to construct a Bott-periodic sequence of interacting symmetry types, we must use a different model.
    \Cref{40_04}, constructing an isomorphism $E_{4,0}\cong E_{0,4}$ of fermionic groups, provides the answer. It implies that alternatively decreasing $\ell$ and increasing $k$, like in our model of the Bott spiral, produces an $8$-periodic sequence of fermionic groups, hence an $8$-periodic sequence of interacting symmetry types matching the Bott clock on the free side.
\end{itemize}

As part of our study of the Bott spiral, we proved some other results that may be of independent interest.
\begin{itemize}
    \item In \cref{MCB_dpin_F2I,smithy_f2i,YR_exm,more_crystalline_F2I}, we give mathematical models for a number of free-to-interacting maps beyond the type-$(\ell,k)$ and type-$m$ examples discussed above, and compare the behavior of our maps against the physics literature. Some of our examples incorporate spatial reflection symmetries through the use of Manjunath--Calvera--Barkeshli's~\cite[(30)]{manjunath_non-perturbative_nodate} version of the fermionic crystalline equivalence principle (\cref{FCEP}). Using our computations in \cref{s:spin_computations}, we show that our free-to-interacting maps agree with the corresponding maps calculated at a physical level of rigor in~\cite{rosch_unwinding_2012, ryu_interacting_2012, qi_new_2013, yao_interaction_2013, gu_interactions_2012, cho_topological_2015, morimoto_breakdown_2015, ryu_interacting_2015, yoshida_correlation_2015, 
    queiroz_dimensional_2016, song_interaction_2017, yoshida_fate_2017, aksoy_stability_2021, manjunath_non-perturbative_nodate}. We interpret this as evidence in favor of \cref{our_F2I_ansatz}, our mathematical model for a class of free-to-interacting maps.
    \item Yu~\cite{Yu95} studied the $\ko$-homology of $(B\Z/2)^{m}$ and showed that it demonstrates $4$-periodic behavior in $m$. In \cref{M_is_EA}, we show this extends to \emph{some} twists of $\ko$-homology of $(B\Z/2)^m$. Specifically, we show that the $\ko$-homology of $((B\Z/2)^{\ell+k})^{\sigma_1 + \dotsb + \sigma_\ell - \sigma_{\ell+1} - \dotsb - \sigma_{\ell+k}}$ has the same $4$-periodicity in $\ell$ and an additional $1$-periodicity in $k$. Here $\sigma_i$ is the tautological line bundle for the $i^{\mathrm{th}}$ factor of $B\Z/2$.

    This $4$-periodicity is related to the $\Z/4$ summand in the Picard group of $Q_1$-locally stably invertible $\cA(1)$-modules (\cref{Q1stab}); see Bruner~\cite{Bru14}. In future work, we will explicate this connection and relate it to a $4$-periodic family of Smith homomorphisms discussed in~\cite{kapustin_fermionic_2015, TY19, hason_anomaly_2020, WWZ20, debray_long_2024, debray_smith_2024}.
    \item In \cref{dpin_decomp}, we compute the dpin bordism groups of any space $X$ in dimensions $7$ and below, generalizing~\cite[Appendices E and F]{kaidi_topological_2020} for the case of a point. Since the worldsheet in Type I string theory has a dpin structure (\textit{ibid.}, \S 6), this could be of independent interest.
    \item In \cref{spin_Q8_bordism}, we compute $\Spin\times_{\set{\pm 1}}Q_8$ bordism groups in all dimensions, generalizing work of Pedrotti~\cite[Theorem 8.0.8]{Ped17} in dimensions $4$ and below. Using this, we compute the image of the discrete class C free-to-interacting map (\cref{disc_F2I_compute}) and show that $\Spin\times_{\set{\pm 1}}Q_8$ bordism is detected by $\KO$-Pontrjagin classes, Buchanan--McKean's $\mathit{KSp}$-characteristic classes~\cite[\S 5.5]{BM23}, and mod $2$ cohomology classes (\cref{signature_rem}).
    \item In \cref{thm:pin_tilde_c_ABP,thm:pinh_ABP}, we state Anderson--Brown--Peterson-type splitting results for the bordism spectra $\MTPin^{\tilde c\pm}$ and $\MTPin^{h\pm}$ into indecomposables, such that Freed--Hopkins' twisted Atiyah--Bott--Shapiro maps~\cite[\S 9.2.2]{freed_reflection_2021} are nonzero on exactly one summand of each spectrum. These theorems will be proven in a separate paper of the first two named authors and Pacheco-Tallaj~\cite{ABP_splittings}, extending work of Anderson--Brown--Peterson~\cite{ABP67} for $\MTSpin$, Anderson--Brown--Peterson (\textit{ibid.}) and Stong~\cite[Chapter XI]{Sto68} for $\MTSpin^c$, and Mills~\cite{Mil24} and Buchanan--McKean~\cite{BM23} for $\MTSpin^h$.
\end{itemize}

\subsection*{Outline}

In \cref{section_fermionic_groups}, we discuss fermionic groups. We define fermionic groups in \cref{sec:fermgrp}, review superalgebras in \cref{ss:superalgebra}, introduce the twisted fermionic group superalgebra $\R^f[G_f]$ in \cref{subsec:twistedgroupsuperalgebras}, and discuss its $K$-theory in \cref{sec:Ktheoryoffermgps}. In \cref{ss:xi_to_Gf}, we describe the twisted spin structure associated to a fermionic group. %

In \cref{section_free_to_interacting_section}, we review the mathematical models for the kinds of topological phases that we study in this paper: free fermions via $K$-theory in \cref{ff_and_k}, and interacting SPTs in terms of bordism in \cref{interacting_IFT}. In \cref{subsec_symm_types_morita_invnce}, we introduce the fermionic groups we use to model the discrete and continuous versions of the ten Altland--Zirnbauer types (\cref{tenfold_table}). %

We define our \term{free-to-interacting maps} in \cref{section_F2I}. We review the Atiyah--Bott--Shapiro (ABS) construction and Smith homomorphisms in \cref{sec_classical_ABS} and \cref{smith_homs_intro}, respectively. Then, in \cref{ss:twABS} and \cref{subsec_generalized_ABS}, we combine these two ideas to define twisted ABS maps, and apply Anderson duality to define our free-to-interacting maps in \cref{ss:F2I}.

With all of the ingredients in place, we provide our model of the Bott spiral in \cref{section_modeling_the_Bott_spiral}. We define ``primed Altland--Zirnbauer classes'' in 
\cref{primed_AZ}, and interpret them as including a spatial reflection symmetry in \cref{CEP}. In \cref{ss:imF2I}, we study the images of the free-to-interacting maps from \cref{ss:F2I} on homotopy groups. In \cref{ss:spiral_model}, we construct the spiral maps $\mathrm{sp}^\phi_{\ell,k}$ and $\mathrm{sp}_{\ell,k}^\psi$, then use them to model the three Bott spirals of Queiroz--Khalaf--Stern: %
the complex Bott spiral (starting with class AIII) in \cref{fig:cplx_Bott_spiral_maps}, the real Bott spiral starting with class $\mathrm{BDI}'$ in \cref{fig:real_spiral_from_BDI_prime}, and the real Bott spiral starting with class CII in \cref{CII_spiral}. 

Our final two sections, \cref{section_computations} and \cref{s:spin_computations}, are devoted to the computations supporting our main results: the computations of (the interesting part of) the free-to-interacting maps and the spiral maps for the discrete Altland--Zirnbauer symmetry types appearing in the Bott spiral. \cref{section_computations} performs an approximation to our computations, in which spin bordism is replaced with $\ko$-homology. \cref{ss:kogen} begins with some generalities, and \cref{ss:EA_type} introduces four key examples and proves \cref{melk_calc}. In \cref{ss:ko_phi_f2i} and \cref{ss:ko_psi}, we prove \cref{compute_F2I,int_Bott_spiral_ko,psicalc}, calculating $\ko$-theoretic approximations to the free-to-interacting and spiral maps. In \cref{spin_computation} we lift to spin bordism, yielding our main computations in \cref{only_first_summand_real,int_Bott_spiral_1,MSpin_psi}. The remaining sections discuss several variants of the free-to-interacting and spiral maps: ``mixed discrete-continuous'' variants in \cref{other_real_AZ_computation}, complex versions in \cref{complex_computation}, and the discrete class C maps in \cref{spinQ8}.

\subsection*{Acknowledgments}
We would especially like to thank Natalia Pacheco-Tallaj and Ryan Thorngren for many discussions that were helpful in writing this paper.
We additionally thank
Omar Antolín Camarena,
Dan Berwick-Evans,
Jonathan Buchanan,
Vladimir Calvera,
Markus Dierigl,
Dan Freed,
Meng Guo,
Zachary Halladay,
Mike Hopkins,
Theo Johnson-Freyd,
Yigal Kamel,
Ralph Kaufmann,
Yu Leon Liu,
Naren Manjunath,
Stephen McKean,
Lukas Müller,
Niccolò Porciani,
Daniel Sheinbaum,
Ethan Torres,
Mayuko Yamashita,
Weicheng Ye,
and
Matthew Yu.
Part of this project was completed while the authors visited the Perimeter Institute for Theoretical Physics for the conference ``Higher Categorical
Tools for Quantum Phases of Matter''; research at Perimeter is supported by the Government of Canada through Industry Canada and by the Province of Ontario through the Ministry of Research \& Innovation. CK was supported by the National Science Foundation under Grant No.\ DGE-2141064 during the early writing of this paper, and is currently
supported by the Simons Foundation and the Killam Trusts.
LS is supported by ERC Consolidator Grant SYMSPEC and
additionally acknowledges the Simons Foundation and the Atlantic Association for Research in the Mathematical Sciences.
\resumetoc

\section{Fermionic groups, \texorpdfstring{$K$}{K}-theory, and bordism}\label{section_fermionic_groups}
    
In this work, we study the relationship between free fermionic phases, which are classified by $K$-theory groups~\cite{kitaev_periodic_2009, freed_twisted_2013}, and interacting fermionic phases, which under the SPT-bordism conjecture \cite{kapustin2014beyondcohomology, kapustin_fermionic_2015, freed_reflection_2021} are classified by Anderson-dual bordism groups. 
We present a unified account of symmetries in both classifications using the language of fermionic groups, which allows us to predict the form of the free-to-interacting maps in \cref{section_free_to_interacting_section}.

\subsection{The definition of a fermionic group}
\label{sec:fermgrp}

\begin{defn}[{Benson~\cite[\S 7]{Ben88}}]\label{defn_ferm_group}
    A \emph{fermionic group} consists of 
    \begin{itemize}
        \item a Lie group $G_f$,
        \item a central element $(-1)^F$ of order two called \emph{fermion parity},
        \item and a homomorphism $\theta \colon G_f \to \Z/2$,
    \end{itemize}   
    such that $\theta((-1)^F) = 0$.

    A morphism of fermionic groups $(G_f, ((-1)^F)_G, \theta_G)\to (H, ((-1)^F)_H, \theta_H)$ is a Lie group homomorphism $\varphi\colon G_f\to H_f$ such that $\varphi(((-1)^F)_G) = ((-1)^F)_H$ and $\theta_G = \theta_H\circ\varphi$. With these morphisms, fermionic groups form a category $\cat{FermGrp}$. %
\end{defn}

There is a central subgroup $\{\pm 1\} \subseteq G_f$ generated by $(-1)^F$. 
Physically, $G_f$ represents the group of symmetries of a given physical systems with fermions.
This group potentially contains time-reversing (or anti-unitary) symmetries as indicated by $\theta$.
We denote by $G_b \coloneqq {G_f}/\set{\pm 1}$ the associated bosonic symmetry group.
\begin{defn}\label{fermionic_twist}
    Let $(-1)^F \in G_f \xrightarrow{\theta} \Z/2$ be a fermionic group.
    The \emph{associated twist} $\tau$ is the element
    \begin{equation}\label{twistdat}
    (\theta,\omega) \in H^1(BG_b; \Z/2) \times H^2(BG_b; \Z/2)
    \end{equation}
    defined as follows.
    \begin{itemize}
        \item The Hurewicz and universal coefficient theorems imply a natural isomorphism $\Hom(G_b, \Z/2)\overset\cong\to H^1(BG_b;\Z/2)$. We use this isomorphism to identify $\theta\colon G_b\to\Z/2$ with the class $\theta$ in~\eqref{twistdat}.
        \item The class $\omega$ in~\eqref{twistdat} classifies the extension
            \begin{equation}
            \shortexact*{\{\pm 1\}}{G_f}{G_b}.
            \end{equation}
    \end{itemize}
\end{defn}

The terminology `twist' comes from twisted cohomology, as $\tau$ also specifies a twist of $G_b$-equivariant $\KO$-theory (see~\cite{DK70}, \cite[\S 3.4.2]{gomi_freed-moore_2021}) and spin bordism of $BG_b$ (see~\cite{HJ20}). 
We will see in later sections how such twisted $G_b$-equivariant cohomology theories classify $G_b$-protected SPT phases.
The twist encodes physical properties, including which elements of $G_f$ act by antiunitary symmetries, and how they mix with fermion parity $(-1)^F$
(e.g.\ $T^2 = (-1)^F$ in the fermionic group $\Pin_1^-$).

Observe that twists are natural in homomorphisms of fermionic groups. More specifically, a homomorphism $G_f \to H_f$ induces a homomorphism $\phi\colon G_b \to H_b$ such that $\phi^*(\omega_G) = \omega_H$ and $\phi^*(\theta_G) = \theta_H$.

\begin{exm}\label{repn_fg}
If $\rho\colon G_b \to\O_n$ is a representation, there are two associated fermionic groups defined by pulling back the $\Pin^\pm$ extensions of $\O_n$ along $\rho$ (see Footnote~\ref{pin_convention}): %
\begin{equation}
\begin{tikzcd}
    1 \ar[r] & \set{\pm 1}\ar[r] \ar[d,equal] & G_f \ar[r] \ar[d]& G_b\ar[r] \ar[d,"\rho"] & 1
    \\
    1\ar[r] & \set{\pm 1}\ar[r] & \Pin_n^\pm \ar[r]& \O_n\ar[r] & 1.
\end{tikzcd}
\end{equation}
For the \pinp
twist, $\theta = w_1(\rho)$ and $\omega = w_2(\rho)$; for the \pinm twist, $\theta = w_1(\rho)$ and $\omega = w_2(\rho) + w_1(\rho)^2$, where $w_i(\rho)$ denotes the $i^{\mathrm{th}}$ Stiefel--Whitney class of the vector bundle over $BG_b$ induced by $\rho$.

The universal case $G_b = \O_n$ and $\rho = \id$ defines fermionic group structures on $\Pin_n^\pm$ with $(\Pin_n^\pm)_b\cong \O_n$.
For $G_b = \SO_n$ and $\rho$ the inclusion, we obtain a fermionic group structure on $\Spin_n$.
\end{exm}

The fermionic tensor product of $G$ and $H$ is a certain fermionic group whose underlying bosonic group is $G_b \times H_b$:
\begin{defn}[{Benson~\cite[\S 7]{Ben88}}]
\label{ferm_tens_defn}
The \emph{fermionic tensor product} $G \ftens H$ of $(G,\theta_G, \omega_G), (H,\theta_H, \omega_H)$ is the set 
\begin{equation}
\frac{G \times H}{\{\pm 1\}}
\end{equation}
equipped with the operation
\begin{equation}
(g_1 \mathbin{\hat{\otimes}} h_1) (g_2 \mathbin{\hat{\otimes}} h_2) = 
\begin{cases}
(-1)^F g_1 g_2 \mathbin{\hat{\otimes}} h_1 h_2, & \text{ if } \theta(h_1) = 1 \text{ and } \theta(g_2) = 1
\\
    g_1 g_2 \mathbin{\hat{\otimes}} h_1 h_2,  & \text{ if } \theta(h_1) = 0 \text{ or } \theta(g_2) = 0.
\end{cases}
\end{equation}
$G\ftens H$ has the structure of a fermionic group, with the following data.
\begin{enumerate}
    \item $\theta(g, h) = \theta_G(g) + \theta_H(h)$ for $g\in G$ and $h\in H$; since $\theta_G$ and $\theta_H$ vanish on the $\{\pm 1\}$ subgroup we quotiented by, this is indeed well-defined.
    \item The central subgroup $\{\pm 1\}\times\{\pm 1\}\subset G\times H$ maps to a central subgroup of $G\ftens  H$ isomorphic to $\{\pm 1\}$ under the quotient map; we define $(-1)^F$ of $G\ftens  H$ to be the generator of this $\{\pm 1\}$ subgroup.
\end{enumerate}
\end{defn}

It is straightforward, if a little tedious, to check the following proposition directly.
\begin{prop}\label{fermionic_tensor_symm_mon}
$\cat{FermGrp}$ is symmetric monoidal with respect to the fermionic tensor product; the unit is $\{\pm1\}$ and the braiding is given by
\begin{equation}
g \mathbin{\hat{\otimes}} h \mapsto 
\begin{cases}
(-1)^F h \mathbin{\hat{\otimes}} g  & \text{ if } \theta(h) = 1 \text{ and } \theta(g) = 1
\\
    h \mathbin{\hat{\otimes}} g,  & \text{ if } \theta(h) = 0 \text{ or } \theta(g) = 0.
\end{cases}
\end{equation}
\end{prop}
\begin{lem}\label{twist_of_product}
Let $G_f$ and $H_f$ be fermionic groups.
\begin{enumerate}
    \item\label{bosonic_product} There is a natural isomorphism $(G_f\ftens H_f)_b\xrightarrow{\cong} G_b\times H_b$ of groups.
    \item %
    The associated twist to $G_f\ftens  H_f$ is
    \begin{equation}\label{fake_Whitney}
        (\theta_G + \theta_H, \omega_G + \theta_G \theta_H + \omega_H).
    \end{equation}
\end{enumerate}
\end{lem}
\begin{proof}
Part~\eqref{bosonic_product} follows directly from \cref{ferm_tens_defn}. The formula for $\theta$ of $G_f\ftens  H_f$ follows from the the natural identification $H^1(BK; \Z/2)\cong \Hom(K, \Z/2)$ of abelian groups, where $K$ is a Lie group, and the fact that we defined $\theta$ of $G_f\ftens  H_f$ by adding $\theta_G$ and $\theta_H$.

The formula for $\omega$ is less trivial. It must be a natural formula in the data of two fermionic groups; the only natural cohomology classes associated to this data are $\theta_G$, $\theta_H$, $\omega_G$, and $\omega_H$ and classes built out of them (e.g.\ products). Moreover, the formula must be symmetric in $G_f$ and $H_f$, because the fermionic tensor product is symmetric monoidal. Therefore there are constants $\textcolor{Rhodamine}{\lambda_1},\textcolor{Orange}{\lambda_2},\textcolor{Cyan}{\lambda_3}\in\Z/2$ such that
\begin{equation}\label{preformula_for_product}
    \omega_{G_f\ftens  H_f} = \textcolor{Rhodamine}{\lambda_1(\omega_G + \omega_H)} + \textcolor{Orange}{\lambda_2(\theta_G^2  + \theta_H^2)} + \textcolor{Cyan}{\lambda_3(\theta_G\theta_H)}.
\end{equation}
It suffices to determine $\textcolor{Rhodamine}{\lambda_1}$, $\textcolor{Orange}{\lambda_2}$, and $\textcolor{Cyan}{\lambda_3}$. Therefore we may without loss of generality pass to cohomology.
We will show $\textcolor{Rhodamine}{\lambda_1 = 1}$, $\textcolor{Orange}{\lambda_2 = 0}$, and $\textcolor{Cyan}{\lambda_3 = 1}$.
\begin{enumerate}
    \item To show $\textcolor{Rhodamine}{\lambda_1 = 1}$, let $G_f$ be any fermionic group with $\theta_G = 0$ but $\omega_G\ne 0$ and let $H_f = \{\pm 1\}$. For example, we could take $G_f = \SU_2$ with the fermionic group structure discussed in~\cite[Example 9]{stehouwer_interacting_2022}. Then $G_f\ftens  H_f\cong G_f$ as fermionic groups, so $\omega_{G_f\ftens  H_f} = \omega_G$.
    \item To show $\textcolor{Orange}{\lambda_2 = 0}$, let $G_f$ be any fermionic group with $\theta_G\ne 0$, $\theta_G^2\ne 0$, and $\omega_G = 0$, and let $H_f = \{\pm 1\}$ again. For example, we could take $G = \Pin_1^+$. Then $G_f\ftens  H_f\cong G_f$ again, and now $\omega_{G_f\ftens  H_f} = 0$ and $\theta_G^2\ne 0$, forcing $\textcolor{Orange}{\lambda_2 = 0}$.
    \item To show $\textcolor{Cyan}{\lambda_3 = 1}$, let $G_f = H_f = \Z/4\times \{\pm1\}$, so that $G_b = H_b = \Z/4$, with $\theta_G,\theta_H\colon\Z/4\to\Z/2$ both equal to the unique nontrivial homomorphism. We also have $\omega_G = \omega_H = 0$. Eilenberg--Mac Lane~\cite{EM45, EM47, Eil49} proved an isomorphism $H^*(B\Z/4;\Z/2)\cong\Z/2[x, y]/(x^2)$ with $\abs x = 1$ and $\abs y = 2$; since $\theta_G$ and $\theta_H$ are nontrivial, they both equal $x$, and we deduce $\theta_G^2 = \theta_H^2 = 0$.

    Thus for this choice of $G_f$ and $H_f$,~\eqref{preformula_for_product} simplifies to \begin{equation}
    \omega_{G_f\ftens  H_f} = \textcolor{Cyan}{\lambda_3}\theta_G\theta_H,
    \end{equation}
    and all we have to do is show that the extension
    \begin{equation}
        \shortexact*{\{\pm 1\}}{G_f\ftens  H_f}{\Z/4\times\Z/4},
    \end{equation}
    which is classified by $\omega_{G_f\ftens H_f}$, is not split. The split extension $\{\pm 1\}\times\Z/4\times\Z/4$ is abelian, so it suffices to show $G_f\ftens H_f$ is not abelian, which can be checked directly from \cref{ferm_tens_defn}. \qedhere
\end{enumerate}
\end{proof}
\begin{rem}
\label{faux_whitney}
It is no coincidence that~\eqref{fake_Whitney} looks like the Whitney sum formula. Indeed, recall that $B\O$ is a grouplike $E_\infty$-space under $\oplus$, and let $B\O/B\Spin$ denote the cofiber, in the $\infty$-category of $E_\infty$-spaces, of the forgetful $E_\infty$-map $B\Spin\to B\O$. Then $B\O/B\Spin$ is again a grouplike $E_\infty$-space, and is equivalent to $B\Z/2\times B^2\Z/2$ as spaces but not as $E_1$-spaces~\cite[\S 1.2.3]{DY23}---the direct sum structure on $B\O$ induces a non-product $E_\infty$-structure on $B\Z/2\times B^2\Z/2$, which on cohomology is the Whitney sum formula for Stiefel--Whitney classes. See also~\cite[Proposition 4.19]{beardsley2023brauer} for a related computation.
\end{rem}
Sometimes fermionic groups arise as (s)pin covers of representations. Recall from Footnote~\ref{pin_convention} our conventions for pin groups, and
recall that the direct sum of inner product spaces induces a smooth homomorphism
\begin{equation}
\label{oplus_defn_orth}
    \oplus\colon\O_{\ell,k}\times\O_{\ell',k'}\longrightarrow \O_{\ell+\ell',k+k'}
\end{equation}
for all $\ell,\ell',k,k'\ge 0$. These maps are associative on a triple product $\O_{\ell,k}\times\O_{\ell',k'}\times\O_{\ell'',k''}$.
\begin{defn}
Let $\cat{Rep}$ denote the symmetric monoidal category specified by the following data.
\begin{subequations}
\begin{enumerate}
    \item An object of $\cat{Rep}$ is a Lie group $G$ and a smooth homomorphism $\rho\colon G\to\O_{\ell,k}$.
    \item A morphism $f$ from $(G, \rho\colon G\to\O_{\ell,k})$ to $(H, \rho'\colon H\to \O_{\ell,k})$ is a smooth group homomorphism $f_1\colon G\to H$ intertwining the maps to $\O_{\ell,k}$ up to conjugation by an element of $\O_{\ell,k}$. If $(\ell,k)\ne (\ell',k')$, there are no morphisms from $(G, \rho\colon G\to\O_{\ell,k})$ to $(H, \rho'\colon H\to \O_{\ell',k'})$.
    \item The unit is the trivial group with its unique representation into $\O_{0,0} = \set 1$.
    \item The monoidal product is the external product. Specifically, the external product of $\rho\colon G\to\O_{\ell,k}$ and $\rho'\colon H\to\O_{\ell',k'}$ is the composition
    \begin{equation}
        G\times H \xrightarrow{\rho \times\rho'} \O_{\ell,k}\times\O_{\ell',k'} \xrightarrow[\eqref{oplus_defn_orth}]{\oplus} \O_{\ell+\ell',k+k'}.
    \end{equation}
    \item The braiding is induced from the usual one in the Cartesian symmetric monoidal structure on the symmetric monoidal category of Lie groups.
\end{enumerate}
\end{subequations}
In addition, let $\cat{Rep}^{\mathrm{cpt}}$ be the full subcategory of $\cat{Rep}$ consisting of compact $G$ with $\rho$ factoring through $\O_\ell\times\O_k$.\footnote{If $G$ is compact, $\rho$ factors through $\O_\ell\times\O_k$ up to conjugation, so the second condition is no loss of generality.}
\end{defn}
\begin{defn}
\label{tildeconj}
For $A\in\O_{\ell,k}$, let $\mathrm{Ad}_A\colon\O_{\ell,k}\to\O_{\ell,k}$ be conjugation by $A$: $\mathrm{Ad}_A(B)\coloneqq ABA^{-1}$.

$A$ has exactly two preimages under $\Pin_{\ell,k}\to\O_{\ell,k}$; since they differ by a central element, they define the same conjugation homomorphism $\Pin_{\ell,k}\to\Pin_{\ell,k}$; call this map $\widetilde{\mathrm{Ad}}_A$.
\end{defn}
\begin{defn}
\label{zeta_defn}
Let $\zeta\colon\cat{Rep}\to\cat{FermGrp}$ be the functor sending $\rho\colon G\to \O_{\ell,k}$ to the pullback $G_f$ in
\begin{subequations}
\begin{equation}
\begin{tikzcd}
	{G_f} & {\Pin_{\ell,k}} \\
	G & {\O_{\ell,k},}
	\arrow[from=1-1, to=1-2]
	\arrow["\pi"', from=1-1, to=2-1]
	\arrow["\lrcorner"{anchor=center, pos=0.125}, draw=none, from=1-1, to=2-2]
	\arrow[from=1-2, to=2-2]
	\arrow["\rho", from=2-1, to=2-2]
\end{tikzcd}
\end{equation}
with $\set{\pm 1}\coloneqq\ker(\pi)$ and $\theta\coloneqq\det\circ\rho\circ\pi\colon G_f\to\set{\pm 1}$.

The definition on morphisms is a little nontrivial.
Let $f\colon (G, \rho_G\colon G\to\O_{\ell,k})\to (H, \rho_H\colon H\to\O_{\ell,k})$ be a morphism in $\cat{Rep}$, meaning $f\colon G\to H$ is a group homomorphism and there is an $A\in\O_{\ell,k}$ such that for all $g\in G$, $\mathrm{Ad}_A(\rho_G(g)) = \rho_H(f(g))$. Thus we have a commutative diagram whose top and bottom faces are pullback squares:
\begin{equation}\begin{tikzcd}[row sep=0.5cm, column sep=0.5cm]
	{G_f} && {\Pin_{\ell,k}} & \\
	& G && {\O_{\ell,k}} \\
	{H_f} && {\Pin_{\ell,k}} \\
	& H && {\O_{\ell,k}.}
	\arrow["{\widetilde\rho_G}", from=1-1, to=1-3]
	\arrow["{\pi_G}"{description}, from=1-1, to=2-2]
	\arrow["\lrcorner"{anchor=center, pos=0.125, rotate=45}, draw=none, from=1-1, to=2-4]
	\arrow[from=1-3, to=2-4]
	\arrow["{\widetilde{\mathrm{Ad}}_A}"{pos=0.2}, from=1-3, to=3-3]
	\arrow["{\rho_G}"{pos=0.3}, from=2-2, to=2-4, crossing over]
	\arrow["{\mathrm{Ad}_A}", from=2-4, to=4-4]
	\arrow["{\widetilde\rho_H}"'{pos=0.8}, from=3-1, to=3-3]
	\arrow["{\pi_H}"{description}, from=3-1, to=4-2]
	\arrow["\lrcorner"{anchor=center, pos=0.125, rotate=45}, draw=none, from=3-1, to=4-4]
	\arrow[from=3-3, to=4-4]
	\arrow["{\rho_H}"', from=4-2, to=4-4]
    \arrow["f"'{pos=0.2}, from=2-2, to=4-2, crossing over]
\end{tikzcd}\end{equation}
\end{subequations}
The maps $f\circ\pi_G\colon G_f\to H$ and $\widetilde{\mathrm{Ad}}_A\circ \widetilde\rho_G\colon G_f\to \Pin_{\ell,k}$ thus commute under the maps to $\O_{\ell,k}$ in the bottom right of the cube, so by the universal property of the pullback they induce a map $\zeta(f)\colon G_f\to H_f$. Since $\mathrm{Ad}_A$ intertwines the determinant and $\widetilde{\mathrm{Ad}}_A$ intertwines multiplication by $(-1)^F\in\Pin_{\ell,k}$, $\zeta(f)$ is indeed a morphism of fermionic groups.
\end{defn}
This is a mixed-signature generalization of \cref{repn_fg}.
\begin{lem}
There is a pullback diagram
\begin{equation}\begin{tikzcd}\label{pin_pullback}
	{\Pin_{\ell,k}\ftens \Pin_{\ell',k'}} & {\Pin_{\ell+\ell',k+k'}} \\
	{\O_{\ell,k}\times\O_{\ell',k'}} & {\O_{\ell+\ell',k+k'}.}
	\arrow[from=1-1, to=1-2]
	\arrow[from=1-1, to=2-1]
	\arrow["\lrcorner"{anchor=center, pos=0.125}, draw=none, from=1-1, to=2-2]
	\arrow[from=1-2, to=2-2]
	\arrow["\oplus", from=2-1, to=2-2]
\end{tikzcd}\end{equation}
natural with respect to the associators for triple products and the braidings on both sides.
\end{lem}
\begin{proof}[Proof sketch]
This diagram and the desired naturality statements follow directly from the definition of the pin group of a vector space equipped with a quadratic form, as in~\cite[Definition I.2.3]{lawson_spin_1989}.
\end{proof}
See also~\cite[\S 3.2]{HJ20}.
\begin{cor}
\label{zetamon}
$\zeta\colon \cat{Rep}\to\cat{FermGrp}$ admits the structure of a symmetric monoidal functor.
\end{cor}
\begin{proof}[Proof sketch]
Compose the pullback square~\eqref{pin_pullback} with the pullback square
\begin{equation}\begin{tikzcd}
	{\zeta(G)\ftens\zeta(H)} & {\Pin_{\ell,k}\ftens \Pin_{\ell',k'}} \\
	{G\times H} & {\O_{\ell,k}\times\O_{\ell',k'}.}
	\arrow[from=1-1, to=1-2]
	\arrow[from=1-1, to=2-1]
	\arrow["\lrcorner"{anchor=center, pos=0.125}, draw=none, from=1-1, to=2-2]
	\arrow[from=1-2, to=2-2]
	\arrow[from=2-1, to=2-2]
\end{tikzcd}
\qedhere\end{equation}
\end{proof}

\begin{defn}
\label{day_conv}
Let $A$ be an $E_\infty$-space. Then there is a symmetric monoidal structure on $\Top_{/A}$, the category of spaces with a map to $A$,
defined as follows:
use the multiplication $\mu\colon A\times A\to A$ to define the ``external sum'' of maps $f_1\colon X\to A$ and $f_2\colon Y\to A$ as follows. The product map $f_1\boxplus f_2\colon X\times Y\to A$ is the composition
\begin{equation}\label{extdir}
    X\times Y\xrightarrow{f_1\times f_2} A\times A\overset\mu\longrightarrow A.
\end{equation} %
\end{defn}
This sum refines to a symmetric monoidal $\infty$-category structure on $\Top_{/A}$, and the functor $\Top_{/A}\to\Top_{/B}$ induced by an $E_\infty$-map $A\to B$ is symmetric monoidal.

In what follows, we use the bar construction model for the classifying space $BG$ of a topological group $G$. With this choice, there is a natural homeomorphism $B(G\times H)\cong BG\times BH$.

Recall that $\cat{Rep}^{\mathrm{cpt}}$ is the full subcategory of $\cat{Rep}$ of representations of \emph{compact} Lie groups valued in $\O_\ell\times\O_k\subset\O_{\ell,k}$.
\begin{defn}\label{xi_defn}
Write $V_j\to B\O_j$ for the tautological bundle.
Let $\xi\colon\cat{Rep}^{\mathrm{cpt}}\to\Top_{/B\O}$ denote the functor sending a representation $\rho\colon G\to\O_\ell\times\O_k$ to the map %
\begin{equation}\label{to_BO_xi}
    BG\xrightarrow{B\rho}B\O_\ell\times B\O_k\xrightarrow{V_\ell - V_k + k-\ell} B\O.
\end{equation}
\end{defn}
We include the $k-\ell$ term in~\eqref{to_BO_xi} so that the virtual bundle has rank zero, hence is classified by $B\O$ instead of $\Z\times B\O$.
\begin{lem}
\label{Rep_to_BO}
Give $B\O$ its direct sum $E_\infty$-structure and $\Top_{/B\O}$ the induced symmetric monoidal structure as in \cref{day_conv}. Then $\xi$ is symmetric monoidal.
\end{lem}
\begin{proof}
This follows essentially by definition of the direct sum $E_\infty$-structure on $B\O$, together with symmetric monoidality of the classifying space functor.
\end{proof}
\begin{lem}
Let $\rho\colon G\to\O_\ell\times\O_k$ be an object of $\cat{Rep}^{\mathrm{cpt}}$ and $f\colon BG\to B\O$ be $\xi(G,\rho)$. Then the twisting data $(\theta,\omega)$ for $\zeta(G)$ equals $(f^*(w_1), f^*(w_2))$.
\end{lem}
\begin{proof}
The equality to prove is natural in $(G,\rho)$, so it suffices to show it in the universal case $\id\colon\O_\ell\times\O_k\to\O_\ell\times\O_k$. Then the Whitney sum formula and \cref{twist_of_product} reduce the claim to checking for the left and right inclusions $\O_\ell\to\O_\ell\times\O_k$, resp.\ $\O_k\to\O_\ell\times\O_k$ respectively. This then boils down to the claims that the twisting data for $\Pin_\ell^-$, resp.\ $\Pin_k^+$, are $(w_1, w_2 + w_1^2)$, resp.\ $(w_1, w_2)$, which is standard.
\end{proof}
    \begin{rem}
\label{rem:BO/BSpin}
    This $E_\infty$-structure on $B\Z/2\times B^2\Z/2$ also arises naturally from the following context.  Let $\sLine_\R$ be the category of real superlines, i.e.\ one-dimensional real supervector spaces with invertible linear maps.
    $\cat{sLine}_\R$ acquires an $E_\infty$-structure $\hat\otimes$ from the fact that it is the maximal Picard groupoid inside the symmetric monoidal category $\sVect_\R$ with its Koszul braiding (see~\eqref{koszul}).
    This makes the inclusion
    \begin{equation}
        B\O/B\Spin \simeq B\Z/2 \times B^2 \Z/2 \hookrightarrow B\Z/2 \times B^2 \R^\times \simeq B\sLine_\R
    \end{equation}
    into an $E_\infty$-map. Like for $B\O/B\Spin$, the equivalence $B\Z/2\times B^2\R^\times\simeq B\sLine_\R$ of spaces is not $E_1$.
\end{rem}

The next definition will play a crucial role in our physical application.
\begin{defn}\label{elk_definition}
The fermionic group $E_{\ell,k}$ is
\begin{equation}
    E_{\ell,k}\coloneqq \underbracket{\Pin_1^+\ftens  \dotsb \ftens  \Pin_1^+}_{\text{$\ell$ copies}} \ftens 
     \underbracket{\Pin_1^-\ftens  \dotsb \ftens  \Pin_1^-}_{\text{$k$ copies}}.
\end{equation}
By \cref{twist_of_product}, $(E_{\ell,k})_b$ is isomorphic to the elementary abelian $2$-group\footnote{Here and throughout this paper, ``$2$-groups'' are in the sense of $p$-groups, i.e.\ finite groups of order $2^s$ for some $s$. We do not use ``$n$-groups'' in the sense of, e.g., \cite{BL04}.}
$(\Z/2)^{\ell+k}$.
\end{defn}
\begin{rem}
\label{altelk}
One may equivalently define $E_{\ell,k}$ as the preimage of the diagonal $(\Z/2)^{\ell+k}\subset\O_{\ell,k}$ under the double cover $\Pin_{\ell,k}\twoheadrightarrow\O_{\ell,k}$, lifting $E_{\ell,k}$ to an object of $\cat{Rep}$.
This definition makes sense for $\ell = k = 0$: in this case, $\O_{\ell,k} = 1$ is the trivial group and so $\Pin_{\ell,k} = \{\pm 1\} = E_{0,0}$. If we regard the empty fermionic product as the monoidal unit, then \cref{elk_definition} also gives $E_{0,0} = \{\pm 1\}$.
\end{rem}

\begin{rem}
\label{twisting_data_Elk}
We record the twisting data for $E_{\ell,k}$, which the reader can verify directly from \cref{twist_of_product}.

Use the Künneth formula to write $H^*((B\Z/2)^{\ell+k};\Z/2)\cong\Z/2[x_1,\dotsc,x_{\ell+k}]$, with $\abs{x_i} = 1$ for all $i$. Under the isomorphism $H^1((B
\Z/2)^{\ell + k};\Z/2)\cong \Hom(\Z/2^{\ell + k}, \Z/2)$, $x_i$ is dual to the $i^{\mathrm{th}}$ standard basis vector in $(\Z/2)^{\ell+k}$. Then the twisting data of $E_{\ell,k}$ is
\begin{equation}
    \theta = \sum_{i=1}^{\ell+k} x_i\qquad\qquad
    \omega = \sum_{i=\ell+1}^{\ell+k} x_i^2 + \sum_{1\le i<j\le \ell+k} x_ix_j.
    \qedhere
\end{equation}
\end{rem}

\subsection{Superalgebras and Clifford algebras}
\label{ss:superalgebra}
In this paper, we frequently work with $\Z/2$-graded objects and require the tensor product to obey the Koszul sign rule with respect to the grading. We summarize the definitions we use; the reader who is familiar with superalgebras and Clifford algebras is welcome to jump to \cref{fg_alg}. 

Fix a field $F$ of characteristic not equal to $2$; in this paper we will only need $F = \R$ and $F = \C$. A \term{super vector space} $V$ over $F$ is a $\Z/2$-graded vector space over $F$. The tensor product $V\mathbin{\hat\otimes} W$ of two super vector spaces $V$ and $W$ is given the following grading: if $v\in V$ and $w\in W$ are homogeneous elements, then $v\mathbin{\hat \otimes} w$ has degree $\abs v + \abs w\in\Z/2$. We use the Koszul sign rule isomorphism $V\mathbin{\hat\otimes} W\xrightarrow{\cong} W\mathbin{\hat\otimes} V$, defined on pure tensors of homogeneous elements by
\begin{equation}\label{koszul}
    v\mathbin{\hat\otimes} w \mapsto (-1)^{\abs v\abs w} w\mathbin{\hat\otimes }v.
\end{equation}
This formula extends uniquely to define an isomorphism on the entire tensor product, and is the symmetry data in a symmetric monoidal category $\sVect_F$ of super vector spaces and grading-preserving $F$-linear maps.
\begin{defn}\hfill
\begin{enumerate}
    \item A \term{superalgebra} $A$ over $F$ is a super vector space that is also an algebra, such that the unit has even grading and the multiplication $A\mathbin{\hat\otimes} A\to A$ preserves the grading.
    \item A \emph{superhomomorphism} $A \to B$ between two superalgebras is a grading-preserving algebra homomorphism.
    \item A \term{supermodule} over a superalgebra $A$ is a super vector space $M$ that is also an $A$-module whose action map preserves the grading.\footnote{This does \emph{not} mean that a superalgebra $A$ acts on a supermodule $M$ by grading-preserving endomorphisms! Instead, it means that if $a\in A$ and $m\in M$ are odd, then $a\cdot m$ has the same grading as $a\mathbin{\hat\otimes} m$ in $A\mathbin{\hat\otimes} M$: even.}
\end{enumerate}
\end{defn}
The tensor product of superalgebras is canonically a superalgebra with the product map defined to extend linearly from the following formula on pure tensors of homogeneous elements:
\begin{equation}
    (v_1 \mathbin{\hat\otimes} w_1)\cdot (v_2\mathbin{\hat\otimes} w_2) \coloneqq (-1)^{\abs{v_2}\abs{w_1}} v_1v_2\mathbin{\hat\otimes} w_1w_2.
\end{equation}
\begin{defn}
The \emph{opposite} of a superalgebra $A\op$ is defined to be equal to $A$ as a graded vector space but with multiplication
\begin{equation}
a{\op} \cdot b{\op} = (-1)^{|a||b|} (b a){\op}.
\end{equation}
\end{defn}
A straightforward computation shows that $(A \mathbin{\hat{\otimes}} B){\op} \cong A{\op} \mathbin{\hat{\otimes}} B{\op}$.

One can thus define superbimodules and module and bimodule homomorphisms to be grading-preserving linear maps that commute with the action maps.

\begin{rem}
\label{rem:Morita2cat}
There is a symmetric monoidal bicategory $\sAlg_F$ whose objects are $F$-superalgebras, whose $1$-morphisms are bimodules, and whose $2$-morphisms are bimodule homomorphisms. 
The symmetric monoidal structure is the tensor product with Koszul sign rule as in~\eqref{koszul}. 
This is called the \term{Morita bicategory of superalgebras}; we refer to \cite[\S 3.2]{davidovichthesis} for the ungraded case and \cite{Claudiathesis} for a general construction of a higher Morita category of algebras in a symmetric monoidal category.
The name of this bicategory comes from the fact that its equivalences between objects are precisely the Morita equivalences of superalgebras.

If $\sAlg^1_F$ is the $1$-category of superalgebras, there is a symmetric monoidal functor $\sAlg^1_F \to \sAlg_F$ given by sending a homomorphism $\phi\colon A \to B$ to the associated $(B,A)$-bimodule with action twisted by $\phi$.
It can be convenient to know whether a functor $\sAlg^1_F \to \mathcal{C}$ factors through $\sAlg_F$ as it will automatically imply Morita invariance on objects.
\end{rem}

\begin{defn}\label{cliff_defn}
Let $F$ be a field whose characteristic is not $2$, and $V$ be a finite-dimensional vector space over $F$. Let $q\colon V\to F$ be a quadratic form.
The \term{Clifford algebra} $\Cl(F, V, q)$ is the $F$-algebra
\begin{equation}\label{cldef}
    \Cl(F,V, q)\coloneqq T(V)/((v^2 - q(v))\cdot 1),%
\end{equation}
where $T(V)$ denotes the tensor algebra. We make the following notational shortcuts.
\begin{itemize}
    \item Let $q_{\ell,k}$ denote the quadratic form on $\R^{\ell+k}$ defined by
    \begin{equation}
        q_{\ell,k}(x_1,\dotsc,x_{\ell+k}) = x_1^2 + \dotsb + x_\ell^2 - x_{\ell+1}^2 - \dotsb - x_{\ell+k}^2.
    \end{equation}
    We will let $\Cl_{\ell,k}\coloneqq\Cl(\R, \R^{\ell+k}, q_{\ell,k})$.
    \item If $k = 0$ in the previous example, we will just write $\Cl_\ell$; if instead $\ell = 0$ we will write $\Cl_{-k}$.
    \item Let $q_m$ denote the quadratic form on $\C^m$ defined by
    \begin{equation}
        q_m(z_1,\dotsc,z_m) \coloneqq z_1^2 + \dotsb + z_m^2.
    \end{equation}
    We write $\Cxl_m\coloneqq \Cl(\C, \C^m, q_m)$.
\end{itemize}
\end{defn}
We give $\Cl(F, V, q)$ the structure of a super vector space by specifying that a product of $n$ elements of $V$ has grading $n\bmod 2$; since the quotient in~\eqref{cldef} is by an ideal spanned by products of even numbers of elements of $V$, this grading, a priori defined on $T(V)$, descends to $\Cl(F, V, q)$ as claimed. Moreover, multiplication preserves the grading and therefore $\Cl(F, V, q)$ is a superalgebra.
There is an isomorphism $\Cl_{\ell,k}\otimes\C \cong \Cl_{k,\ell}\otimes\C\cong \Cxl_{\ell+k}$ of superalgebras, since multiplying generators by $i$ changes the sign of their square.
\begin{lem}[{Atiyah--Bott--Shapiro~\cite[Proposition 1.6]{atiyah_clifford_1963}}]
\label{clifftens}
Let $V$ and $q$ be as in \cref{cliff_defn}, and suppose $V = V_1\oplus V_2$. Then the inclusions $V_1\to V$ and $V_2\to V$ extend to a natural isomorphism of superalgebras
\begin{equation}
    \Cl(F, V_1, q|_{V_1}) \mathbin{\hat\otimes}_F \Cl(F, V_2, q|_{V_2}) \overset\cong\longrightarrow \Cl(F, V, q).
\end{equation}
\end{lem}
In particular, for $m_1,m_2,n_1,n_2\ge 0$, $\Cl_{m_1,n_1}\mathbin{\hat\otimes} \Cl_{m_2,n_2}\cong \Cl_{m_1+m_2,n_1+n_2}$.
We also have $\Cl_{m,n}{\op} \cong \Cl_{n,m}$.

\subsection{Twisted group superalgebras}
\label{subsec:twistedgroupsuperalgebras}

Every fermionic group has a canonical group superalgebra associated to it:

\begin{defn}[{Benson~\cite[\S 7]{Ben88}}]
\label{fg_alg}
    Let $G_f$ be a finite fermionic group. 
    The \emph{fermionic group algebra of $G_f$} is the real superalgebra
    \begin{equation}
    \R^f[G_f] \coloneqq \frac{\R[G_f]}{((-1)^F + 1)}
    \end{equation}
    graded by $\theta$. 
\end{defn}
\begin{rem}\label{twist_gp_alg}
Let $(\theta,\omega)$ be the twisting data for a fermionic group $G_f$. Let $\widetilde\omega$ be a cocycle\footnote{The cocycle condition is necessary in order for the multiplication in~\eqref{twprodform} to be associative. An alternative approach is to allow all cochains and construct nonassociative algebras; for example, Albuquerque--Majid~\cite{AM99, AM00} construct the octonion algebra as a twisted group $\R$-algebra for $\Z/2\times\Z/2\times\Z/2$ with a cochain that is not a cocycle.} whose cohomology class is $\omega$, and without loss of generality assume $\widetilde\omega$ is
\term{unital}, meaning $\widetilde\omega(e, g) = \widetilde\omega(g, e) = 0$, where $e\in G_b$ is the identity. Then $\R^f[G_f]$ as defined in \cref{fg_alg} is isomorphic as superalgebras to vector space $\R[G_b]$ with the multiplication modified to satisfy the formula
\begin{equation}\label{twprodform}
    g_1 \cdot_{\R^f[G_f, \widetilde\omega]} g_2 \coloneqq (-1)^{\widetilde\omega(g_1,g_2)} g_1g_2.
\end{equation}
As $G_b$ is a natural basis for $\R^f[G_b]$, this formula extends uniquely to define a product on the entire vector spae. One can check this product is associative and that $e\in G_b$ is the unit for it; this latter fact uses that $\widetilde\omega$ is unital. The $\Z/2$-grading is defined in the same way as in \cref{fg_alg}.
\end{rem}
\begin{rem}
\label{rem:groupC*alg}
    The major reason we restrict to finite groups is that the group algebras of infinite groups are not suitable for our application.
    However, we expect many of our considerations to generalize at least to compact Lie groups if we replace the fermionic group algebra by a twisted group $C^*$-algebra.
\end{rem}
\begin{exm}
\label{pin1_to_cliff}
We calculate $\R^f[\Pin_1^\pm]\cong\Cl_{\pm 1}$ using \cref{twist_gp_alg}. This computation is also done a different way by Stolz~\cite[Proposition 8.3]{Sto98} and Albuquerque--Majid~\cite{albuquerque2002clifford}.

Since $(\Pin_1^\pm)_b\cong\O_1\cong\Z/2$, we need to compute $\Z/2$-valued cocycles on $\Z/2$. For any $n\ge 0$, $H^n(B\Z/2;\Z/2)\cong\Z/2$. In general, if $M$ carries the trivial $G$-action, the group of cocycles $Z^1(G;M)$ is the group of homomorphisms $G\to M$; thus we may represent the nontrivial element $x\in H^1(B\Z/2;\Z/2)$ by $\id\colon \Z/2\to\Z/2$. Thus $x^2\in H^2(B\Z/2;\Z/2)$ is represented by the $2$-cocycle $\id\smile\id\colon \Z/2\times\Z/2\to\Z/2$ with the formula
\begin{equation}\label{an_explicit_cocycle}
    (\id\smile\id)(a, b) \coloneqq a\cdot b.
\end{equation}
The bundle $\sigma\to B\Z/2$ is a nontrivial line bundle, so $w_1(\sigma)\ne 0$, which forces $w_1(\sigma) = x$, and $w_2(\sigma) = 0$. Thus $\Pin_1^+$ has $\theta = x$ and $\omega=  0$. That is, we do not modify the multiplication in $\R[\O_1]$:
\begin{equation}\label{pin1_p_cliff}
    \R^f[\Pin_1^+]\cong \R[\O_1]\cong\R[a]/(a^2 - 1),
\end{equation}
with $\Z/2$-grading odd on $a$ and even on scalars. This is $\Cl_1$ by definition.

For $\Pin_1^-$, we modify the multiplication by the formula in~\eqref{an_explicit_cocycle}. This cocycle is only nonzero when evaluated on $(1,1)$, so only has the effect of multiplying $a^2$ by $(-1)^{(\id\smile\id)(1,1)} = -1$. Since $\theta$ is the same as it was for $\Pin_1^+$, we again have $a$ odd and scalars even; thus
\begin{equation}
    \R^f[\Pin_1^-] \cong \R[a]/(a^2 + 1)\cong\Cl_{-1}.
    \qedhere
\end{equation}
\end{exm}
\begin{exm}[{Stolz~\cite[Proposition 8.3]{Sto98}}]
\label{ex:complexfermgp}
Consider the fermionic group $C \coloneqq \{\pm 1, \pm i\} \cong \Z/4 $ with $\theta$ trivial and $(-1)^F = -1 \in C$. This fermionic group has $G_b\cong\Z/2$ and $\omega$ nontrivial, so following a similar line of reasoning as in \cref{pin1_to_cliff} one learns that as superalgebras, $\R^f[C]\cong\C$ in purely even grading. $\C$ and $\Cl_{-1}$ are isomorphic as algebras, but not as superalgebras, as the gradings do not match.
\end{exm}
Maschke's theorem generalizes readily to the setting of twisted group superalgebras.

\begin{lem}[{Ganter--Kapranov~\cite[Example 3.5.1]{GK14}}]
\label{super_maschke}
For all finite fermionic groups $G_f$, $\R^f[G_f]$ and $\R^f[G_f]\otimes\C$ are semisimple.
\end{lem}

\begin{prop}\label{gpalg_tensor}
The functor $G_f\mapsto \R^f[G_f]$ from $\cat{FermGrp}^{\mathrm{finite}}$ to the category $\sAlg_k^1$ of superalgebras and super algebra homomorphisms is symmetric monoidal.
\end{prop}
\begin{proof}
    A fermionic group homomorphism $\phi\colon G_f \to H_f$ functorially induces a grading-preserving algebra homomorphism $\R^f[G_f] \to \R^f[H_f]$ since $\phi((-1)^F_G) = (-1)^F_H$.
    The nontrivial assertion is therefore the symmetric monoidality.
    There is a unique linear map
    \begin{equation}
    \R^f[G_f] \mathbin{\hat\otimes} \R^f[H_f] 
\longrightarrow \R^f[G_f \ftens H_f]
    \end{equation}
    extending the assignment $g \otimes h \mapsto g \ftens h$.
    This map is well-defined since $g (-1)^F \ftens h$ and $g \ftens (-1)^F h$ %
    both get sent to $-g\otimes h$.
    By dimension counting the linear map is an isomorphism.
    It is grading-preserving by definition of the grading $\theta_G + \theta_H$ on the fermionic tensor product and the graded tensor product of vector spaces.
    It follows directly from the definition of the fermionic tensor product that we have defined an algebra map.
    The fact that the functor is braided is immediate for the same reason.
\end{proof}
\begin{rem}
    It follows by \cref{rem:Morita2cat,gpalg_tensor} that there is also a symmetric monoidal functor $\cat{FermGrp} \to \sAlg_k$ into the Morita bicategory of superalgebras.
\end{rem}
The real Clifford algebras $\Cl_{\ell,k}$ are twisted group algebras of fermionic groups with underlying bosonic groups $(\Z/2)^{\ell+k}$.
\begin{cor}[Albuquerque--Majid~\cite{albuquerque2002clifford}]
\label{elk_alg}
With $E_{\ell,k}$ as in \cref{elk_definition}, there is a canonical isomorphism of superalgebras
\begin{equation}
    \R^f[E_{\ell,k}] \xrightarrow{\cong} \Cl_{\ell, k}.
\end{equation}
\end{cor}
\begin{proof}
This follows by combining \cref{clifftens}, \cref{pin1_to_cliff}, and \cref{gpalg_tensor}.
\end{proof}
See also  Thiang~\cite[Remark 6.4]{Thiang_2015}, Abłamowicz~\cite{ablamowicz2016cliffordalgebrasrelatedfinite}, and Ichikawa--Tachikawa~\cite[Lemma 2.7]{IT23}.

Explicitly, $E_{\ell,k}$ is the extraspecial $2$-group
    \begin{equation}
    \begin{aligned}
    \langle e_1, \dots, e_{\ell+k}, (-1)^F &: e_1^2 = \dotsb = e_\ell^2 = 1, e_{\ell+1}^2 = \dotsb = e_{\ell+k}^2 = (-1)^F,
    \\
    &e_i e_j = (-1)^F e_j e_i,~ i \neq j, (-1)^F e_i = e_i (-1)^F, ((-1)^F)^2 = 1  \rangle,
    \end{aligned}
    \end{equation}
    isomorphic to the subset of $\Cl_{\ell,k}$ generated by the set $\{-1, e_1,\dots, e_{\ell+k}\}$ as a group under multiplication. The grading $\theta$ agrees with the grading on $\Cl_{\ell,k}$.

\begin{rem}
\label{rem:Pic KO}
Recall from \cref{rem:BO/BSpin} that the interesting $E_\infty$-structure on $B\Z/2 \times B^2\Z/2$ can be interpreted algebraically using the Picard $1$-groupoid of real superlines.
The maximal Picard $2$-groupoid $\Pic(\sAlg_\R)$ of the symmetric monoidal Morita bicategory of real superalgebras $\sAlg_\R$ also inherits an $E_\infty$-structure. As spaces, but again not as $E_\infty$-spaces, $\pi_{\leq 2} \Pic(\KO) \simeq \Z/8 \times B \Z/2 \times B^2 \Z/2$ \cite[Theorem 5.27]{beardsley2023brauer}.
Here $\Z/8 \cong \pi_0 \Pic(\sAlg_\R)$ is the super Brauer group (also called Brauer--Wall group) of the real numbers~\cite{wall1964graded}.
The connected component of the basepoint of this space is equivalent as $E_\infty$-spaces to $B\O/B\Spin$, corresponding to the fact that the automorphisms of the trivial algebra $\R$ are given by $\sLine_\R$.
See \cite[Section 5.2]{gepner2021brauer} for a general definition of the Brauer spectrum over a ring.
\end{rem}

\begin{rem}
\label{rem:donovankaroubi}
    Using the diagonal map $X \to X \times X$ and the $E_\infty$-space $\pi_{\leq 2} \Pic(\KO)$, we obtain an $E_\infty$-structure on the set of maps $X \to \Z/8 \times B\Z/2 \times B^2\Z/2$. 
Beardsley--Luecke--Morava~\cite{beardsley2023brauer} show that induced group structure on 
\begin{equation}
\Z/8 \times H^1(X;\Z/2) \times H^2(X;\Z/2),
\end{equation}
which is not in general the direct product structure,
is naturally isomorphic to the (real) graded Brauer group of the space $X$ introduced in Donovan--Karoubi~\cite{DK70}.
\end{rem}

\subsection{The \texorpdfstring{$K$}{K}-theory of fermionic group algebras}
\label{sec:Ktheoryoffermgps}

The classification of free fermion phases protected by symmetry modeled by a fermionic group $G_f$ is in terms of the $K$-theory of its fermionic group algebra.
In the case that $G_f$ is \textit{finite}, this $K$-theory is given by the \textit{Clifford module quotient group} $\CMQ_0(\R^f[G_f])$.

Let $A$ be a real superalgebra and let $\Mod_A$ be the category of finitely generated projective graded modules over $A$.
The set $\pi_0 (\Mod_A)$ of isomorphism classes is a monoid under $\oplus$. 
Following the ideas of \cite{atiyah_clifford_1963}, we define %
\begin{defn}
\label{CMQ}
The \emph{Clifford module quotient} group of $A$ is the quotient of monoids
\begin{equation}
\label{eq:CMQ}
\CMQ_0(A) \coloneqq \frac{\pi_0(\Mod_{A})}{\pi_0(\Mod_{A \mathbin{\hat{\otimes}} \Cl_{-1}})}.
\end{equation}
    More generally, $\CMQ_n(A)$ is defined as $\CMQ_0(Cl_{-n} \mathbin{\hat{\otimes}} A)$. 
\end{defn}

Even though $\pi_0(\Mod_{A})$ is not a group, it can be shown that $\CMQ_n(A)$ is a group.

\begin{rem}
For infinite-dimensional trivially graded algebras $A$, \cref{CMQ} 
agrees with $K$-theory
in degree zero. 
Indeed, in that case a graded $A$-module is nothing but a pair $(M_1,M_2)$ of $A$-modules.
It extends to a graded $A \mathbin{\hat{\otimes}} \Cl_{-1}$-module if and only if $M_1 \cong M_2$.
Therefore the map
\begin{equation}
(M_1, M_2) \mapsto M_1 - M_2
\end{equation}
defines an isomorphism from $\CMQ_0(A)$ to the Grothendieck group of the monoid $\pi_0(\Mod_A^{ug})$ of ungraded modules.
\end{rem}

\begin{rem}\label{rem_CMQ_insufficient_infinite_dim}
If $A$ is a finite-dimensional $C^*$-superalgebra, then $\CMQ_n(A)$ models the $K$-theory of $A$ for all $n \in \mathbb{N}$.
    However, for infinite-dimensional algebras this is not the case. 
    For example, if $A = C(S^1, \C)$ with trivial grading, a short computation shows that $\CMQ_1(A) = 0$; see \cite[Example 3.61]{stehouwer_k-theory_nodate}.
    On the other hand, we have
    \begin{equation}
    K_1(C(S^1)) \cong \KU^1(S^1) \cong \KU^1(\pt) \oplus \widetilde{\KU}{}^1(S^1) \cong \Z,
    \end{equation}
    where $K_n(A)$ denotes $K$-theory of the (ungraded) $C^*$-algebra $A$ in degree $n$.
\end{rem}

\begin{rem}
    If $A$ is finite-dimensional and semisimple, $\Mod_A$ is the category of finite-dimensional graded modules over $A$.
    Indeed, finitely generated modules are exactly the finite-dimensional modules. These are automatically projective because $A$ is semisimple.
\end{rem}

\begin{rem}
\label{rem:opconvention}
    Another convention sometimes seen in the literature is to quotient by graded $A$-modules that extend to $\Cl_{+1} \mathbin{\hat{\otimes}} A$-modules.
    These conventions are related by taking the opposite superalgebra.
\end{rem}

We will mainly be interested in the case where the algebra is a real Clifford algebra $\Cl_{\ell,k}$ or a complex Clifford algebra $\Cxl_k$.
It follows from \cite[Theorem 11.5]{atiyah_clifford_1963} that 
\begin{equation}
\label{eq:shiftthing}
K_p (\Cl_{\ell,k}) \cong \KO_{p - \ell + k}(\pt) \quad\text{and}\quad K_p (\Cxl_{k}) \cong \KU_{p-k}(\pt).
\end{equation}

Our focus in this paper is on the fermionic group algebras of the fermionic group $E_{\ell,k}$, whose $K$-theory reduces to the non-equivariant setting via \cref{eq:shiftthing}.
However, 
we would like to make a connection between the $K$-theory of finite fermionic group algebras and other definitions of equivariant $K$-theory to offer another perspective and illustrate the more general usage.

\begin{rem}\label{prop:equivKthvsfermgpKth}
    Let $(G_f,\theta,\omega)$ be a finite fermionic group. (In particular, this means that $G_b$ is a finite group.)
    We will restrict to the case where the twist arises from an orthogonal $G_b$-representation $V$ of rank $r_V$, meaning that $w_1(V) = \theta \in H_{G_b}^1(\mathrm{pt};\Z/2)\cong H^1(BG_b;\Z/2)$ and $w_2(V) = \theta \in H_{G_b}^2(\mathrm{pt};\Z/2)\cong H^2(BG_b;\Z/2)$. Let $S^V$ denote the one-point compactification of $V$ as a $G_b$-space, and let $\KO_{G_b}$ denote the $G_b$-equivariant $K$-theory functor. Then there is a group isomorphism
    \begin{equation}
    \label{KCl_is_Keq}
        K_0(\R^f[G_f]\mathbin{\hat\otimes} \Cl_{0,p}) \cong \KO_{G_b}^{-p}(S^{V-r_V}) = \KO_{G_b}^{-p-V+r_V}(\mathrm{pt}).
    \end{equation}
    There is a similar statement in the complex case.
    
    See e.g.\ \cite{karoubi_algebres_1968} for further discussion and Gomi~\cite[Theorem 3.19]{gomi_freed-moore_2021} for a proof. In this paper, we only consider the case where $G_b$ is an elementary abelian $2$-group. In this case, all twists can be realized by orthogonal $G_b$-representations and the corresponding twisted equivariant $\KU$- and $\KO$-cohomology groups of a point have been computed by Hu--Kriz~\cite{HK06} and Balderrama~\cite{balderrama_equivariant_2022} (see also Karoubi~\cite{karoubi_equivariant_2002}), allowing for a proof of~\eqref{KCl_is_Keq} for these $G_b$ by direct comparison.
\end{rem}

\begin{exm}\label{Clifford_twists_ex}
As a special case of the previous remark, take the finite fermionic group to be $G_f = E_{1,0}$.
Then the twist is realized by the orthogonal $G_b = \Z/2$-representation $V = \sigma$ (see \cref{pin1_to_cliff}).
Using the isomorphism of superalgebras $\R^f[G_f] \cong \Cl_1$ from~\eqref{pin1_p_cliff} and the periodicity of $\KO$-theory with respect to tensoring with Clifford algebras, we get
\begin{equation}
\KO^{-p+1}(\pt) \cong \KO_{p-1}(\R) \cong K_{0}(\R^f[E_{1,0}] \mathbin{\hat{\otimes}} \Cl_{0,p}) \cong \KO^{-p}_{\Z/2}(S^{\sigma-1}) \cong \KO_{\Z/2}^{-p - \sigma+1}(\pt).
\qedhere
\end{equation}
\end{exm}

The next result follows from \cref{elk_alg} and~\eqref{eq:shiftthing}:
\begin{prop}\label{K-theory_of_Elk}
    If $G_f = E_{\ell,k}$, then
\begin{equation}\label{cliff_susp_iso}
K_p(\R^f[E_{\ell,k}]) \cong K_p (\Cl_{\ell,k}) \cong \KO_{p - \ell + k}(*).
\end{equation}
\end{prop}
This can be upgraded to an equivalence of spectra.

Following \cref{K-theory_of_Elk}, $(1,1)$-periodicity \cite{atiyah_clifford_1963} takes the form
\begin{equation}
K_p (\Cl_{\ell,k}) \cong K_p (\Cl_{\ell-1,k-1}).
\end{equation}
The following example is a special case:
\begin{exm}\label{Cl_1_twist}
    Consider the fermionic group $E_{1,1}$ generated by $e$, $f$, and $(-1)^F$ with 
    \begin{equation}
    e^2 = 1, ~ f^2 = (-1)^F, ~ ((-1)^F)^2 = 1,~  ef = (-1)^F fe, ~ e (-1)^F = (-1)^F e, ~ f (-1)^F = (-1)^F f.
    \end{equation}
    This group is abstractly isomorphic to the symmetry group of a square with rotation $f$ and reflection $e$, but as a fermionic group $e$ and $f$ are time-reversing, so the fermionic group algebra is $\R^f[E_{1,1}]\cong \Cl_{1,1}\cong M_{1|1}(\R)$, the algebra of $2\times 2$ real matrices with off-diagonal matrices in odd grading. This algebra is Morita equivalent to $\mathbb R$ and thus $K_p(\R^f[E_{1,1}]) \cong K_p(\mathbb R) \cong \KO_p(*)$, consistent with \cref{K-theory_of_Elk}.
\end{exm}

To accommodate the complex symmetry classes, we will also consider the $K$-theory of the complexified fermionic group algebra, $K(\R^f[G_f]\otimes_\R\C)$.
In particular, for $G_f = E_{\ell,k}$ we obtain the $K$-theory of complex Clifford algebras
\begin{equation}
K(\R^f[E_{\ell,k}] \otimes_\R \C) = K(\Cl_{\ell,k} \otimes_\R \C) \cong \KU_{k-\ell}(\pt),
\end{equation}
by \cref{eq:shiftthing}.
Over $\C$, $K$-theory becomes $2$-periodic, related to the fact that %
\begin{equation}
\Cl_{1,0} \otimes_\R \C \cong \Cl_{0,1} \otimes_\R \C \quad \text{ and } \quad \Cl_{2,0} \otimes_\R \C \cong \End(\C^{1|1}).
\end{equation}
There is an analog of tensoring with the complex numbers at the fermionic group level, by extending the group $\mathbb Z/2$ of norm one scalars in $\mathbb R$ to the group $\U_1$ of norm one scalars in $\mathbb C$.
Namely, given a fermionic group $G_f$, we can consider $G_f \ftens \U_1$, where for $\U_1 = \Spin_2$ we take $\theta$ to be trivial and $(-1)^F = -1$ nontrivial.
The group $G_f \ftens \U_1$ then fits into a short exact sequence
\begin{equation}
\label{chargeextension}
    \shortexact*{\U_1}{G_f\ftens\U_1}{G_b}.
\end{equation}
Central extensions of $G_b$ by $\U_1$ are classified by $H^3(BG_b;\Z)$; under this classification, \eqref{chargeextension} corresponds to the Bockstein of
$\omega \in H^2(BG_b; \Z/2)$.

One can make analogs of \cref{defn_ferm_group,fg_alg} in the complex case. The details are similar, so we will be brief.
\begin{defn}[{\cite[\S 5.2]{Thiang_2015}}]
\label{complex_fg_alg}
Let $\theta \colon K \to \Z/2$ be a $\Z/2$-graded Lie group with a distinguished central subgroup isomorphic to $\U_1$.
    The \emph{complex fermionic group algebra} $\C^f[K]$ is the quotient of the group algebra of $K$ where we identify the $\U_1 \subseteq K$ with the $\U_1 \subseteq \C$ in the scalars of the algebra.
    We consider it as a $\Z/2$-graded algebra under $\theta$.
 \end{defn}
 Thiang's definition in \textit{loc.\ cit.} is more general than what appears here.

For the special case \cref{chargeextension}, \cref{complex_fg_alg} specializes to $\R^f[G_f] \otimes_\R \C$.

\begin{rem}\label{remark_charge}
    Physically, we can think of $G_f \ftens \U_1$ as ``$G_f$ plus an additional $\U_1$-charge $Q$.''
    Even though we will not discuss fermionic group $C^*$-algebras of infinite fermionic groups here, one can think of modules over $\R^f[G_f] \otimes_\R \C$ as ``representations of $G_f \ftens \U_1$ of unit charge.'' See \cite[Section 5.3]{neutralluuk} for further discussion.
\end{rem} 

\begin{rem}
\label{rem:freed-mooregroup}
    Even more generally, it can be physically relevant to look at non-central extensions, where $\theta$ does not need to be induced by a homomorphism $\phi\colon K \to \Z/2$ satisfying $z k = k z^{\phi(h)}$ for all $k \in H_f$ and $z \in \U_1 \subseteq K$.
    In more physical language, these groups can contain symmetries that anti-commute with the charge operator.
    Such symmetries are called \term{charge conjugation operators} in high energy physics and are related to particle-hole symmetries in condensed matter~\cite{zirnbauer2021particle}.
    Freed--Moore~\cite{freed_twisted_2013} apply this group-theoretic structure to SPT phases, where they call it an \term{extended QM symmetry class}.
\end{rem}

\subsection{Fermionic group twists of spin bordism}
\label{ss:xi_to_Gf}

The data of a fermionic group $G_f$ also defines a variant of the notion of spin structure. This variant is the tangential structure present on spacetime in a fermionic field theory with $G_f$-symmetry. We first go over the generalities, then specialize to several examples that will be helpful in our analysis of the Bott spiral.

\begin{defn}
Let $X$ be a space, $\theta \in H^1(X; \Z/2)$, and $\omega\in H^2(X;\Z/2)$. An \term{$(X, \theta, \omega)$-twisted spin structure} on a vector bundle $V\to M$ is data of a map $f\colon M\to X$ and trivializations of the classes $w_1(V) - f^*(\theta)$ and $w_2(V) + w_1(V)^2 - f^*(\omega)$. Two $(X, \theta, \omega)$-twisted spin structures are equivalent if they lie in the same path component of such data. %
\label{def:twistedspinstructure}
\end{defn}
We will consider bordism classes of manifolds equipped with $(X,\theta, \omega)$-twisted spin structures. Bordism groups of such manifolds are modules over the spin bordism ring.
This notion of twisted spin structure was studied by B.L.\ Wang~\cite[Definition 8.2]{Wan08} in the special case $\theta = 0$, and is closely related to Kreck's notion of a normal $1$-smoothing in~\cite[\S 2]{kreck_surgery_1999}.
\begin{exm}[Vector bundle twists]\label{vb_twisted_spin}
Suppose there is a vector bundle $E\to X$ with $w_1(E) = \theta$ and $w_2(E) = \omega$. Then the Whitney sum formula implies an $(X, \theta, \omega)$-twisted spin structure on $V\to M$ is equivalent to the data of the map $f\colon M\to X$ and a spin structure on $V\oplus f^*(E)$. In this case, we will call $(X, \theta, \omega)$-twisted spin bordism a \term{vector bundle twist} of spin bordism.
We also will also refer to such twists as $(X,V)$-twists.

We learned this notion from Hason--Komargodski--Thorngren~\cite[\S 4.1]{hason_anomaly_2020};
see also 
MacAlpine~\cite{MA66} and Stolz~\cite[\S 2.9]{Sto98} for related but different notions.
\end{exm}
\begin{rem}\label{nonVB}
Given $(X, \theta, \omega)$, it is not always possible to find a vector bundle $E\to X$ with $w_1(E) = \theta$ and $w_2(E) = \omega$: see the discussion in~\cite{GKT89, RWG14, TJF19, Kuhn20, Speyer22}.
This is typically not a problem in practice: all twisted spin structures appearing in~\cite{Cam17, Ped17, BC18, guo_fermionic_2020, debray_invertible_2021, freed_reflection_2021} can be realized as vector bundle twists, and the same is true for all twists appearing in this paper. See~\cite{DY23, DY24} for an example of a twisted spin structure that cannot be realized as a vector bundle twist, applied to an anomaly cancellation question in supergravity. 
\end{rem}
\begin{exm}
Let $G_f$ be a fermionic group, and, as in \cref{fermionic_twist}, let $G_b\coloneqq G_f/\{\pm 1\}$ and $(\theta, \omega)$ be the associated twist. Then the data $(BG_b, \theta, \omega)$ defines a notion of twisted spin structure which we call a \term{$G_f$-twisted spin structure}.
\end{exm}
Thus a $G_f$-twisted spin structure is a vector bundle twist if $G_f$ is in the essential image of $\zeta\colon\cat{Rep}\to\cat{FermGrp}$ (\cref{zeta_defn}).
\begin{defn}[{Stolz~\cite[\S 2.6]{Sto98}}]
\label{FG_to_xi}
Given a fermionic group $G_f$ and an integer $d\ge 0$, define the Lie group
\begin{equation}
    H_d(G_f)\coloneqq (\Pin^+_d \ftens  G_f)_{\mathit{ev}},
\end{equation}
where ``$(\bl)_{\mathit{ev}}$'' means to take $\theta^{-1}(0)$, the even subgroup. The double cover $\Pin_d^+\to\O_d$ induces a map $H_d(G_f)\to \O_d$ given by killing the $G_f$ factor of the fermionic tensor product.

Taking the colimit, we also allow $d = \infty$, defining a topological group $H(G_f)$ with a map to $\O$; apply the classifying space functor and we obtain a tangential structure in the sense of Lashof~\cite{Las63}, which we call an \term{$H(G_f)$-structure}. 
\end{defn}
\begin{prop}[Stehouwer~\cite{stehouwer_interacting_2022}]
\label{FG_twist_Hd}
Let $G_f$ be a fermionic group with associated twist $(\theta, \omega)$. The following are equivalent data on a vector bundle $V\to M$.
\begin{enumerate}
    \item A $(BG_b, \theta, \omega)$-twisted spin structure.
    \item An $H(G_f)$-structure.
\end{enumerate}
\end{prop}
Stehouwer does not state the result in exactly this way, but it follows from~\cite[Proposition 14]{stehouwer_interacting_2022}, the definition of the pullback, and the Whitney sum formula.

The maps $H_d(G_f)\to\O_d$ commute with the inclusions $H_d(G_f)\to H_{d+1}(G_f)$ (induced by $\Pin_d^+\hookrightarrow\Pin_{d+1}^+$) and $\O_d\hookrightarrow\O_{d+1}$, so there is a notion of bordism of manifolds with $G_f$-twisted spin structures on their tangent bundles. Lashof~\cite[Theorem C]{Las63} shows that the Pontrjagin--Thom map identifies the corresponding bordism groups with the homotopy groups of the Thom spectrum $\mathit{MTH}(G_f)$.

One can check using the Whitney sum formula that if $M$ is a spin manifold and $N$ has a $G_f$-twisted spin structure, then $M\times N$ has a canonical $G_f$-twisted spin structure. This refines to define the structure of an $\MTSpin$-module spectrum on $\mathit{MTH}(G_f)$, and we obtain a spectrum-level lift of \cref{FG_twist_Hd}.
\begin{prop}[{Shearing~\cite[Lemma 10.18]{DDHM23}}]
\label{shearing}
Suppose there is a rank-$r$ virtual vector bundle $V\to BG_b$ with $w_1(V) = \theta$ and $w_2(V) = \omega$. Then there is an equivalence of $\MTSpin$-modules, natural in the data of $G_f$ and $V$,
\begin{equation}
    \mathit{MTH}(G_f)\overset\simeq\longrightarrow \MTSpin\wedge (BG_b)^{V - r}.
\end{equation}
\end{prop}
Typically the vector bundle $V$ is far from unique. Furthermore, the homotopy type of $\MTSpin\wedge (BG_b)^{V-r}$ only depends on $w_1(V)$ and $w_2(V)$. See~\cite[Theorem 1.39]{debray_invertible_2021}.
\begin{exm}\label{pin_tangential_example}
Consider the two fermionic groups $G_f= \Pin_1^\pm$. As we showed in \cref{pin1_to_cliff}, for both of these groups, $G_b\cong\Z/2$ and $\theta$ is nontrivial, but $\omega = 0$ for $\Pin_1^+$ and $\omega = \theta^2$ for $\Pin_1^-$. \cite[Proposition 1.26]{ADKPSS24} constructs isomorphisms $H_d(\Pin_1^\pm)\cong \Pin_d^\mp$ for all $d$ including $d = \infty$.

Let $\sigma\to B\Z/2$ denote the tautological line bundle. Then $w_1(\sigma)\ne 0$ and $w_2(\sigma) = 0$, so by \cref{shearing}, $\Pin_1^+$-twisted spin bordism groups are the homotopy groups of $\MTSpin\wedge (B\Z/2)^{\sigma-1}$. The identification $H(\Pin_1^+)\cong \Pin^-$, together with \cref{FG_twist_Hd}, implies the following equivalence of $\MTSpin$-modules, first recorded by Peterson~\cite[\S 7]{Pet68}:
\begin{subequations}
\begin{equation}
    \MTPin^-\overset\simeq\longrightarrow \MTSpin\wedge (B\Z/2)^{\sigma-1}.
\end{equation}
The Whitney sum formula implies that if $V$ is either of the virtual bundles $3\sigma$ or $-\sigma$, then $w_1(V) \ne 0$ and $w_2(V) = w_1(V)^2$, so the Stiefel--Whitney classes of $V$ match the associated twist of $\Pin_1^-$. Thus, following a similar line of reasoning as we did for $\Pin_1^+$, we deduce that $\Pin_1^-$-twisted spin structures are the homotopy groups of
\begin{equation}
    \MTPin^+\simeq \MTSpin\wedge (B\Z/2)^{1-\sigma}\simeq \MTSpin\wedge (B\Z/2)^{3\sigma-3}.
\end{equation}
This is a theorem of Stolz~\cite[\S 8]{stolz_exotic_1988}.
\end{subequations}
\end{exm}
The most important examples in this paper are the fermionic groups defined in terms of the extraspecial $2$-groups in \cref{elk_definition}. The following proposition will play an important role. 
\begin{prop}\label{interacting_symm_mon}
Let $M\colon \cat{Top}_{/B\O}\to\cat{Spectra}$ denote the functor sending a rank-zero virtual vector bundle $V\to X$ to its Thom spectrum $X^V$. Then the composition
\begin{equation}
\label{spinthom}
    \cat{Rep}^{\mathrm{cpt}} \xrightarrow[\eqref{xi_defn}]{\xi} \cat{Top}_{/B\O} \xrightarrow{M} \cat{Spectra} \xrightarrow{\wedge\MTSpin} \cat{Mod}_{\MTSpin}
\end{equation}
is a symmetric monoidal functor of $\infty$-categories.
\end{prop}
\begin{proof}
Since~\eqref{spinthom} comes as a composition of three functors, it suffices to show each of the three is symmetric monoidal. For $\xi$, we did this in \cref{Rep_to_BO}. For $M$, this is Lewis' theorem~\cite[\S IX.7.4]{LMSM}; the $\infty$-categorical version is due to Ando--Blumberg--Gepner~\cite[Theorem 1.6]{ABG18}. For the base change $\wedge\MTSpin$, this is in~\cite[\S 4.5.3]{HA} given an $E_\infty$-ring structure on $\MTSpin$, such as the one constructed by Joachim~\cite{Joa04}.
\end{proof}
The upshot is that, for $G_f$ in the essential image of $\zeta$, we have a description of $\mathit{MTH}(G_f)$ compatible with the fermionic tensor product, strengthening \cref{FG_twist_Hd,shearing}. In light of \cref{nonVB}, implying that $\zeta$ is not essentially surjective, it would be nice to lose the hypothesis of being in its essential image. Thus we make the following conjecture.
\begin{conj}
\label{direct_SM_conj}
With notation as in \cref{interacting_symm_mon},
there is a symmetric monoidal functor $\mathit{TW}\colon\cat{FermGrp}\to\Top_{/B\O/B\Spin}$ of $\infty$-categories such that
\begin{subequations}
\begin{enumerate}
    \item $\mathit{TW}$ of a fermionic group $G_f$ with twisting data $(\theta,\omega)$ is homotopy equivalent to the map
    \begin{equation}
        BG_b\xrightarrow{(\theta,\omega)} B\Z/2\times B^2\Z/2\xrightarrow[\eqref{faux_whitney}]{\simeq} B\O/B\Spin.
    \end{equation}
    \item The following diagram of symmetric monoidal functors commutes.
\begin{equation}
\begin{tikzcd}
	{\cat{Rep}^{\mathrm{cpt}}} & {\cat{Top}_{/B\O}} & {\cat{Spectra}} \\
	{\cat{FermGrp}} & {\cat{Top}_{/B\O/B\Spin}} & {\cat{Mod}_{\MTSpin}.}
	\arrow["\xi", "{\eqref{xi_defn}}"', from=1-1, to=1-2]
	\arrow["\zeta"', "{\eqref{zeta_defn}}", from=1-1, to=2-1]
	\arrow["M", from=1-2, to=1-3]
    \arrow["q", from=1-2, to=2-2]
	\arrow["{\wedge\MTSpin}", from=1-3, to=2-3]
	\arrow["{\mathit{TW}}"', color={BrickRed}, from=2-1, to=2-2]
	\arrow["T"', from=2-2, to=2-3]
\end{tikzcd}
\end{equation}
Here $q$ is induced from the usual map $B\O\to B\O/B\Spin$ and $T\colon\cat{Top}_{/B\O/B\Spin}\to \cat{Mod}_{\MTSpin}$ is the $\MTSpin$-module Thom spectrum functor: see, e.g., \cite[Theorem 1.13]{DY23}, based on~\cite[Lemma IV.2.6]{MQRT77} and~\cite{ABGHR14a, ABGHR14b}.
\end{enumerate}
\end{subequations}
\end{conj}

\begin{defn}\label{MElk_defn}
For $k,\ell\ge 0$, recall the fermionic group $E_{\ell,k}$ from \cref{elk_definition} and let\footnote{The notation $\ME$ is meant to represent a Thom spectrum (``$M$'') for an elementary abelian group (``$E$'').} 
\begin{equation}
    \ME_{\ell,k}\coloneqq ((B\Z/2)^{\sigma-1})^{\wedge \ell} \wedge ((B\Z/2)^{1-\sigma})^{\wedge k}.
\end{equation}
By \cref{FG_twist_Hd}, an $MTH(E_{\ell,k})$-structure on a manifold $M$ is equivalent to a spin structure on the virtual bundle $TM + L_1 + \dots + L_\ell - L_{\ell+1} - \dots - L_{\ell+k}$ for a collection of line bundles $L_i$ over $M$.
We will call such a manifold a \emph{spin-$(\ell,k)$ manifold}.

In the same way, we define a \term{\spinc-$(\ell,k)$-manifold} to be an $H(E_{\ell,k}\ftens\Spin_2)$-structure. Building on the equivalence of \spinc structures on $V$ and data of a complex line bundle $E$ and a spin structure on $V\oplus E$, one can show that a \spinc-$(\ell,k)$ structure is naturally equivalent to a \spinc structure on $TM + L_1 + \dots + L_\ell - L_{\ell+1} - \dots - L_{\ell+k}$, where the $L_i$ are real line bundles as before.
\end{defn}

\begin{lem}\label{lem_shearing_Elk}
There is an $\MTSpin$-module equivalence
\begin{equation}
    \mathit{MTH}(E_{\ell,k})\overset\simeq\longrightarrow \MTSpin\wedge\ME_{\ell,k}. 
\end{equation}
\end{lem}
\begin{proof}
This follows by combining \cref{pin_tangential_example,interacting_symm_mon}.
\end{proof}
We elaborate on the case $\ell= k = 1$, which we began in \cref{Cl_1_twist} and which will be a running example throughout this paper.

\begin{defn}[{Kaidi--Parra-Martinez--Tachikawa~\cite[\S 6.1]{kaidi_topological_2020}}]
\label{dpin_defn}
There is an isomorphism $H^*(B\Z/2\times B\Z/2;\Z/2)\cong\Z/2[x, y]$ with $\abs x =\abs y = 1$ specified by asking that, under the canonical isomorphisms
\begin{equation}
    H^1(B\Z/2\times B\Z/2; \Z/2) \overset\cong\longrightarrow
    \Hom(\pi_1(B\Z/2\times B\Z/2), \Z/2)\overset\cong\longrightarrow
    \Hom(\Z/2\times \Z/2, \Z/2),
\end{equation}
$x(1, 0) =1$, $x(0, 1) = 0$, $y(1, 0) = 0$, and $y(0, 1) = 1$. A \term{dpin} 
structure on a vector bundle $V$ is a $(B\Z/2\times B\Z/2, x, xy)$-twisted spin structure.
\end{defn}

\begin{lem}\label{dpinequiv}
There is a natural homotopy equivalence between the spaces of dpin structures and of $E_{1,1}$-twisted spin structures on any vector bundle $V$.
\end{lem}
\begin{proof}
We have $(E_{1,1})_b\cong\Z/2\times\Z/2$. By \cref{twist_of_product}, the associated twist is $(x+y, x^2+xy)$, so $E_{1,1}$-twisted spin structures are $(B\Z/2\times B\Z/2, x+y, x^2+xy)$-twisted spin structures. %

There is an automorphism $\varphi$ of $\Z/2\times\Z/2$ that sends $(1, 0)\mapsto (0, 1)$ and $(0, 1)\mapsto (1, 1)$, and whose value on the remaining elements is uniquely specified by the stipulation that $\varphi$ is a group homomorphism. It follows from our construction of $x$ and $y$ that $\varphi^*(x) = x+y$ and $\varphi^*(y) = x$, so if we pull back the twisting data defining a dpin structure along $\varphi$, we get the twisting data of an $E_{1,1}$-twisted spin structure. Thus, precomposing the map to $B\Z/2\times B\Z/2$ with $B\varphi$ or $B(\varphi^{-1})$ turns dpin structures into $E_{1,1}$-twisted spin structures and vice versa.
\end{proof}
We will discuss spin-$(1,1)$ structures and dpin structures interchangeably throughout the rest of this paper, often leaving the automorphism $\varphi$ implicit.

Finally, we discuss the complex case.
To define twisted \spinc bordism associated to a fermionic group, we use the map of spectra $\MTSpin\to\MTSpin^c$, which refines the fact that spin structures induce \spinc structures.\footnote{In terms of obstructions,  if $w_2=0$, then $W_3=\beta(w_2) =0$ as well.} This map gives $\MTSpin^c$ the structure of an $E_\infty$-$\MTSpin$-algebra, guaranteeing that the constructions in this section are natural. The necessary modifications are:
\begin{enumerate}
    \item Exchange $\Pin_d^c$ for $\Pin_d^+$ in \cref{FG_to_xi}.
    \item Note that the fermionic groups $\Pin_1^\pm$ both give rise to the same twisted \spinc structure, namely \pinc structure.
    This is because the Lie groups $\Pin_d^\pm\times_{\set{\pm 1}}\U_1$ are isomorphic with the isomorphism commuting with the structure map to $\O_d$.
    Alternatively, one can think about this in terms of group algebras: $\R[\Pin_1^\pm]\cong\Cl_{\pm 1}$, and the complexifications of these two superalgebras are isomorphic: if $e^2 = 1$, then $(ie)^2 = -1$. We will see more algebraic consequences of this isomorphism in \S\ref{complex_computation}.
\end{enumerate}

\begin{rem}
Under the hood, 
there is a more general notion of twisted \spinc structure defined in terms of a map to $B\O/B\Spin^c\simeq K(\Z/2, 1)\times K(\Z, 3)$ (see~\cite{FW99, Dou06, DY23}), and the twisting data $(\theta, \omega)$ associated to a fermionic group only manifests through $\theta$ and the integral Bockstein of $\omega$. So different choices of $\omega$ with the same Bockstein lead to equivalent notions of twisted \spinc bordism, as happened for $\Pin_1^\pm$.
Up to homotopy, $BG_b \to B\O/B\Spin^c$ is equivalent to a group homomorphism $\theta\colon G_b \to \Z/2$ and an isomorphism class of an extension of $G_b$ by $\U_1$; compare with \cref{rem:freed-mooregroup}.

Similarly to \cref{rem:Pic KO}, there is a larger $E_\infty$-space $\pi_{\leq 3}\Pic(\KU) \cong \Z/2 \times B\Z/2 \times B^3 \Z$, and the connected component of the basepoint is equivalent as $E_\infty$-spaces to $B\O/B\Spin^c$.
Analogously to \cref{rem:donovankaroubi}, Beardsley--Luecke--Morava~\cite{beardsley2023brauer} show that the induced group structure on 
\begin{equation}
\Z/2 \times H^1(X;\Z/2) \times H^3(X;\Z)
\end{equation}
is naturally isomorphic to the complex graded Brauer group of $X$ as defined by Donovan--Karoubi~\cite{DK70}.

    We also saw in \cref{rem:freed-mooregroup} that there is a more general notion of twists classified by $H^1(BG_b; \Z/2) \times H^3(BG_b; \Z_\phi)$, where $H^3(BG_b; \Z_\phi)$ is twisted cohomology for a fixed homomorphism $\phi\colon G_b \to \Z/2$.
    We expect these to give Real-equivariant twists for the genuine $\Z/2$-equivariant spectrum $\Spin^c$ given by complex conjugation; see also \cite{halladay2024realABS}.
    We will not need this level of generality in this paper.
\end{rem}

\begin{prop}\label{cpx_collapse}
Let $R$ be one of $\MTSpin^c$, $\ku$, or $\KU$. Then there is a canonical $R$-module equivalence $R\wedge (B\Z/2)^{\sigma-1}\simeq R\wedge (B\Z/2)^{1-\sigma}$.
\end{prop}

We are not sure where \cref{cpx_collapse} was first proven: it appears to be well-known but not written down. It follows from the fact that $2\sigma$ is complex, hence $R$-oriented: see~\cite[Example 6.13]{debray_smith_2024}.

\begin{defn}\label{MEm_defn}
    Let $\ME_m\coloneqq\ME_{m,0}$.
\end{defn}

\begin{cor}\label{spinc_simplification}
With $R$ as in \cref{cpx_collapse}, there is a canonical $R$-module isomorphism $R\wedge\ME_{\ell, k}\xrightarrow{\simeq} R\wedge\ME_{\ell+k}$.
\end{cor}

\section{Homotopical classifications of SPT phases}\label{section_free_to_interacting_section}
    In this section, we discuss the classifications of both free and interacting SPTs, in preparation for defining maps between them in \cref{ss:F2I}.
We first introduce free fermion Hamiltonians and their $K$-theoretic classification in \cref{ff_and_k}, then discuss invertible field theories as a model for interacting SPT phases in \cref{interacting_IFT}. In \cref{subsec_symm_types_morita_invnce}, we introduce and discuss the fermionic groups we will use to model the ten Altland--Zirnbauer classes.

\subsection{Free fermion phases and \texorpdfstring{$K$}{K}-theory}
\label{ff_and_k}

Let $A$ be a real $C^*$-algebra with a $\Z/2$-grading. Physically, $A$ represents the algebra of (not-necessarily internal) symmetries, and the grading distinguishes time-preserving and time-reversing symmetries.
According to the physics definition, the set of free fermion phases protected by $A$-symmetry are the connected components of the space of free fermion systems---often represented by Hamiltonian operators---with $A$-symmetry.
In \cite{neutralluuk} one of us argued for the following ansatz.

\begin{ansatz}[{\cite[ Definition 3.7]{neutralluuk}}]\label{neutral_luuk_ansatz}
    The set of free fermionic SPT phases protected by $A$ is the set of equivalence classes of triples $[M,H_1,H_2]$, where $M$ is a finitely generated $A$-module and $H_i \colon M \to M$ are operators such that $H_i a = (-1)^{|a|} a H_i$ and $H_i^2 = -1$.
\end{ansatz}

The set of SPT phases is then given an abelian group structure by stacking (see \cite[Definition 3.5]{neutralluuk}).

\begin{rem}
    For the interested reader, we quickly run through the motivation for the above ansatz. 
    \begin{enumerate}
        \item If we model neutral\footnote{i.e.\ phases without charge conservation symmetry.} %
        free fermions in the Bogoliubov--de Gennes framework, a free Hamiltonian $H$ is a \emph{skew}-adjoint operator on the real space $M$ of Majorana operators. One should think of this operator as $i$ times a self adjoint operator acting on $M \otimes_\R \C$.
        \item An $A$-symmetry manifests itself as an action on $M$. Because of the extra factor of $i$ in the Hamiltonian, the operator $H$ commutes with time-preserving symmetries and anti-commutes with time-reversing symmetries.
        \item We shift the Fermi energy to zero so that gapped systems correspond to invertible Hamiltonians.
        \item The space of all gapped Hamiltonians is predicted to be homotopy equivalent to the space of flattened Hamiltonians, which are the Hamiltonians $H$ satisfying $H^2 = -1$.
        \item We replace individual Hamiltonians by pairs of Hamiltonians, which we think of as formal differences of two systems. Mathematically, this implements the group completion necessary to obtain $K$-theory, while physically this kills fragile phases \cite{po_fragile_2018}.
    \end{enumerate}
    We refer to \cite{neutralluuk} for details.
\end{rem}

The following theorem is a generalization and formalization of the arguments in \cite{kitaev_periodic_2009}.\footnote{\label{Kth_cite}Kitaev's proposal has been generalized to many settings, using many different formalisms for $K$-theory. See, for example, \cite{bellissard2005k, heinzner2005symmetry,
freed_twisted_2013,
Thiang_2015,
kellendonk2017c,
alldridge2020bulk,
gomi_freed-moore_2021}.}

\begin{thm}[{\cite[Theorem 3.35]{neutralluuk}}]\label{neutralluuk_main_thm_citation}
Assuming \cref{neutral_luuk_ansatz},
    The group of free fermion SPT phases protected by $A$ is isomorphic to $K_2(A)$.
\end{thm}

The general idea of the proof is to observe that the group of SPT phases described in \cref{neutral_luuk_ansatz} is a version of the Karoubi model for the $K$-theory of $\Z/2$-graded $C^*$-algebras~\cite{karoubi_algebres_1968}.

Now, let $G_f$ be a finite fermionic group and $d$ be an integer representing the spatial dimension.
It is argued in \cite{neutralluuk} that the corresponding symmetry algebra is given by $A = C_f^*(\R^d \times G_f)$, where $C_f^*$ denotes the fermionic group ($C^*$-)algebra.\footnote{Because the
group $\R^d \times G_f$ is infinite,
using $\R^f[\R^d\times G_f]$ does not suffice: among other things, one has to complete, and that is why $C_f^*(\R^d\times G_f)$ appears here.}
This allows one to prove a relationship with the $K$-theory of the fermionic group algebra:

\begin{thm}[{\cite[Remark 4.34]{neutralluuk}}]
\label{free_classification_ansatz}
Free fermion SPT phases with finite fermionic symmetry $G_f$ are classified by $K_{2-d}(\R^f[G_f])$.     
\end{thm}

\begin{rem}
    The reason for the choice of algebra is as follows.
    \begin{enumerate}
    \item $M$ should have fermionic symmetry $G_f$, i.e.\ be a module over the fermionic group algebra.
        \item $M$ should also have a $\Z^d$ translation symmetry coming from the lattice of atoms, which we can model by an action of the group algebra $C^*(\Z^d)$ on $M$.
        \item If we ignore weak phases---phases protected by the discrete translation symmetry---we should impose $\R^d$-symmetry instead of $\Z^d$-symmetry, to allow for continuous translations. See \cite{ADKPSS24} for more on this perspective on weak phases.
        \item Strong phases in dimension $d$ should thus be symmetric for both $\R^d$ and $G_f$, meaning that we have a $C^*_f(\R^d \times G_f)$-action.
        \qedhere
    \end{enumerate}
\end{rem}

When $G_f = E_{\ell,k}$ we obtain the following $8$-periodic expression by using \cref{K-theory_of_Elk}.

\begin{cor}
\label{Elk_free_phases}
Free fermion SPT phases protected by $E_{\ell,k}$-symmetry in spatial dimension $d$ are classified by $\KO_{2-d-\ell+k}(\pt)$. 
\end{cor}

The $8$-periodicity above is related to the $8$ real symmetry classes in the tenfold way.
We provide more details on how we model symmetry types in \cref{subsec_symm_types_morita_invnce}, but give a few examples now:

\begin{exm}[Class D]
    The case of no symmetry, $G_f = E_{0,0}$, models class D systems.
    For example, for $d = 2$, it is predicted (see Kitaev~\cite{kitaev_periodic_2009}) that the $p+ip$ topological superconductor represents a generator of $\KO_{2-d}(\pt) = \KO_0(\pt) = \Z$.
\end{exm}

\begin{exm}[Class BDI]
\label{TRS_Maj_exm_K-theory}
    The symmetry algebra $G_f = E_{1,0}$ consists of fermion parity $(-1)^F$ and a single time-reversal symmetry $T$ with $T^2 = 1$.
    For $d = 1$, the time-reversal-symmetric Majorana chain~\cite{kitaev_unpaired_2001} is believed to represent a generator of $\KO_{2-1-d}(\pt) = \KO_0(\pt) \cong \Z$.
    Any number of stacked copies of the chain is not adiabatically connected to the trivial phase via symmetric, quadratic terms \cite{fidkowski_effects_2010}. 
\end{exm}

We will deal with the remaining two complex symmetry classes by using the complex version of fermionic group algebras given in \cref{complex_fg_alg}.

\begin{thm}
\label{free_classification_ansatz_complex}
Suppose that $G_f$ is a fermionic group and $\U_1 \subseteq G_f$ is a central subgroup such that $G_f/\U_1$ is finite.
Free (charged) fermion SPT phases with fermionic symmetry $G_f$ are classified by $K_{2-d}(\C^f[G_f])$. 
For $G_f = E_{\ell,k} \ftens \U_1$ this gives the $2$-periodic expression $\KU_{2-d-\ell+k}(\pt) \cong \KU_{d+\ell+k}(\pt)$. 
\end{thm}

In the above theorem, the $\U_1$ subgroup should be interpreted as electric charge.

\begin{exm}[Class A]
    For class A, we have that $G_f = E_{0,0} \ftens \U_1 = \U_1$ is  charge conservation. %
    The quantum Hall insulator is predicted to represent a generator for $d = 2$ of $K_{2-d}(\C^f[G_f]) = \KU_{0}(\pt) \cong \Z$~\cite{kitaev_periodic_2009}.
\end{exm}
The $K$-groups of a $C^*$-superalgebra $A$ are Morita invariant, so \cref{neutralluuk_main_thm_citation} has the consequence that the classification of free fermion phases with symmetry given by a fermionic group $G_f$ depend only weakly on $G_f$ through the Morita class of $\R^f[G_f]$. See~\cite[\S 2.3]{stehouwer_interacting_2022} and~\cite[Remark 3.34]{neutralluuk}. Work of Müssnich--Vasconcellos Vieira~\cite{mussnich_weakly_2026} suggests this Morita invariance phenomenon extends into the weakly interacting regime as well. 

For example, by \cref{elk_alg}, $\R^f[E_{\ell,k}]\cong\Cl_{\ell,k}$, which is Morita equivalent to $\R^f[E_{\ell',k'}]$ if and only if $\ell-k\equiv \ell'-k'\bmod 8$. This gives many examples of nonisomorphic fermionic groups whose groups of free fermion phases in every dimension are isomorphic. Later, in \cref{only_one_long_summand}, we will prove that the classifications of reflection-positive IFTs for these fermionic groups are in general nonisomorphic.

As we discussed in the introduction, this is an important concern when describing Altland--Zirnbauer classes, which behave more like Morita equivalence classes than isomorphism classes of fermionic groups or superalgebras.

\subsection{Interacting SPT phases and invertible field theories}
\label{interacting_IFT}
In the physics literature, the codomain of a free-to-interacting map is the space of all gapped invertible topological phases of matter of a given dimension and symmetry type. Providing a rigorous mathematical construction of this space is a difficult open question, but it is conjectured that 
gapped invertible phases are classified by their low energy limits, which are invertible field theories (IFTs).

\begin{ansatz}[Freed--Hopkins \cite{freed_reflection_2021}]\label{LEFT_ansatz}
Interacting SPT phases of a given symmetry type are classified by 
deformation classes of reflection-positive, fully extended invertible field theories whose tangential structure corresponds to the symmetry type of the SPTs.
\end{ansatz}
In this paper, we will take this low-energy approximation ansatz as given.
See~\cite{Fre19, freed_reflection_2021} for further discussion on this ansatz and~\cite{freed_reflection_2021, Cam17, KT17, BC18, WG18, Gaiotto_SPT, freed_invertible_2019, ABK21, debray_invertible_2021, BCHM22, ADKPSS24} for supporting computational evidence.
We view our results in this work, and our ongoing program of constructing homotopical free-to-interacting maps, as further tests of the ansatz.

By results of Freed--Hopkins~\cite{freed_reflection_2021} and Grady~\cite{Gra23}, reflection-positive IFTs are classified in terms of homotopy theory. So, once we admit the ansatz, we have access to computational methods.
In the rest of this subsection, we give some details on the classification.

Let $\xi\colon B\to B\O$ be a stable tangential structure.
Recall that an ($n$-dimensional, $\xi$-structured) topological field theory is a symmetric monoidal functor
\begin{equation}
    \alpha\colon \cat{Bord}_n^\xi\longrightarrow \cat C,
\end{equation}
where $\cat{Bord}_n^\xi$ is the bordism category of $n$-manifolds with $\xi$-structure, symmetric monoidal under disjoint union, and $\cat C$ is some symmetric monoidal category, often $\cat{sVect}_\C$. One may also take $\cat{Bord}_n^\xi$ and $\cat C$ to be symmetric monoidal higher categories. Given two TFTs $\alpha$ and $\beta$, their tensor product $\alpha\otimes\beta$ is defined by the formula \begin{equation}
    (\alpha\otimes\beta)(X) \coloneqq \alpha(X) \otimes \beta(X)
\end{equation}
for $X$ an object, morphism, or higher morphism.
\begin{defn}[{Freed--Moore~\cite[Definition 5.7]{FM06}}]
Let $\boldsymbol 1\colon\cat{Bord}_n^\xi\to\cat C$ denote the constant functor valued at the monoidal unit of $\cat C$ and the identity morphism. A TFT $\alpha$ is \term{invertible} if there is another TFT $\beta$ such that $\alpha\otimes\beta\simeq\boldsymbol 1$. 
\end{defn}
Equivalently, the image of $\alpha$ is contained in $\otimes$-invertible objects and composition-invertible (higher) morphisms. 
For example, if $\cat C = \cat{sVect}_\C$, the tensor invertible objects are the one-dimensional vector spaces.

Freed--Hopkins--Teleman~\cite{FHT10} and Schommer-Pries~\cite{schommer_pries_invertible_2024} classify invertible TFTs in terms of the Madsen--Tillmann spectrum $\Sigma^n \mathit{MT\xi}_n$,\footnote{Rovi--Schoenbauer~\cite{rovi_relating_2022} and Kreck--Stolz--Teichner (unpublished; see~\cite{StolzTalk}) have also proven related classification theorems for invertible TFTs that are not stated in terms of Madsen--Tillmann spectra.} where $\xi_n \colon B_n \to B\O_n$ is the unstable tangential structure associated to $\xi$.
For applications to unitary quantum field theory, we
will want slightly more structure: a \term{reflection-positive} IFT,\footnote{Reflection positivity appears because it is the Wick-rotated version of unitarity. See~\cite[\S 3]{freed_reflection_2021}.} defined for invertible topological field theories by Freed--Hopkins~\cite{freed_reflection_2021} and in the invertible, non-topological setting by Grady--Pavlov~\cite[\S 5]{GP21} (see also~\cite{GP20, Gra25}).\footnote{For unitarity and/or reflection positivity in the topological, noninvertible setting, see~\cite{JF17, MS23, DAGGER24, CFHPS24, Ste24}.} %

To state the classification of reflection-positive IFTs, we introduce a few more spectra.
\begin{defn}[Brown--Comenetz~\cite{BC76}]
\label{ICx_defn}
Let $A$ be an injective abelian group. The \term{Brown--Comenetz dual of the sphere (over $A$)}, written $I_A$, is defined to represent
the generalized cohomology theory
\begin{equation}
    (I_A)^n(X) \coloneqq \Hom(\pi_n(X), A).
\end{equation}
\end{defn}
In the homotopy theory literature, $A$ is most often taken to be $\Q/\Z$, following Brown--Comenetz. In the mathematical physics literature, $A = \C^\times$ is more common. If $A$ is a field of
characteristic $0$, there is a weak equivalence $I_A\xrightarrow{\simeq} HA$.
\begin{defn}[Anderson~\cite{And69}]\label{anddual}
The \term{Anderson dual of the sphere}, denoted $I_\Z$, is the fiber of the map $H\C\simeq I_\C\to I_{\C^\times}$ induced by
the exponential map $\exp\colon\C\to\C^\times$.\footnote{Some authors define $I_\Z$ in a different but equivalent way, e.g.\ using another characteristic zero field in place of $\C$. The resulting spectra are all canonically equivalent to $I_\Z$ as defined in \cref{anddual}.}
\end{defn}
Using this definition, one can show that $I_\Z$ satisfies the following universal property: for any spectrum $X$, there is a natural short exact sequence
\begin{equation}
\label{IZproperty}
\shortexact{\Ext(\pi_{n-1}(X), \Z)}{[X, \Sigma^n I_\Z]}{\Hom(\pi_n(X), \Z)},
\end{equation}
and this characterizes $I_\Z$ up to homotopy equivalence. Equation \eqref{IZproperty} splits, but not naturally, inducing a
(non-canonical) isomorphism from $[X, \Sigma^nI_\Z]$ to the direct sum of the torsion summand of $\pi_{n-1}(X)$ and the
free summand of $\pi_n(X)$.%

Now we return to the classification of reflection-positive IFTs.
\begin{thm}[{Freed--Hopkins~\cite[Theorem 1.1]{freed_reflection_2021}, Grady~\cite[Theorem 1]{Gra23}}]
\label{IFT_class}
Let $G_f$ be a fermionic group.
There is an isomorphism, natural in $G_f$, from the abelian group of deformation classes of $n$-dimensional reflection-positive IFTs on manifolds with $H(G_f)$-structure to $[\mathit{MTH}(G_f), \Sigma^{n+1}I_\Z]$.
\end{thm}
\begin{rem}
\Cref{IFT_class} holds in more generality. Indeed, for \emph{any} tangential structure $\xi\colon B\to B\O$, using the construction in~\cite[\S 7]{DAGGER24} one can define a notion of reflection-positive $\xi$-structured IFT analogous to Freed--Hopkins' definition~\cite[\S 8]{freed_reflection_2021}, and Lukas Müller observed that Freed--Hopkins' proof immediately generalizes to this setting.

Though we do not need this extra generality in this paper, it is useful for applications of the general version of \cref{IFT_class} to string theory, such as those in~\cite{FH21b, tachikawa_topological_2022, yonekura_heterotic_2022, dierigl_discrete_2023, tachikawa_anderson_2023, TY23, basile_global_2024, basile_anomaly_2024, debray_CHL_2024, debray_IIA_2024, johnson_freyd_periodicity_2024, tachikawa_discrete_2024, basile_non_supersymmetric_2025, kaidi_non_supersymmetric_2025, saito_cancelling_2025, dierigl_modified_2026}, which study reflection-positive IFTs on manifolds with various notions of twisted string structure.
\end{rem}
\begin{notn}\label{IFT_notn}
    We will often denote the group $[\mathit{MT\xi}, \Sigma^{n+1}I_\Z]$ by $\mho_\xi^{n}$. 
    The IFTs classified by this group have partition functions defined on $n$-manifolds, where $n$ corresponds to the spacetime dimension of the physical theory. 
    We will often prefer to focus instead on the spatial dimension, which we indicate by $d$; these are related by $n=d+1$.
\end{notn}
As we discussed in \cref{ss:xi_to_Gf}, for a fermionic theory with a symmetry given by a fermionic group $G_f$, the tangential structure of the corresponding IFTs is the structure $H(G_f)$ constructed in~\cite[\S 2.6]{Sto98} (here \cref{FG_to_xi}).
\begin{rem}
The classification of topological field theories $Z\colon\cat{Bord}_n^\xi\to\cat{C}$ depends on the choice of codomain $\cat{C}$, and this remains true when restricting to invertible TFTs. Freed--Hopkins~\cite[\S 5.3]{freed_reflection_2021} argue that $I_\Z$ is the universal target for classifying deformation classes of invertible IFTs, which is why it appears as the codomain in \cref{IFT_class}. See also~\cite[\S\S 6.8, 6.9, 9.4]{Fre19}, as well as~\cite{freed_vienna_notes, debray_arf-brown_2018, freed_fully_2026} for constructions of example codomain ($d+1$)-categories $\cat C$ in dimensions $d\le 2$ justifying this choice.
\end{rem}

\begin{exm}[2d \pinm IFTs]
\label{pinm_IFT_exm}
Let $G_f = \Pin_1^+$, so that the spacetime tangential structure is \pinm, as we established in \cref{pin_tangential_example}. To classify reflection-positive \pinm IFTs, by \cref{IFT_class} we compute $[\MTPin^-, \Sigma^{n+1}I_\Z]$, which by~\eqref{IZproperty} can be deduced from $\pi_*(\MTPin^-)$, the bordism groups of manifolds with \pinm structures on their tangent bundles. 

Now we specialize to $n = 2$. Anderson--Brown--Peterson~\cite[Theorem 5.1]{ABP69} showed $\Omega_2^{\Pin^-}\cong\Z/8$ and $\Omega_3^{\Pin^-}$ is torsion,\footnote{The result in~\cite{ABP69} is quite general; see~\cite[Footnote 6]{DK24} for how to extract the case $n= 2$, as well as Giambalvo~\cite[Theorem 3.4(b)]{Gia73}, Kirby--Taylor~\cite[Lemma 3.6]{KT90}, and Campbell~\cite[Theorem 6.4]{Cam17} for additional calculations of $\Omega_2^{\Pin^-}\cong\Z/8$.}
so the group of 2d \pinm reflection-positive IFTs is also isomorphic to $\Z/8$.
The field theory $\alpha_\text{ABK}$ whose partition is the Arf--Brown--Kervaire invariant~\cite{Bro71,KT90} is a generator of $\Hom(\Omega_2^{\Pin^-}, \C^\times)\cong\Z/8$ ~\cite[\S 5]{debray_arf-brown_2018}.
This IFT {and its role as the low-energy field theory of the time-reversal symmetric Majorana chain of \cref{TRS_Maj_exm_K-theory}} is discussed in detail in~\cite{debray_arf-brown_2018};
see also~\cite{Ste16, Tur20, MS23} for more general results on 2d \pinm TFTs.
\end{exm}

\subsection{Fermionic groups and symmetry types}\label{subsec_symm_types_morita_invnce}

We have seen so far that fermionic groups are useful for describing the symmetries of SPT phases; indeed, their definition is very close to descriptions in the physics literature.
In this section, we construct the fermionic groups we will use to model the tenfold way in the rest of the paper.
However, there is a major subtlety: we need to distinguish between a discrete and a continuous version.

Mathematically, there are two natural families of ten fermionic groups to consider: the \emph{discrete tenfold way} given (except for class C) by certain $E_{\ell,k}$ groups, and the \emph{continuous tenfold way}, 
which gives rise to the internal symmetry groups
used in \cite{freed_reflection_2021}.
These groups can each be 
constructed from the ten real superdivision algebras as follows.

By the classification of super division algebras $A$, we know that if $A$ is not the quaternions, then it is a Clifford algebra on the vector space $\mathbb{F}^{\ell+k}$ with bilinear form of signature $(\ell,k)$, where $\mathbb{F} = \R$ or $\C$.
For our application,
not all values of $\ell$ and $k$ can occur, but we will allow $\ell = k = 0$.
We will first discuss the Clifford algebra case, returning to the case $A=\mathbb{H}$ momentarily.

\begin{defn}[Discrete and Continuous Forms]
\label{disc_cont_defn}
For both of the following cases, the fermionic group structure is given by $(-1)^F = -1$ and $\theta$ coming from the $\Z/2$-grading of the Clifford algebra.
\begin{itemize}
    \item Pick an orthonormal basis $\mathcal{B} \coloneqq \{e_1, \dots, e_{\ell+k}\}$ of the underlying vector space $\mathbb{F}^{\ell+k}$.
    The subgroup of $(\F^{\ell+k})^\times$ generated by the basis $\mathcal{B}$ together with the norm-$1$ scalars is called the \emph{discrete form} 
    of the fermionic group in this symmetry class.
    \item The subgroup of $(\F^{\ell+k})^\times$ generated by \textit{all} homogeneous norm-$1$ vectors
    of the $\Z/2$-graded vector space $\mathbb{F}^{\ell} \oplus \mathbb{F}^k$ is the group $\Pin_{\ell,k}^{\mathbb{F}}$, which we call the \emph{continuous form}.
\end{itemize}
\end{defn}

Observe that the discrete form is always contained in the continuous form.
In the seven relevant real cases, for which we will we take
\begin{equation}
    (\ell,k) = (0,0), (0,1), (0,2), (0,3), (1,0), (2,0), (3,0),
\end{equation}
this procedure gives the fermionic groups $E_{\ell,k} \subseteq \Pin_{\ell,k}$, while in the two complex cases it gives $\Spin^c_1$ and $\Pin^c_1$.
Finally, for the 
remaining 
exceptional case $A = \mathbb{H}$, we take the quaternion subgroup $Q_8 \subseteq \SU_2 \subseteq \mathbb{H}$ generated by $i,j$ and $k$ and equip it with the fermionic group structure
$(-1)^F = -1$ and $\theta = 0$, to get the discrete form, and likewise choose $\SU_2$ with the same $\theta$ and $(-1)^F$ for the continuous form.

We refer the reader to \cref{tenfold_table} for a translation to the usual Altland--Zirnbauer class notation.
Additionally, we clarify the relationship with Freed--Hopkins~\cite{freed_reflection_2021}, whose symmetry types depend on a parameter $s$. 
In all cases in the table, the parameter $s$ and the choice of $\ell,k$ in $E_{\ell,k}$ are related by $s=\ell-k$.
Also, the super division algebra $A$ is the fermionic group algebra of the 
relevant
discrete form fermionic group. 

For the classes D, BDI, DIII, A, and AIII, the discrete and continuous forms are equal.
For the remaining 5 real classes (AI, AII, C, CI, CII), 
the inclusion $E_{\ell,0}\subset \Pin_\ell^+$ (resp. $E_{0,k}\subset \Pin_k^-$) is no longer an isomorphism for $\ell >1$ (resp. $k>1$), and likewise $Q_8\subset\SU_2$ is not an isomorphism.

\begin{exm}[class AI]
Consider the case $A = \Cl_2$, corresponding to class AI.
The discrete class AI symmetry group $E_{2,0}$ consists of two anticommuting time-reversal symmetries $T_1,T_2$ such that $T_1^2 = T_2^2 = 1$.
Setting $g \coloneqq T_1 T_2$, this is equivalent to a single time-preserving symmetry with $g^2 = (-1)^F$ and a single anticommuting time-reversal symmetry $T_1$.
It is also physically reasonable to interpret the unitary symmetry $g = iQ$ as charge and to enlarge the symmetry group ``generated by'' $g$ to a whole continuous $\U_1$. 
The total resulting internal symmetry group is $\Pin_2^+$, the continuous class AI symmetry group.
\end{exm}

The distinction between continuous and discrete symmetry classes is important. 
In general, the discrete and continuous fermionic groups for a given Altland--Zirnbauer class give rise to completely different twisted spin bordism groups, and hence also completely different groups of IFTs.
\begin{thm}
\label{discrete_ne_cts}
In Altland--Zirnbauer classes AI, AII, C, CI, and CII, the twisted spin bordism groups associated to the discrete and continuous fermionic groups in \cref{tenfold_table} are not isomorphic. For classes AI, AII, and C, they are not even rationally isomorphic.
\end{thm}
In the remaining five classes (A, AIII, BDI, D, DIII), the discrete and continuous fermionic groups are equal, as we discussed above. Nevertheless, even in these cases there are other physically motivated choices for a fermionic group representing a given symmetry class giving rise to nonisomorphic twisted spin bordism groups: see~\cite[\S 4]{stehouwer_interacting_2022} for an example in class D.
\begin{proof}[Proof of \cref{discrete_ne_cts}]
We compute the twisted spin bordism groups corresponding to our discrete fermionic groups in \cref{melk_calc}, modulo certain $\Z/2$ summands (see \cref{rem:bosonicsummand}), except for class C; we do the class C computation in \cref{spin_Q8_bordism} (see~\eqref{explicit_Q8}).\footnote{In this proof, we only need $\Omega_4^{\Spin\times_{\set{\pm 1}}Q_8}$, which was computed earlier by Pedrotti~\cite[Theorem 8.0.8]{Ped17}.}
Freed--Hopkins~\cite[\S 9.3]{freed_reflection_2021} compute the twisted spin bordism groups corresponding to the continuous fermionic groups in degrees $4$ and below, and using their Adams charts (\textit{ibid.}, Figure 5) and the Anderson--Brown--Peterson splitting~\cite{ABP67} (reviewed in \cref{ABP_thm}), one can extend these computations to degrees $\le 11$. To finish the proof, directly compare the results of these two computations. Explicitly, there are nonnegative integers $w_{\mathrm{AI}}$, $w_{\mathrm{AII}}$, $w_{\mathrm{CI}}$ and $w_{\mathrm{CII}}$ such that\footnote{One can show that $w_{\mathrm{AI}} = 2$, $w_{\mathrm{AII}} = 1$, $w_{\mathrm{CI}}= 8$, and $w_{\mathrm{CII}} = 6$ by explicitly working out the Adams spectral sequences for these four cases. We do not need the specific numbers, so do not go into detail.}
\begin{equation}
\begin{alignedat}{2}
    \Omega_2^{\Pin^{\tilde c-}} &\cong\Z \qquad\qquad\qquad\qquad\qquad\quad && \Omega_2^{E_{2,0}} \cong (\Z/2)^{\oplus w_{\mathrm{AI}}}\\
    \Omega_2^{\Pin^{\tilde c+}} &\cong\Z  && \Omega_2^{E_{0,2}} \cong (\Z/2)^{\oplus w_{\mathrm{AII}}}\\
    \Omega_4^{\Spin^h} &\cong\Z^2  &&\Omega_4^{\Spin\times_{\set{\pm 1}}Q_8} \cong \Z\\
    \Omega_8^{\Pin^{h+}} &\cong\Z/32\oplus\Z/8\oplus (\Z/2)^{\oplus 3}  && \Omega_8^{E_{3,0}} \cong \Z/32\oplus (\Z/2)^{\oplus w_{\mathrm{CI}}}\\
    \Omega_6^{\Pin^{h-}} &\cong\Z/16\oplus\Z/4\oplus\Z/2  && \Omega_6^{E_{0,3}} \cong \Z/16\oplus (\Z/2)^{\oplus w_{\mathrm{CII}}},
\end{alignedat}
\end{equation}
corresponding to classes AI, AII, C, CI, and CII, respectively. %
\end{proof}

For our physical application, discrete symmetry types suffice, simplify computations, and in fact illuminate a version of Bott periodicity, as we discuss further in \cref{section_modeling_the_Bott_spiral}.

\begin{table}[h!]
\begin{center}
    Complex Symmetry Types
\end{center}
\begin{tabular}{r c c c c}
\toprule
$s$ & AZ class & $G_f$ & $\C^f[G_f]$ & $H^c(s) = H(G_f)$\\
\midrule
$0$ & A & $\U_1$ & $\C$ & $\Spin^c$\\
$1$ & AIII & $\U_1\times\Z/2$ & $\C\ell_1$ & $\Pin^c$\\
\bottomrule
\end{tabular}

\vspace{3mm}

\begin{center}
    Real Symmetry Types
\end{center}
\begin{tabular}{r c c c c l l}
\toprule
$s$ & AZ class & Continuous $K_f$ & Discrete $G_f$ & $\R^f[G_f]$ &  $H(s) = H(K_f)$ & $H(G_f)$\\
\midrule
$-3$ & CII & $\Pin^-_3$ & $E_{0,3}$ & $\Cl_{-3}$ &  $\Pin^{h-}\coloneqq \Pin^-\times_{\set{\pm 1}}\SU_2$ & $\Spin$-$(0,3)$\\
$-2$ & AII & $\Pin^-_2$ & $E_{0,2}$&  $\Cl_{-2}$ &  $\Pin^{\tilde c+}\coloneqq \Pin^+\ltimes_{\set{\pm 1}} \U_1$ & $\Spin$-$(0,2)$\\
$-1$ & DIII & $\Pin^-_1$ & $E_{0,1}$ & $\Cl_{-1}$
 & $\Pin^+$ & $\Spin$-$(0,1)$\\
$0$ & D & $\Spin_1$ & $E_{0,0}$&  $\R$
 & $\Spin$& $\Spin$\\
$1$ & BDI & $\Pin^+_1$& $E_{1,0}$ & $\Cl_1$
 & $\Pin^-$& $\Spin$-$(1,0)$\\
$2$ & AI & $\Pin^+_2$ & $E_{2,0}$& $\Cl_2$ & $\Pin^{\tilde c-}\coloneqq \Pin^-\ltimes_{\set{\pm 1}} \U_1$& $\Spin$-$(2,0)$\\
$3$ & CI & $\Pin^+_3$ & $E_{3,0}$& $\Cl_3$  & $\Pin^{h+}\coloneqq \Pin^+\times_{\set{\pm 1}}\SU_2$ & $\Spin$-$(3,0)$ \\
$4$ & C & $\SU_2$ & $Q_8$ & $\Cl_4$
& $\Spin^h\coloneqq \Spin\times_{\set{\pm 1}}\SU_2$& $\Spin\times_{\set{\pm 1}}Q_8$\\ 
\bottomrule
\end{tabular}
\caption{
Here we list symmetry types corresponding to the tenfold way of fermionic insulators and superconductors. The first subtable lists the complex classes, and the second subtable lists real classes. 
In the complex cases, discrete and continuous symmetry types agree, so we make no distinction.
In the real cases, we list the two versions separately.
In all cases, the continuous classes agree with those studied by Freed--Hopkins~\cite[Tables (9.24), (9.25)]{freed_reflection_2021}, and for these we include their notation with the parameter $s$ and symmetry type $H(s)$ for comparison. See also \cite[Table 1]{stehouwer_interacting_2022}.
\\
The fermionic group algebra $\R^f[G_f]$ of the discrete fermionic group $G_f$ is used to compute free SPT phases (see \cref{ff_and_k}) in the relevant class.
The corresponding twisted spin structure groups $H(G_f)$ and $H(K_f)$ are used to compute interacting SPT phases (see \cref{interacting_IFT}). In physics language, $K_f$ and $G_f$ are internal symmetry groups, and $H(\bl)$ defines the tangential structure present on spacetime.
}
\label{tenfold_table}
\end{table}

The fermionic group $Q_8$ is an outlier, as it is is not isomorphic as a fermionic group\footnote{As groups, $Q_8$ is isomorphic to $E_{0,2}$, but this isomorphism does not preserve $\theta$; $Q_8$ does not have any time-reversing symmetries.} to one of the form $E_{\ell,k}$.
However, it is related to the groups $E_{\ell,k}$ in the following way:
\begin{lem}\label{class_C_prime}
    There are isomorphisms $s_+\colon E_{0,3} \overset\cong\to Q_8 \ftens E_{1,0}$ and $s_-\colon E_{3,0} \overset\cong\to Q_8 \ftens E_{0,1}$.
\end{lem}
\begin{proof}
Let $f_1,f_2,f_3$ be the canonical odd generators of $E_{0,3}$ and $f$ the generator of $E_{1,0}$.
    We define a map $Q_8 \ftens E_{1,0} \to E_{0,3}$ on generators as follows:
    \begin{equation}\label{double_phi}
    \begin{aligned}
        i \ftens 1 &\mapsto f_1 f_2, \qquad 
        j \ftens 1 \mapsto f_1 f_3, \qquad
        k \ftens 1 \mapsto f_2 f_3,
        \\
        1 \ftens f &\mapsto f_1 f_2 f_3.
    \end{aligned}
    \end{equation}
    Note that $f_1 f_2 f_3$ is odd and the $f_i f_j$ are even.
    A direct computation shows that the $f_i f_j$ satisfy the quaternion group relations.
    It follows that $(f_1 f_2 f_3)^2 = f_1^2 (f_2 f_3)^2 = 1$.
    One also checks that $f_1 f_2 f_3$ commutes with $f_i f_j$ as is required from the fermionic tensor product.
    Therefore, this map is a well-defined group homomorphism.
    By checking surjectivity and comparing cardinalities, one can verify it is a bijection.
    Showing that $E_{3,0} \cong Q_8 \ftens E_{0,1}$ is completely analogous; if $f_i^2 = 1$ instead, then we still have $(f_i f_j)^2 = -1$, but now $(f_1 f_2 f_3)^2 = -1$.
\end{proof}
Tensoring the above isomorphisms with $E_{0,1}$ and $E_{1,0}$ respectively, we obtain:
\begin{cor}
\label{40_04}
There is an isomorphism of fermionic groups $E_{0,4}\cong E_{4,0}$.
\end{cor}
Similarly, there are isomorphisms of fermionic groups $E_{\ell,k+4} \cong E_{\ell+4,k}$ for nonzero $\ell,k$, and as a result there are also equivalences of spectra
\begin{equation}\label{4exchange}
    \mathit{MTH}(E_{\ell,k+4}) \simeq \MTSpin \wedge \ME_{\ell,k+4}  \simeq \MTSpin \wedge \ME_{\ell+4,k} \simeq \mathit{MTH}(E_{\ell+4,k}),
\end{equation}
which will play an important role in \cref{spiral_maps_physics_section}.

\begin{rem}
    When we apply the fermionic group algebra functor to the isomorphism $E_{\ell,k+4} \cong E_{\ell+4,k}$, we recover the familiar fact that there is an isomorphism (not just a Morita equivalence) of $\Z/2$-graded algebras $\Cl_{\ell,k+4}\cong \Cl_{\ell+4,k}$.
\end{rem}
\begin{rem}
\label{non_VB_isom}
The second half of \cref{class_C_prime} is equivalent to the assertion that there is a group isomorphism $\varphi\colon (E_{3,0})_b\xrightarrow{\cong}(Q_8\ftens E_{0,1})_b$ which pulls back the twisting data of $Q_8\ftens E_{0,1}$ to the twisting data of $E_{3,0}$. Such a map $\varphi$ can be extracted from its double cover~\eqref{double_phi}; in the rest of this remark, we make that choice.

The twisting data for $E_{\ell,k}$ comes from the virtual vector bundle which is the sum of $\ell$ copies of $\sigma$ and $k$ copies of $-\sigma$, so it is natural to wonder whether there is a virtual vector bundle $V\to B(Q_8)_b$ and an isomorphism $\overline\varphi\colon V\boxplus (-\sigma)\cong \sigma^{\boxplus 3}$ lifting $\varphi$ in the sense that taking the first two Stiefel--Whitney classes of the map $\overline\varphi$ recovers the effect of $\varphi$ on twisting data.

Even though the twisting data on $Q_8\ftens E_{0,1}$ can be realized by a vector bundle twist, it turns out no such $\overline\varphi$ can exist! To prove this, use the Whitney sum formula to calculate the Stiefel--Whitney classes of $V$, assuming it exists, and obtain a contradiction. This is a different problem from the existence of non-vector-bundle twists (\cref{nonVB}), and will complicate our definition of the spiral map $\mathrm{sp}^\psi$ in \cref{ss:spiral_model}.
\end{rem}
In the above discussion, we saw how multiple fermionic groups (discrete or continuous) can reasonably be associated to symmetry classes, leading to different interacting SPT classifications.
Another common way in which two distinct fermionic groups are considered to be in `the same symmetry class' is when their respective fermionic group algebras are \textit{Morita equivalent}.
In that case these symmetries generate identical free SPT classifications, but in general inequivalent interacting SPT classifications~\cite{stehouwer_interacting_2022}.
This is a crucial aspect of the free-to-interacting maps we define in \cref{section_F2I} and of our choice of fermionic groups in \cref{primed_AZ}; see specifically \cref{Morita_variance_rmk_2}.

\begin{exm}\label{E11_vs_1}
    The fermionic group algebra $\R^f[E_{1,1}]\cong\Cl_{1,1}$ is Morita equivalent to $\R$ (\cref{elk_alg}), thus the $K$-theory of this fermionic group coincides with that of the trivial fermionic group: $K_i(\Cl_{1,1})\simeq\KO_i$. 
    In particular, the classification of $E_{1,1}$-protected free fermion phases is the same as $E_{0,0}$-protected (i.e.\ class D) free phases by \cref{free_classification_ansatz}.
    However, this does \emph{not} carry over for twisted spin bordism---the groups are different. In general, they do not even agree rationally. See \cref{tab:SpinvsDPin}.
    \end{exm}

    \begin{table}[h!]
\begin{tabular}{l c c}
\toprule
$d$ & $\Omega^{\text{Spin}}_d$ & $\Omega^{\text{DPin}}_d$
\\\midrule
0   & $\Z$                     & $\Z/2$                   \\
1   & $\Z/2$                   & $\Z/2$                   \\
2   & $\Z/2$                   & $(\Z/2)^2$               \\
3   & $0$                      & $\Z/8$                   \\
4   & $\Z$                     & $(\Z/2)^2$               \\
5   & $0$                      & $0$                      \\
6   & $0$                      & $(\Z/2)^2$  
\\
\bottomrule
\end{tabular}
\caption{Comparing low-dimensional spin bordism groups~\cite{milnor_spin_1963} with dpin bordism groups \cite[Theorem F.1]{kaidi_topological_2020}.}
\label{tab:SpinvsDPin}
\end{table}

Exactly the same thing as in \cref{E11_vs_1} happens for $E_{k,k}$, whose group algebra $\Cl_{k,k}$ (\cref{elk_alg}) is Morita trivial, but whose spin-$(k,k)$ bordism groups in general depend on $k$ (see \cref{melk_calc}). All of these different notions of twisted spin bordism in some sense correspond to class D free fermionic phases.

\section{Defining the free-to-interacting map}
\label{section_F2I}

As we discussed in the introduction, one of our main goals is to produce homotopy-theoretic models of \textit{free-to-interacting maps}, which relate the group of free fermion SPT phases with the group of interacting SPT phases for a specified symmetry type.

Physically, one can study a free Hamiltonian representing a free fermion phase and compute the order of the group of free fermion SPTs that it generates by stacking copies of the phase and checking whether it can be trivialized via free terms. When one passes to the interacting classification, new terms become available with which to deform the Hamiltonian, which can change the order of the group of phases it generates. In fact, for our examples of interest, quartic interaction terms in the Hamiltonian will always suffice to trivialize the phase \cite{queiroz_dimensional_2016}.

A free-to-interacting map should model this process by taking the deformation class of a free Hamiltonian to the interacting deformation class of the same Hamiltonian.
When distinct free phases can be adiabatically connected by interacting terms, we observe that the free-to-interacting map should have a kernel. Conversely, intrinsically-interacting phases, which have no free analog, should form the cokernel of a free-to-interacting map.

Following the approach of Freed--Hopkins \cite[Sections 9, 10]{freed_reflection_2021}, we model free-to-interacting maps as maps from $K$-theory groups, which classify free SPTs by \cref{neutral_luuk_ansatz,neutralluuk_main_thm_citation}, to groups of deformation classes of invertible field theories, which classify interacting SPTs according to \cref{LEFT_ansatz}.
Each element of the $K$-theory group in the source represents a free fermion phase, while each element in the group of deformation classes of invertible field theories in the target represents an interacting fermion phase.

On the free side, there are many approaches in the literature connecting free fermion Hamiltonian models to representatives of $K$-theory classes, including, for example, \cite{bellissard2005k, heinzner2005symmetry,
freed_twisted_2013,
Thiang_2015,
kellendonk2017c,
alldridge2020bulk,
gomi_freed-moore_2021,
neutralluuk}. In \cref{ff_and_k}, we discussed such a model using Karoubi triples. On the interacting side, relating lattice models to invertible field theories is a more subtle problem and an active area of research: see for example~\cite{debray_arf-brown_2018, Fre19, debray_low_energy_2020, fidkowski_exactly_2020, fidkowski_disentangling_2020, bourne_classification_2021, freed_reflection_2021, ogata_index_2021, sopenko_index_2021, freed_topological_2022, ogata_invariant_2022, ogata_classification_2022, spodynieko_Hall_2023, artymowicz_mathematical_2024, sopenko_index_2024, feng_onsiteability_2025, fidkowski_QCA_2025, freed_discrete_2025,
feng_Pauli_2026, kapustin_topological_2025, kubota2025stable, sopenko_reflection_2025, shirley_anomaly_2025, feng_higher_2026, sun_Clifford_2025, else_anomalies_2026}.
Thus we rely on \cref{LEFT_ansatz} to justify our approach. We view the existence of mathematical free-to-interacting maps as supporting evidence for the homotopical approach to SPT phase classification, and one of our main goals in this paper is to add to this body of evidence.

In this section, we define a homotopical free-to-interacting map for a broader collection symmetry types than in prior work, including the fermionic groups $E_{\ell,k}$.
We compute the image of free-to-interacting maps of type $E_{\ell,k}$ in \cref{section_computations}, which we then compare with the physical computations of Queiroz--Khalaf--Stern  \cite{queiroz_dimensional_2016} in \cref{section_modeling_the_Bott_spiral}.

\subsection{Classical Atiyah--Bott--Shapiro orientation}\label{sec_classical_ABS}
The heart of the free-to-interacting map is a natural transformation between spin bordism and $\KO$-theory called the Atiyah--Bott--Shapiro (ABS) orientation. Atiyah--Bott--Shapiro \cite{atiyah_clifford_1963} provided formulas for the Thom class in $\KO$-theory (resp. $\KU$-theory) of a spin (resp.\ \spinc) vector bundle.
The choice of Thom classes defines an orientation, which gives a map of spectra $\MTSpin\to \KO$ (resp.\ $MT\Spinc\to \KU$) \cite{ABP67,hopkins_algebraic_2002}.

In \cite{freed_reflection_2021}, Freed--Hopkins generalized the ABS maps from spin and \spinc bordism to the Thom spectra $\mathit{MTH}(K_f(s))$ corresponding to the \emph{continuous} tenfold way fermionic groups $K_f(s)$ (see \cref{disc_cont_defn,tenfold_table}).
We first recall their construction; then, we explain our generalization for twists by pairs of vector bundles.

In the following, $X$ will be a locally compact Hausdorff space. For finite $X$, taking $A=C_0(X)$ and regarding $A$ as a purely even superalgebra recovers the setting of \cref{sec:Ktheoryoffermgps}.

\begin{defn}
    Let $V\to X$ be a real Euclidean vector bundle over $X$, so that each fiber is equipped with a metric $h_x$. The \emph{Clifford bundle} $\Cl(V)$ of $V$ is the algebra bundle whose fiber at $x$ is the Clifford algebra $\Cl(V_x,-q_x)$, where $q_x(v) \coloneqq h_x(v, v)$.
\end{defn}
\begin{defn}
    A \emph{graded Clifford module} of $V\to X$ is a $\Z/2$-graded real vector bundle $E\to X$ equipped with actions
    $V\otimes_\R E^0\to E^1$ and $V\otimes_\R E^1\to E^0$
    such that for  each $v\in V$ and $e\in E$,
    \begin{equation}
        v(v(e)) = - \lVert v \rVert^2 e.
    \end{equation}
\end{defn}
\begin{rem}
If $X$ is compact Hausdorff, graded Clifford modules for $V\to X$ are equivalent data to finitely generated, projective modules over the $C^*$-algebra $\Gamma(X; \Cl(V))$. This follows from the fact that the Serre--Swan theorem~\cite{swan_vector_1962}, that taking sections defines an equivalence of categories from vector bundles on $X$ to finitely generated projective $C(X)$-algebras, in fact refines to a monoidal equivalence by Conlon~\cite[\S 7.5]{conlon_differentiable_1993} and Faurot--Ferrer~\cite[Theorem A.7]{faurot_homotopy_2026}.
\end{rem}

\begin{exm}\label{lambda_V_class}
    Let $V\to X$ be a vector bundle with a quadratic form. 
    The exterior power bundle $\Lambda^* V$ admits the structure of a $\Z/2$-graded $\Cl(V)$-module: its graded pieces are the even and odd powers $E^0 = \Lambda^{\text{ev}}V$, $E^1 = \Lambda^{\text{odd}} V$ and the action of $V$ is given by
    \begin{subequations}
    \begin{align}
        V\otimes_\R \Lambda^{*}(V) &\longrightarrow \Lambda^* V \\
        v\otimes e &\mapsto v\wedge e - (v\,\neg\,e).
    \end{align}
    \end{subequations}
    Here $\neg\,e$ is the adjoint of $\wedge\, e$ with respect to the bilinear form on $\Lambda^*(V)$ induced from the quadratic form on $V$.
\end{exm}

In \eqref{eq:CMQ}, we defined the Clifford module quotient of a graded algebra by considering graded modules modulo those that extend to the action of one more positive Clifford generator. 
For the vector bundle setting, let $V\to X$ be a rank-$k$  vector bundle with quadratic form $q_V$ and let $E$ be a graded $\Cl(V)$-module. Equip $V\oplus \underline{\R}$ with the quadratic form $q_V\oplus \id_{\underline{\R}}$.
Then if $E$ extends to a $\Cl(V\oplus \underline{\R})$ module, each fiber $E_x$ admits the action of $\Cl(V_x)\mathbin{\hat{\otimes}}\Cl_{-1}$, by \cref{clifftens}.
As a special case of \eqref{eq:CMQ}, we thus get:
\begin{defn}
    Let $V$ and $V\oplus \underline{\R}$ be as above, and let $\pi_0(\cat{Mod}_{\Cl(V)})$ be the commutative monoid
    of isomorphism classes of finite-dimensional graded $\Cl(V)$-modules under $\oplus$.
    The Clifford module quotient\footnote{This corresponds to the group Atiyah--Bott--Shapiro call $A(V)$~\cite[\S 11]{atiyah_clifford_1963}.} is the abelian group
    \begin{equation}
        \CMQ^0(V) \coloneqq 
        \mathrm{coker}\,(\pi_0(\cat{Mod}_{\Cl(V\oplus \underline{\R})}) \to \pi_0(\cat{Mod}_{\Cl(V)})).
    \end{equation}
     where the map of monoids is induced by the inclusion $V\hookrightarrow V\oplus \underline{\R}$. 
    More generally, define $\CMQ^{-n}(V)$ as $\CMQ^0(V\oplus \underline{\R}^n)$, where the trivial bundle $\underline{\R}^n$ is equipped with the standard quadratic form.
\end{defn}
In the complex case, there is an analogous group built from $\Cxl(V)$-modules, where $V$ is a complex vector bundle with a Hermitian metric.

As noted in \cref{rem_CMQ_insufficient_infinite_dim}, the Clifford module quotient group does not in general give a model for $K$-theory. Instead, there is a homomorphism
$\CMQ^0(V)\to \widetilde{\KO}{}^0(X^V)$, which Atiyah--Bott--Shapiro call the \term{Euler characteristic map}~\cite[Definition 8.1]{atiyah_clifford_1963}. To discuss it, we will use the L-form of $K$-theory classes in the case of a 2-term complex, meaning that we specify classes in $\KO^0(X,Y)$ by a pair of vector bundles $V,W\to X$ and a map $s\colon V\to W$ between them, written $[V,W;s]$, 
where $s$ restricts to an isomorphism over $Y$. 
See e.g.\ \cite[\S 7]{atiyah_clifford_1963}, \cite[Definition 9.23]{lawson_spin_1989} for the details of this model.

\begin{defn}
    Let $V\to X$ be a real vector bundle, let $D(V)$ and $S(V)$ be the disk bundle and sphere bundle, resp., and let $\pi\colon D(V)\to X$ be the projection. The Euler characteristic map makes the assignment
    \begin{equation}
    \begin{aligned}
        \chi_V\colon \CMQ^0(V) &\to \widetilde{\KO}{}^{0}(X^V) \\
        E=E^0\oplus E^1 &\mapsto [\pi^* E^1,\pi^* E^0;\sigma] 
    \end{aligned}
    \end{equation}
    where $\sigma$
    is given by multiplication by minus the vector in the base. 
    That is, $\sigma|_{v\in V}\colon e \mapsto -v\cdot e$ for $e\in \Cl(V)|_v$. 
\end{defn}
Here, we used that $\widetilde{\KO}{}^{0}(X^V) \cong \KO^0(D(V),S(V))$.
See e.g.\ \cite[\S8, \S9]{atiyah_clifford_1963} for more details. The complex case is analogous.

A complex vector bundle $V$ always admits a spin$^c$ structure, meaning that it is oriented for $K$-theory. We next recall the Clifford module construction for the $K$-theory Thom class from \cite[\S11, \S12]{atiyah_clifford_1963}.

\begin{thm}[{\cite[Theorem (12.3)]{atiyah_clifford_1963}}]\label{K_Thom_class}
    Let $V$ be a rank-$2k$ \spinc vector bundle. Then there is a class $u_V^{KU} \in \widetilde{KU}^{2k}(X^V)$ such that 
    $\widetilde{\KU}{}^*(X^V)$ is a free $\KU^*(X)$-module of rank one generated by $u^{\KU}_V$.
\end{thm}

In particular, for each degree $i$   
    there is an isomorphism
    \begin{equation}
        \KU^i(X) \xrightarrow{-\cdot u^{\KU}_V} \widetilde{\KU}{}^{i+2k}(X^V).
    \end{equation}

\begin{rem}
    We choose to put the Thom class in degree $2k$, but many sources discuss this class as living in degree zero, using periodicity of $KU$-theory.
\end{rem}

\begin{rem}
\label{spinc_thom} 

Let $V$ be a rank-$k$ complex vector bundle.
Then $V$ admits a spin$^c$ structure, meaning an isomorphism
\begin{equation}
        V\cong P_{\Spin^c}\times_{\Spin_{2k}^c}\R^{2k}
\end{equation}
for $P_{\Spin^c}$ a principal $\Spin^c_{2k}$ bundle.
    From this bundle and the Clifford module $\Lambda^*(\C^k)$, we may form the Clifford module $P_{\Spin^c}\times_{\Spin^c_{2k}}\Lambda^*(\C^k)$, using the action of $\Spin^c_{2k}\subset \C\ell_{2k}^\times$.
    Then there is an equivalence
    $\chi_{V}(\Lambda^*(V)) = \chi_{V}(P_{\Spin^c}\times_{\Spin^c_{2k}}\Lambda^*(\C^k))$, so we may use either construction to form the Thom class. See \cite[Proposition 11.6]{atiyah_clifford_1963}.
\end{rem}

Complex vector bundles have canonical \spinc structures and even (real) rank, so are particular cases of the above construction. To specify the $\KU$-theory orientation of an \textit{odd-rank} \spinc vector bundle $V$, one can stabilize to the even-rank bundle $V\oplus\underline{\R}$, whose Thom space is naturally homotopy equivalent to $\Sigma X^V$, and use the suspension isomorphism on $\KU$-theory to obtain the Thom class over $X^V$.

For real $\KO$-theory, the construction is analogous, except that the representation theory of real Clifford modules is richer than that of complex ones. In this case, we will need to stabilize to a rank that is a multiple of $8$ by adding trivial bundles.

\begin{thm}[{\cite[Theorem (12.3)]{atiyah_clifford_1963}}]\label{KO-Thom-class-8k}
    Let $V\to X$ be a rank-$8k$ spin vector bundle, let $\lambda^k \in \Mod(\Cl_{8k})$ be an irreducible graded module, and let $P_{\Spin}$ be a principal $\Spin_{8k}$-bundle such that $V\cong P_{\Spin}\times_{\Spin_{8k}} \R^{8k}$ as spin vector bundles.
    Write $\beta\in \KO^{-8}$ for the real Bott class. Define
    \begin{equation}
        u^{\KO}_V \coloneqq \chi_V (P_{\Spin}\times_{\Spin_{8k}} \lambda^k){\cdot \beta^{-k}} \in \widetilde{\KO}{}^{{8k}}(X^V).
    \end{equation}
    Then $\widetilde{\KO}{}^*(X^V)$ is a free $\KO^*(X)$-module of rank one generated by $u^{\KO}_V$. 
\end{thm}
In particular, for each degree $i$   
    there is an isomorphism
    \begin{equation}
        \KO^i(X) \xrightarrow{-\cdot u^{\KO}_V} \widetilde{\KO}{}^{i+8k}(X^V).
    \end{equation}

See \cite[Cor.\ (6.6), Section 11]{atiyah_clifford_1963}.

Above, we discussed the construction of Thom classes for spin$^c$ and spin vector bundles, but we will ultimately use maps of spectra.
Returning to the complex case,
to specify the orientation map $\MTSpin^c\to \KU$ it suffices to provide classes in the $KU$-theory of the Thom spaces that make up the spectrum $\MTSpin^c$ that are compatible with the structure maps of the spectrum.
In more detail,
recall that the spaces in the spectrum $\MTSpin^c$ are the Thom spaces $M\Spin^c_n = \mathrm{Th}(B\Spin_n^c; V^{\Spinc}_n)$,\footnote{A priori, this process builds a different spectrum, $M\Spin^c$; $\MTSpin^c$ is the Thom spectrum of $-V^{\Spin^c}\to B\Spin^c$. The Pontrjagin--Thom theorem identifies the homotopy groups of $M\Spin^c$, resp.\ $\MTSpin^c$, with the bordism groups of manifolds with \spinc structures on their stable normal bundles, resp.\ stable tangent bundles. Fortunately, there is an equivalence $M\Spin^c\simeq\MTSpin^c$: see~\cite[Remark 3.3]{morava_cobordism_2012}. This uses the well-known fact that the map $B\Spin^c\to B\O$ is $A_\infty$, in fact $E_\infty$. See~\cite[Example 2.58]{debray_IIA_2024} for a non-original reference. The story for $M\Spin\simeq\MTSpin$ is completely analogous. For more general twisted spin structures, though, including spin-$(\ell,k)$ structures, this equivalence does not generalize.} where $V^\Spinc_n$ is the pullback of the universal spin$^c$ bundle along the map $\Spin^c_n \to \Spinc$. We write $\KU_n$ for the classifying space for the functor $\KU^{-n}$.
To produce a collection of compatible maps $M\Spin^c_n\to \KU_n$, we will take the classifying maps of a set of $\KU$-theory classes. Specifically, in each dimension $n$, we take the class $u_{V_n^\Spinc}^{\KU}$, which gives the Thom class in $\KU$-theory by \cref{spinc_thom}:
\begin{equation}
u^{\KU}_{V_n^{\Spinc}} \in \widetilde{\KU}{}^0(M\Spin^c_n) ~ \rightsquigarrow ~ M\Spin^c_n\to \KO_n.
\end{equation}
See e.g.\ \cite{joachim_twisted_1997} for more details on the map of spectra. The real case is analogous.

\subsection{Smith homomorphisms}\label{smith_homs_intro}

Smith homomorphisms are maps on bordism groups induced by zero section maps between Thom spaces. They were first introduced by Conner--Floyd~\cite[Theorem 26.1]{ConnerFloyd}, and the full theory was worked out in~\cite{debray_smith_2024}. See there for further background, references, and examples.

Smith homomorphisms have the effect of twisting the tangential structure on manifolds between their domains and codomains, which allows them to capture certain forms of symmetry breaking, as studied in \cite{hason_anomaly_2020, COSY, kobayashi_anomaly_2021, DNT24, copetti_anomalies_2025, debray_long_2024, jones_charge_2025, manjunath_anomalous_2025} and as we will discuss in \cref{smith_symm_br}.
We will introduce them as induced from maps of Thom spectra, following~\cite[\S 3.2]{debray_smith_2024}.

Let $X$ be a space and let $W\to X$ be a $k$-dimensional real vector bundle. Taking the zero section of $W$ defines a map $X_+\hookrightarrow X^W$ from the base space into the Thom space. Similarly, if $V$ is a (possibly virtual) vector bundle on $X$, the inclusion $V\hookrightarrow V\oplus W$ also defines a map on Thom spectra:
\begin{equation}\label{smith_map_of_spectra_specific}
    e_W\colon X^V \longrightarrow X^{V+W}.
\end{equation}
In the case that $V$ is an actual vector bundle, the map descends to Thom spaces.

\emph{Smith homomorphisms} are induced from maps $e_V$ as in~\eqref{smith_map_of_spectra_specific} by smashing with a bordism spectrum $MT\xi$ and taking homotopy, then using the Pontrjagin--Thom theorem to identify the resulting groups with bordism groups.

\begin{defn}\label{smith_hom_defn}
The \term{Smith homomorphism} in $\xi$-bordism in degree $n$ associated to~\eqref{smith_map_of_spectra_specific} is the induced map to~\eqref{smith_map_of_spectra_specific} on $\Omega_n^\xi$:
\begin{equation}
        \sm_W\colon \Omega^\xi_n(X^V) \to \Omega^\xi_n(X^{V+W}) \cong \Omega^\xi_{n-k}(X^{V+W-k})
    \end{equation}
On the right side we used the suspension isomorphism in bordism to normalize degrees.
\end{defn}

\begin{lem}[{\cite[Proposition 3.17, Corollary 4.39]{debray_smith_2024}}]
The Smith homomorphism associated to the data above can be equivalently described as
\begin{enumerate}
    \item the map sending 
 the class of a closed $n$-manifold $[M]$ to the bordism class $[N]$, where $N\subset M$ is the $(n-k)$-dimensional submanifold defined as the intersection of the zero section of $W$ (restricted to $M$) with a generic section $s$, or
 \item the map sending a class $[M]$ to the Poincaré dual of the cobordism\footnote{By ``cobordism'' we mean the generalized cohomology theory associated to the spectrum $\mathit{MT\xi}$, whose homology theory is called bordism. This terminology is due to Atiyah~\cite{atiyah_bordism_1961}.} Euler class 
$e^{\mathit{MT\xi}}(f^*W)$.
\end{enumerate}

\end{lem}

To type-check description (1), first
invoke the preimage theorem to ensure that $N=s^{-1}(0)$ is a closed $(n-k)$-manifold, assuming $s$ is transverse to the zero section, which is true for generic $s$. Hason--Komargodski--Thorngren~\cite[\S 4.2]{hason_anomaly_2020} show that, in that case, the bordism class of $N$ is independent of the choice of $s$.
By definition, $M$ is a closed $n$-manifold with an $(X, V)$-twisted spin structure, so there is a map $f\colon M\to X$ and a spin structure on $TM\oplus f^*V$. Equip $N$ with the map $g\colon N\hookrightarrow M\xrightarrow{f} X$. The normal bundle $\nu$ to the embedding of $N$ is isomorphic to $g^* W$ by construction,
so the exact sequence
\begin{equation}
    0 \to TN \to TM|_N \to \nu \to 0
\end{equation}
implies that $TN$ carries a $(X, V\oplus W)$-twisted spin structure.

See~\cite[\S3, \S7]{debray_smith_2024} for a general discussion and an extensive collection of examples. The examples relevant to this article are the Smith maps for which $X = B\Z/2$, $V$ and $W$ are direct sums of some number of copies of the universal bundle $\sigma\to B\Z/2$, and $\xi$ is spin or \spinc structure, which are discussed in (\textit{ibid.}, Examples 7.8, 7.13, and 7.37) as well as the references therein.

\begin{exm}\label{1st_Z2_Smith_exm}
    Let $X=B\Z/2$, $V=0$, and $W=\sigma$. The zero section map $0\hookrightarrow \sigma$ defines a map of spectra
    \begin{equation}\label{BZ2_Smith}
        e_\sigma\colon (B\Z/2)_+ \to (B\Z/2)^\sigma.
    \end{equation}
    Taking spin bordism yields homomorphisms
    \begin{equation}\label{first_smith_isom}
        \sm_\sigma\colon \Omega^\Spin_n(B\Z/2) \to \Omega^\Spin_{n-1}((B\Z/2)^{\sigma-1})\cong \Omega_{n-1}^{\Pinm}.
        \qedhere
    \end{equation}
The isomorphism $\Omega^\Spin_{n-1}((B\Z/2)^{\sigma-1})\cong \Omega_{n-1}^{\Pinm}$ was first observed in~\cite[\S 7]{Pet68}; it can also be proven using the shearing isomorphism of \cref{shearing}.
Restricting the domain of \eqref{first_smith_isom} to \textit{reduced} spin bordism defines an isomorphism for all $n$~\cite{ABP69}.

The previous map fits into a family of four types of Smith homomorphisms for $X=B\Z/2$, $W=\sigma$, and $V=n\sigma$. Since four copies of any bundle, and in particular $4\sigma$, is spin, the twisted spin tangential structures appearing in iterated Smith homomorphisms forms a periodic family depending only on $n\bmod 4$. The four symmetry types, starting from $V=0$, go as follows:
\begin{equation}\label{spin_sigma_symm_types}
    \Spin\times \Z/2 ~ \rightsquigarrow ~ \Pinm ~ \rightsquigarrow ~  \Spin\times_{\set{\pm 1}}\Z/4 ~ \rightsquigarrow ~ \Pinp ~ \rightsquigarrow ~ \dots
\end{equation}
See~\cite{kapustin_fermionic_2015, TY19, hason_anomaly_2020, WWZ20, debray_long_2024, debray_smith_2024} for more on this $4$-periodic family of Smith homomorphisms.
\end{exm}

\begin{exm}\label{spinc_Z2_Smith}
    Similarly, we may take spin$^c$ bordism of \eqref{BZ2_Smith} to get Smith homomorphisms
    \begin{equation}\label{spinc_smith_eq}
        \sm_{\sigma}\colon \Omega^{\Spinc}_n(B\Z/2) \to \Omega^{\Pin^c}_{n-1}.
    \end{equation}
    Like in the previous example, \cref{shearing} can be used to show $\Omega_n^{\Spin^c}((B\Z/2)^{\sigma-1})\cong \Omega_n^{\Pin^c}$, a theorem of Bahri--Gilkey~\cite{BG87,BG87a}. Bahri--Gilkey also showed that, if we use \emph{reduced} \spinc bordism on the left-hand side of~\eqref{spinc_smith_eq}, $e_\sigma$ is an isomorphism for all $n$.

    Meanwhile, starting with $V=\sigma$ defines a map
    \begin{equation}
        e_{\sigma}\colon (B\Z/2)^{\sigma} \to (B\Z/2)^{2\sigma}.
    \end{equation}
    For any vector bundle $V$, $V\oplus V$ has a complex structure, hence a \spinc structure. Therefore there is a Thom isomorphism $\Omega_*^{\Spin^c}((B\Z/2)^{2\sigma-2})\cong\Omega_*^{\Spin^c}(B\Z/2)$, and we get Smith homomorphisms
    \begin{equation}
        \sm_{\sigma}\colon \Omega^{\Pin^c}_n \to \Omega^{\Spinc}_{n-1}(B\Z/2).
    \end{equation}
    In general, these are not isomorphisms.
    
    Overall, we get a 2-periodic family of twisted spin$^c$ structures:
    \begin{equation}
        \Spinc\times\Z/2 ~ \rightsquigarrow ~ \Pinc \rightsquigarrow \dots
    \end{equation}
    This family of Smith homomorphisms also appears in~\cite{BG87a,hambleton_fundamental_2013,debray_long_2024,debray_smith_2024}.
\end{exm}
In the next section, we will use Smith homomorphisms to give explicit descriptions of certain twisted ABS maps. Later, in \cref{smith_symm_br}, we will explain the role that \textit{Anderson-dualized} Smith homomorphisms play in symmetry breaking, motivating our construction in \cref{spiral_maps_of_IFTs}.

\subsection{Twisted ABS maps}
\label{ss:twABS}
Freed--Hopkins~\cite[\S 9.2.2]{freed_reflection_2021} define twisted ABS maps for the ten continuous symmetry types $H(s)$ and $H^c(s)$ from \cref{disc_cont_defn} (see \cref{tenfold_table} for this notation). We start off by describing their constructions in the real case with the parameter $|s|\leq 3$.

\begin{defn}
\label{lambda_V_defn}
    Let $V\to B\O_s$ be the universal vector bundle. Using the Clifford module construction of \cref{lambda_V_class}, define
    \begin{equation}
        \lambda_V \coloneqq \chi_V([\Lambda^{\text{ev}}(V),\Lambda^{\text{odd}}(V);\sigma_V]),
    \end{equation}
    where $\sigma_V$ is the map sending $(v, w)\in V\times \Lambda^* V$ to $(v,{v\wedge w + v\mathbin \lrcorner w})$ in the other graded piece of $V\times \Lambda^* V$.
\end{defn}

\begin{rem}
    If $V$ were complex, the class $\lambda_V$ described above would give the $\KU$-theory Thom class of $V$ (see \cref{K_Thom_class}), since any complex vector bundle is also spin$^c$, and thus oriented for $\KU$-theory. However, the universal bundles over $B\O_s$ are not spin, and thus do not have Thom isomorphisms in $\KO$-theory.\footnote{For refinements to equivariant $\KO$-theory, see~\cite{DK70, FHT11, gomi_freed-moore_2021,berwick2024power}.}
\end{rem}
The classifying map of $\lambda_V$ can be described as a map $B\O_s^V \to \KO$. We may desuspend $s$ times so that the domain is the Thom spectrum of a rank-zero virtual bundle:
\begin{equation}\label{shift_lambda}
    \Sigma^{-s}\lambda_V \colon B\O_s^{V-s} \to \Sigma^{-s}\KO.
\end{equation}

Freed--Hopkins use this class to define the twisted ABS maps for $s=1,2,3$.

\begin{defn}[{\cite[Proposition 10.27]{freed_reflection_2021}}]\label{positive_s_ABS}
    Let $s=1$, $2$, or $3$, and let $V\to B\O_s$ be the universal bundle. Then $H(s)$-structures can be identified with $(B\O_s, V)$-twisted spin structures (this is implicit in Freed--Hopkins~\cite[\S 10]{freed_reflection_2021}), so by \cref{shearing}, $\mathit{MTH}(s)\simeq\MTSpin\wedge (B\O_s)^{V-s}$~\cite[(10.2)]{freed_reflection_2021}.
    
    The twisted ABS map for $\mathit{MTH}(s)$ is the composition
    \begin{equation}
        \mathit{MTH}(s)\simeq MT\Spin\wedge (B\O_s)^{V-s} \xrightarrow{{\ABS}_0^r\wedge \Sigma^{-s}\lambda_V} \KO\wedge \Sigma^{-s} \KO \to \Sigma^{-s}\KO.
    \end{equation}
\end{defn}
where $\ABS_0\colon \MTSpin \to \KO$ is the classical ABS map described in \cref{sec_classical_ABS}.

See~\cite[Proposition 10.27]{freed_reflection_2021} for a comparison to an index-theoretic definition. We will mostly focus on the cases $s =1$ in \cref{positive_s_ABS} (and $s = -1$ in \cref{negative_s_ABS}, below). For these, see also the more recent discussion in \cite{freed_index_2024}.

\begin{exm}\label{1_lam_defn}
     For $s=1$, the twisted ABS map is
     \begin{equation}
         \ABS_{s=+1}\colon MT\mathrm{Pin}^-\simeq MT\Spin\wedge (B\Z/2)^{\sigma-1} \xrightarrow{\ABS_0\wedge \lambda_{\sigma}} \Sigma^{-1}\KO. 
         \qedhere
     \end{equation}
\end{exm}

For $s<0$, there is a different construction. In the following, for $V$ the universal bundle over $B\O_{\abs s}$, we will use $e_V$ to denote the zero section map
\begin{equation}
    e_V\colon (B\O_{\abs s})^{-V} \to (B\O_{\abs s})_+.
\end{equation}
In what follows, denote by $\underline{1}$ the classifying map for the trivial line bundle, in this case over $B\O_{|s|}$.

\begin{defn}[{\cite[Proposition 10.24]{freed_reflection_2021}}]\label{negative_s_ABS}
For $s=-1$, $-2$, or $-3$, the twisted ABS map for $\mathit{MTH}(s)$ is the composition
\begin{equation}\label{minus_s_ABS_eq}
    \mathit{MTH}(s) \simeq MT\Spin\wedge (B\O_{|s|})^{|s|-V} \xrightarrow{ABS_0\wedge e_V} \KO\wedge \Sigma^{|s|} (B\O_{|s|})_+ \xrightarrow{\id\wedge \underline{1}} \Sigma^{|s|} \KO.
\end{equation}
Like for positive $s$, Freed--Hopkins~\cite[(10.2)]{freed_reflection_2021} prove $\mathit{MTH}(s) \simeq MT\Spin\wedge (B\O_{|s|})^{|s|-V}$, implicitly identifying $H(s)$-structures with $(B\O_{\abs s}, -V)$-twisted spin structures.

On homotopy, \eqref{minus_s_ABS_eq} factors as
\begin{equation}\label{minus_s_factor}
    \Omega_n^{H(s)} \xrightarrow{\sm_V} \Omega_{n+s}^\Spin(B\O_{-s}) \xrightarrow{c} \Omega^\Spin_{n+s} \xrightarrow{\ABS_0} \KO_{n+s},
\end{equation}
where $V\to B\O_{-s}$ is the universal bundle, $\sm_V$ is the Smith homomorphism, and $c$ is the map forgetting the data of the auxiliary $B\O_{-s}$-bundle.
\end{defn}
The factorization~\eqref{minus_s_factor} was used in~\cite[Appendix A]{ADKPSS24} to evaluate the map $\ABS_{-s}$ on a set of generators for $\Omega_4^{\Pin^{\tilde c+}}$.
\begin{defn}[{\cite[Proposition 10.24]{freed_reflection_2021}}]\label{spinh_ABS_map}
Let $\mathit{KSp} = \Sigma^4\KO$ denote the quaternionic $K$-theory spectrum. There is a unique homotopy class of maps $\psi\colon (B\SO_3)^{3-V}\to\mathit{KSp}$ whose composition with the map $\mathbb{HP}^1\to (B\SO_3)^{3-V}$ is in the homotopy class of the standard generator of $\widetilde{\mathit{KSp}}{}^0(\mathbb{HP}^1)$ under the diffeomorphism $\mathbb{HP}^1\cong S^4$. %

For $s = 4$, the twisted ABS map for $\mathit{MTH}(s)$ is the composition
\begin{equation}\label{spinh_ABS_eqn}
    \mathit{MTH}(4) \simeq\MTSpin\wedge (B\SO_3)^{3-V} \xrightarrow{\ABS_0 \wedge \psi} \KO\wedge\Sigma^{-4}\KO \to \Sigma^{-4}\KO.
\end{equation}
\end{defn}
The $s = 4$ twisted ABS map has also been studied by Hu~\cite[\S 6.1]{hu_invariants_2023} and Buchanan--McKean~\cite{BM23}.

\begin{exm}\label{-1_lam_defn}
    For $s=-1$, the twisted ABS map is
    \begin{equation}
        \ABS_{s=-1}\colon MT\mathrm{Pin}^+\simeq MT\Spin\wedge (B\Z/2)^{1-\sigma} \xrightarrow{\ABS_0\wedge e_\sigma} \Sigma \KO\wedge (B\Z/2)_+ \xrightarrow{\id\wedge \underline{1}} \Sigma \KO.
        \qedhere
    \end{equation}
\end{exm}

\begin{rem}\label{sign_of_s_and_ABS}
    These definitions are asymmetric: the generalized ABS maps constructed by Freed--Hopkins for $s>0$ and $s<0$ are different, and they are carefully chosen to land in the desired degrees. This in particular uses more information than just the fermionic group of symmetries.
    
    For example, for $G_f = \Pin_1^-$, corresponding by \cite[Proposition 1.26]{ADKPSS24} to \pinp structure, \cref{-1_lam_defn} used the equivalence $\MTPin^+\simeq \MTSpin\wedge (B\Z/2)^{1-\sigma}$. Using the Whitney sum formula, one can show that $4\sigma-4$ is spin, so that (see, e.g., \cite[Theorem 1.39]{debray_invertible_2021} or \cite[\S 6]{debray_long_2024}) there is also an $\MTSpin$-module equivalence $\MTPin^+\simeq\MTSpin\wedge (B\Z/2)^{3\sigma-3}$.\footnote{In fact, this is how Stolz~\cite[\S 8]{stolz_exotic_1988} first decomposed $\MTPin^+$; the $1-\sigma$ description came later.}
    Thus we could have defined a map
    \begin{equation}\label{wrong_pinp}
        MT\Spin \wedge (B\Z/2)^{3\sigma-3} \xrightarrow{\ABS_0\wedge \lambda_{3\sigma}} \Sigma^{-3} \KO
    \end{equation}
    but this choice of target is not appropriate
    for our free fermion theories of interest---we land in class CII, whereas Freed--Hopkins place \pinp structures in class DIII~\cite[(9.2.5)]{freed_reflection_2021}. The corresponding free fermion phases have different classifications in most dimensions, and therefore the choice~\eqref{wrong_pinp} would contradict the established classification of these free fermion phases.
\end{rem}

\begin{exm}
The Klein bottle $K$ has four \pinp structures; two of them are bounding and two are nonbounding, and either of the two nonbounding ones generates $\Omega_2^{\Pin^+}\cong\Z/2$~\cite[Proposition 3.9]{KT90}. In this example, we compute $\ABS_1(K,\mathfrak p)$ for each \pinp structure $\mathfrak p$ of $K$.

The Klein bottle is a circle bundle over the circle, explicitly described as the quotient of $S^1\times [0,1]$ by the reflection $(z,0)\sim (\overline z, 1)$. If $t$ denotes the coordinate on $[0,1]$, then the vector field $\partial_t$ on $S^1\times [0,1]$ descends to $K$ and is nonvanishing, so $TK\cong \sigma\oplus\underline{\R}$ for a real line bundle $\sigma\to K$, which is the pullback of the Möbius bundle on the base $S^1$. Thus $w_1(K)$ is the pullback of the top-degree class on the base, meaning that the inclusion $j\colon S^1\to K$ of any fiber is a representative of the Poincaré dual of $w_1(K)$, which can be identified with the intersection of a generic section and the zero section. The vector field $\partial_t$ also trivializes the normal bundle to $j$, so any choice of \pinp structure $\mathfrak p$ on $K$ induces a \pinp structure on the fiber, hence (after choosing once and for all an orientation on the fiber) a spin structure $\mathfrak s_{\mathfrak p}$.

The composition
\begin{equation}
\label{most_of_twisted_ABS_pinp}
    \Omega_2^{\Pin^+} \xrightarrow{\sm_\sigma} \Omega_1^\Spin(B\Z/2) \xrightarrow{c} \Omega_1^\Spin,
\end{equation}
in which the map $c$ forgets the principal $\Z/2$-bundle, sends $(K, \mathfrak p)\mapsto (S^1, \mathfrak s_{\mathfrak p})$. Kirby-Taylor (\textit{ibid.}) show that $\mathfrak s_{\mathfrak p}$ is nonbounding if and only if $\mathfrak p$ is: \eqref{most_of_twisted_ABS_pinp} is an isomorphism $\Z/2\xrightarrow{\cong}\Z/2$.

The last step in the computation of the twisted ABS map is to apply $\ABS_0$, which is an isomorphism $\Omega_1^\Spin\to\KO_1$. That is:
\begin{itemize}
    \item If $\mathfrak p$ is one of the two nonbounding \pinp structures on $K$, $\ABS_1(K,\mathfrak p) = \ABS_0(S_{\mathit{nb}}^1) = \eta$, the nonzero element of $\KO_1\cong\Z/2$.
    \item If $\mathfrak p$ is one of the two bounding structures on $K$, $\ABS_1(K,\mathfrak p) = \ABS_0(S_b^1) = 0$.
\end{itemize}
Since $\Omega_3^{\Pin^+}$ is torsion~\cite{Gia73}, this also implies that the Anderson dual of $\ABS_1$ is an isomorphism in degree $2$, as originally shown by Freed--Hopkins~\cite[Corollary 9.83]{freed_reflection_2021}. In light of \cref{LEFT_ansatz}, we interpret this isomorphism as saying that the nontrivial free phase in one spatial dimension, the class DIII superconductor, survives the incorporation of interactions and generates the group of interacting phases.
\end{exm}

\subsection{Generalization for Spin-\texorpdfstring{$({\ell,k})$}{(l, k)}}\label{subsec_generalized_ABS}
We now generalize the ABS orientation to assign $\KO$- and $\KU$-classes to certain twisted spin manifolds, including spin-$(\ell,k)$ and spin$^c$-$(\ell,k)$ manifolds (see \cref{MElk_defn}). This will be the main ingredient in the generalization of the free-to-interacting map to spin-$(\ell,k)$ and \spinc-$(\ell,k)$ theories.
Our generalization combines the techniques for the twisted ABS maps discussed above in \cref{positive_s_ABS,negative_s_ABS}, i.e.\ the cases $s\ne 4$.

\begin{defn}\label{defn_generalized_ABS}
    Let $G_1$ and $G_2$ be compact Lie groups. Let $V_1\to BG_1$ and $V_2\to BG_2$ be vector bundles of ranks $r_1$, resp.\ $r_2$. 
    The $(V_1,V_2)$-\textit{twisted real ABS map} is the composition
    \begin{equation}
\label{generalized_ABS_defn_actual}
    \begin{tikzcd}
        MT\Spin\wedge (BG_1)^{V_1-r_1}\wedge (BG_2)^{\,r_2-V_2} \ar[d,"\ABS_0^r\wedge \lambda_{V_1}\wedge  e_{V_2}"] \\
        \KO\wedge \Sigma^{-r_1}\KO \wedge \Sigma^{r_2}(BG_2)_+ \ar[d,"\id \wedge \underline{1}"] \\
        \KO\wedge \Sigma^{-r_1}\KO\wedge \Sigma^{r_2} \KO \ar[d,"\mu"] \\
        \Sigma^{r_2-r_1} \KO,
    \end{tikzcd}
\end{equation}
where $\mu$ is the multiplication induced by the ring spectrum structure on $\KO$.

Analogously, we define the $(V_1,V_2)$\textit{-twisted complex ABS map} to be the composition
\begin{equation}
\label{generalized_ABS_defn_complex}
    \begin{tikzcd}
        MT\Spin^c\wedge (BG_1)^{V_1-r_1}\wedge (BG_2)^{\,r_2-V_2} \ar[d,"\ABS_0^c\wedge \lambda_{V_1}\wedge e_{V_2}"] \\
        \KU\wedge \Sigma^{-r_1}\KU \wedge \Sigma^{r_2}(BG_2)_+ \ar[d,"\id\wedge \underline{1}"] \\
        \KU\wedge \Sigma^{-r_1}\KU\wedge \Sigma^{r_2} \KU \ar[d,"\mu"] \\
        \Sigma^{r_2-r_1} \KU.
    \end{tikzcd}
\end{equation}
\end{defn}

\begin{rem}
    We refer to these maps as twisted ABS \textit{maps} rather than a twisted ABS \textit{orientations} in order to follow the usual convention that an orientation is multiplicative, i.e.\ a map of ring spectra. 
    For $X$ a spectrum, the spectrum $\MTSpin\wedge X$ need not be a ring spectrum, but only an $\MTSpin$-module spectrum. For example, when $s\ne 0$, one can show that $\mathit{MTH}(s)$ is not a ring spectrum.
\end{rem}

\begin{rem}\label{localization}
    
    Just as the twisted ABS map \cref{negative_s_ABS} factors through $e_V$, as in~\eqref{minus_s_factor}, the $(V_1,V_2)$-twisted ABS maps factor through the composition
    \begin{equation}\label{localization_eqn}
    \begin{tikzcd}
        \Omega_n^\Spin((BG_1)^{V_1-r_1}\wedge (BG_2)^{\,r_2-V_2}) \ar[r,"\sm_{V_2}"] & \Omega_{n-r_2}^\Spin(BG_1^{V_1-r_1}\wedge BG_2) \ar[r,"c"] & \Omega_{n-r_2}^\Spin((BG_1)^{V_1-r_1})
    \end{tikzcd}
    \end{equation}
    where $\sm_{V_2}$ is a Smith homomorphism, which cuts out a submanifold, and $c$ is the crush map, which forgets the $G_2$-bundle. There is a similar factorization in the complex case.

    We may interpret this factorization as a localization of the twisted index to a codimension-$r_2$, $(BG_1,V_1)$-twisted spin submanifold. In this way, we recover some results of Fast--Ochanine \cite{fast_ko_2004},
    Hayashi \cite{hayashi_localization_2014}, and Miyazawa \cite{miyazawa_localization_2025} for various choices of $G_1$, $G_2$, $V_1$, and $V_2$.
\end{rem}
\begin{lem}[{\cite[Lemma 2.3]{Ati61a}}]
\label{thomsmash}
There is a homotopy equivalence~
\begin{equation}\label{external_internal_products}
    (BG_1)^{V_1-r_1}\wedge (BG_2)^{\,r_2-V_2} \simeq 
    B(G_1\times G_2)^{(V_1-r_1)\boxplus(r_2-V_2)}.
\end{equation}
\end{lem}
For our application to free-to-interacting maps, the choice of $G_1,G_2$ and the pair of bundles will be determined by the symmetries of the system. In this work, we will elaborate further only for the special case of the Bott spiral. In this case, we model the symmetries of the relevant SPTs using spin-$(\ell,k)$-structures, so we specialize to $G_1 = (\Z/2)^\ell$, $G_2 = (\Z/2)^k$, $V_1 = \sigma^{\boxplus \ell}$, and $V_2 = \sigma^{\boxplus k}$. Recall the spectrum $\ME_{\ell,k}$ from \cref{MElk_defn}; then, by \cref{thomsmash}, $(BG_1)^{V_1-r_1}\wedge (BG_2)^{\,r_2-V_2}\simeq\ME_{\ell,k}$.

\begin{defn}\label{defn:gen_ABS}
    The twisted ABS map $\ABS_{\ell,k}$ is the composition
\begin{equation}
\label{generalized_ABS_defn_spin_lk}
    \begin{tikzcd}
        MT\Spin\wedge \ME_{\ell,k} \ar["{\eqref{MElk_defn}}", r,equals] &
        MT\Spin\wedge ((B\Z/2)^{\sigma-1})^{\wedge \ell}\wedge ((B\Z/2)^{1-\sigma})^{\wedge k} \ar[d,"\ABS_0\wedge \lambda_\sigma^{\wedge \ell}\wedge e_\sigma^{\wedge k}"] \\
        & \KO\wedge  (\Sigma^{-1}\KO)^{\wedge \ell}\wedge (\Sigma(B\Z/2)_+)^{\wedge k} \ar[d,"\id \wedge \underline{1}"] \\
        & \KO\wedge \Sigma^{-\ell}\KO\wedge \Sigma^k \KO \ar[d,"\mu"] \\
        & \Sigma^{k-\ell} \KO.
    \end{tikzcd}
\end{equation}
\end{defn}

\begin{rem}
\label{rem:same_map}
By definition, $\ABS_{1,0}$ (resp.\ $\ABS_{0,1}$) from \cref{defn:gen_ABS} coincides with Freed--Hopkins' twisted ABS map in the case $s = 1$ (resp.\ $s = -1$): see \cref{1_lam_defn,-1_lam_defn}.
\end{rem}

\begin{exm}\label{ABS11_ex}
    Return to the case of $\MTSpin\wedge \ME_{1,1} \simeq MT\mathrm{DPin}$. Then $\ABS_{1,1}$ is the composition
\begin{equation}
    \begin{tikzcd}
        MT\mathrm{DPin} \ar[r,phantom,"\simeq"] & MT\Spin\wedge (B\Z/2)^{\sigma-1}\wedge (B\Z/2)^{1-\sigma} \ar[d,"\ABS_0\wedge \lambda_\sigma\wedge e_\sigma"] \\
        & \KO\wedge \Sigma^{-1}\KO\wedge \Sigma(B\Z/2)_+ \ar[d,"\id\wedge \underline{1}"] \\
        & \KO\wedge \Sigma^{-1}\KO\wedge \Sigma \KO \ar[d,"\mu"] \\
        & \KO.
    \end{tikzcd}
\end{equation}

    On homotopy, we get a map 
    \begin{equation}
        \ABS_{1,1} \colon \Omega^{E_{1,1}}_d \longrightarrow \KO_{d}.
    \end{equation}
    For degree reasons, this is the zero map except for $d \equiv 0,1,2,4\bmod 8$. When $d\equiv 0\bmod 4$, $\KO_d\cong\Z$ but $\Omega_d^{E_{1,1}}$ is torsion, by \cref{spinbord_torsion}, so the map is also $0$. This leaves $d\equiv 1,2\bmod 8$.

    For $d = 1$, we have $\Omega_1^{E_{1,1}}\cong\Z/2$~\cite[Theorem F.1]{kaidi_topological_2020}. By \cref{MElk_defn}, an $E_{1,1}$-structure on a manifold $M$ is a pair of line bundles $L_1,L_2\to M$ and a spin structure on $TM + L_1 - L_2$; thus $\RP^1$ with $L_1 = L_2 = \sigma$ admits an $E_{1,1}$ structure, and this represents the nonzero element of $\Omega_1^{E_{1,1}}$, as it is detected by the bordism invariant $\int w_1(L_1)$. The Smith homomorphism sends $\RP^1\mapsto\RP^0$ with some $\Z/2$-bundle, and $c$ forgets the bundle, giving us $\RP^0\cong\pt$, the nonzero element of $\Omega_0^{\Pin^-}$. Freed--Hopkins~\cite[Corollary 9.85]{freed_reflection_2021} showed that $\Sigma I_\Z$ of $\ABS_{s = 1}$ is an isomorphism in degree $0$, and since the domain and codomain of $\ABS_{s = 1}$ are both torsion, this implies $\ABS_{s =1}$ is an isomorphism in degree $0$. Thus $\ABS_{1,1}$ sends $\RP^1\mapsto \eta\in\KO_1$.

    Because $\ABS_{1,1}$ is a map of $\MTSpin$-modules, it commutes with the action of any element of $\Omega_*^\Spin$. Using this, we can immediately conclude that in all degrees $d\equiv 1,2\bmod 8$, $d>0$, $\ABS_{1,1}$ is surjective.
    \begin{itemize}
        \item A \term{Bott manifold} is a closed spin $8$-manifold whose image under the Atiyah--Bott--Shapiro map is the Bott class $\beta\in\KO_8$. Freed--Hopkins~\cite[\S 5.3]{FH21b} construct a Bott manifold $B$.
        Thus if $d = 8k+1$, $\ABS_{1,1}(B^k\times\RP^1) = \beta^k \eta$, which is the unique nonzero element of $\KO_d$.
        \item The same argument works in degrees $8k+2$ as soon as we show that $\ABS_{1,1}$ is surjective in degree $2$. For this, we use that $\ABS_{1,1}$ must commute with $\eta\in\Omega_1^\Spin$. The nonbounding spin circle $S_{\mathit{nb}}^1$ represents $\eta$ in $\Omega_1^\Spin$, so this tells us $\ABS_{1,1}$ sends $\RP^1\times S_{\mathit{nb}}^1$ to $\eta^2$, which is the unique nonzero element of $\KO_2$---and likewise, $B^k\times\RP^1\times S_{\mathit{nb}}^1\mapsto \beta^k\eta^2$, the unique nonzero element of $\KO_{8k+2}$.
    \end{itemize}
    One can show that $\RP^2$ has an $E_{1,1}$ structure with which it is linearly independent to $\RP^1\times S_{\mathit{nb}}^1$ in $\Omega_2^{E_{1,1}}\cong\Z/2\oplus\Z/2$, given by the line bundles $(\sigma, \underline{\R})$. Said differently, there are two $E_{1,1}$ structures on $\RP^2$ induced from its two \pinm structures. Choose either of these two $E_{1,1}$ structures. We will see that the Smith homomorphism kills the bordism class of this manifold, implying $\ABS_{1,1}(\RP^2) = 0$.
    
    To compute the Smith homomorphism $\sm_\sigma\colon \Omega^{E_{1,1}}_2 \to \Omega_1^{\Pinm}(B\Z/2)$, we may use the Smith long exact sequence of bordism groups \cite{debray_smith_2024}; specifically, per \cite[(7.1)]{debray_smith_2024} with $k=-1$, there is a fiber sequence
    \begin{equation}
        \mathbb{S} \to (B\Z/2)^{1-\sigma} \xrightarrow{e_\sigma} \Sigma(B\Z/2),
    \end{equation}
    which, after smashing with $MT\mathrm{Spin}\wedge (B\Z/2)^{\sigma-1}\simeq MT\Pinm$, becomes
    \begin{equation}
        MT\Pinm \to MT\Spin\wedge ME_{1,1} \xrightarrow{\sm_\sigma} MT\Pinm\wedge \Sigma(B\Z/2).
    \end{equation}
    Taking homotopy groups yields a long exact sequence of bordism groups. One can work out the groups and homomorphisms in this long exact sequence in low degrees, using the $E_{1,1}$ bordism groups computed in~\cite[Appendix F]{kaidi_topological_2020} and the groups $\Omega_*^{\Pin^-}(B\Z/2)$ computed in low degrees in~\cite[Theorem 3.2.1]{Her17} and \cite[Theorem 16]{guo_fermionic_2020}; we leave this as an exercise for the reader. However, for our goal of computing on $\RP^2$, we do not need to do this---by exactness, if the $E_{1,1}$ structure on a closed $d$-manifold $M$ was induced from a \pinm structure, then $\mathrm{sm}_\sigma(M) = 0\in\Omega_{d-1}^{\Pin^-}(B\Z/2)$. Since this is how we constructed the $E_{1,1}$ structure on $\RP^2$, we conclude its image under the Smith homomorphism---and therefore also under $\ABS_{1,1}$---is $0$.
\end{exm}

Now we turn to \spinc-$(\ell,k)$ structures. By \cref{spinc_simplification}, in these cases the two indices $\ell,k$ collapse to just $m \coloneqq\ell+k$, so it suffices to consider twisted ABS maps with domain $\MTSpin^c\wedge\ME_m$, where $\ME_m = ((B\Z/2)^{\sigma-1})^{\wedge m}$ (\cref{MEm_defn}).

\begin{defn}\label{defn:gen_ABS_cplx}
    The twisted ABS map $\ABS^c_{m}$ is the composition
\begin{equation}
\label{generalized_ABS_defn_Bott_spiral_cplx}
    \begin{tikzcd}
        MT\Spin^c\wedge ((B\Z/2)^{\sigma-1})^{\wedge m}\ar[d,"\ABS_0^c\wedge \lambda_\sigma^{\wedge m}"] \\
        \KU\wedge \Sigma^{-m}\KU \ar[d,"\mu"] \\
        \Sigma^{-m} \KU.
    \end{tikzcd}
\end{equation}
\end{defn}
At this point, we have defined twisted ABS maps for all continuous symmetry types and all discrete symmetry types from \cref{disc_cont_defn}, except for the class C discrete case. We will use the continuous ABS map defined by Freed--Hopkins (\cref{spinh_ABS_map}) to define the discrete analog.
\begin{defn}
\label{Q8_twABS}
Let $\rho\colon Q_8\inj\SU_2$ be the map of fermionic groups induced by the inclusion of $\{\pm 1, \pm i, \pm j, \pm k\}$ into the norm-one quaternions, where both fermionic groups have trivial map to $\Z/2$ and $(-1)^F = -1$. Then, the discrete class C twisted ABS map is the composition
\begin{equation}
    \ABS^\delta_{s=4}\colon \mathit{MT}(\Spin\times_{\set{\pm 1}} Q_8) \xrightarrow{\rho} \MTSpin^h \xrightarrow{\eqref{spinh_ABS_eqn}} \Sigma^{-4}\KO.
\end{equation}
\end{defn}
\begin{rem}
\label{ABS_through_ko}
Because the maps $\ABS_{\ell,k}$ defined in \cref{defn:gen_ABS} are $\MTSpin$-module maps, and the codomain $\Sigma^{k-\ell}\KO$ acquires its $\MTSpin$-module structure through a sequence of $E_\infty$-ring maps $\ABS_0\colon \MTSpin\to\ko$ followed by $\ko\to\KO$ (i.e.\ the ABS map factors through $\ko$), the maps $\ABS_{\ell,k}$ naturally factor as
\begin{equation}
    \ABS_{\ell,k}\colon \MTSpin\wedge\ME_{\ell,k} \xrightarrow{\ABS_0\wedge_{\MTSpin}\id} \ko\wedge\ME_{\ell,k} \xrightarrow{abs_{\ell,k}} \Sigma^{k-\ell}\KO,
\end{equation}
where the maps $\ABS_{\ell,k}$ are $\ko$-module maps. In the same way, the map $\ABS_m^c$ from \cref{defn:gen_ABS_cplx} factors through a $\ku$-module map $\ku\wedge ((B\Z/2)^{\sigma-1})^m\to\Sigma^{-m}\KU$. %
\end{rem}

\subsection{Free-to-interacting maps from ABS maps}
\label{ss:F2I}

To get from ABS maps to maps between groups of free fermion theories and deformation classes of invertible field theories, we Anderson-dualize. Recall from \cref{anddual} that the Anderson dual to the sphere spectrum is the spectrum $I_\Z$ defined as the fiber of the \emph{exponential map} $I_\C \to I_{\C^\times}$. By \cref{IFT_class}, invertible field theories are classified by Anderson-dual $\xi$-bordism: maps from $MT\xi$ to a shift of $I_\Z$.
Meanwhile, free fermion theories are classified by $K$-theory, as discussed in \cref{ff_and_k} for the case of finite group superalgebra symmetries. This classification may be recast in terms of Anderson-dual $K$-theory using the Anderson self-duality of $K$-theory:

\begin{thm}[{\cite[Theorem
4.16]{And69}}]\label{and_KO}
    There are equivalences $\varphi\colon I_\Z \KO \xrightarrow{\simeq} \Sigma^4 \KO$ and $\varphi^c\colon I_\Z \KU \xrightarrow{\simeq} \KU$ of $\KO$, resp.\ $\KU$-modules.
\end{thm}
Because $-1\colon\KO\to\KO$ is a $\KO$-module equivalence not homotopic to the identity, $\varphi$ and $\varphi^c$ are not unique. Choose one of each.\footnote{It is possible to choose $\varphi$ and $\varphi^c$ to be compatible with respect to the realification and complexification maps between $\KO$ and $\KU$, which follows from the fact that \cref{and_KO} refines to a shifted Anderson self-duality of $\mathit{KR}$~\cite[Theorem 8.2]{heard_k_2014}.} The specific choice is not important for what follows, but we need to make \emph{a} choice.

For proofs of \cref{and_KO}, see Anderson's unpublished notes (\textit{ibid.}), or any of Yosimura~\cite[Theorem 4]{Yos75}, Freed--Moore--Segal~\cite[Proposition B.11]{FMS07a}, Heard--Stojanoska~\cite[Theorem 8.1]{heard_k_2014}, Ricka~\cite[Corollary
5.8]{Ric16}, Greenlees--Meier~\cite[Example 1.5]{greenlees_gorenstein_2017}, Greenlees--Stojanoska~\cite[Example 6.1]{greenlees_anderson_2018}, or Hebestreit--Land--Nikolaus~\cite[Example 2.8]{HLN20}. (Some of these proofs only show an equivalence of spectra.) Heard--Stojanoska~\cite[Theorem 8.2]{heard_k_2014} lifted \cref{and_KO} to a proof of shifted Anderson self-duality for $\KR$-theory, with additional proofs given by Ricka~\cite[Theorem 5.4]{Ric16} and Greenlees--Meier~\cite[Example 1.5]{greenlees_gorenstein_2017}.

\begin{exm}
    In our case of interest for the Bott spiral, free theories in dimension $d$ with $\Cl_{\ell,k}$-symmetry are classified by the group $K_{2-d}(\Cl_{\ell,k}) \cong \KO^{d+\ell-k-2}(\pt)$ (see \cref{Elk_free_phases}). There is an identification $I_\Z \Sigma^{-d-\ell+k+2} \KO\simeq \Sigma^{d+\ell-k+2}\KO$.    
\end{exm}

We may express Anderson self-duality as a map of spectra called the Pfaffian.
\begin{defn}[{\cite[\S 9.2.5]{freed_reflection_2021}, \cite[\S 11.3]{Fre19}}]\label{dualizing_class}
    Consider the map $u\colon\Sph\to \KO$ representing $1\in \pi_0(\KO)$. By applying the Anderson self-duality $\varphi$ from \cref{and_KO}, $u$ induces a map $u^\vee\colon \KO\to\Sigma^{-4}I_\Z$.
    The \term{real Pfaffian} is the map $\mathrm{Pfaff}\colon \KO \to \Sigma^4 I_\Z$ obtained from $\Sigma^{8}(u^\vee)\colon\Sigma^8\KO\to\Sigma^4 I_\Z$ by Bott periodicity.%
\footnote{Freed--Hopkins define their map in differential cohomology. This is important for non-topological invertible field theories: see Freed~\cite[Lecture 9]{Fre19}. We will not need this level of generality in this paper, hence work with ordinary generalized cohomology.}
\end{defn}
The Pfaffian is so-called because for a Dirac operator on a spin manifold, this map conjecturally gives the determinant line or Pfaffian line of the Dirac operator.
See \cite[\S 9.2.5]{freed_reflection_2021} for further physical discussion.

The free-to-interacting maps for the tenfold way, proposed by Freed--Hopkins, are the following.
\begin{defn}[{Freed--Hopkins~\cite[Conjecture 9.70]{freed_reflection_2021}}]\label{FH_F2I_maps}
    Let $\ABS_s\colon \mathit{MTH}(s) \to \Sigma^{-s}\KO$ be the twisted ABS map of~\eqref{positive_s_ABS}, \eqref{negative_s_ABS}, or \eqref{spinh_ABS_eqn}, depending on $s$. The free-to-interacting map for symmetry type $H(s)$ is the dualized map
    \begin{equation}
    \label{FH_f2i_map}
        \Sigma^{s-2} \KO\xrightarrow[\eqref{and_KO}]{\simeq}
        \Sigma^2 I_\Z (\Sigma^{-s} \KO) \xrightarrow{I_\Z \ABS_s} \Sigma^2 I_\Z \mathit{MTH}(s).
    \end{equation}
    Equivalently, for a fixed degree $d$, thought of as the spatial dimension, the free-to-interacting map is the map sending a free theory $x\in \KO^{d+s-2}(\pt)$ to the class of the composition
    \begin{equation}
        \mathit{MTH}(s) \xrightarrow{\ABS_s \wedge x} \Sigma^{-s} \KO \wedge \Sigma^{d+s-2}\KO \xrightarrow{ \mu} \Sigma^{d-2} \KO \xrightarrow{\mathrm{Pfaff}} \Sigma^{d+2} I_\Z
    \end{equation}
    in $[MTH(s),\Sigma^{d+2} I_\Z]$.
\end{defn}
\begin{ansatz}[{Freed--Hopkins~\cite[\S 9.2.6]{freed_reflection_2021}}]
\label{FH_F2I_ansatz}
\hfill
\begin{enumerate}
    \item Under the identifications in \cref{neutralluuk_main_thm_citation,IFT_class} identifying the groups of $d$-dimensional free fermion Hamiltonians of Altland--Zirnbauer type $s$, resp.\ reflection-positive IFTs on $H(s)$-manifolds with $\KO^{d+s-2}$, resp.\ $\mho_{H(s)}^{d+2}$, \eqref{FH_f2i_map} is the homomorphism assigning to a free fermion Hamiltonian its low-energy effective field theory.
    \item The same is true mutatis mutandis with $H(\U_1\ftens K_f(s))$ in place of $H(s)$, $\KU$ in place of $\KO$, and Freed--Hopkins' complex analog of~\eqref{FH_f2i_map}, written explicitly in~\cite[(1.57)]{ADKPSS24}, in place of~\eqref{FH_f2i_map}.
\end{enumerate}
\end{ansatz}
Freed--Hopkins~\cite[\S 9.3]{freed_reflection_2021} provide computational evidence supporting \cref{FH_F2I_ansatz}. \cite[Ansatz 1.73, \S 3]{ADKPSS24} generalizes \cref{FH_F2I_ansatz} to weak symmetry-protected topological phases and provides additional supporting evidence.
\begin{exm}\label{F2I_TRS_Majorana}
    As we discussed in \cref{TRS_Maj_exm_K-theory}, the time-reversal-symmetric Majorana chain~\cite{kitaev_unpaired_2001} is a $1$-dimensional free fermion phase with an $E_{1,0}$ symmetry, i.e.\ fermion parity and a time-reversal symmetry squaring to $1$. It is nontorsion as a free fermion phase with respect to stacking. Fidkowski--Kitaev~\cite{fidkowski_effects_2010, fidkowski_topological_2011}\footnote{See also~\cite{turner_topological_2011, GW14, you_topological_2014, kapustin_fermionic_2015, witten_fermion_2016, debray_arf-brown_2018, kobayashi_pin_2019, inamura_nonlocal_2020, Tur20, freed_reflection_2021, turzillo_duality_2024} for additional derivations of this or closely related facts.} argued that, if we allow deformations by Hamiltonians with quartic terms, eight copies of this phase can be deformed to the trivial phase, thus giving the first example of a free fermion phase whose order as a free versus as an interacting phase differs.

    To support \cref{FH_F2I_ansatz}, we would like for the corresponding free-to-interacting map to be a surjective map $\Z\to\Z/8$.
    
    Mathematically, this map is modeled by the case $s=1$ of \cref{FH_F2I_maps} on homotopy in degree $1$:
    \begin{equation}
        \KO^0 \cong \Z \xrightarrow{I_\Z \ABS_{+1}} \mho^2_{\Pinm} \cong \Z/8.
    \end{equation}
    Freed--Hopkins computed \cite[Corollary 9.83]{freed_reflection_2021} that this map is a reduction modulo eight, reproducing the physical computation.
    The low-energy TFT describing the interacting Majorana chain is the Arf--Brown--Kervaire IFT of \cref{pinm_IFT_exm}, and the deformation class of this IFT generates the group $\mho^2_{\Pinm}$~\cite{debray_arf-brown_2018, freed_reflection_2021}.
\end{exm}

\begin{exm}
    Consider the case where $n=4$ and $s=-2$, where the ABS map has domain $\Omega_4^{\Pin^{\tilde c+}}$ and codomain $\KO_2$. Theories on manifolds with pin$^{\tilde c+}$ structures model physical theories with both a time reversal symmetry and particle number conservation, and include the famous topological insulators studied by Fu--Kane \cite{fu_topological_2007} and Fu--Kane--Mele \cite{fu_topological_2007-2}. Freed--Hopkins~\cite[Theorem 9.87]{freed_reflection_2021} computed the free-to-interacting maps in low dimensions for this symmetry group.
    With Antolín Camarena, Pacheco-Tallaj, and Sheinbaum, we computed the ABS map in degree four on the level of manifold generators in \cite[Appendix A]{ADKPSS24}, with an application toward free-to-interacting maps for theories with discrete translation symmetry. 
    
    We found that the map
    \begin{equation}
        \Omega_4^{\Pin^{\tilde c+}}\cong (\Z/2)^2 \xrightarrow{\ABS_{-2}} \KO_2\cong \Z/2
    \end{equation}
    sends the bordism generators $\RP^4$ and $\CP^2$ with appropriate pin$^{\tilde c+}$ structures to $0\in \KO_2$, while the class of $\CP^1\times \CP^1$ with pin$^{\tilde c+}$ structure induced from its complex structure is sent to the generator of $\KO_2$, which is dual to $\eta^2\in \KO^{-2}$. 
    This means that the invertible field theory that arises from the unique nontrivial free theory has a nontrivial partition function when evaluated on the spacetime manifold $\CP^1\times\CP^1$.
\end{exm}

Equipped with the generalized ABS map \eqref{defn:gen_ABS}, we may now define the free-to-interacting map for the symmetry types we built from the fermionic groups $E_{\ell,k}$ from \cref{elk_definition}.

\begin{defn}\label{Elk_F2I}
The free-to-interacting map of type $(\ell,k)$ is
\begin{equation}\label{eqn:Elk_F2I}
    F2I_{\ell,k} \coloneqq 
    I_\Z\ABS_{\ell,k}\colon 
    \Sigma^{-k+\ell-2}\KO \simeq \Sigma^2 I_\Z( \Sigma^{k-\ell} \KO) \xrightarrow{I_\Z \ABS_{\ell,k}} \Sigma^2 I_\Z (MT\Spin\wedge \ME_{\ell,k}).
\end{equation}
\end{defn}
By \cref{elk_alg,K-theory_of_Elk,lem_shearing_Elk}, the domain of $F2I_{\ell,k}$ is $\KO$-module equivalent to $K(\R^f[E_{\ell,k}])$ and the codomain is $\MTSpin$-module equivalent to $\mathit{MTH}(E_{\ell,k})$. Therefore, in light of \cref{FH_F2I_ansatz}, we interpret $F2I_{\ell,k}$ as modeling the map sending a free fermion theory with $E_{\ell,k}$ symmetry to its low-energy effective field theory: see \cref{our_F2I_ansatz} below. A similar consideration applies to the type-$m$ free-to-interacting maps below.

On homotopy, this is the map
    \begin{equation}\label{F2I_Elk_on_pi*}
        F2I_{\ell,k}\colon \KO^{d+\ell-k-2}(\pt) \to [MT\Spin\wedge\ME_{\ell,k}, \Sigma^{d+2}I_\Z]
    \end{equation}
    taking 
    $x\in KO^{d+\ell-k-2}(\pt)$ to the homotopy class of the composition
    \begin{equation}
\label{general_F2I_defn}
\begin{tikzcd}[column sep=0.5em]
     MT\Spin\wedge ((B\Z/2)^{\sigma-1})^{\wedge \ell}\wedge ((B\Z/2)^{1-\sigma})^{\wedge k} \ar[r,phantom,"\simeq"] & MT\Spin\wedge ((B\Z/2)^{\sigma-1})^{\wedge \ell}\wedge ((B\Z/2)^{1-\sigma})^{\wedge k} \wedge \mathbb{S} \ar[d,"\ABS_{\ell,k}\wedge x"] \\
     & \Sigma^{k-\ell}\KO \wedge \Sigma^{d+\ell-k-2} \KO
     \ar[d,"\text{mult}"] \\
     & \Sigma^{d-2} \KO
     \ar[d,"\text{Pfaff}"] \\
     & \Sigma^{d+2} I_\Z.
\end{tikzcd}
\end{equation}
We will state the physical interpretation of this map in \cref{our_F2I_ansatz}.

\begin{rem}
\label{phi_defn_remark}
We compute these free-to-interacting maps in \cref{only_first_summand_real}.
The result will ultimately be determined by the restriction of the twisted ABS map $\ABS_{\ell,k}$ to the $\ME_{\ell,k}$ factor.
\end{rem}

For the complex cases, we dualize the complex $m$-twisted ABS map $\ABS^c_m$ of \cref{defn:gen_ABS_cplx}. Bott periodicity implies the fourfold shift in \cref{complex_F2I_map_defn} below is unnecessary, just as in \cref{and_KO}; we maintain this shift to give a unified account in the real and complex cases.

\begin{defn}\label{complex_F2I_map_defn}
    The free-to-interacting map
    of type $m$ is %
    \begin{equation}
        F2I_m \coloneqq I_\Z \ABS_m^c\colon \Sigma^{m-2}\KU \xrightarrow[\eqref{and_KO}]{\simeq} \Sigma^2 I_\Z(\Sigma^{-m} \KU) \xrightarrow{I_\Z \ABS_m^c}  \Sigma^2 I_\Z (\MTSpin^c\wedge \ME_m).
    \end{equation}
\end{defn}

On homotopy groups, this is the map
\begin{equation}
    \KU^{d+m-2}(\pt) \to [\MTSpin^c\wedge \ME_m, \Sigma^{d+2} I_\Z]
\end{equation}
sending $x\in \KU^{d+m-2}$ to the homotopy class of the composition
\begin{equation}
    \begin{tikzcd}
        \MTSpin^c\wedge \ME_m \wedge \mathbb{S} \ar[d, "\ABS_m^c\wedge x"] \\
        \Sigma^{-m} \KU \wedge \Sigma^{d+m-2} \KU \ar[d,"\mu"] \\
        \Sigma^{d-2} \KU \ar[d,"\mathrm{Pfaff}"] \\
        \Sigma^{d+2} I_\Z.
    \end{tikzcd}
\end{equation}
\begin{ansatz}
\label{our_F2I_ansatz}
\hfill
\begin{enumerate}
    \item Under the identifications in \cref{neutralluuk_main_thm_citation,IFT_class} identifying the groups of $E_{\ell,k}$-symmetric $d$-dimensional free fermion Hamiltonians with $\KO^{d+s-2}$ and reflection-positive IFTs on spin-$(\ell,k)$-manifolds with $\mho_{\ell,k}^{d+2}$, resp., the map $F2I_{\ell,k}$ from \eqref{eqn:Elk_F2I} is the homomorphism assigning to a free fermion Hamiltonian its low-energy effective field theory.
    \item The same is true mutatis mutandis with $\spinc$-$(\ell,k)$ in place of spin-$(\ell,k)$, $\mho_{\Spin^c\text{-}(\ell,k)}^{d+2}$ in place of $\mho_{\ell,k}^{d+2}$, $\KU$ in place of $\KO$, and the map $F2I_m^c$ from \cref{complex_F2I_map_defn} in place of $F2I_{\ell,k}$, where $m = \ell+k$.
\end{enumerate}
\end{ansatz}

\begin{rem}
\label{Q8_F2I}
As usual, discrete class C must be handled separately. Apply Anderson duality to the map $\ABS_{s = 4}^\delta$ from \cref{Q8_twABS} to obtain a map $F2I_{s=4}^\delta\colon\Sigma^{2}\KO\to \Sigma^2 I_\Z(\mathit{MT}(\Spin\times_{\set{\pm 1}}Q_8))$. Following \cref{our_F2I_ansatz} we interpret $F2I_{s=4}^\delta$ as the map sending a discrete class C free fermion theory to the interacting field theory it describes.
\end{rem}

\begin{rem}\label{F2I_from_general_ABS_map}
    In a completely analogous way, we may also define free-to-interacting maps for any twisted ABS map of the form in \cref{defn_generalized_ABS} by applying Anderson duality, and we predict analogously to \cref{our_F2I_ansatz} that these describe the process of obtaining a low-energy IFT from a free fermion Hamiltonian for a more general class of symmetries.
\end{rem}

\begin{rem}
\label{IZ_ABS_through_ko}
By applying Anderson duality to \cref{ABS_through_ko}, we see that the maps $F2I_{\ell,k}$, as well as $F2I_{s=4}^\delta$, factor as the composition of a map $\mathit{f2i}_{\ell,k}\colon \Sigma^{-k+\ell-2}\KO\to \Sigma^2 I_\Z(\ko\wedge\ME_{\ell,k})$, followed by the Anderson dual of the ABS map, and analogously in the complex case.
A similar factorization exists for the general case mentioned in the previous remark.
\end{rem}

\begin{rem}
\label{MCB_dpin_F2I}
Manjunath--Calvera--Barkeshli~\cite[\S IV.I]{manjunath_non-perturbative_nodate} use methods quite different from the ones in this paper to study a free-to-interacting map for 2d phases with symmetry given by the fermionic group $G_f$ with $G_b = \Z/2\times\Z/2$ and twisting data $(x, x^2+xy)$, where $x$ and $y$ are as in \cref{dpin_defn}. Similarly to the proof of \cref{dpinequiv}, one can use an automorphism of $\Z/2\times\Z/2$ to prove that $G_f\cong E_{1,1}$ as fermionic groups, so we can check whether our map $F2I_{1,1}$ matches Manjunath--Calvera--Barkeshli's.

Manjunath--Calvera--Barkeshli (\textit{ibid.}) argue that in 2d, the free-to-interacting map is a surjection $\Z\twoheadrightarrow\Z/8$. By \cref{only_first_summand_real} and the computation $\mho^3_{1,1}\cong\Z/8$ in~\cite[Appendix E]{kaidi_topological_2020}, our model for this free-to-interacting map, $\pi_{-2}(F2I_{1,1})\colon\KO^0\to\mho_{\mathrm{DPin}}^3$, is also a surjection $\Z\twoheadrightarrow\Z/8$. Thus our model agrees with Manjunath--Calvera--Barkeshli's result, providing evidence supporting \cref{our_F2I_ansatz}.
\end{rem}
\begin{rem}\label{smithy_f2i}
Ryu~\cite[\S 3.1]{ryu_interacting_2015} and Rosch~\cite{rosch_unwinding_2012} study free-to-interacting maps for 2-, resp.\ 1-dimensional phases with symmetries given by the fermionic groups $\{\pm 1\}\times\Z/2$, resp.\ $\Pin_1^+\times\Z/2$, where $\theta$ vanishes on the $\Z/2$ factor. One can directly check that these correspond to the tangential structures $\Spin\times\Z/2$, resp.\ $\Pin^-\times\Z/2$ (see~\cite[Proposition 1.26]{ADKPSS24}). Ryu, resp.\ Rosch provide physical arguments that the free-to-interacting maps are surjective maps of the form $\Z^2\twoheadrightarrow\Z/8$, resp.\ $\Z\twoheadrightarrow\Z/4$.

Though we have not defined free-to-interacting maps for these symmetries, our computations suggest the following models for these maps:
\begin{subequations}
\label{Z2_F2I_maps}
\begin{gather}
    \label{ryu_map}
    \Sigma^{-2}\KO \xrightarrow[\eqref{eqn:Elk_F2I}]{\Sigma^{-1}I_\Z\ABS_{1,0}} \Sigma I_\Z\MTPin^- \xrightarrow[\eqref{first_smith_isom}]{I_\Z e_\sigma} \Sigma^2 I_\Z(\MTSpin\wedge (B\Z/2)_+)\\
    \label{rosch_map}
    \Sigma^{-1}\KO \xrightarrow[\eqref{eqn:Elk_F2I}]{\Sigma^{-1}I_\Z\ABS_{2,0}} \Sigma I_\Z(\MTSpin\wedge\ME_{2,0}) \xrightarrow[\eqref{first_smith_isom}]{I_\Z e_\sigma} \Sigma^2 I_\Z(\MTPin^-\wedge (B\Z/2)_+).
\end{gather}
\end{subequations}
In~\eqref{ryu_map}, $e_\sigma$ is smashed with $\MTSpin$ before applying $I_\Z$; in~\eqref{rosch_map}, $e_\sigma$ is smashed with $\MTPin^-$.

Applying $\pi_{-2}$, resp.\ $\pi_{-1}$ to these maps, we obtain maps $\KO^0\to\mho_\Spin^3(B\Z/2)$ and $\KO^0\to\mho_{\Pin^-}^2(B\Z/2)$. By construction of $e_\sigma$, the images of these maps lie in the reduced versions of these Anderson-dualized bordism groups; see~\cite[\S 7.1]{debray_smith_2024}. There are isomorphisms $\widetilde\mho_\Spin^3(B\Z/2)\cong\Z/8$~\cite{MM76} and $\widetilde\mho_{\Pin^-}^2(B\Z/2)\cong\Z/4$ (\cite[Theorem 3.2.1]{Her17} or \cite[Theorem 16]{guo_fermionic_2020}), and Freed--Hopkins' computation of $F2I_{1,0}$~\cite[Corollary 9.85]{freed_reflection_2021} and our computation in \cref{only_first_summand_real} of $I_\Z\ABS_{1,1}$, together with the fact that $I_\Z e_\sigma$ is an isomorphism onto the reduced $I_\Z\mathit{MTH}$-cohomology of $B\Z/2$ (see \cref{1st_Z2_Smith_exm} and~\cite[\S 7.1]{debray_smith_2024}), implies that on $\pi_{-2}$, resp.\ $\pi_{-1}$, the maps in~\eqref{Z2_F2I_maps} are surjections $\Z\twoheadrightarrow \Z/8$, resp.\ $\Z\twoheadrightarrow \Z/4$, onto the reduced Anderson-dualized bordism groups. This matches Ryu and Rosch up to the second $\Z$ free fermion summand of the former, which we interpret as Ryu allowing fermions to transform in a larger class of representations of the symmetry group.

In both cases, the reduced classification is a strict subset of the total classification, as $\mho_{\Pin^-}^2(\pt)$ and $\mho_\Spin^3(\pt)$ are nonzero~\cite[Corollaries 9.81 and 9.85]{freed_reflection_2021}; following \cref{F2I_from_general_ABS_map} we interpret these as representing intrinsically interacting SPT phases that Ryu and Rosch's methods do not detect.
\end{rem}

\section{Modeling the Bott spiral}\label{section_modeling_the_Bott_spiral}

We now apply the mathematical framework of the previous sections to the physical situation
of the \textit{Bott spiral}
of Queiroz--Khalaf--Stern \cite{queiroz_dimensional_2016}.
In contrast to the ``Bott clock'' or periodic table of free fermion phases in the tenfold way, where the classification of free SPTs ($\Z$, $\Z/2$, or $0$) repeats periodically in the dimension and symmetry type, the Bott spiral consists of interacting SPTs whose classification order grows with the dimension. For example, Kitaev's Bott clock~\cite{kitaev_periodic_2009} reports that the groups of free fermion phases in class CII in dimensions $1$ and $9$ are both isomorphic to $\Z$. But Queiroz--Khalaf--Stern report that the corresponding interacting phases form a $\Z/2$ in dimension $1$ and $\Z/32$ in dimension $9$.

There are two key differences between the Bott spiral and the Bott clock. First, the phases in the Bott spiral are interacting; specifically, they are interacting SPTs that are adiabatically connected to free fermion models. %
This means that in our homotopical model, these phases should lie in the image of free-to-interacting maps.
Second, the symmetry types in the Bott spiral are a mild generalization of those in the tenfold way: half of the Altland--Zirnbauer classes considered are \textit{primed}, meaning that an additional $E_{1,1}$-symmetry is present.

Queiroz--Khalaf--Stern computed the classification of SPTs in three different spirals: starting from spatial dimension $d=1$ in either class BDI, CII, or AIII, since these are the three cases in which the classification of free SPTs is a $\Z$. %
Their methods included a form of \textit{dimensional reduction} that relates an SPT in spatial dimension $d$ with an SPT in dimension $d+1$ with a modified symmetry type, interchanging primed and unprimed classes as one goes up the spiral.
This approach allowed them to reduce to dimension 0, where they studied explicit interacting Hamiltonians and computed their classification orders via a similar approach to that of Fidkowski--Kitaev in the original class BDI case \cite{fidkowski_effects_2010}.
Their results are as follows.

\begin{physres}[{\cite[(1)]{queiroz_dimensional_2016}}]\label{QKS_result}
    Consider the minimal chiral Dirac Hamiltonian in spatial dimension $d$ whose dimensional reduction to $d=1$ is a generator of the group of free SPT phases in class BDI, CII, or AIII.
    The subgroup of interacting SPTs generated by the interacting class of this Hamiltonian is a cyclic group $\Z/n$, where
    \begin{subequations}
    \begin{equation}\label{QKS_first_spiral}
        n = \begin{cases}
            4\cdot 2^{\lfloor \frac{d-1}{2} \rfloor}\cdot \mu & \text{for class BDI in $d=1$} \\
            2\cdot 2^{\lfloor \frac{d-1}{2} \rfloor}\cdot \mu & \text{for class CII in $d=1$} \\
            4\cdot 2^{\lfloor \frac{d-1}{2} \rfloor} & \text{for class AIII in $d=1$}
        \end{cases}
    \end{equation}
    and
    \begin{equation}\label{QKS_complex_spiral}
        \mu = \begin{cases}
            2 & \text{for class BDI, D$'$, or DIII} \\
            1 & \text{otherwise.}
        \end{cases}
    \end{equation}
    \end{subequations}
\end{physres}

\begin{exm}
Let us spell out \cref{QKS_result} a bit, following~\cite[Figure 1]{queiroz_dimensional_2016}.
\begin{enumerate}
    \item If one dimensionally reduces to class BDI in $d = 1$, \eqref{QKS_first_spiral} gives a $\underline{\Z/8}$ classification of $d = 1$ BDI phases, a $\underline{\Z/8}$ of $d = 2$ $\mathrm{D'}$ phases, a $\underline{\Z/16}$ of 3d DIII phases, a $\Z/8$ of 4d $\mathrm{AII}'$ phases, a $\Z/16$ of 5d CII phases, a $\Z/16$ of 6d $\mathrm{CII}'$ phases, a $\Z/32$ of 7d CI phases, and so on.
    \item If one dimensionally reduces to class CII in $d = 1$, \eqref{QKS_first_spiral} gives a $\Z/2$ of 1d CII phases, a $\Z/2$ of 2d $\mathrm{C'}$ phases, a $\Z/4$ of 3d CI phases, a $\Z/4$ of 3d CI phases, a $\underline{\Z/16}$ of 5d BDI phases, a $\underline{\Z/16}$ of 6d $\mathrm{D'}$ phases, a $\underline{\Z/32}$ of 7d DIII phases, and so on.
    \item If one dimensionally reduces to class AIII in $d = 1$, \eqref{QKS_complex_spiral} gives a $\Z/4$ of 1d AIII phases, a $\Z/4$ of 2d $\mathrm{A'}$ phases, a $\Z/8$ of 3d AIII phases, a $\Z/8$ of 4d $\mathrm{A'}$ phases, a $\Z/16$ of 5d AIII phases, a $\Z/16$ of 6d $\mathrm{A'}$ phases, a $\Z/32$ of AIII phases, and so on.
\end{enumerate}
In these examples, underlined groups correspond to $\mu = 2$, and the remaining cases to $\mu =1$.
\end{exm}
If we ignore the effect of $\mu$, which distinguishes triplet superconductors, this classification doubles in size in every other dimension step. In particular, this is true for the AIII series, which does not use $\mu$.

One of our goals is to reproduce this result and explain it by alternative methods, providing a test of our homotopical ansatz.
To do so, we use \cref{FH_F2I_ansatz,our_F2I_ansatz} to model the passage from a free fermion theory to the corresponding interacting theory as the action of the free-to-interacting maps of \cref{Elk_F2I,complex_F2I_map_defn}. We model the interacting theory through the (deformation class of) its low-energy reflection-positive IFT, following \cref{LEFT_ansatz}.
Then, to determine the group of interacting SPTs connected to these free fermion models, it suffices to compute the image of the free-to-interacting map.
Our mathematical version of \cref{QKS_result} is given in \cref{cplx_F2I_image_cor,real_F2I_image_cor}.

Additionally, motivated by work of Hason--Komargodski--Thorngren \cite[\S 5]{hason_anomaly_2020}, who noted the relationship between the order of the large 2-torsion summands in the 4-periodic family of Smith homomorphisms of \cref{1st_Z2_Smith_exm}, we also define an invertible field theory model for the \textit{dimensional reduction} process. To avoid overloading the term,\footnote{See \cref{rem:dim_red_terminology}.} we will henceforth call the maps of invertible field theories modeling this process \textit{spiral maps} (\cref{spiral_maps_of_IFTs}). We actually define two different maps for this process, whose application depends on whether we begin in a primed phase or unprimed phase.

Consider the complex cases; i.e.\ the spiral starting from class AIII. As an immediate corollary of \cref{QKS_result}, the interacting classification order doubles at every even-to-odd dimension step. For this reason, we desire that our complex spiral maps are a multiplication by two at every even-to-odd dimension step and an isomorphism in the other cases. In \cref{complex_bott_spiral_computation_corollary}, we observe that this is the case, and in \cref{real_bott_spiral_computation_corollary} we observe the real analog.

\subsection{Primed Altland--Zirnbauer classes}\label{primed_AZ}

The classification of free fermion phases in a tenfold way class in a certain dimension corresponds to one of the groups in the $\KO$-theory or $\KU$-theory of a point. For simplicity, we first consider the cases ($\KU^0$, $\KO^0$, and $\KO^{-4}$) where this group is $\Z$.

Tenfold way free fermion phases classified by $\Z$ in even dimensions host chiral edge modes, which are gapless.
In order to successively produce lower-dimensional \textit{gapped} theories,
Queiroz--Khalaf--Stern define \textit{primed} Altland--Zirnbauer classes, generalizing the definitions of class D$'$ and class DIII+$\mathcal{R}$ put forward in \cite{ryu_interacting_2012, qi_new_2013,yao_interaction_2013}, and related to Song--Schnyder's extended Altland--Zirnbauer-style classification in~\cite[\S II.A.1]{song_interaction_2017}.
To physically produce a primed Altland--Zirnbauer class, one stacks two systems in the same Altland--Zirnbauer class with opposite invariants. Normally, these invariants would add and the phases would cancel, but to prevent this one imposes an additional symmetry formed from an antiunitary symmetry combined with a reflection symmetry.

Correspondingly, to form the symmetry group of a primed class,
one starts with the symmetry group of the original class and enlarges it by adding generators $\mathcal{T}$ and $\mathcal{R}$.
Here
$\mathcal T$ is an internal symmetry that is antiunitary, squares to the identity, and commutes with the rest of the symmetry group, while $\mathcal R$ is a unitary symmetry that squares to $(-1)^F$ and also commutes with the rest of the symmetry group.
Our mathematical model for primed phases encodes the additional symmetry generators $\mathcal{T}$ and $\mathcal{R}$ as the generators of the fermionic group $E_{1,1}$ of \cref{Cl_1_twist}: %
\begin{defn}\label{G_f_prime}
    Let $G_f$ be a fermionic group. Define $G_f' \coloneqq G_f \ftens E_{1,1}$.
\end{defn}
In our mathematical model, if $G_f$ describes a physical symmetry type, the \textit{primed} symmetry type will be described by $G_f'$.

\begin{rem}\label{Morita_variance_rmk_2}
Recall the discussion of the failure of Morita invariance for interacting SPTs from \cref{subsec_symm_types_morita_invnce}.
    While the $\Cl_{1,1}$ periodicity of $\KO$-theory (\cref{K-theory_of_Elk}) ensures that the free classifications of primed and unprimed phases are isomorphic, the same is not true for the interacting classifications. 
    The tangential structure $H(G_f\hat{\times} E_{1,1})$ is measurably different from $H(G_f)$: as explained in \cref{E11_vs_1}, the corresponding bordism groups and, dually, the groups of IFTs, are \textit{not} the same.
\end{rem}

\begin{exm}
The simplest example of a primed class is class D$'$. For both the discrete and continuous forms of Altland--Zirnbauer class D (see \cref{disc_cont_defn,tenfold_table}), $G_b = 1$ and $G_f = E_{0,0} = \Spin_1 = \Z/2$ with trivial grading and nontrivial $(-1)^F$, so for class D$'$ we have $G_f'\cong E_{1,1}$. Recall from \cref{Cl_1_twist} that 
the fermionic group algebra $\R^f[E_{1,1}]\cong \Cl_{1,1}\cong M_{1|1}(\mathbb R)$ is Morita equivalent to $\mathbb R$, which is the fermionic group algebra corresponding to class D.
Correspondingly, 
$K_p(\R^f[E_{1,1}]) \cong K_p(\mathbb R) \cong \KO_p$, and the classification of free class D$'$ phases is the same as that for free class D phases.

On the interacting side, there is a clear difference: $E_{1,1}$ corresponds to spin-$(1,1)$ structures (or dpin structures by \cref{dpinequiv}), while $\R$ corresponds to spin structures. For example, in spatial dimension 2, interacting class D superconductors are classified by $\mho_{\Spin}^3 \cong \Z$~\cite[Corollary 9.81]{freed_reflection_2021}, while interacting class D$'$ superconductors are classified by $\mho^3_{\DPin} \cong \Z/8$~\cite[Appendix E]{kaidi_topological_2020}.\footnote{See~\cite[\S 4]{stehouwer_interacting_2022} for another fermionic group, $\Pin_2^+$ with $\theta$ trivial, giving rise to the same class D classification for free fermion phases and groups of IFTs not isomorphic to the groups of spin nor dpin IFTs.}
\end{exm}

\begin{exm}
    In \cref{subsec_symm_types_morita_invnce}, we discussed how our discrete model for class C was exceptional because it was not of the form $E_{\ell,k}$. Thanks to \cref{class_C_prime}, however, 
    the discrete class $\mathrm C'$ fermionic group is isomorphic to $E_{0,4}$ and $E_{4,0}$.
\end{exm}

\begin{exm}
    The only complex primed class we will need to consider is class A$'$. As for class D, we have $G_f'=E_{1,1}$, but now the complex fermionic group algebra is $\C^f[E_{1,1}] \cong \Cxl_2$, oblivious to the signature of the quadratic form. 
    From \cref{cpx_collapse}, we also know that the Thom spectrum of the primed symmetry type simplifies as
    \begin{equation}
        MT\Spinc\wedge \ME_{1,1} \simeq MT\Spinc\wedge \ME_2.
    \end{equation}
    In \cref{dpin_c}, we show that $\mho_{\Spin^c}^3(\ME_2)\cong\Z/4$, but $\mho_{\Spin^c}^3\cong\Z^2$~\cite[Corollary 9.89]{freed_reflection_2021}.
\end{exm}

\subsubsection{Crystalline equivalence principle for primed classes}\label{CEP}
According to Queiroz--Khalaf--Stern, primed classes may be defined assuming that $\mathcal{R}$ is an internal unitary $\Z/2$-symmetry (following D$'$ as defined by Qi \cite{qi_new_2013} and Ryu--Zhang~\cite{ryu_interacting_2012}), or alternatively a spatial $\Z/2$ reflection symmetry (following DIII+$\mathcal{R}$ as defined by Yao--Ryu \cite{yao_interaction_2013}). In this section, we show that these two definitions agree, using 
the \term{crystalline equivalence principle} (CEP).
This principle, introduced by Thorngren--Else~\cite{thorngren_gauging_2018}, posits an equivalence between the classifications of invertible phases respecting a $G$-symmetry acting on space and those of invertible phases with a purely internal $G$-symmetry. The details of the internal symmetry depend on the specific $G$-action on space, how it mixes with time-reversal or other internal symmetries, etc. In particular, for fermionic phases, through the work of many groups of researchers~\cite{freed_invertible_2019, guo_fermionic_2020, mao_mirror_2020, zhang_construction_2020, debray_invertible_2021, sheinbaum_crystallographic_2021, zhang_crystalline_2021, cheng_rotation_2022, huang_effective_2022, zhang_construction_2022, zhang_real_space_2022, manjunath_non-perturbative_nodate, zhang_anomalous_2023, kobayashi_topological_2024, lee_crystalline_2024, manjunath_characterization_2024, ren_stacking_2024, soldini_interacting_2024, sheinbaum_CEP_2025, stockall_generalized_2025, cheng_boundary_2024, liu_real_space_2026, lee_connection_2024, manjunath_crystalline_2026}, it has become clear that the fermionic groups acting on the spatial and internal sides have the same $G_b$ but different associated twists.

Of the various formulations of the fermionic crystalline equivalence principle in the above references, we found the proposal of Manjunath--Calvera--Barkeshli~\cite[\S III.A]{manjunath_non-perturbative_nodate} most applicable to the problem at hand.

Let $\Euc_d\coloneqq \R^d\rtimes\O_d$ denote the group of isometries of $d$-dimensional Euclidean space and $p\colon\Euc_d\to\O_d$ denote the quotient by the normal subgroup of translations.
The data that goes into Manjunath--Calvera--Barkeshli's ansatz is a (not necessarily compact) fermionic group $G_f$ and a group homomorphism $\lambda\colon G_b\to\Euc_d$, where $d$ is the dimension of space. This data captures both how the symmetry group acts on space (through $\lambda$) as well as how it mixes with fermion parity (through $G_f\to G_b$).
\begin{defn}[{Manjunath--Calvera--Barkeshli~\cite[(30)]{manjunath_non-perturbative_nodate}}]
\label{int_defn}
Let $G_f$ be a fermionic group with associated bosonic group $G_b$ and $\lambda\colon G_b\to\Euc_d$ be a homomorphism. Let $(\theta, \omega)$ be the associated twist of $G_f$. Define another fermionic group $G_f^{\mathrm{int}}$ to have the bosonic group $G_b$ and the associated twist $(\theta^{\mathrm{int}}, \omega^{\mathrm{int}})$ given by the following formulas:
\begin{subequations}\label{int_twisting_data}
\begin{align}
    \theta^{\mathrm{int}} &= \theta + (p\circ\lambda)^*(w_1)\\
    \omega^{\mathrm{int}} &= \omega + (p\circ\lambda)^*(w_1)\cdot\theta^{\mathrm{int}} + (p\circ\lambda)^*(w_2).
\end{align}
\end{subequations}
\end{defn}
This defines $G_f^{\mathrm{int}}$ only up to isomorphism, but that will suffice for our applications in this paper.
\begin{ansatz}[{Fermionic CEP of Manjunath--Calvera--Barkeshli~\cite[\S III.A]{manjunath_non-perturbative_nodate}}]
\label{FCEP}
Given a fermionic group $G_f$ and a group homomorphism $\lambda\colon G_b\to\Euc_d$, the abelian group of $d$-dimensional invertible interacting fermionic phases with symmetry group $G_f$ acting on space by $\lambda$ is isomorphic to the group of $d$-dimensional interacting invertible phases with symmetry group $G_f^{\mathrm{int}}$.
\end{ansatz}
Where this overlaps in scope with the proposals of~\cite{freed_invertible_2019, debray_invertible_2021}, the three proposals make equivalent predictions. In general, their predictions are in line with what has been computed in the physics literature by other methods. Compare for example the computations in~\cite{debray_invertible_2021,zhang_construction_2022}.

Let us start from the perspective that $\mathcal{R}$ is a reflection symmetry.
If $\rho\colon\Z/2\to\Euc_d$ is the inclusion of a reflection $\mathcal{R}$ and $\mathrm{proj}_3\colon A\times B\times C\to C$ is the projection onto the third factor, then the spatial symmetry data is that of the group $G_f\ftens E_{1,1}$ acting spatially by
\begin{equation}
\label{QKS_spatial_group}
    \lambda = \rho\circ \mathrm{proj}_3\colon (G_f\ftens E_{1,1})_b = G_b\times \Z/2\times\Z/2\xrightarrow{\mathrm{proj}_3}  \Z/2 \xrightarrow{\rho} \Euc_d.
\end{equation}
\begin{prop}
\label{primed_equals_primed}
If $(G_f, \lambda)$ are as in~\eqref{QKS_spatial_group} above, then there is an isomorphism of fermionic groups $G_f^{\mathrm{int}}\cong G_f'$.
\end{prop}
In other words, the fermionic group is agnostic as to whether $\mathcal{R}$ is spatial or internal.
\begin{proof}
Both $G_f^{\mathrm{int}}$ and $G_f'$ have bosonic group isomorphic to $G_b\times\Z/2\times\Z/2$, so it suffices to use \cref{twist_of_product} and~\eqref{int_twisting_data} to compute their associated twists. Specifically, if $(\theta,\omega)\in H^1(BG_b;\Z/2)\times H^2(BG_b;\Z/2)$ is the associated twist for $G_f$, and $x_1$, resp.\ $x_2$ represent the classes in $H^1(B\Z/2\times B\Z/2;\Z/2)$ corresponding to the first, resp.\ second factors of $\Z/2$, then \cref{twist_of_product} implies the associated twist of $G_f'$ is
\begin{subequations}
\begin{equation}\label{primetwist}
    (\theta + x_1 + x_2,\, \omega + \theta(x_1 + x_2) + x_1^2)
\end{equation}
and~\eqref{int_twisting_data} implies the associated twist for $G_f^{\mathrm{int}}$ is
\begin{equation}\label{inttwist}
    (\theta + x_1, \,\omega + \theta x_1 + x_1^2 + x_2^2).
\end{equation}
\end{subequations}
Though $\eqref{primetwist}\ne\eqref{inttwist}$, they differ by an automorphism of $\Z/2\times\Z/2$. Specifically, the automorphism defined by extending $(1, 0)\mapsto (1,1)$ and $(0,1)\mapsto (0,1)$ $\Z/2$-linearly sends $x_1\mapsto x_1+x_2$ and $x_2\mapsto x_2$, and one can check that this carries the data~\eqref{inttwist} to the data~\eqref{primetwist}. Therefore this automorphism lifts to an isomorphism of fermionic groups $G_f^{\mathrm{int}}\xrightarrow{\cong} G_f'$.
\end{proof}

\begin{exm}
\label{YR_exm}
Return to class D$'$.
By the discussion above, this also corresponds to the spatial fermionic symmetry generated by commuting operators $\mathcal T$ and $\mathcal R$, with $\mathcal T^2 = 1$ and $\mathcal R^2 = (-1)^F$, such that $\mathcal T$ is antiunitary and $\mathcal R$ is a reflection.

Qi~\cite{qi_new_2013}, Ryu--Zhang~\cite{ryu_interacting_2012}, and Yao--Ryu~\cite{yao_interaction_2013} studied (2+1)d SPT phases with this fermionic symmetry data; they found a $\Z$ classification of free fermion phases, which maps surjectively onto a $\Z/8$ group of interacting phases. These SPTs were subsequently studied by several collaborations, including~\cite{gu_interactions_2012, cho_topological_2015, morimoto_breakdown_2015, ryu_interacting_2015, queiroz_dimensional_2016, song_interaction_2017, yoshida_fate_2017, aksoy_stability_2021, manjunath_non-perturbative_nodate}, who reproduced Yao--Ryu's results by different methods.  Thus we can give another derivation of this $\Z/8$ classification using the CEP to identify it with $\mho^3_{\mathrm{DPin}} \cong \mho^3_{1,1}$, which was calculated to be $\Z/8$ in~\cite[Appendix E]{kaidi_topological_2020}.
We also get a mathematical model for the class $\mathrm{D'}$ free-to-interacting map: first apply $F2I_{1,1}\colon \Sigma^{-2}\KO\to \Sigma^2 I_\Z(\MTSpin\wedge\ME_{1,1})$, then apply the fermionic CEP to identify the codomain with the spectrum classifying phases in class $\mathrm{D'}$. By our computations in \cref{only_first_summand_real}, this is indeed a surjection $\Z\twoheadrightarrow \Z/8$ in dimension $2$.
\end{exm}
\begin{exm}
\label{more_crystalline_F2I}
Generalizing \cref{YR_exm}, there are several more examples in the physics literature of computations of free-to-interacting maps for a symmetry group $G_f$ that acts on space as above such that under the fermionic CEP (\cref{FCEP}), $G_f^{\mathrm{int}}\cong E_{\ell,k}$ for some $(\ell,k)\ne (0,0)$. Therefore in these cases we may mathematically model the free-to-interacting map as $F2I_{\ell,k}$ followed by the CEP identification, like we did in \cref{YR_exm}. This includes the following examples.
\begin{enumerate}
    \item\label{BDI_R++} Cho--Hsieh--Morimoto--Ryu~\cite[\S IV.D]{cho_topological_2015} and Song--Schnyder~\cite{song_interaction_2017} study the free-to-interacting map for class ``$\mathrm{BDI}+R_{++}$,'' given by a time-reversal symmetry $\mathcal T$ and a spatial reflection symmetry $\mathcal R$ that commute and both square to $1$. Using methods different from this paper's and from each other, Cho--Hsieh--Morimoto--Ryu compute the free-to-interacting map in 2d and Song--Schnyder compute it in all dimensions. They conclude that the free-to-interacting map has image $\Z/2^{4k+3}$ in dimensions $d = 8k+2$, $\Z/2^{4k+4}$ in dimensions $d = 8k+6$, $\Z/2$ in dimensions $d = 8k$ and $8k+1$, and $0$ otherwise. Under \cref{FCEP}, we get $G_f^{\mathrm{int}} \cong E_{1,1}$ again, so just like in \cref{YR_exm}, \cref{only_first_summand_real} computes that our free-to-interacting map $F2I_{1,1}$ has the same behavior as the one in in~\cite{cho_topological_2015,song_interaction_2017}. These authors do not discuss the cokernel of the free-to-interacting map.
    \item Morimoto--Furusaki--Mudry~\cite[\S IV.B]{morimoto_breakdown_2015} and Yoshida--Furusaki~\cite[\S III]{yoshida_correlation_2015} consider three-dimensional phases in class ``$\mathrm{AII}+R$,'' arguing that the corresponding free-to-interacting map is a surjection $\Z\surj\Z/8$. In more detail, they consider the continuous class AII fermionic group $\Pin_2^-$ together with a spatial reflection symmetry $R$ squaring to $(-1)^F$ and commuting with the internal $\Pin_2^-$ symmetry. Passing through the fermionic CEP (\cref{FCEP}), one obtains the tangential structure $\Pin^{\tilde c+}\ftens E_{1,0}$. We model the corresponding free-to-interacting map by the ``mixed discrete-continuous F2I map'' defined in \cref{cts_AZ_f2i}; roughly speaking, it is the map obtained by combining Freed--Hopkins' free-to-interacting map for $\Pin^{\tilde c+}$~\cite[\S 9.2]{freed_reflection_2021} (\cref{FH_F2I_maps}, case $s = -2$) with our $F2I_{1,0}$ (\cref{Elk_F2I}):
    \begin{equation}\label{AIIR_map}
        F2I_{1,0,-2}\colon \Sigma\KO \xrightarrow{\eqref{eqn:mixed_F2I}} \Sigma^2 I_\Z(\MTPin^{\tilde c+}\wedge\ME_{1,0}).
    \end{equation}
    Our model for the free-to-interacting map in~\cite{morimoto_breakdown_2015,yoshida_correlation_2015} is $\pi_{-3}$ of~\eqref{AIIR_map}. In \cref{gen_AZ_f2i}, we show $\pi_{-3}F2I_{1,0,-2}$ is a map $\Z\to\Z/8\oplus (\Z/2)^s$, for some $s>0$, which surjects onto the $\Z/8$ summand. Thus our model for this free-to-interacting map agrees with Morimoto--Furusaki--Mudry and Yoshida--Furusaki's; they do not discuss the cokernel.\footnote{By studying the Anderson--Brown--Peterson-style splitting for $\MTPin^{\tilde c+}$ in \cref{thm:pin_tilde_c_ABP} more carefully, one can prove that $s = 3$, predicting a $(\Z/2)^{\oplus 3}$ of intrinsically interacting invertible phases for this collection of symmetries.}

    Tin telluride ($\mathrm{SnTe}$) is predicted to be in class $\mathrm{AII}+R$: see~\cite{hsieh_SnTe_2012, dziawa_SnTe_2012, tanaka_SnTe_2012, xu_SnTe_2012, okada_SnTe_13} and~\cite[\S IV.C]{morimoto_breakdown_2015}. It would be interesting to determine whether the behavior of this class's free-to-interacting map could be studied experimentally.
    \item Song--Schnyder~\cite{song_interaction_2017} studied the free-to-interacting map in all dimensions for many symmetry classes including spatial reflections or twofold rotations. Seven of these correspond under the fermionic CEP (\cref{FCEP}) to free-to-interacting maps newly defined in our paper: $\mathrm{BDI}+R_{++}$ as studied above in part~\eqref{BDI_R++}, and also $\mathrm{AIII}+R_+$, $\mathrm{AI}+R_+$, $\mathrm{DIII}+R_{++}$, $\mathrm{AII}+R_+$, $\mathrm{CII}+R_{++}$, and $\mathrm{CI}+R_{++}$. Each of has an internal symmetry group given by the continuous fermionic group $K_f(s)$ (see \cref{tenfold_table}) from the corresponding Altland--Zirnbauer class, together with a spatial reflection commuting with the internal symmetry and squaring to $1$. Under \cref{FCEP}, the corresponding internal symmetry group is $K_f(s)\ftens \Pin_1^-$, so we can model Song--Schnyder's free-to-interacting maps by the maps we defined in \cref{Elk_F2I,complex_F2I_map_defn,cts_AZ_f2i}. In order left-to-right, then top-to-bottom, the six remaining examples are modeled by
    \begin{equation}
    \begin{alignedat}{2}
        F2I_2\colon& \KU\to\Sigma^2 I_\Z(\MTSpin^c\wedge\ME_2) \qquad\qquad &
            F2I_{0,1,-2}\colon &\Sigma^3\KO\to \Sigma^2 I_\Z(\MTPin^{\tilde c-}\wedge\ME_{0,1})\\
        F2I_{0,2}\colon &\Sigma^4\KO\to \Sigma^2 I_\Z(\MTSpin\wedge\ME_{0,2}) &
            F2I_{0,1,2}\colon &\Sigma^{-1}\KO\to \Sigma^2 I_\Z(\MTPin^{\tilde c+}\wedge\ME_{0,1})\\
        F2I_{0,1,-3}\colon & \Sigma^{-1}\KO\to \Sigma^2 I_\Z(\MTPin^{h-}\wedge\ME_{0,1}) &
            F2I_{0,1,3}\colon & \KO\to \Sigma^2 I_\Z(\MTPin^{h+}\wedge\ME_{0,1}).
    \end{alignedat}
    \end{equation}
    The singly-indexed maps $F2I_m$ are the ones from \cref{complex_F2I_map_defn}; the doubly-indexed maps $F2I_{\ell,k}$ are from \cref{Elk_F2I}, and the triply-indexed maps $F2I_{\ell,k,s}$ are from \cref{cts_AZ_f2i}.
    
    In \cref{only_first_summand_real,gen_AZ_f2i,cplx_F2I_image_thm}, we computed these maps, and comparing those results with~\cite[Table 1]{song_interaction_2017}, we see complete agreement in the cases we consider.

    We can also model Song--Schnyder's $\mathrm{AIII}+R_-$ example, consisting of an internal $\Pin_1^c$ symmetry together with a spatial reflection squaring to $1$ and \emph{anti}commuting with time-reversal, in a manner similar to that of \cref{smithy_f2i}. After applying the fermionic CEP, we obtain the tangential structure $\Pin^c\times\Z/2$, so following~\eqref{Z2_F2I_maps}, we can define a free-to-interacting map to be the composition
    \begin{equation}\label{AIIIR-}
        \Sigma^{-1}\KU \xrightarrow[\eqref{complex_F2I_map_defn}]{\Sigma^{-1}I_\Z\ABS_2} \Sigma I_\Z(\MTSpin^c\wedge\ME_2) \xrightarrow[\eqref{first_smith_isom}]{I_\Z e_\sigma} \Sigma^2 I_\Z(\MTPin^c\wedge (B\Z/2)_+)
    \end{equation}
    together with the identification in the fermionic CEP. Combining our computation of $F2I_2$ in \cref{cplx_F2I_image_thm} with a similar argument to the one in \cref{smithy_f2i}, one can show that on $\pi_{-d}$, corresponding to $d$-dimensional phases, the image of~\eqref{AIIIR-} consists of $\Z/2^{k+2}$ if $d = 2k+1$ and is $0$ if $d$ is even, matching what Song--Schnyder obtain. As usual, they do not discuss the cokernels of their free-to-interacting maps.
\end{enumerate}
There are also a number of examples in the literature, including some of those studied in~\cite{isobe_theory_2015,song_interaction_2017,lee_connection_2024}, where applying the fermionic CEP (\cref{FCEP}) produces one of the ten continuous fermionic symmetry groups $K_f(s)$ (see \cref{tenfold_table}), and Freed--Hopkins' corresponding free-to-interacting map~\cite[\S 9.2]{freed_reflection_2021} (\cref{FH_F2I_maps}) produces isomorphic domain and image groups in all dimensions to the example. As these are accounted for by previous literature, we do not go into the details.
\end{exm}
Despite these examples, in general, work of Sheinbaum--Antolín Camarena~\cite{sheinbaum_CEP_2025} shows that one must be careful when discussing the CEP in the context of the free-to-interacting map.

\subsection{Image of the free-to-interacting map for primed phases}
\label{ss:imF2I}

Equipped with our model for primed symmetry types and with the free-to-interacting maps for $E_{\ell,k}$- and $E_m$-type twists (\cref{Elk_F2I,complex_F2I_map_defn}, resp.), we can now describe our mathematical version of \cref{QKS_result}: the interacting classification order $n$ is given by the order of the image of the free-to-interacting map of the appropriate symmetry type.
We compute these orders in \cref{section_computations} and \cref{s:spin_computations}, and present a summary of our results here.

First, consider the real cases. The following is a corollary of \cref{compute_F2I} and \cref{only_first_summand_real}.
Note that $d$ should be interpreted as the spatial dimension of the initial free fermion SPT, while $\ell$ and $k$ set the symmetry type to be associated with the fermionic group $E_{\ell,k}$.
The variable $i$ tracks $\ell\bmod 4$, while $\widetilde{m}$ is an auxiliary variable used to simplify the theorem statement; essentially, $\widetilde{m}$ tracks how many times we have gone fully around the spiral.

Recall the definition of $\ldeg(\ell, k)$ from \cref{ldeg_defn}.
\begin{cor}\label{real_F2I_image_cor}
    Let $\widetilde m\coloneqq \lfloor(d-\ell-k+i+1)/8\rfloor$, where $i\in\{0,1,2,3\}$ with $i\equiv \ell\bmod 4$. If $d\ge \ldeg(\ell, k)-2$, the image of the type-$E_{\ell,k}$ free-to-interacting map
    \begin{equation}
        F2I_{\ell,k}\colon \KO^{d+\ell-k-2} \to \mho^{d+2}_{\Spin}(\ME_{\ell,k})
    \end{equation}
    is
    \begin{equation}
    \mathrm{im} \, (F2I_{\ell,k}) \cong
        \begin{cases}
            \Z/2^{4+4\widetilde m-i} & d\equiv \ell+k-2i+2\bmod 8 \\
            \Z/2^{5+4\widetilde m-i} & d \equiv\ell+k-2i+6\bmod 8 \\
            \Z/2 & d\equiv \ell+k-2i\bmod 8 \text{ or } \ell+k-2i+1\bmod 8.
        \end{cases}
    \end{equation}
    If $d<\ldeg(\ell, k)-2$, $F2I_{\ell,k}$ vanishes in degree $d$.
\end{cor}
We give many more details in \cref{section_computations} and \cref{s:spin_computations}.
Here, note that the first two cases describe the breakdown of a $\Z$'s worth of free phases into a $\Z/2^{n}$'s worth of interacting phases, which are the cases considered in \cite{queiroz_dimensional_2016}.
We also computed the third case, for which the domain of the free-to-interacting map is isomorphic to $\Z/2$ instead. In these cases, the free-to-interacting map is actually injective.

\begin{rem}
We don't compute the entirety of $\mho_\Spin^{d+2}(\ME_{\ell,k})$---as we describe in \cref{rem:bosonicsummand}, we leave out some IFTs that do not correspond to interesting fermionic SPTs, which substantially simplifies the computation. As we show in \cref{compute_F2I}, the image of the free-to-interacting map never intersects these uninteresting IFTs nontrivially, so for the purposes of modeling the Bott spiral, this suffices:
Queiroz--Khalaf--Stern's results do not fully characterize the interacting classification of phases, but only those phases connected to free fermion models (i.e.\ those in the image of the F2I map).
\end{rem}

To illustrate the theorem statement, we will consider a couple of examples.

\begin{exm}
    Return to the running example: class BDI superconductors. As discussed in \cref{subsec_symm_types_morita_invnce}, the usual choice of fermionic group for the symmetry class BDI is $E_{1,0}$. Setting $\ell=1$, $k=0$, and setting $i=1$ to match $\ell\bmod 4$, we can read off the results. When $d\equiv-1,0\bmod 8$, the image is isomorphic to $\Z/2$, while when $d= 1$, for example, we observe the $\Z/2^3 =\Z/8$ interacting classification generated by the TRS Majorana chain (see e.g.\ \cite{fidkowski_effects_2010}).

    This result recapitulates the computation of \cite[Corollary 9.85]{freed_reflection_2021} and the physics references cited therein for $d=-1$ to $d=4$, since indeed our free-to-interacting map for $E_{1,0}$ matches their \pinm free-to-interacting map by definition. Although it is not directly relevant for condensed matter applications, our computation holds for arbitrarily high dimensions $d$. Above dimension $d=4$, we see for example $\Z/2$ images in $d=7,8$, a $\Z/2^7 = \Z/128$ in $d=9$, and so on. We see the $16$-fold increase in the classification at each increase in eight dimensions for dimensions $1, 5\bmod 8$.
\end{exm}

\begin{exm}\label{ex:Dpin_F2I_im}
    Consider now class D$'$ superconductors, for which we set $\ell=k=1$ and $i=1$. Then, $\widetilde{m}=\lfloor d/8 \rfloor$ and we can read off that the image of $F2I_{1,1}$ is a $\Z/2$ in spatial dimensions 0 and 1, a $\Z/8$ in $d=2$, zero in the next three dimensions, and a $\Z/16$ in $d=6$. For $d\equiv 2,6\bmod 8$, there is an increase in the classification order by an order of $16$ every eight dimension steps.

    This example is related to the previous one. In particular, there is an isomorphism between the $\Z/8$ classification of class D$'$ interacting phases in $d=2$ (studied e.g.\ by \cite{qi_new_2013,yao_interaction_2013,queiroz_dimensional_2016}) and the $\Z/8$ classification of class BDI interacting phases in $d=1$; the latter is a \textit{dimensional reduction} of the first. We discuss this more in the next subsection.
\end{exm}

The complex cases follow a similar but simpler pattern.
The following is a corollary of \cref{cplx_F2I_image_thm}, which we prove in \cref{ss:comp_spinc}.

\begin{cor}\label{cplx_F2I_image_cor}
    The image of the type-$E_m$ free-to-interacting map
    \begin{subequations}
    \begin{equation}
        F2I_m\colon \KU^{d+m-2} \to \mho^{d+2}_{\Spinc}(\ME_m)
    \end{equation}
    is
    \begin{equation}
        \mathrm{im}\,(F2I_m^c) = 
        \begin{cases}
        \Z/2^{2+(d-m)/2}, & d-m \text{\rm{} even, }d-m\ge -2 \\
        0, & \text{\rm otherwise}.
        \end{cases}
    \end{equation}
    \end{subequations}
\end{cor}

\begin{exm}
    Consider class AIII, for which we take $m=1$. \Cref{cplx_F2I_image_cor} implies that in odd dimensions $d\ge -1$, the free-to-interacting map has domain $\Z$ and surjects onto a $\Z/2^{(d+3)/2}$ summand: a $\Z/2$ for $d = -1$, $\Z/4$ for $d = 1$, $\Z/8$ for $d =3$, etc. For $d$ even the free-to-interacting map is trivial in degree $d$. Looking at the homotopy groups of $\MTSpin^c\wedge\ME_1\simeq\MTPin^c$ computed by Bahri--Gilkey~\cite[Theorem 0.2(b)]{BG87a}, we see that this free-to-interacting map is surjective in degrees $2$ and below: $\mho_{\Spin^c}^3(\ME_1)\cong\Z/8\oplus\Z/2$, and the second summand is not in the image of $F2I_1$. In higher dimensions, the cokernel is even larger.

    This free-to-interacting map coincides with Freed--Hopkins' free-to-interacting map $\Sigma\KU\to \Sigma^2 I_\Z(\MTPin^c)$~\cite[\S 9.2.4]{freed_reflection_2021},
    akin to the real case from \cref{rem:same_map}. Thus \cref{cplx_F2I_image_cor} recovers as a special case Freed--Hopkins' computation of their free-to-interacting map in low degrees~\cite[Corollary 9.91]{freed_reflection_2021}; as usual their $n$ is our $d+1$. The computations for $m>1$ are new.
\end{exm}

\begin{rem}
    Consider the cases connected to the class AIII insulator in $d=1$. Queiroz--Khalaf--Stern argue that the interacting classification order relies on the dimension of the minimal massless Dirac Hamiltonian with chiral symmetry, which is $\lfloor \frac{d}{2} \rfloor$ for the operator acting in $d$ spatial dimensions.

    To unpack this a bit, consider the algebra $\Cxl_d\otimes \Cxl_1$, which is physically represented by the gamma matrices and a chiral symmetry operator.
    The complex spinor representation is the minimal-dimensional representation of $(\Cxl_d\otimes\Cxl_1)_{\mathit{ev}}\cong \Cxl_d$.
    
    For $d=2k$, there is an isomorphism of superalgebras $\Cxl_{2k}\cong \End(\C^{2^{k-1}\mid 2^{k-1}})$, which as an ungraded algebra is isomorphic to the matrix algebra $M_{2^k}(\C)$. %
    The minimal-dimensional representation is given by acting as matrices on $\C^{2^{k-1}\mid 2^{k-1}}$.
    For $d=2k+1$, there is an isomorphism $\Cxl_{2k+1}\cong \Cxl_{2k}\otimes\Cxl_1 \cong \End(\C^{2^{k-1}\mid 2^{k-1}})\oplus \End(\C^{2^{k-1}\mid 2^{k-1}})$, and the minimal-dimensional representation acts on $\C^{2^{k-1}\mid 2^{k-1}}$. %
\end{rem}

\subsection{Spiraling out}
\label{ss:spiral_model}
In this section, we develop an invertible field theory model for the dimensional reduction procedure used by Queiroz--Khalaf--Stern to derive their classification.
With this model, we not only recover the physical picture, but also clarify the role of Bott periodicity on the interacting side, recognizing the 16-fold multiplication that occurs going fully around the spiral as being induced by taking a product with a Bott manifold.

Here, we define the spiral maps of IFTs and describe their behavior---the actual computation of these maps appears in \cref{section_computations}.
We begin by providing the motivation for our construction of the spiral maps: Smith homomorphisms and symmetry breaking.

\subsubsection{Smith homomorphisms and symmetry breaking}\label{smith_symm_br}

In \cref{smith_homs_intro}, we defined Smith homomorphisms as the homomorphisms of bordism groups induced by the zero section map of a given vector bundle. Anderson-dualizing such maps produces group homomorphisms of (deformation classes of) invertible field theories, which by the ansatz discussed in \cref{interacting_IFT} model interacting SPTs.

In this section, we follow the discussion in \cite[\S 8]{debray_smith_2024} and \cite[\S III.B]{debray_long_2024}. See also \cite{hason_anomaly_2020,COSY} for more detail.

The next definition follows \cite{debray_smith_2024} and \cite[\S4.2]{hason_anomaly_2020}:
\begin{defn}\label{W_defect_map_defn}
    Let $W \to X$ be a vector bundle of rank $k$, $V \to X$ a virtual vector bundle, $X^V\hookrightarrow X^{V\oplus W}$ be the map $e_W$ induced by the zero section of $W$ as in \cref{smith_map_of_spectra_specific}, and let $\xi$ be a stable tangential structure. The \textit{defect map} corresponding to this data is the map 
    \begin{equation}\label{defect_map}
    \mathrm{Def}_W\colon [MT\xi\wedge X^{V\oplus W}, \Sigma^{d+2} I_\Z] \to [MT\xi\wedge X^V, \Sigma^{d+2} I_\Z].
\end{equation}
    induced by smashing $e_V$ with $MT\xi$ and applying $[-,\Sigma^{d+2} I_\Z]$.
Rewritten using \cref{IFT_notn} and using the suspension isomorphism on the source, this is
\begin{equation}\label{defect_map_clearer_notation}
    \mathrm{Def}_W \colon \mho_\xi^{d+2-k}(X^{V\oplus W-k}) \to \mho_\xi^{d+2}(X^V).
\end{equation}
\end{defn}

\begin{rem}
In physical applications, $X$ will often come in the form $BG$ for $G$ a compact Lie group, and $W$ will often be an associated vector bundle to some representation of $G$. To describe invertible field theories modeling of real (resp.\ complex) fermionic SPTs, we will often take $\xi = \Spin$ (resp.\ $\xi=\Spinc$), $G = G_b$, and $V$ inducing the desired internal fermionic symmetry group through the procedure described in \cref{repn_fg}.
\end{rem}

This map of invertible field theories models a map between the anomaly of a $d$-dimensional QFT and the anomaly of a $(d-k)$-dimensional \textit{$W$-defect} theory. According to the theory of \textit{anomaly inflow} (see e.g.\ \cite{freed_anomalies_2014,witten_anomaly_2019,thorngren_anomalous_2021}), the anomalies of these QFTs are described by SPTs in one dimension higher, which following \cref{LEFT_ansatz} we will mathematically model using invertible field theories.
So, the anomaly of the $W$-defect QFT, which is modeled by a (deformation class of a) $(d-k+1)$-dimensional invertible field theory, lives in the source of $\mathrm{Def}_W$, while the image of the map classifies the anomalies of the possible bulk $d$-dimensional QFTs that can admit the source $W$-defect. When $k=1$, such defects are called domain walls \cite{solitons-jackiw-rebbi, hason_anomaly_2020}, while vortices \cite{thouless, volovik-superfluid} and hedgehogs are examples in $d=2$ and $d=3$, resp. %

Physically, one begins with the bulk $d$-dimensional theory, often specified by a Hamiltonian or Lagrangian, whose symmetry type is encoded by $X$, $V$, and $\xi$. Then, one modifies the theory by adding and then condensing a charged order parameter, whose symmetry charge (i.e.\ representation) is encoded by the bundle $W$.
The submanifold of spacetime where the order parameter vanishes corresponds to the spacetime of the defect theory, and, for a given starting bulk spacetime manifold $M$, to the image of the bordism class $[M]$ under the homomorphism $\mathrm{sm}_W$. Ground states on this defect generally possess a different symmetry from those of the original theory---the symmetry of the original theory has been spontaneously broken. 

Mathematically, the modification to the symmetry in the theory is achieved by twisting the tangential structures on spacetime by the bundle $W$, and this is exactly what the Smith homomorphism achieves.
For more details on the creation of $W$-defects, see e.g.\ \cite[\S 4]{hason_anomaly_2020} and \cite[\S III.B]{debray_long_2024}, where they are called $\rho$-defects. 

\begin{exm}[{\cite[\S IV.C]{debray_long_2024}}]
    Consider our running example, the time-reversal symmetric Majorana chain, whose low energy field theory corresponds to the theory $\alpha_{\mathit{ABK}}$ of \cref{pinm_IFT_exm}. This theory generates the group $\mho^2_\Pinm \cong \Hom(\Omega_2^\Pinm, \C^\times)\cong \Z/8$. 
    The Anderson-dual map to the Smith homomorphism of \cref{1st_Z2_Smith_exm} is    
    the defect map
    \begin{equation}
        \mathrm{Def}_\sigma\colon \mho^2_\Pinm \cong \Z/8 \to 
 \Z/8\oplus \Z \cong \mho^3_{\Spin}(B\Z/2).
    \end{equation}
    This turns out to be given by inclusion into the first factor, meaning that $\alpha_{\mathit{ABK}}$ arises as the defect theory to the theory that generates the $\Z/8$ factor in $\mho^3_{\Spin}(B\Z/2)$. The symmetry type of this theory, $\Spin\times\Z/2$, encodes fermions with a unitary $\Z/2$ internal symmetry $U$ that squares to the identity.

    Under the anomaly inflow hypothesis, this computation shows that under this form of symmetry breaking, the anomaly of the 2d fermions with unitary $\Z/2$ symmetry is sent isomorphically to the anomaly of the domain wall fermions, while the gravitational anomaly (corresponding to the $\Z$ factor) is forgotten. 
\end{exm}

For an example concerning $\mho_{\Pin^+}^4\cong\Z/16$, which is known at the physics level of rigor to be generated by the low-energy theory of a Majorana fermion, and a more involved anomaly matching computation, see~\cite[III.C.3]{debray_long_2024}.

\begin{rem}
    In this section, we described how Anderson-dualized Smith homomorphisms model defect maps of SPTs. In fact, one can extend \cref{defect_map} to an entire long exact sequence of groups of deformation classes of invertible field theories, which was studied in \cite{debray_long_2024,debray_smith_2024}. The physics processes corresponding to the other two maps in the sequence involve \textit{residual anomalies}, which determine when it is possible to construct a $W$-defect, and the pumping of \textit{Berry phases}.
\end{rem}

\subsubsection{The spiral maps of interacting SPTs}\label{spiral_maps_physics_section}

Queiroz--Khalaf--Stern only considered the dimensional reductions of free theories, incorporating interactions \textit{after} the reduction was performed. 
However, 
inspired by the Smith homomorphism models for symmetry breaking for SPT phases discussed in the previous section (\ref{smith_symm_br}), and following a suggestion of Hason--Komargodski--Thorngren \cite[\S 5]{hason_anomaly_2020}, we define two maps of Anderson-dual bordism groups that model analogs of dimensional reduction on the interacting side.
To avoid conflict with other uses of the term \textit{dimensional reduction} for field theories,\footnote{Again, see \cref{rem:dim_red_terminology}.} we will refer to our mathematical interacting versions of Queiroz--Khalaf--Stern's dimensional reduction maps as \textit{spiral maps}.

As observed by Hason--Komargodski--Thorngren, the dimensional reduction procedures of Queiroz--Khalaf--Stern are quite similar to the $W$-defect creation described in \cref{smith_symm_br}. However, there are at least two confounding aspects at play.

The first aspect to notice is that a Smith homomorphism simply does not have the correct type signature.
A map modeling Queiroz--Khalaf--Stern's procedure should raise the dimension (of IFTS) by 1\footnote{So that the dual map (on bordism classes of manifolds) \textit{lowers} dimension by 1.} and change the tangential structure by one hour of the Bott clock. However, no Smith homomorphism recovers this situation in our model. More precisely, our choice of model for primed phases involves twisting not just over $\Z/2$, but over products $(\Z/2)^{\ell+k}$: for this reason, our model for the Bott spiral map must compare groups of twisted spin IFTs for twists over two different base spaces (e.g. twists over $B\Z/2$ versus over $B(\Z/2\times\Z/2)\simeq B\Z/2\times B\Z/2$).
Since Smith homomorphisms modify the twist but not the base space, they alone do not suffice for this application.

A second aspect of the problem is that our maps for Queiroz--Khalaf--Stern's process should have two different behaviors: they should be either isomorphisms or multiplication by two. In the case of spiraling up from class AIII in $d=1$, for example, each map starting from an unprimed phase should be an isomorphism, while each map starting from a primed phase should be multiplication by two.

With the above in mind, we define two different spiral maps, each involving a Smith homomorphism component. The other components will be the following:

\begin{defn}\label{crush_map_defn}
    Let $X$ be a space. The \textit{crush map}
    $ c_X\colon \Sigma_+^\infty X\to\Sph $
    is 
    $\Sigma_+^\infty$ applied to the unique map $X\to\pt$. 
    When $X$ is clear from context, we will write $c$ instead of $c_X$.
\end{defn}

\begin{defn}\label{diag_map_defn}
    The \textit{diagonal map} is the map
    $$ \Delta\colon (B\Z/2)_+ \to (B\Z/2)^{\sigma-1}\wedge (B\Z/2)^{1-\sigma} = \ME_{1,1} $$
    induced by the diagonal map $B\Z/2\to B\Z/2\times B\Z/2$. 
\end{defn}
Note that for this definition to parse we use that the bundle $(\sigma - 1) \boxplus (1 - \sigma)$ over $B\Z/2 \times B\Z/2$ pulls back to the bundle $(\sigma - 1) \oplus (1 - \sigma) = 0$ over $B\Z/2$.

\begin{rem}
The crush map $\Sigma^\infty B\Z/2 \to \mathbb{S}$ is induced by the unique homomorphism $\Z/2\times\{\pm 1\}\twoheadrightarrow \{\pm 1\}$ of fermionic groups.
Meanwhile, let $e_1,e_2$ be the generators of the fermionic group $E_{1,1}$. The diagonal map is induced by the map of fermionic groups $\Z/2 \times\Spin_1\to E_{1,1}$ sending the nontrivial element of $\Z/2$ to the product $e_1e_2$. 
Multiplied with the identity on a fermionic group $G_f$, $\Delta$ induces a map $G_f\times\Z/2 \to G_f'$.
\end{rem}

Now we are ready to define the two spiral maps.
Recall from \cref{MElk_defn} that for nonnegative integers $\ell$ and $k$, $\ME_{\ell,k} = ((B\Z/2)^{\sigma-1})^{\wedge\ell} \wedge ((B\Z/2)^{1-\sigma})^{\wedge k}$, and recall from \cref{smith_map_of_spectra_specific} that $e_\sigma$ is the map of spectra induced by taking the zero section of the bundle $\sigma\to B\Z/2$, which in turn induces Smith homomorphisms of the form in \cref{smith_hom_defn}. %

\begin{defn}\label{spiral_maps_of_spectra}
    For $\ell \geq 0$ and $k > 0$, let $\phi_{\ell,k}$ be the composition
    \begin{equation}\label{spiral_1_defn}
        \phi_{\ell,k}\colon \ME_{\ell,k} = \ME_{\ell,k-1}\wedge (B\Z/2)^{1-\sigma} 
        \xrightarrow{\id\wedge e_\sigma} 
        \ME_{\ell,k-1} \wedge \Sigma (B\Z/2)_+
        \xrightarrow{\id\wedge c}
        \Sigma\ME_{\ell,k-1},
    \end{equation}
    and let $\psi_{\ell,k}$ be the composition
    \begin{equation}\label{spiral_2_defn}
        \psi_{\ell,k}\colon \ME_{\ell,k} = \ME_{\ell,k-1}\wedge (B\Z/2)^{1-\sigma}
        \xrightarrow{\id\wedge e_\sigma}
        \ME_{\ell,k-1}\wedge \Sigma (B\Z/2)_+
        \xrightarrow{\id\wedge \Delta}
        \ME_{\ell,k-1}\wedge \Sigma \ME_{1,1} = \Sigma\ME_{\ell+1,k}.
    \end{equation}
\end{defn}

Note that $\phi_{\ell,k}$ changes $(\ell,k)$ by $(0,-1)$ while $\psi_{\ell,k}$ changes $(\ell,k)$ by $(1,0)$.

We will also want maps $\psi_{\ell,0}$ for $\ell\ge 3$ in order to model the Bott spiral. Since \cref{spiral_maps_of_spectra} does not immediately generalize to these values of $(\ell,k)$, we give a different definition.

Recall the functor $\zeta\colon\cat{Rep}\to\cat{FermGrp}$ (\cref{zeta_defn}) sending a representation $\rho\colon G\to\O_{\ell,k}$ to its pin cover, as well as the functor from \cref{interacting_symm_mon} sending $(G,\rho)$ to $\MTSpin\wedge (BG)^{\rho-r}$, where $r$ is the rank of the virtual representation induced by $\rho$. By \cref{zetamon,interacting_symm_mon}, both of these functors are symmetric monoidal.

The fermionic groups $E_{\ell,k}$ and $Q_8$ are in the essential image of $\zeta$: for $E_{\ell,k}$, this is in \cref{altelk}, and for $Q_8$, use $G = \Z/2\times\Z/2$ acting as the dihedral symmetry $\mathrm D_2$ (in Schönflies notation) as a subgroup of $\SO_3\subset\O_3\times\O_0$. Thus by \cref{zetamon}, \cref{interacting_symm_mon}, and \cref{lem_shearing_Elk}, there is an $\MTSpin$-module equivalence
\begin{equation}
\label{mdefn__}
    m\colon\mathit{MTH}(Q_8\ftens E_{\ell-3,k-1}) \xrightarrow{\simeq} \mathit{MTH}(Q_8)\wedge_{\MTSpin}\MTSpin\wedge\ME_{\ell-3,k-1}\simeq \mathit{MTH}(Q_8)\wedge\ME_{\ell-3,k-1}.
\end{equation}
\begin{defn}
\label{using_Q8}
Let $\ell\ge 3$. \Cref{class_C_prime} gives an isomorphism of fermionic groups $s_-\colon Q_8\ftens E_{0,1}\overset\cong\to E_{3,0}$; by tensoring with $\id_{E_{\ell-3,k}}$, we get an isomorphism of fermionic groups $s_{\ell,k}\colon Q_8\ftens E_{\ell-3,k+1}\overset\cong\to E_{\ell,k}$. Define $\overline s_{\ell,k}$ to be the following composition, which is an equivalence of $\MTSpin$-modules:
\begin{equation}
    \overline s_{\ell,k}\colon \MTSpin\wedge\ME_{\ell,k} \xrightarrow[\simeq]{\eqref{lem_shearing_Elk}} \mathit{MTH}(E_{\ell,k}) \xrightarrow[\simeq]{s_{\ell,k}^{-1}} \mathit{MTH}(Q_8\ftens E_{\ell-3,k+1}) \xrightarrow[\simeq]{m} \mathit{MTH}(Q_8)\wedge \ME_{\ell-3,k+1},
\end{equation}
where $m$ is as in~\eqref{mdefn__}.
\end{defn}
\begin{rem}
This is an example where a positive answer to \cref{direct_SM_conj} would simplify our construction, by allowing us to avoid the auxiliary data of representations in defining $m$.
\end{rem}
\begin{defn}
\label{Q8_psi}
Let $\ell\ge 3$. Define $\overline\psi_{\ell,0}$ to be the following composition:
\begin{equation}
   \overline\psi_{\ell,0}\colon \MTSpin\wedge\ME_{\ell,0} \xrightarrow[\simeq]{\overline s_{\ell,0}} \mathit{MTH}(Q_8)\wedge \ME_{\ell-3,1} \xrightarrow{\id\wedge\psi_{\ell-3,1}} \mathit{MTH}(Q_8)\wedge \Sigma\ME_{\ell-2,1} \xrightarrow[\simeq]{\overline s_{\ell+1,1}{}^{-1}} \Sigma\MTSpin\wedge\ME_{\ell+1,1}.
\end{equation}
For $k>0$ and $\ell\ge 0$, define $\overline\psi_{\ell,k}\coloneqq \id_{\MTSpin}\wedge\psi_{\ell,k}\colon\MTSpin\wedge\ME_{\ell,k}\to \Sigma\MTSpin\wedge\ME_{\ell+1,k}$ and define $\overline\phi_{\ell,k}\coloneqq \id_{\MTSpin}\wedge\phi_{\ell,k}$ analogously.
\end{defn}
\begin{rem}
We would have preferred to define $\overline\psi_{\ell,0}$ as $\id_{\MTSpin}$ smashed with a map of Thom spectra, like in \eqref{spiral_2_defn}. To do so, we would need
an equivalence of Thom spectra $(B(Q_8)_b)^V\wedge \ME_{\ell-3,1}\overset\simeq\to \ME_{\ell,0}$, for some rank-zero virtual vector bundle $V$, ideally induced by an isomorphism of vector bundles. This is obstructed by \cref{non_VB_isom}.
\end{rem}
By taking Anderson duals, we obtain two maps of groups of (deformation classes of) invertible field theories. In other words, we take the Anderson-dual bordism cohomology of the previous two maps to obtain our maps of interest. For simplicity, we state the following definitions for the real case, using twists of spin. The definitions for spin$^c$ are analogous.
\begin{defn}[Spiral maps]\label{spiral_maps_of_IFTs}
    Let
    \begin{equation}
        \mathrm{sp}^\phi_{\ell,k} \colon [\MTSpin\wedge\Sigma\ME_{\ell,k-1}, \Sigma^{d+2} I_\Z]
        \to [\MTSpin\wedge\ME_{\ell,k}, \Sigma^{d+2} I_\Z]
    \end{equation}
    be the map induced by applying $[\bl, \Sigma^{d+2}I_\Z]$ to $\overline\phi_{\ell,k}$,
    and let
    \begin{equation}
        \mathrm{sp}^\psi_{\ell,k} \colon [\MTSpin\wedge\Sigma\ME_{\ell+1,k}, \Sigma^{d+2} I_\Z] \to [\MTSpin\wedge\ME_{\ell,k}, \Sigma^{d+2} I_\Z]
    \end{equation}
    be the map induced in the same way from $\overline \psi_{\ell,k}$.
    Rewritten using \cref{IFT_notn} and the suspension isomorphism, the maps are of the form
    \begin{equation}
        \mathrm{sp}^\phi_{\ell,k} \colon \mho_{\Spin}^{d+1}(\ME_{\ell,k-1})
        \to
        \mho_{\Spin}^{d+2}(\ME_{\ell,k})
    \end{equation}
    and
    \begin{equation}
        \mathrm{sp}^\psi_{\ell,k} \colon 
        \mho_{\Spin}^{d+1}(\ME_{\ell+1,k})
        \to
        \mho_{\Spin}^{d+2}(\ME_{\ell,k}).
    \end{equation}
\end{defn}
We compute the relevant portion of each of these maps in \cref{section_computations}. The computation for $\mathrm{sp}^\phi_{\ell,k}$ appears in \S\S\ref{ss:ko_phi_f2i}, \ref{spin_computation}, and \ref{complex_computation}, while that for $\mathrm{sp}^\psi_{\ell,k}$ appears in \S\S\ref{computing_spiral_2_map}, \ref{spin_computation}, \ref{other_real_AZ_computation}, and \ref{complex_computation}. %

\begin{rem}[Dimension check]\label{dimension_check_rem}
Observe that, as desired, each map increases the dimension on invertible field theories by one. Just like the $W$-defect map of \cref{W_defect_map_defn}, these maps increase dimension, while the physical process of condensing to a defect decreases the dimension; each of the spiral maps is meant to run from the defect SPT to the one-dimension-higher bulk SPT.
\end{rem}

\begin{rem}[Physical Intuition]
The guiding intuition for the definitions of our two maps, other than the fact that they behave computationally as we desire, is that the primed-to-unprimed phase physical process involves adding and condensing a charged order parameter, while the unprimed-to-primed physical process first requires doubling to form the primed phase, then forming a defect.
In the definition of $\mathrm{sp}^\phi$, we view $e_\sigma$ as forming the defect, as in the Smith homomorphism story reviewed in \cref{smith_homs_intro}, and we view the crush map $c$ as forgetting an auxiliary $\Z/2$-gauge field.
In the definition of $\mathrm{sp}^\psi$, we view $\Delta$ as forming the primed phase in class A$'$ from two copies of the phase in class A, and then $e_\sigma$ as concentrating to the defect submanifold.
We will not attempt to connect more closely to the Hamiltonian picture in this paper, instead pursuing a complementary picture via \cref{LEFT_ansatz}.
\end{rem}
As part of our computations, we will sometimes want to approximate $\mathrm{sp}^\phi_{\ell,k}$ and $\mathrm{sp}^\psi_{\ell,k}$ using spectra other than $\MTSpin$.
\begin{defn}\label{spiral_1_map_E_version}
For a spectrum $E$, $\ell\ge 0$, and $k>0$, let $\spint_{\ell,k}(E)\colon \Sigma I_\Z(E\wedge \ME_{\ell,k-1})\to I_\Z(E\wedge\ME_{\ell,k})$, resp $\mathrm{sp}_{\ell,k}^\psi(E)\colon I_\Z(E\wedge\ME_{\ell+1,k})\to I_\Z(E\wedge\ME_{\ell,k})$ be the maps defined in the same way as $\spint_{\ell,k}$, resp.\ $\mathrm{sp}_{\ell,k}^\psi$, except with $\MTSpin$ replaced with $E$.

If $k = 0$ and $\ell\ge 3$, we must assume $E$ is an $\MTSpin$-module to define $\mathrm{sp}_{\ell,0}^\psi(E)$. In this case, define $\mathrm{sp}_{\ell,0}^\psi(E)$ analogously, except applying $E\wedge_{\MTSpin}$ to $\overline\psi_{\ell,k}$, then taking the Anderson dual.
\end{defn}

\begin{exm}\label{d_pin_spiral_exm}
    We return to the running example between symmetry types dpin and pin$^-$ for an example of the map $\mathrm{sp}^\phi$.
    Let us specialize to dimension $d=3$ and consider the dual map on bordism. Recall from \cref{MElk_defn} that a spin-$(\ell,k)$ structure on a manifold $M$ is equivalent to data of real line bundles $L_1,\dotsc,L_{\ell+k}\to M$ and a spin structure on $TM + L_1 + \dotsb + L_\ell - L_{\ell+1} - \dotsb - L_{\ell+k}$.
    On manifold representatives, the map $\phi_{1,1}$ in bordism is given by~\cite[Lemma F.15]{kaidi_topological_2020}
\begin{equation}
\begin{tikzcd}[row sep=small]
    \Omega^{\mathrm{DPin}}_{3} \cong \Z/8 \ar[r] & \Z/8 \cong \Omega^{\Pin^-}_{2} \\
    (\RP^3,L_1,L_2) \ar[r,mapsto] & (\RP^2,L_1)
\end{tikzcd}
\end{equation}
where the line bundles are such that $T\RP^3 + L_1 - L_2$ is spin and such that $T\RP^2+L_1$ is spin. Recall that $T\RP^n\simeq (n+1)\sigma-1$, and that for any vector bundle $V$, $V^{\oplus 4}$ is spin, to see that we can choose $L_1 = L_2 =\sigma$. 

The dual map on (groups of deformation classes of) IFTs is also an isomorphism:%
\begin{equation}
\begin{tikzcd}
    \mho_{\Pinm}^2 \cong \Z/8 
    \ar[r,"{\mathrm{sp}^\phi_{1,1}}","\cong"'] & \mho_{\DPin}^3 \cong \Z/8.
\end{tikzcd}
\end{equation}
Physically, we interpret
a generator of the domain $\Z/8$ as the
class BDI superconductor in spatial dimension $d=1$, created as a line defect in the class D$'$ superconductor in spatial dimension $d=2$.
\end{exm}

Alternately iterating these two spiral maps, we construct our mathematical models for entire Bott spirals of phases.
We begin our discussion with the complex case.
The physical story that we wish to reproduce is the following.
In \cite[Figure 1 (c)]{queiroz_dimensional_2016}, the complex Bott spiral begins from the base case of the $d=1$ class AIII topological insulator, which generates a $\Z/4$ of interacting SPT phases. It is a dimensional reduction of the $d=2$ class A$'$ topological insulator, which generates a $\Z/4$ of interacting phases, which in turn is a dimensional reduction of the $d=3$ class AIII insulator, which generates a $\Z/8$.
Continuing upward, with each dimension increase by one, we exchange classes AIII and A$'$, and at every even-to-odd dimension step, the order of the interacting SPT grows by a factor of two.

\begin{figure}[h!]
    \centering
\includegraphics{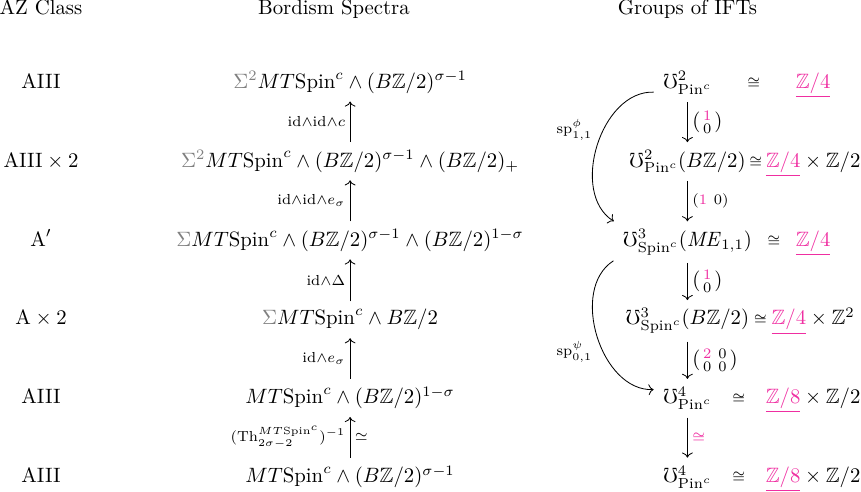}
    \caption{This figure shows the sequence of interacting SPT phases in the complex version of the Bott spiral, starting with the $d=1$ class AIII insulator, which is predicted to generate the group $\mho_{\Pinc}^2\cong\Z/4$ in the top right. The leftmost column indicates the Altland--Zirnbauer class of the SPT at each step. The middle column encodes spectra of the form $\MTSpin^c\wedge \ME_{\ell,k}$ and the maps between them. The last column indicates the $I_\Z$-cohomology of the middle column; i.e.\ the groups of (deformation classes of) invertible field theories that provide our model for the groups of interacting SPTs.
    The spatial dimension of the SPT increases as the rightmost column is read downward; see \cref{dimension_check_rem}.
    \textcolor{Rhodamine}{Pink, underlined text} indicates the image of the free-to-interacting map and maps restricted to those summands.
    The maps $\spint_{\ell,k}$ and $\mathrm{sp}^\psi_{\ell,k}$ are defined in \cref{spiral_maps_of_IFTs} and are the composites of the vertical maps shown, while the last map, in the bottom right, is the Thom isomorphism in $\mho_{\Spinc}^*$-cohomology. Bahri--Gilkey~\cite[Theorems 1 and 2]{BG87} compute the bordism groups corresponding to $\mho_{\Pin^c}^*$ and $\mho_{\Spin^c}^*(B\Z/2)$; \cite[Theorem 4.57]{debray_invertible_2021} does the same for $\mho_{\Pin^c}^2(B\Z/2)$, and \cref{dpin_c} computes $\mho_{\Spin^c}^3(\ME_{1,1})$.
    }
    \label{fig:cplx_Bott_spiral_maps}
\end{figure}

In our IFT model, we begin (in the top right of \cref{fig:cplx_Bott_spiral_maps}) with the group $\mho_{\Pinc}^2\cong\Z/4$~\cite{BG87a, BG87}, which under \cref{LEFT_ansatz} classifies SPTs with \pinc symmetry in spatial dimension 1. We may rewrite, using the Smith isomorphism of \cref{spinc_Z2_Smith}, $\mho_{\Pinc}^2\cong \mho_{\Spinc}^2((B\Z/2)^{\sigma-1}) = \mho_{\Spinc}^2(\ME_{1,0})$.
The spiral map
\begin{equation}\label{ME2_c_spiral}
    \mathrm{sp}^\phi_{{1,1}} \colon \mho_{\Spinc}^2(\ME_{1,0}) \to \mho^3_{\Spinc}(\ME_{1,1})
\end{equation}
is an isomorphism $\Z/4\xrightarrow{\cong} \Z/4$ (see \cref{dpin_c} for the isomorphism $\mho_{\Spin^c}^3(\ME_{1,1})\cong\Z/4$). Thus~\eqref{ME2_c_spiral} maps a generator of the source---predicted to be the $d=1$ class AIII insulator---to a generator of the target---the $d=2$ class A$'$ insulator, modeled as an invertible field theory on spin$^c$-(1,1) manifolds.
The intermediate step shown in \cref{fig:cplx_Bott_spiral_maps}, dual to the crush map, is the addition of an auxiliary $\Z/2$-bundle on the IFT side, which we interpret as the formation of two copies of the initial class AIII SPT.

The next spiral map,
\begin{equation}
    \mathrm{sp}^\psi_{0,1} \colon \mho^3_{\Spinc}(\ME_{1,1}) \to \mho_{\Spinc}^4(\ME_{0,1}) \cong \mho_{\Pinc}^4\cong\Z/8\oplus\Z/2,
\end{equation}
sends any generator of the $\Z/4$ in the source to twice a generator of the $\Z/8$ summand in the classification of $d=3$ class AIII insulators, missing the extra $\Z/2$ summand.

To continue up the spiral, we use \cref{cpx_collapse} to reset from $(\ell,k) = (0,1)$ to $(\ell,k) = (1,0)$ using the Thom isomorphism for $2\sigma$ in $\mho_\Spinc^*$-cohomology. In other words, because $2\sigma$ is a complex vector bundle, it admits a spin$^c$ structure, allowing us to exchange the $\Z/2$ twist by $1-\sigma$  with the twist by $2\sigma-2 +1-\sigma = \sigma-1$.
Equivalently, we are using that $E_{0,1}\ftens \U_1 \cong E_{1,0}\ftens \U_1$, a complexified version of \cref{40_04}.
To continue to arbitrary higher dimensions, we may then iterate $\mathrm{sp}_{1,1}^\phi$, $\mathrm{sp}_{0,1}^\psi$, and the Thom isomorphism.

The physics we wish to describe is only 
that of the interacting SPTs connected to free fermion models---the image of the free-to-interacting map---not the entire groups of invertible field theories. By the results of our computation (\cref{cplx_F2I_image_thm}), we know that the image of the free-to-interacting map is a large two-torsion summand---called the long summand in \cref{section_computations}.
Combining this with our computation of the spiral maps (\cref{ku_int_Bott,complex_psi}), we have the following.

\begin{cor}\label{complex_bott_spiral_computation_corollary}
The spiral maps restricted to the image of the free-to-interacting map (\cref{complex_F2I_map_defn}) take the form
\begin{equation}
    \begin{tikzcd}[row sep=small]
\mathrm{sp}_{1,1}^\phi|_{\mathrm{im}\, F2I^c} \colon \Z/2^{2+(d-1)/2} \ar[r,"\cong"] & \Z/2^{2+(d-1)/2} \\
\qquad \qquad \mho_{\Pin^c}^{d+1} \ar[u,phantom,sloped,"\supset",shift right = 5ex] & \mho_\Spinc^{d+2}(\ME_{1,1}) \ar[u,phantom,sloped,"\supset"]
    \end{tikzcd}
\end{equation}
for $d\geq -1$ odd
and
\begin{equation}
    \begin{tikzcd}[row sep=small]
\mathrm{sp}_{0,1}^\psi|_{\mathrm{im}\, F2I^c} \colon \Z/2^{1+d/2} \ar[r,"\times 2"] & \Z/2^{2+d/2} \\
\qquad \qquad \mho_\Spinc^{d+1}(\ME_{1,1}) \ar[u,phantom,sloped,"\supset",shift right = 5ex] & \mho_{\Pin^c}^{d+2} \ar[u,phantom,sloped,"\supset"]
    \end{tikzcd}
\end{equation}
for $d\geq -2$ even. 
\end{cor}
In other words, as we continue up the spiral, we indeed see the doubling of the invariant at every other dimension step in our invertible field theory model. That is, our model of the Bott spiral beginning in 1d class AIII is in agreement with Queiroz--Khalaf--Stern's from \cref{QKS_result}.

The real cases---spiraling up from class BDI and class CII, resp., in $d=1$ \cite[Figure 1 (a), (b)]{queiroz_dimensional_2016}---are similar, but require more care in our choices of fermionic group assignments to symmetry classes. As in \cref{subsec_symm_types_morita_invnce}, we will use discrete fermionic group models for the Altland--Zirnbauer classes, consisting (except class C) of $E_{\ell,k}$'s. And as explained in \cref{primed_AZ}, for the primed classes (except C$'$), we will simply form the fermionic tensor product with $E_{1,1}$. For class C$'$, we must invoke \cref{class_C_prime} to see that $E_{4,0} \cong Q_8\ftens E_{1,1}$ or $E_{0,4}\cong Q_8\ftens E_{1,1}$ are both valid choices. See \cref{tab:ferm_grp_models_primed_AZ} for the collection of all primed and unprimed Altland--Zirnbauer classes appearing in the real spirals.

\begin{table}[h!]
    \centering
    \begin{tabular}{ccccccccc}
    \toprule
         C$'$ & CII & AII$'$ & DIII & D$'$ & BDI & AI$'$ & CI & C$'$ \\
         \midrule
         $E_{0,4}$ & $E_{0,3}$ & $E_{1,3}$ & $E_{0,1}$ & $E_{1,1}$ & $E_{1,0}$ & $E_{3,1}$ & $E_{3,0}$ & $E_{4,0}$ \\
    \bottomrule
    \end{tabular}
    \caption{(Discrete) fermionic group models for the symmetry types appearing in the Bott spiral, expanding from \cref{tenfold_table} to include primed classes. Class C$'$ is listed twice to emphasize periodicity.}
    \label{tab:ferm_grp_models_primed_AZ}
\end{table}

As in the complex case, by alternately applying the two different spiral maps, we can draw out spirals of phases. When the dimension increases from $2m-1$ to $2m$, we get an isomorphism on the subgroup of IFTs in the image of the free-to-interacting map. Going from dimension $2m$ to $2m+1$, the map is multiplication by $2$. We computed these results in
\cref{int_Bott_spiral_1,MSpin_psi,special_case_psi_ko}.
This replicates \cref{QKS_result} if we omit the parameter $\mu$.

Recall the definition of $\ldeg(\ell, k)$ from \cref{ldeg_defn}.
\begin{cor}\label{real_bott_spiral_computation_corollary}
Let $\widetilde{m}$ and $i$ be as in  \cref{real_F2I_image_cor}, and suppose $d\equiv \ell+k-2i+2\bmod 4$ and $d\ge \ldeg(\ell, k)-2$. If $d\equiv \ell+k-2i+2\bmod 8$, set $c(d)\coloneqq 0$; if $d\equiv \ell+k-2i+6\bmod 8$, set $c(d)\coloneqq 1$.
The spiral maps restricted to the image of the free-to-interacting map (\cref{Elk_F2I}) take the form
\begin{equation}
    \begin{tikzcd}[row sep=small]
\mathrm{sp}_{\ell,k}^\phi|_{\mathrm{im}\, F2I_{\ell,k-1}} \colon \Z/2^{4+4\widetilde{m}+c(d)-i} \ar[r,"\cong"] & \Z/2^{4+4\widetilde{m}+c(d)-i} \\
\qquad \qquad \mho_{\Spin}^{d+1}(\ME_{\ell,k-1}) \ar[u,phantom,sloped,"\supset",shift right = 5ex] & \mho_\Spin^{d+2}(\ME_{\ell,k}) \ar[u,phantom,sloped,"\supset"]
    \end{tikzcd}
\end{equation}
and
\begin{equation}
    \begin{tikzcd}[row sep=small]
\mathrm{sp}_{\ell,k}^\psi|_{\mathrm{im}\, F2I_{\ell+1,k}} \colon \Z/2^{4+4\widetilde{m}+c(d)-i} \ar[r,"\times 2"] & \Z/2^{5+4\widetilde{m}+c(d)-i}. \\
\qquad \qquad \mho_\Spin^{d+2}(\ME_{\ell+1,k}) \ar[u,phantom,sloped,"\supset",shift right = 5ex] & \mho_{\Spin}^{d+3}(\ME_{\ell,k}) \ar[u,phantom,sloped,"\supset"]
    \end{tikzcd}
\end{equation}
\end{cor}
Thus our models for the Bott spirals starting in 1d classes BDI and CII are in \emph{nearly} complete agreement with Queiroz--Khalaf--Stern's \cref{QKS_result}. We discuss the differences below in \cref{mu2_strikes_back}.

We find for the real cases that it is actually more illuminating to draw an entire grid of possible sequences of spiral maps. We do so in \cref{fig:grid_for_real_spirals}, identifying symmetry types for spin-$(\ell,k)$ IFTs according to \cref{tab:ferm_grp_models_primed_AZ} plus following the rule that increasing $(\ell,k)$ by $(1,1)$ adds a prime. For example, the fermionic group $E_{2,2}$ models a class we call D$''$.

\begin{rem}
\label{mu2_strikes_back}
Our analysis of the interacting Bott spiral clarifies an aspect of Queiroz--Khalaf--Stern's analysis that is invisible from the free fermion side of the problem. Specifically, Queiroz--Khalaf--Stern use the parameter $\mu$ to double the size of the groups of interacting SPTs in classes BDI, D$'$, and DIII, and thus obtain their desired interacting classification in a slightly different way from the behavior of the other five real Altland--Zirnbauer classes. In general, the Bott spiral computations suggest that when one increases the dimension by $2$, the order of the image of the free-to-interacting map doubles. This suggests that, for example, in $d = 1$ class BDI, there is a $\Z/4$ classification of interacting phases. However, Queiroz--Khalaf--Stern change their parameter $\mu$ to obtain a $\Z/8$ of phases, matching the classification in, e.g., \cite{fidkowski_effects_2010, fidkowski_topological_2011, turner_topological_2011, GW14, you_topological_2014, kapustin_fermionic_2015, witten_fermion_2016, debray_arf-brown_2018, kobayashi_pin_2019, inamura_nonlocal_2020, Tur20, freed_reflection_2021, turzillo_duality_2024}.

In our model for the Bott spiral, using $\mu$ in this way corresponds to removing a prime: the spiral we most naturally obtain includes the discrete Altland--Zirnbauer types $\mathrm{BDI}'$, $\mathrm{D}''$, and $\mathrm{DIII}'$, corresponding to $(\ell, k) = (2, 1)$, $(2, 2)$, and $(1, 2)$ respectively. The classification of $\mathrm{BDI}'$ phases in 1d is $\Z/4\subset \mho_{2,1}^2$ (\cref{real_bott_spiral_computation_corollary}), matching the expected behavior of the Bott spiral but not Queiroz--Khalaf--Stern's modification. The same factor of $2$ discrepancy occurs in classes BDI vs $\mathrm{BDI}'$ in all dimensions $4m+1$, $\mathrm{D}'$ vs $\mathrm{D}''$ in dimensions $4m+2$, and DIII vs $\mathrm{DIII}'$ in dimensions $4m+3$.

Which of these two approaches correctly models the Bott spiral? We use the spiral maps $\mathrm{sp}^\phi_{\ell,k}$ and $\mathrm{sp}^\psi_{\ell,k}$ to answer this question. We defined the spiral maps to model the processes Queiroz--Khalaf--Stern describe to pass between different entries in the Bott spiral. Therefore we believe \emph{the correct model for the Bott spiral is one in which it makes sense to alternate the spiral maps $\mathrm{sp}^\phi$ and $\mathrm{sp}^\psi$ through a full period of the spiral.}
As the spiral maps are defined on groups of IFTs, this ansatz is visible only on the interacting side of the free-to-interacting correspondence. %

Using this, we argue that the Bott spiral must include $\mathrm{BDI}'$, $\mathrm{D}''$, and $\mathrm{DIII}'$ instead of BDI, $\mathrm{D}'$, and DIII. For the sake of contradiction, begin the spiral in dimension $d$ and class BDI. Then, applying $\mathrm{sp}_{1,1}^\phi$, we land in dimension $d+1$ class $\mathrm{D'}$. Next, apply $\mathrm{sp}_{0,1}^\psi$ to land in $(d+2)$-dimensional class DIII. In \cref{fig:grid_for_real_spirals,CII_spiral}, we color these three steps in \textcolor{Orange}{orange}.

So far there is no contradiction. By \cref{real_bott_spiral_computation_corollary}, if $d = 8j+1$ we have a $\Z/2^{4j+3}$ summand in $\mho_{1,0}^{8j+1}$ (class BDI), a $\Z/2^{4j+3}$ summand in $\mho_{1,1}^{8j+2}$ (class $\mathrm{D'}$), and a $\Z/2^{4j+4}$ summand in $\mho_{0,1}^{8j+3}$ (class DIII), in agreement with the $\mu = 2$ case of~\cite[Figure 1(a)]{queiroz_dimensional_2016}. The case $j = 0$ specializes to the $\Z/8$ of 1d \pinm IFTs (\cref{pinm_IFT_exm}), $\Z/8$ of 2d dpin IFTs, and $\Z/16$ of 3d \pinp IFTs. Likewise, $d = 8j+5$ reproduces the $\mu = 2$ cases of~\cite[Figure 1(b)]{queiroz_dimensional_2016}. \Cref{real_bott_spiral_computation_corollary} implies that the maps $\mathrm{sp}^\phi_{1,1}$, resp.\ $\mathrm{sp}^\psi_{0,1}$ are an isomorphism, resp.\ multiplication by $2$ on those summands, as predicted.

Next we apply $\mathrm{sp}^\phi_{0,2}$ to land in $(d+3)$-dimensional class AII. This is a slight problem, as we expected class $\mathrm{AII}'$, and the groups of IFTs continue to be off by a factor of $2$. There does not appear to be a physically meaningful map from classes DIII or AII to $\mathrm{AII}'$. But then we have a bigger problem: since $\mathrm{sp}^\psi$ lowers the value of $\ell$, it is undefined on class AII, for which $\ell = 0$. Therefore if we started the spiral in class BDI, we cannot continue past here. By contrast, if we began with $\mathrm{BDI}'$, then after three steps we are in class $\mathrm{AII}'$ in accordance with Queiroz--Khalaf--Stern's prediction, and $\ell = 1$, so we can continue the spiral. We interpret this as evidence that the correct model for the Bott spiral uses classes $\mathrm{BDI}'$, $\mathrm{D}''$, and $\mathrm{DIII}'$ instead of BDI, $\mathrm{D}'$, and DIII, and therefore has a slightly different classification of IFTs in those cases.
\end{rem}

\begin{figure}[h!]
\[\begin{tikzcd}[column sep=3.3em, row sep = 2.7em]
	{\textcolor{black}{\underbracket{\mho_{3,0}^i}_{\text{$\mathrm{CI}$ }}}} & {\underbracket{\mho^{i+1}_{3,1}}_{\textrm{AI}'}} & {\underbracket{\mho^{i+2}_{3,2}}_{\textrm{BDI}''}} & {\underbracket{\mho^{i+3}_{3,3}}_{\textrm{D}'''}} & {\underbracket{\mho^{i+4}_{3,4}}_{\textrm{DIII}'''}} & {\underbracket{\mho^{i+5}_{3,5}}_{\textrm{AII}'''}} & \cdots \\
	{\underbracket{\mho^{i+1}_{2,0}}_{\textrm{AI}}} & {\textcolor{MidnightBlue}{\underbracket{\mho^{i+2}_{2,1}}_{\textrm{BDI}'}}} & {\textcolor{MidnightBlue}{\underbracket{\mho^{i+3}_{2,2}}_{\textrm{D}''}}} & {\underbracket{\mho^{i+4}_{2,3}}_{\textrm{DIII}''}} & {\underbracket{\mho^{i+5}_{2,4}}_{\textrm{AII}''}} & {\underbracket{{\mho^{i+6}_{2,5}}}_{\textrm{CII}''}} & \cdots \\
	{\textcolor{Orange}{\underbracket{\mho^{i+2}_{1,0}}_{\textrm{BDI}}}} & {\textcolor{Orange}{\underbracket{\mho^{i+3}_{1,1}}_{\textrm{D}'}}} & {\textcolor{MidnightBlue}{\underbracket{\mho^{i+4}_{1,2}}_{\textrm{DIII}'}}} & {\textcolor{MidnightBlue}{\underbracket{\mho^{i+5}_{1,3}}_{\textrm{AII}'}}} & {\underbracket{\mho^{i+6}_{1,4}}_{\textrm{CII}'}} & {\underbracket{\mho^{i+7}_{1,5}}_{\textrm{C}''}} & \cdots \\
	& {\textcolor{Orange}{\underbracket{\mho^{i+4}_{0,1}}_{\textrm{DIII}}}} & {\underbracket{\mho^{i+5}_{0,2}}_{\textrm{AII}}} & {\textcolor{MidnightBlue}{\underbracket{\mho^{i+6}_{0,3}}_{\textrm{CII}}}} & {\textcolor{MidnightBlue}{\underbracket{\mho^{i+7}_{0,4}}_{\textrm{C}'}}} & {\underbracket{\mho^{i+8}_{4,1}}_{\textrm{CI}'}} & \cdots \\
	&&&& {\textcolor{MidnightBlue}{\underbracket{\mho^{i+8}_{3,0}}_{\textrm{CI}}}} & {\textcolor{MidnightBlue}{\underbracket{\mho^{i+9}_{3,1}}_{\textrm{AI}'}}} & \cdots \\
	&& {} &&  & \vdots
	\arrow["\mathrm{sp}^{\phi}_{3,1}", from=1-1, to=1-2]
	\arrow["\mathrm{sp}^{\phi}_{3,2}", from=1-2, to=1-3]
	\arrow["\mathrm{sp}^{\psi}_{2,1}", from=1-2, to=2-2]
	\arrow["\mathrm{sp}^{\phi}_{3,3}", from=1-3, to=1-4]
	\arrow["\mathrm{sp}^{\psi}_{2,2}", from=1-3, to=2-3]
	\arrow["\mathrm{sp}^{\phi}_{3,4}", from=1-4, to=1-5]
	\arrow["\mathrm{sp}^{\psi}_{2,3}", from=1-4, to=2-4]
	\arrow["\mathrm{sp}^{\phi}_{3,5}", from=1-5, to=1-6]
	\arrow["\mathrm{sp}^{\psi}_{2,4}", from=1-5, to=2-5]
	\arrow["\mathrm{sp}^{\phi}_{3,6}", from=1-6, to=1-7]
	\arrow["\mathrm{sp}^{\psi}_{2,5}", from=1-6, to=2-6]
	\arrow["\mathrm{sp}^{\phi}_{2,1}", from=2-1, to=2-2]
	\arrow["\mathrm{sp}^{\phi}_{2,2}", color={MidnightBlue}, from=2-2, to=2-3]
	\arrow["\mathrm{sp}^{\psi}_{1,1}", from=2-2, to=3-2]
	\arrow["\mathrm{sp}^{\phi}_{2,3}", from=2-3, to=2-4]
	\arrow["\mathrm{sp}^{\psi}_{1,2}", color={MidnightBlue}, from=2-3, to=3-3]
	\arrow["\mathrm{sp}^{\phi}_{2,4}", from=2-4, to=2-5]
	\arrow["\mathrm{sp}^{\psi}_{1,3}", from=2-4, to=3-4]
	\arrow["\mathrm{sp}^{\phi}_{2,5}", from=2-5, to=2-6]
	\arrow["\mathrm{sp}^{\psi}_{1,4}", from=2-5, to=3-5]
	\arrow["\mathrm{sp}^{\phi}_{2,6}", from=2-6, to=2-7]
	\arrow["\mathrm{sp}^{\psi}_{1,5}", from=2-6, to=3-6]
	\arrow["\mathrm{sp}^{\phi}_{1,1}", color={Orange}, from=3-1, to=3-2]
	\arrow["\mathrm{sp}^{\phi}_{1,2}", from=3-2, to=3-3]
	\arrow["\mathrm{sp}^{\psi}_{0,1}", color={Orange}, from=3-2, to=4-2]
	\arrow["\mathrm{sp}^{\phi}_{1,3}", color={MidnightBlue}, from=3-3, to=3-4]
	\arrow["\mathrm{sp}^{\psi}_{0,2}", from=3-3, to=4-3]
	\arrow["\mathrm{sp}^{\phi}_{1,4}", from=3-4, to=3-5]
	\arrow["\mathrm{sp}^{\psi}_{0,3}", color={MidnightBlue}, from=3-4, to=4-4]
	\arrow["\mathrm{sp}^{\phi}_{1,5}", from=3-5, to=3-6]
	\arrow["\mathrm{sp}^{\psi}_{0,4}", from=3-5, to=4-5]
	\arrow["\mathrm{sp}^{\phi}_{1,6}", from=3-6, to=3-7]
	\arrow["\mathrm{sp}^{\psi}_{4,1}\circ{\eqref{4exchange}}", from=3-6, to=4-6]
	\arrow["\mathrm{sp}^{\phi}_{0,2}", from=4-2, to=4-3]
	\arrow["\mathrm{sp}^{\phi}_{0,3}", from=4-3, to=4-4]
	\arrow["\mathrm{sp}^{\phi_{0,4}}", color={MidnightBlue}, from=4-4, to=4-5]
	\arrow["\mathrm{sp}^{\phi}_{4,1}\circ{\eqref{4exchange}}", from=4-5, to=4-6]
	\arrow["{\textcolor{MidnightBlue}{\mathrm{sp}^{\psi}_{3,0}\circ{\eqref{4exchange}}}}", draw={MidnightBlue}, from=4-5, to=5-5]
	\arrow["\mathrm{sp}^{\phi}_{4,2}", from=4-6, to=4-7]
	\arrow["{\mathrm{sp}^{\psi}_{3,1}}", from=4-6, to=5-6]
	\arrow["\mathrm{sp}^{\phi}_{3,1}"', color={MidnightBlue}, from=5-5, to=5-6]
	\arrow["\mathrm{sp}^{\phi}_{3,2}", from=5-6, to=5-7]
	\arrow["\mathrm{sp}^{\psi}_{2,1}", color={MidnightBlue}, from=5-6, to=6-6]
\end{tikzcd}\]

    \caption{In this figure, we abbreviate $\mho^i_{\ell,k}\coloneqq \mho^i_{\Spin}(\ME_{\ell,k})$.
    This grid shows the groups of (deformation classes of) invertible field theories on spin-$(\ell,k)$ manifolds that can be reached by repeatedly applying the spiral maps of \cref{spiral_maps_of_IFTs} along with identifications resulting from \cref{40_04}.
    Physically, these represent the groups of interacting SPTs able to be reached by the creation of a $\Z/2$-symmetry defect, referred to as dimensional reduction in \cite{queiroz_dimensional_2016}.
    We discuss the \textcolor{MidnightBlue}{blue} and \textcolor{Orange}{orange} zigzags in the main text.
    }
    \label{fig:grid_for_real_spirals}
\end{figure}

\begin{lem}
    The diagram in \cref{fig:grid_for_real_spirals} commutes.
\end{lem}
\begin{proof}
    Zero section maps are natural in maps of vector bundles.
    More precisely, if $f \colon X \to Y$ is a map, $V \to Y$ is a virtual vector bundle and $W \to Y$ an honest vector bundle, then
    \begin{equation}
    \begin{tikzcd}
        X^{f^* V} \ar[d] \ar[r] & X^{f^* V \oplus f^* W} \ar[d]
        \\
        Y^V \ar[r] & Y^{V \oplus W}
    \end{tikzcd}
    \end{equation}
    commutes.
\end{proof}

In the grid, there are several spirals we can draw out.\footnote{Or, \textit{unravel}, so to speak.} The spiral according to \cite[Figure 1 (a)]{queiroz_dimensional_2016} with $\mu$ fixed to 1 corresponds to the zigzag---drawn in \textcolor{MidnightBlue}{blue}---beginning with $\Z/4\subset \mho^2_\Spin(\ME_{2,1})$, whose generator we call the 1-dimensional class BDI$'$ superconductor.
Meanwhile, to match a version with $\mu=2$, we can start instead from $\mho^2_\Spin(\ME_{1,0}) \cong \Z/8$, whose generator corresponds to the conventional class BDI superconductor. We draw the three steps of this zigzag, corresponding to the three steps of the Bott spiral that Queiroz--Khalaf--Stern consider with $\mu = 2$, in \textcolor{Orange}{orange}. As we discussed in \cref{mu2_strikes_back}, the orange zigzag cannot continue---we cannot employ the identification of \cref{40_04} until at least one index reaches 4. For this reason, we find the \textcolor{MidnightBlue}{blue} zigzag to be more descriptive.

\begin{figure}[h!]
\[\begin{tikzcd}[column sep=small,row sep=scriptsize]
	{\text{BDI}'} & {\text{D}''} & {\text{DIII}'} & {\text{AII}'} & {\text{CII}} & {\text{C}'} & {\text{CI}} & {\text{AI}'} & {\text{BDI}'} \\
	{\mho^2_{2,1}} & {\mho^3_{2,2}} & {\mho^4_{1,2}} & {\mho^5_{1,3}} & {\mho^6_{0,3}} & {\mho^7_{0,4}\cong\mho^7_{4,0}} & {\mho^8_{3,0}} & {\mho^9_{3,1}} & {\mho^{10}_{2,1}} & \cdots \\
	{\Z/4} & {\Z/4} & {\Z/8} & {\Z/8} & {\Z/16} & {\Z/16} & {\Z/32} & {\Z/32} & {\Z/64} & \cdots
	\arrow["\phi", from=2-1, to=2-2]
	\arrow["\psi", from=2-2, to=2-3]
	\arrow["\phi", from=2-3, to=2-4]
	\arrow["\psi", from=2-4, to=2-5]
	\arrow["\phi", from=2-5, to=2-6]
	\arrow["\psi", from=2-6, to=2-7]
	\arrow["\phi", from=2-7, to=2-8]
	\arrow["\psi", from=2-8, to=2-9]
	\arrow[from=2-9, to=2-10]
	\arrow["\cong", from=3-1, to=3-2]
	\arrow["{\times 2}", from=3-2, to=3-3]
	\arrow["\cong", from=3-3, to=3-4]
	\arrow["\times2", from=3-4, to=3-5]
	\arrow["\cong", from=3-5, to=3-6]
	\arrow["\times2", from=3-6, to=3-7]
	\arrow["\cong", from=3-7, to=3-8]
	\arrow["{\times 2}", from=3-8, to=3-9]
	\arrow[from=3-9, to=3-10]
    \arrow["\subset", phantom, sloped, from=3-1, to=2-1]
    \arrow["\subset", phantom, sloped, from=3-2, to=2-2]
    \arrow["\subset", phantom, sloped, from=3-3, to=2-3]
    \arrow["\subset", phantom, sloped, from=3-4, to=2-4]
    \arrow["\subset", phantom, sloped, from=3-5, to=2-5]
    \arrow["\subset", phantom, sloped, from=3-6, to=2-6]
    \arrow["\subset", phantom, sloped, from=3-7, to=2-7]
    \arrow["\subset", phantom, sloped, from=3-8, to=2-8]
    \arrow["\subset", phantom, sloped, from=3-9, to=2-9]
\end{tikzcd}\]
    \caption{As in \cref{fig:grid_for_real_spirals}, we abbreviate $\mho^i_{\Spin}(\ME_{\ell,k})$ by $\mho^i_{\ell,k}$. We also abbreviate $\mathrm{sp}^\phi_{\ell,k}$ by $\phi$ and $\mathrm{sp}^\psi_{\ell,k}$ by $\psi$.
    This sequence of maps is the blue zigzag from \cref{fig:grid_for_real_spirals} starting in degree 2, modeling the spiral of SPT phases starting from class BDI$'$ in spacetime dimension 2. In this figure, we also label the symmetry types and fill in the results of our computations (see \cref{section_computations,s:spin_computations}).
    }
    \label{fig:real_spiral_from_BDI_prime}
\end{figure}

As our last extended example, we show in \cref{CII_spiral} how our formalism recovers Queiroz--Khalaf--Stern's third Bott spiral, which starts in 1d in (discrete) class CII~\cite[Figure 1(b)]{queiroz_dimensional_2016}. The subtlety we found in \cref{mu2_strikes_back} returns: as we argued there, this Bott spiral includes classes $\mathrm{BDI}'$, $\mathrm D''$, and $\mathrm{DIII}'$ (corresponding to $(\ell, k) = (2, 1)$, $(2, 2)$, and $(1, 2)$ respectively) rather than BDI, $\mathrm D'$, and DIII. This corresponds to changing the value of $\mu$ in Queiroz--Khalaf--Stern's Bott spiral.

\begin{figure}[h!]
\centerline{\includegraphics{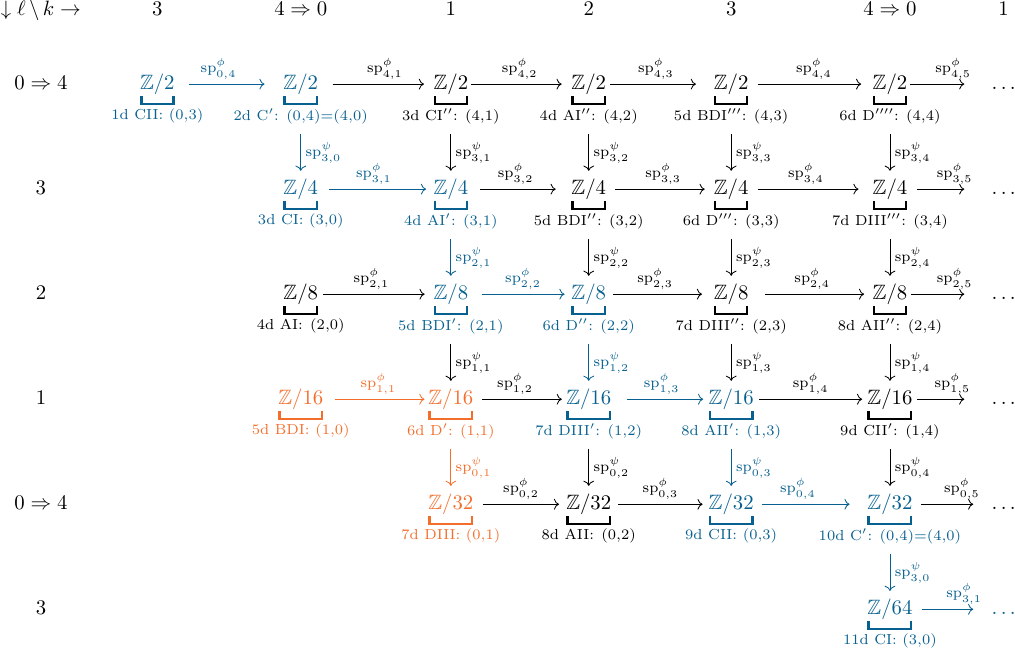}}
\caption{The third Bott spiral: as in~\cite[Figure 1(b)]{queiroz_dimensional_2016}, this spiral starts in dimension $1$ in (discrete) Altland--Zirnbauer class CII. The bracket under each entry displays the (spatial) dimension $d$, the discrete Altland--Zirnbauer class, and the values $(\ell, k)$; the group is the long summand $\Z/2^N\subset \mho_{\ell,k}^{d+1}$. These groups, and the effects of the spiral maps $\mathrm{sp}_{\ell,k}^\phi$ and $\mathrm{sp}_{\ell,k}^\psi$, are given in \cref{real_bott_spiral_computation_corollary}. The orange entries in the lower left correspond to the case $\mu = 1$, as discussed in \cref{mu2_strikes_back}.
}
\label{CII_spiral} 
\end{figure}

\begin{rem}[Discrete versus continuous models for Altland--Zirnbauer classes]
\label{mixed_DC_spiral}
Unlike for classes BDI and AIII, we have two fermionic group models for class CII: the discrete choice $E_{0,3}$ and the continuous choice $\Pin_3^-$. For both of these choices, the group of IFTs in (spatial) dimension $1$ in the image of the free-to-interacting map is isomorphic to $\Z/2$: see \cref{only_first_summand_real} for the discrete case and~\cite[Corollary 9.103]{freed_reflection_2021} for the continuous case. In \cref{CII_spiral} we chose the discrete symmetry group. This is because, if we start with the continuous symmetry group instead, we soon enough run into a problem defining $\mathrm{sp}^\psi$.
See \cref{the_wrong_CII_spiral} for a chart of these spiral maps. Similar issues occur when beginning a spiral with any of the continuous symmetry groups not isomorphic to their discrete counterparts.

The first step of the pin\textsuperscript{$h-$} Bott spiral is not the issue: apply $\mathrm{sp}^\phi$ as usual, and it is an isomorphism on the long summands. We called this map $\mathrm{sp}_{0,1}^\phi(\MTPin^{h-})$ in \cref{spiral_1_map_E_version}. This, and the other spiral maps we use in this remark, are computed in \cref{remaining_real_int_Bott_thm,MSpinHs_psi}.

After this, it is unclear how to apply $\mathrm{sp}^\psi$, because we do not seem to have an $\ME_{1,0}$ factor. Fortunately, though, by~\cite[(3.11)]{DK24} there are $\MTSpin$-module equivalences
\begin{equation}\label{C_I_II_III}
    \MTPin^{h-} \wedge\ME_{0,1} \simeq \MTSpin^h\wedge\ME_{1,1}\simeq \MTPin^{h+}\wedge\ME_{1,0}.
\end{equation}
With this identification, we can apply $\mathrm{sp}^\psi_{0,0}(\MTPin^{h+})$, and land in $\mho_{\Pin^{h+}}^4\cong\Z/4$~\cite[Theorem 9.97]{freed_reflection_2021}, then apply $\mathrm{sp}^\phi_{0,1}(\MTPin^{h+})$ and land in $\mho_{\Pin^{h+}}^5(\ME_{0,1})$, which also contains a $\Z/4$ long summand. However, now we are stuck: as far as we can tell, there is no way to rearrange $\MTPin^{h+}\wedge\ME_{0,1}$ to obtain an $\ME_{1,0}$ factor, hence no way to apply $\mathrm{sp}^\psi$ and continue the spiral. We interpret this fact as evidence that the correct model for the Bott spiral uses the discrete fermionic groups, not the continuous ones.
\end{rem}

\begin{figure}[h!]
\[%
\begin{tikzcd}[row sep=1em]
	{\mho_{\Pin^{h-}}^2} & {\mho^3_{\Pin^{h-}}(\ME_{0,1}) \cong \mho^3_{\Pin^{h+}}(\ME_{1,0})} & {\mho^4_{\Pin^{h+}}} & {\mho^5_{\Pin^{h+}}(\ME_{0,1})} & {?} \\
	{\Z/2} & {\Z/2} & {\Z/4} & {\Z/4} & {?}
	\arrow["\phi", from=1-1, to=1-2]
	\arrow["\subset", phantom, sloped, from=2-1, to=1-1]
	\arrow["\psi", from=1-2, to=1-3]
	\arrow["\supset", phantom, sloped, from=1-2, to=2-2]
	\arrow["\phi", from=1-3, to=1-4]
	\arrow["\supset", phantom, sloped, from=1-3, to=2-3]
	\arrow[dashed, from=1-4, to=1-5]
	\arrow["\subset", phantom, sloped, from=2-4, to=1-4]
	\arrow["\cong", from=2-1, to=2-2]
	\arrow["{\times 2}", from=2-2, to=2-3]
	\arrow["\cong", from=2-3, to=2-4]
	\arrow[dashed, from=2-4, to=2-5]
\end{tikzcd}\]
\caption{If one tries to model the Bott spiral starting with \emph{continuous} class CII in dimension $1$, the first several maps in the Bott spiral work out the same as in the discrete case, but once we get to dimension $5$, there is no map $\mathrm{sp}^\psi$ that would allow us to continue the spiral. Here $\phi$ and $\psi$ mean the corresponding $\mathrm{sp}^\phi$ and $\mathrm{sp}^\psi$ maps smashed with $\MTPin^{h\pm}$. We discuss this further in \cref{mixed_DC_spiral}, where we interpret this as evidence that, when modeling the Bott spiral, the correct symmetry types to use to represent Altland--Zirnbauer classes are the discrete ones.
The groups of IFTs in this spiral are computed in~\cite[Theorem 9.97]{freed_reflection_2021} and \cref{gen_AZ_f2i}. The spiral maps are computed in \cref{remaining_real_int_Bott_thm,MSpinHs_psi}. The isomorphisms $\mho_{\Pin^{h-}}^*(\ME_{0,1})\cong\mho_{\Spin^h}^*(\ME_{1,1})\cong\mho_{\Pin^{h+}}^*$ appear in~\eqref{C_I_II_III}.
\label{the_wrong_CII_spiral}
}
\end{figure}

A theme of this paper has been the breakdown of Bott periodicity as we allow interactions, the tension between the ``Bott clock'' of free tenfold way SPTs and the ``Bott spiral'' of related interacting SPTs.
Now, we can finally describe what remains of Bott periodicity on the interacting side. Instead of an isomorphism after eight steps around the spiral---meaning, eight steps in zigzag within \cref{fig:grid_for_real_spirals}---we find a multiplication by 16 on (deformation classes of) groups of IFTs.
In fact, the dual process to the multiplication by 16 is another multiplication by 16 on the bordism side. On the level of manifold representatives, it is given by multiplication by a real \textit{Bott manifold}, such as the one constructed in~\cite[\S 5.3]{FH21b}, specifically.
As we mentioned in \cref{ABS11_ex}, a real Bott manifold is any closed spin manifold whose bordism class maps to the real Bott class $\beta\in \KO_8$ under the Atiyah--Bott--Shapiro map.
We see that Atiyah--Bott--Shapiro, which was the central ingredient in our free-to-interacting maps (\cref{ss:F2I}), is also the underlying reason that we observe the interacting spiral. %

\section{Computations over $\ko$}\label{section_computations}
    Let $\ko$ denote the connective cover of the real $KO$-theory spectrum. For a few different reasons, $\ko$ is a reasonable approximation to the spin bordism spectrum $\MTSpin$: in degrees $7$ and below, they are equivalent, and in general $\ko$ is a summand of $\MTSpin$~\cite{ABP67}. Therefore, we will separate our computations into two steps: in this section, we study the free-to-interacting maps and the spiral maps with $\MTSpin$ replaced with $\ko$. Then, in \S\ref{s:spin_computations}, we lift our computations to $\MTSpin$.
\begin{subequations}
Our main results in this section are:
\begin{enumerate}
    \item \Cref{melk_calc}, computing $\pi_*(\ko\wedge\ME_{\ell,k})$ modulo simple $2$-torsion \term{Whitney summands}, which do not play a role in the Bott spiral (see \cref{rem:bosonicsummand}).
    \item \Cref{compute_F2I}, computing the image of the $\ko$-module analog of the free-to-interacting map that we constructed in \cref{IZ_ABS_through_ko}:
    \begin{equation}
        \mathit{f2i}_{\ell,k}\colon \KO^{d+\ell-k-2}\to (I_\Z\ko)^{d+2}(\ME_{\ell,k}),
    \end{equation}
    for all $\ell,k$ not both equal to $0$.
    \item \Cref{int_Bott_spiral_ko}, computing the $\ko$ version of $\mathrm{sp}^\phi$ (see \cref{spiral_1_map_E_version}),
    \begin{equation}
    \spint_{\ell,k}(\ko)\colon (I_\Z\ko)^{d+1}(\ME_{\ell,k})\to
	(I_\Z\ko)^{d+2}(\ME_{\ell,k+1}),
    \end{equation}
    again modulo Whitney summands, for all $\ell,k$ not both equal to $0$.
    \item \Cref{psicalc}, computing the $\ko$ version of $\mathrm{sp}^\psi$ (\cref{spiral_1_map_E_version}) modulo Whitney summands:
    \begin{equation}
    \mathrm{sp}^{\psi}_{\ell,k}(\ko) \colon
    (I_\Z \ko)^{d+1}(\ME_{\ell+1,k}) \to (I_\Z\ko)^{d+2}(\ME_{\ell,k}),\end{equation}
    for all $\ell,k>0$. (We will tackle $\mathrm{sp}^\psi_{\ell,0}(\ko)$, where $\ell\ge 3$, in \cref{other_real_AZ_computation}.)
\end{enumerate}
\end{subequations}
Our computation is entirely homotopy-theoretic, which has advantages and disadvantages. Our computations build on and generalize previous work computing the $\ko$-theory of $BV$ for $V$ an elementary abelian $2$-group~\cite{Oss89, Yu95, JW97, Bru99, BG03, BG10, Bru14, Mal11, Sie13, Pow14a, Pow14b, Ric14,  BMSS15, Cam17}: we show that these computations generalize to \emph{some} twists of $\ko$-theory over $BV$, giving many new computations of twisted $\ko$-homology for these spaces. Taking mod $2$ cohomology, we obtain a connection between the isomorphism of fermionic groups $E_{4,0}\cong E_{0,4}$ from \cref{40_04}, the $4$-torsion in the $Q_1$-local Picard group of $\cA(1)$~\cite{Yu95, Bru14}, and the $4$-periodicity in Smith families over $\MTSpin$ and $\ko$, studied by~\cite{kapustin_fermionic_2015, TY19, hason_anomaly_2020, WWZ20, debray_long_2024, debray_smith_2024} and reviewed in~\cite[Examples 6.14 and 7.8]{debray_smith_2024}. We plan to elaborate on this connection in future work.

On the other hand, our computations make use of technical tools from homotopy theory, many of which are uncommon in the mathematical physics literature. For this reason, we would be interested in learning whether other methods could be used to compute the spiral and F2I maps of spectra that we defined in \cref{Elk_F2I,complex_F2I_map_defn,spiral_maps_of_IFTs}.

We begin in \S\ref{ko_generalities} with some groundwork and utility lemmas for our computations. Then, in \S\ref{ss:EA_type}, we introduce our key computational examples and use them to prove \cref{melk_calc}, calculating the interesting part of $\pi_*(\ko\wedge\ME_{\ell,k})$. In \S\ref{ss:ko_phi_f2i} we calculate $\mathrm{sp}^\phi_{\ell,k}(\ko)$ and $\mathit{f2i}_{\ell,k}$. Finally, in \S\ref{ss:ko_psi}, we calculate $\mathrm{sp}_{\ell,k}^\psi(\ko)$ assuming $\ell,k>0$.%

\subsection{Generalities for the computation over \texorpdfstring{$\ko$}{ko}}
\label{ko_generalities}

\label{ss:kogen}
Our computational results are founded on a simple description of $H^*(\ME_{\ell,k};\Z/2)$ that is $(4, 1)$-periodic in $(\ell, k)$, again modulo a trivial piece. This is a twisted version of a continuing program
understanding $H^*(BV;\Z/2)$, $\ku_*(BV)$, and $\ko_*(BV)$ for $V$ an elementary abelian $2$-group, as described above.

Recall that there is a ring isomorphism
\begin{equation}
\ko_*\cong \Z[\eta,a,\beta]/(\eta^3, 2\eta, \eta a, a^2 - 4\beta) \quad \abs\eta = 1, \abs a = 4, \abs\beta = 8.
\end{equation}
The class $\beta$ is called the (real) \term{Bott class}, as it implements Bott periodicity in periodic $\KO$-theory. For any space or spectrum $X$, $\KO_*(X)\cong \ko_*(X)[\beta^{-1}]$, and more generally, for any $\ko$-module $M$, there is a natural isomorphism
\begin{equation}\label{bott_on_modules}
    \pi_*(M)[\beta^{-1}] \xrightarrow{\cong} \pi_*(\KO\wedge_\ko M).
\end{equation}
The Eilenberg--Mac Lane spectrum $H\Z/2$ has a unique $\ko$-algebra structure via the $E_\infty$-ring map $\ko\to H\Z\to H\Z/2$, which is the composition of $0$-truncation and reduction mod $2$ (see for example~\cite[Remark 2.17]{AAR23}). Whenever we refer to $H\Z/2$ as a $\ko$-module, we always mean the module structure induced by this algebra structure. Since $\beta$ acts trivially on $H\Z/2$ for degree reasons, $(H\Z/2)[\beta^{-1}]\simeq 0$.
\begin{defn}
\label{EA_type}
Let $M$ be a bounded below $\ko$-module of finite type, so that there is a $\ko$-module
splitting
\begin{equation}
	M\simeq \overline M\vee \bigvee_i \Sigma^{k_i}H\Z/2
\end{equation}
over some (necessary countable) index $i$, such that $\overline M$ has no $\Sigma^m H\Z/2$ summands. This
specifies $\overline M$ uniquely up to $\ko$-module equivalence.
We say $M$ is \term{of elementary abelian (EA) type} if
\begin{enumerate}
	\item $\beta$ acts injectively on $\pi_*(\overline M)$, and
	\item there is an $m\in\Z$ and a $\KO_*$-module isomorphism
	\begin{equation}
		\pi_*(M)[\beta^{-1}]\overset\cong\longrightarrow\widetilde{\KO}_{*+m}(\RP^\infty).
	\end{equation}
\end{enumerate}
\end{defn} 
\begin{rem}
\label{big_KO_RP}
For convenience, we recall $\widetilde{\KO}_*(\RP^\infty)$ from~\cite[Theorem 2]{BB96} (see also~\cite[\S 4]{RS95}). Let $\mu_{2^\infty}$ denote the abelian group of all
$n^{\mathrm{th}}$ roots of unity as $n$ ranges over all powers of $2$; this group is sometimes denoted
$\Z/2^\infty$.
Then 
\begin{equation}
    \widetilde{\KO}_m(\RP^\infty) = \begin{cases}
        0 & m = 0,4,5,6\bmod 8 \\
        \Z/2 & m = 1,2\bmod 8 \\
        \mu_{2^\infty} & m=3,7\bmod 8.
    \end{cases}
\end{equation}
The action of $\eta\colon
\widetilde{\KO}_n(\RP^\infty)\to\widetilde\KO_{n+1}(\RP^\infty)$ is injective in degrees $1,2\bmod 8$ and vanishes
in all other degrees; in particular, it hits $-1\in\mu_{2^\infty}$ in $\widetilde\KO_3(\RP^\infty)$. The action
of $a$ is uniquely determined by the relation $a^2 = 4\beta$ and the requirement that $\beta$ acts invertibly.
\end{rem}
\begin{defn}
\label{def:summands}
Let $M$ be a $\ko$-module of EA-type. We say a $\Z$-module summand $N\subset \pi_*(M)$ is
\begin{enumerate}
	\item a \term{long summand} if the image of $N$ under the map $\psi\colon \pi_*(M)\to\pi_*(M)[\beta^{-1}]$ is
	nonzero and contained in a $\mu_{2^\infty}$ summand,
	\item a \term{short summand} if the image of $N$ under $\psi$ is nonzero and contained in a $\Z/2$ summand, and
	\item a \term{Whitney summand}\footnote{The name ``Whitney summand'' is inspired by Conner--Floyd's use of \term{Whitney numbers}~\cite[\S 17]{ConnerFloyd} to refer to integrals of mod $2$ cohomology classes of $B\O\times X$ as bordism invariants $\Omega_*^\O(X)\to\Z/2$, generalizing Stiefel--Whitney numbers that only use the cohomology of $B\O$. Our Whitney summands are detected by Whitney numbers in this sense, hence the name.} if $\psi(N) = 0$.
\end{enumerate}
\end{defn}
For each $n$, $\pi_n(M)$ admits a direct-sum decomposition where each summand is long, short, or Whitney.
\begin{rem}[Ignoring Whitney summands]
\label{rem:bosonicsummand}
Above, we foreshadowed that we will give a \emph{mostly} complete description of the interacting Bott spiral map
and the free-to-interacting map. ``Mostly'' here means that $\ko\wedge \ME_{\ell,k}$ is a $\ko$-module of EA-type (\cref{M_is_EA}) and we will be ignoring Whitney summands. 

Our justification for ignoring these pieces is that for our application, namely studying fermionic phases in the Bott spiral, these pieces are not relevant. For example, these pieces do not contribute to the image of the free-to-interacting map. Recall that fermionic phases
are believed to correspond to invertible field theories on manifolds with twisted spin structures. The
IFTs dual to Whitney summands are trivial in that they do not depend on the twisted spin structure, and therefore
the corresponding phases are fermionic only in a trivial sense. See~\cite[\S 3.3]{debray_invertible_2021}.
\end{rem}

\begin{rem}\label{rem:ignoring_Whitney_summands}
We would be tempted to refer to the summands we ignore as \textit{bosonic}, since they do not depend on the choice of twisted spin structure.
However,
there are also IFTs dual to classes in \textit{non}-Whitney summands that do not depend on a choice of twisted spin structure. These theories are multiples or compactifications\footnote{Keeping track of the tangential structure in a general compactified theory can be technical: see Schommer-Pries~\cite[\S 9]{SP18} for a comprehensive analysis and~\cite{greene_compactification_2025} for some explicitly worked-out examples from a physics point of view. We only need the compactification of a twisted spin theory on a spin manifold; in this case things are simpler, because the product of a spin manifold and an $(X, V)$-twisted spin manifold has a canonical $(X, V)$-twisted spin structure. Moreover, Yamashita--Yonekura~\cite[\S 7.3]{YY23} show that for invertible theories, compactification of this sort is a formal consequence of the $\MTSpin$-module structure on the spectra classifying twisted spin IFTs. See also Tachikawa--Yamashita~\cite[\S 2.2.6]{TY23}.}
of other IFTs which do depend on a choice of twisted spin structure; hence, if we are working with a direct-sum decomposition or studying the $\Omega_*^\Spin$-module structure on groups of IFTs for the purpose of computations, these IFTs 
will be relevant.

An illustrative example occurs in two-dimensional \pinm bordism. Recall from \cref{pinm_IFT_exm} that
the group of deformation classes of $(1+1)$d reflection-positive IFTs of \pinm manifolds is isomorphic to $\Z/8$.\footnote{The group of $(1+1)$d \pinm IFTs, without requiring reflection positivity data, is isomorphic to $\C^\times\times\Z/4$. See~\cite[(5.1)]{RW14}, \cite[Examples 4.17 and 4.22]{debray_arf-brown_2018}, and~\cite[Example 5.38]{MS23}, as well as~\cite[Proposition 5.5]{hoekzema_SKK_2025} for a generalization to $(d+1)$d for all odd $d$.} One of the four generators of this group is a theory $\alpha_{\mathit{ABK}}$ whose partition function is the \term{Arf--Brown--Kervaire invariant}~\cite{Bro71,KT90} $\mathit{ABK}\colon\Omega_2^{\Pin^-}\xrightarrow{\cong}\mu_8$, where $\mu_8$ denotes the group of $8^{\mathrm{th}}$ roots of unity in $\C^\times$~\cite[\S 5]{debray_arf-brown_2018}.

There are two \pinm structures on $\RP^2$, and their Arf--Brown--Kervaire invariants are $e^{\pm i\pi/4}\in\mu_8$~\cite[\S 3]{KT90}, so $\alpha_{\mathit{ABK}}$ depends on the choice of \pinm structure. The same argument applies to any power of $\alpha_{\mathit{ABK}}$ except for $\alpha_{\mathit{ABK}}^{\otimes 4n}$---and indeed, for any closed \pinm $2$-manifold $\Sigma$~\cite[Lemma 3.6]{KT90}, 
\begin{equation}
    \mathrm{ABK}(\Sigma)^{4} = (-1)^{\chi(\Sigma)}\in\set{\pm 1}.
\end{equation}
So $\alpha_{\mathit{ABK}}^{\otimes 4}$ is ``bosonic'' in that it does not depend on the \pinm structure of spacetime!

``Surprisingly bosonic'' IFTs such as $\alpha_{\mathit{ABK}}^{\otimes 4}$ occur whenever there is a class in the $E_\infty$-page of the Adams spectral sequence in filtration $0$ (i.e.\ along the $x$-axis in the standard drawing convention); see~\cite[\S 8.4]{FH21b} or~\cite[\S 3.3]{debray_invertible_2021}. This often occurs in long summands, as can be seen in the Ext calculations in \cref{N1_exm,N2_exm,N3_exm,N0_exm}, drawn in the right-hand sides of \cref{N1_figure,,N2_figure,,N3_figure,,N0_figure}. For example, four-dimensional \pinp bordism is isomorphic to $\Z/16$~\cite[\S 2]{Gia73} and eight times any generating IFT $\alpha_\eta$ does not depend on the \pinp structure. One consequence of this appears in work of Barkeshli--Hsin--Kobayashi~\cite{BHK24} on the higher-group symmetry of the fermionic toric code: where one might expect a $\Z/16$ symmetry defined by stacking with $\alpha_\eta$ before gauging the $\Z/2$ symmetry, the lack of dependence of $\alpha_\eta^{\otimes 8}$ on the \pinp structure reduces the symmetry group to $\Z/8$. See (\textit{ibid.}, \S 2.1.1).
\end{rem}

In the examples below, we will briefly use the Adams spectral sequence in the form constructed by Baker--Lazarev~\cite{BL01}; we recommend Beaudry--Campbell~\cite{BC18} for additional background.
\begin{defn}\label{Hko_defn}
For a $\ko$-module $M$, let
\begin{equation}
    H_\ko^*(M)\coloneqq\pi_{-*}\mathrm{Map}_\ko(M, H\Z/2).
\end{equation}
\end{defn}
Let $\cA(1)\coloneqq\ang{\Sq^1, \Sq^2}$.
\begin{lem}[{Baker~\cite[Theorem 5.1]{Bak20}}]
\label{A1_Hko}
$H_\ko^*(H\Z/2)\cong\cA(1)$.
\end{lem}
By composition of maps, $H_\ko(H\Z/2)$ is an algebra over $\Z/2$; with this structure \cref{A1_Hko} is an algebra isomorphism. For a $\ko$-module $M$, composition of maps makes $H_\ko^*(M)$ into an $H_\ko^*(H\Z/2) \cong\cA(1)$-module.
\begin{lem}
\label{Hko_of_ko}
For any spectrum $X$, there is a natural $\cA(1)$-module isomorphism $H_\ko^*(\ko\wedge X)\cong H^*(X;\Z/2)$, where the $\cA(1)$-action on $H^*(X;\Z/2)$ is obtained by restricting the $\cA$-action.
\end{lem}
\begin{proof}
By the tensor-hom adjunction, there is a natural equivalence of spectra $\mathrm{Map}_\ko(\ko\wedge X, H\Z/2)\simeq\mathrm{Map}_\Sph(X, H\Z/2)$; then take $\pi_{-*}$.
\end{proof}
\begin{thm}[{Baker--Lazarev~\cite{BL01}}]
\label{BL}
Let $M$ and $M'$ be $\ko$-modules. Then there is a spectral sequence of Adams type, natural in $M$ and $N$, with signature
    \begin{equation}\label{genAdams}
    E_2^{s,t} = \Ext_{\cA(1)}^{s,t}(H_\ko^*(M), H_\ko^*(M')) \Longrightarrow \pi_{t-s}\mathrm{Map}_\ko(M', M)_2^\wedge.
\end{equation}
If $M'$ is connective and $M$ is a CW $\ko$-module of finite type, this spectral sequence converges.
\end{thm}
We call this the \term{(Baker--Lazarev) Adams spectral sequence}.

What Baker--Lazarev prove is quite a bit more general, but we only need this special case. The finiteness and CW hypotheses in \cref{BL} hold in all examples considered in this paper.

Usually one plugs in $M' = \ko$ to \cref{BL} so that this spectral sequence computes the $2$-completed homotopy groups of $M$: using the natural homotopy equivalence $\mathrm{Map}_R(R\wedge X, Y)\simeq \mathrm{Map}_{\Sph}(X, Y)$ with $R = \ko$, $Y = M$, and $X = \Sph$,
\begin{equation}
    \pi_*\mathrm{Map}_\ko(\ko, M)_2^\wedge \cong \pi_*\mathrm{Map}_\Sph(\Sph, M)_2^\wedge = \pi_*(M)_2^\wedge.
\end{equation}
\begin{rem}
The Baker--Lazarev Adams spectral sequence is additive in the sense that a $\ko$-module splitting $M\simeq M_1\vee M_2$ induces a direct-sum splitting on $H_\ko^*$ and on $\Ext$, and in fact a splitting of the entire Adams spectral sequence into the respective Adams spectral sequences for $M_1$ and $M_2$.
\end{rem}
\begin{cor}
\label{usuAdamscor}
Let $X$ be a CW complex with finitely many cells in each dimension Then there is a convergent spectral sequence of Adams type, natural in $X$, with signature
\begin{equation}\label{usuAdams}
    E_2^{s,t} = \Ext_{\cA(1)}^{s,t}(H^*(X;\Z/2), \Z/2) \Longrightarrow \pi_*(\ko\wedge X)_2^\wedge = \ko_*(X)_2^\wedge.
\end{equation}
\end{cor}
\begin{proof}
Starting from~\eqref{genAdams}, plug in $M' = \ko$ and $M = \ko\wedge X$, then apply $\mathrm{Map}_R(R\wedge X, Y)\simeq \mathrm{Map}_{\Sph}(X, Y)$ as above. To get the $E_2$-page, use that by \cref{Hko_of_ko}, $H_\ko^*(M)\cong H^*(X;\Z/2)$.
\end{proof}
\Cref{usuAdamscor} is the version one more commonly encounters, often proven via a change-of-rings theorem, such as in~\cite{BC18}. We will need a few $\ko$-modules that do not to our knowledge factor as $\ko\wedge X$, so we must use the more general~\cref{BL}.

Throughout this section, $\Ext(N)$ means $\Ext_{\cA(1)}(N, \Z/2)$ if not otherwise specified.

\begin{figure}[h!]
\includegraphics{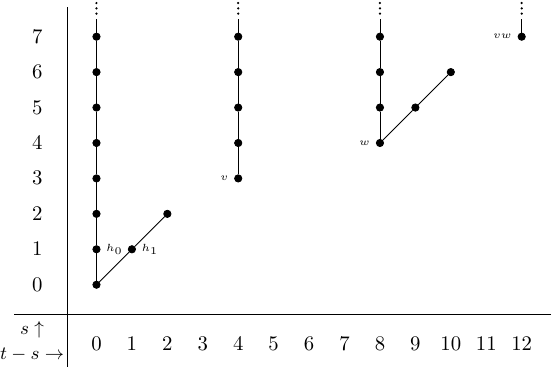}
\caption{The $\Z^2$-graded $\Z/2$-algebra $\E\coloneqq\Ext_{\cA(1)}^{*,*}(\Z/2, \Z/2)$ described in~\cref{A1Extprop}.}
\label{E_figure}
\end{figure}

Let $\E\coloneqq\Ext(\Z/2)$. This is the Adams $E_2$-page for $\ko$, and the $E_\infty$-ring structure on $\ko$ induces a $\Z^2$-graded $\Z/2$-algebra structure on $\E$; see~\cite[\S 4.2]{BC18} for an algebraic description of the product.
\begin{prop}[{Liulevicius~\cite[Theorem 3]{Liu62}}]
\label{A1Extprop}
There is an isomorphism of $\Z^2$-graded $\Z/2$-algebras
\begin{equation}\label{liuthm}
    \E\coloneqq\Ext_{\cA(1)}(\Z/2, \Z/2)\cong \Z/2[h_0, h_1, v, w]/(h_0h_1, h_1^3, vh_1, h_0^2w-v^2)
\end{equation}
with $h_0\in\Ext^{1,1}$, $h_1\in\Ext^{1,2}$, $v\in\Ext^{3,7}$, and $w\in\Ext^{4,12}$.
\end{prop}
Thus $\E$ acts on the $E_2$-page of the Baker--Lazarev Adams spectral sequence of any $\ko$-module. This action commutes with all differentials, and is useful for solving extension problems: $h_0$ lifts to multiplication by $2$, $h_1$ to $\eta$, $v$ to $a$, and $w$ to $\beta$.
See \cref{E_figure} for a picture of $\E$.

The following theorem is surely well-known, but we were unable to find it in the literature, so we have included a proof.
\begin{thm}[Margolis' theorem for $\ko$-modules]
\label{margolis}
Let $M$ be a bounded below $\ko$-module of finite type. Any isomorphism of $\cA(1)$-modules
\begin{subequations}
\begin{equation}
    \overline \vp\colon H_\ko^*(M)\overset\cong\longrightarrow \overline N\oplus F,
\end{equation}
where $F$ is a free $\cA(1)$-module, lifts to a $\ko$-module splitting: there is a $\ko$-module $N$ with $H_\ko^*(N)\cong\overline N$ and a $\ko$-module equivalence
\begin{equation}
    \vp\colon N\vee\bigvee_i \Sigma^{n_i} H\Z/2 \overset\simeq\longrightarrow M
\end{equation}
such that $H_\ko^*\paren{\bigvee \Sigma^{n_i}H\Z/2}\cong F$ and the pullback map of $\vp$ on $H_\ko^*$ is equal to $\overline\vp$.
\end{subequations}
\end{thm}
\begin{proof}
Since $M$ has finite type, the rank of $F$ is countable, so we may induct starting with the lowest-degree summand in $F$, possible because $M$ is bounded below. Therefore in the rest of the proof we will assume $F$ is rank one: $F\cong\Sigma^m \cA(1)$.

It suffices to produce $\ko$-module maps $f\colon \Sigma^m H\Z/2\to M$ and $g\colon M\to\Sigma^m H\Z/2$ with $g\circ f \simeq \id$. To do this, run two instances of the Baker--Lazarev Adams spectral sequence (\eqref{BL}): one for $\mathrm{Map}_\ko(M, \Sigma^m H\Z/2)$ and one for $\mathrm{Map}_\ko(\Sigma^m H\Z/2, M)$. Since $H_\ko^*(H\Z/2)\cong\cA(1)$, as we noted above, the two $E_2$-pages are
\begin{equation}\label{marg_E2}
    \Ext_{\cA(1)}^{s,t}(H_\ko^*(M), \Sigma^m \cA(1))\qquad\text{ and }
    \qquad \Ext_{\cA(1)}^{s,t}(\Sigma^m \cA(1), H_\ko^*(M)).
\end{equation}
Margolis proved that $\cA(1)$ is both projective and injective as a graded $\cA(1)$-module~\cite[Theorem 12.5, Proposition 12.8, Theorem 12.9]{Mar83} (``graded'' is important: see~\cite{Sal23}), so both $E_2$-pages in~\eqref{marg_E2} vanish for $s > 0$, and for $s = 0$, they coincide with graded Hom. This implies these Adams spectral sequences collapse for degree reasons, and that any $\cA(1)$-module homomorphism $H_\ko^*(M)\to\Sigma^m \cA(1)$ (resp.\ $\Sigma^m\cA(1)\to H_\ko^*(M)$) lifts to a $\ko$-module morphism $M\to \Sigma^m H\Z/2$ (resp.\ $\Sigma^m H\Z/2\to M$) and this lift can be chosen to preserve composition up to $\ko$-module homotopy equivalence. Thus it suffices to exhibit maps $f$ and $g$ with $g\circ f\simeq\id$ at the level of $\cA(1)$-modules, where they are the maps splitting the rank-one free summand off of $H_\ko^*(M)$.
\end{proof}
\begin{defn}
\label{stab_iso}
A \term{stable isomorphism} of $\cA(1)$-modules is an $\cA(1)$-module map $f\colon M\to N$ whose kernel and cokernel are free over $\cA(1)$.
\end{defn}
\begin{lem}[{\cite[Proposition 14.1]{Mar83}}]
If $f\colon M\to N$ is a stable isomorphism, then there are free $\cA(1)$-modules $F$ and $F'$ and an extension of $f$ to an isomorphism $\overline f\colon M\oplus F\overset\cong\to N\oplus F'$.
\end{lem}
\Cref{margolis} thus implies that a stable isomorphism on $H_\ko^*$ may affect Whitney summands, but leaves everything we are actually interested in alone.
\begin{rem}\label{origmarg}
Margolis' original theorem~\cite{Mar74} is essentially the same result but with $\Sph$ in place of $\ko$ and $H^*(\bl;\Z/2)$ in place of $H_\ko^*$. Using this, one can use a change-of-rings theorem to prove a weaker version of \cref{margolis}: one requires $M\simeq\ko\wedge X$, and the splitting is only as spectra, not $\ko$-modules. See~\cite[Theorem 3.22]{debray_invertible_2021}.
\end{rem}
\subsection{Examples of \texorpdfstring{$\ko$}{ko}-modules of EA-type}
\label{ss:EA_type}
In this subsection, we will give several examples of $\ko$-modules of EA-type; we will use these examples to first see that $\ko\wedge\ME_{\ell,k}$ is of EA-type, and then in \cref{melk_calc} compute $\pi_*(\ko\wedge\ME_{\ell,k})$ modulo Whitney summands.

Each of our examples is of the form $\ko\wedge X$ for some $X$; we will also compute $H^*(X;\Z/2)$ as a module over the Steenrod subalgebra $\cA(1)\coloneqq \ang{\Sq^1, \Sq^2}$, finding four $\cA(1)$-modules $N_i$ that will play a central role in our computation. They have been studied in detail before by, e.g., Yu~\cite{Yu95} and Bruner~\cite{Bru14}; however, we warn the reader that our modules differ from theirs by suspensions.
This allows us to display the second page of the Adams spectral sequence. In each of these cases the spectral sequence collapses at $E_2$ without
any extension problems, furnishing a proof that $\ko\wedge X$ is in fact of EA-type.
\begin{exm}
\label{N1_exm}
It will be little surprise to the reader that $\ko\wedge\RP^\infty$ is of EA-type. For degree reasons we will find
it slightly more convenient to work with $\ME_{1,0} = (B\Z/2)^{\sigma-1}$, which is equivalent to
$\Sigma^{-1}\RP^\infty$ (see, e.g.,~\cite[Lemma 2.6.5]{Koc96}).

Let $N_1\coloneqq H^*((B\Z/2)^{\sigma-1};\Z/2)$. We draw this module in \cref{N1_figure}, left. $\Ext_{\cA(1)}(N_1)$ was computed by Gitler--Mahowald--Milgram~\cite[\S 2]{GMM68} to be the $\E$-module
\begin{subequations}
\begin{equation}
	\Ext(N_1)\cong \E\set{\alpha, a_i\mid i\ge 1}/\mathcal R_1
\end{equation}
where $\alpha\in\Ext^{0,0}$ and $a_i\in\Ext^{0,4i-2}$, and the submodule $\mathcal R_1$ of relations is
\begin{equation}
    \mathcal R_1 = (h_0\alpha,\ v\alpha,\ h_1^2\alpha - h_0^2 a_1,\ 
    h_0^4 a_2,\
    h_1a_i,\ va_i - h_0^3a_{i+1},\
    wa_i - h_0^4 a_{i+2} \text{ for all } i
	).
\end{equation}
\end{subequations}
We draw this in \cref{N1_figure}, right; all
differentials vanish either for degree reasons or because they must commute with $h_0$ or $h_1$, and there are no
extension problems left after taking the $\E$-module structure into account. Therefore as
$\ko_*$-modules,
\begin{subequations}
\begin{equation}
	\ko_*(\ME_{1,0})\cong \ko_*\set{\overline\alpha, \overline a_i\mid i\ge 1}/\overline{\mathcal R}_1
\end{equation}
where $\abs{\overline \alpha} = 0$ and $\abs{\overline a_i} = 4i-2$, and the submodule $\overline{\mathcal R}_1$ of relations is
\begin{equation}
    \overline{\mathcal R}_1 = (2\overline\alpha,\ a\overline\alpha,\ \eta^2\overline\alpha - 4\overline a_1,\ 16\overline a_2,\
    \eta\overline a_i,\ a\overline a_i - 8\overline
		a_{i+1},\
        \beta \overline a_i - 16\overline a_{i+2}\text{ for all } i).
\end{equation}
\end{subequations}
These relations imply $\ko\wedge\ME_{1,0}$ is
of EA-type, and that $\overline a_i$ generates
a cyclic summand of order $2^{4j+3}$ (if $i = 2j+1$) or $2^{4j}$ (if $i = 2j$); these are the long summands. 
The classes
$\beta^n\overline\alpha $ and $\eta\beta^n\overline\alpha $, $n\ge 0$, generate short summands. There are no
Whitney summands.

Thus the same conclusions are true for $\widetilde{\ko}_*(\RP^\infty)$ with the degrees shifted upwards by $1$. This recovers preexisting computations of $\widetilde\ko_*(\RP^\infty)$: Mahowald~\cite[Lemma 7.3]{Mah82} computes the groups $\widetilde{\ko}_*(\RP^\infty)$, and Bruner--Mira--Stanley--Snaith~\cite[\S 3.1]{BMSS15} give the $\ko_*$-module structure. \qedhere
\begin{figure}[h!]
\begin{subfigure}[c]{0.29\textwidth}
\includegraphics{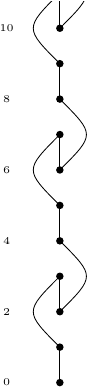}
\end{subfigure}
\begin{subfigure}[c]{0.7\textwidth}
\includegraphics{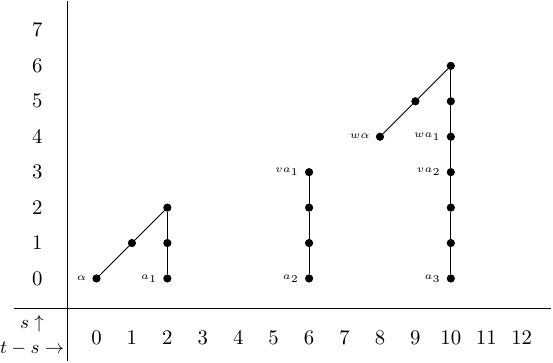}
\end{subfigure}
\caption{Left: the $\cA(1)$-module $N_1\coloneqq H^*((B\Z/2)^{\sigma-1};\Z/2)$. This is also $\widetilde
H^*(\RP^\infty;\Z/2)$ shifted down in degree by $1$. Right: $\Ext(N_1)$, the $E_2$-page of the Adams spectral
sequence for $\ko\wedge (B\Z/2)^{\sigma-1}$. All differentials and extension questions are trivial. We discuss $N_1$ and its Adams spectral sequence in \cref{N1_exm}.}
\label{N1_figure}
\end{figure}
\end{exm}
\begin{exm}
\label{N2_exm}
Let $A_4$ denote the alternating group on four letters. Then $H^*(BA_4;\Z/2)\cong\Z/2[u, v, w]/(u^3+v^2+w^2+vw)$
with $\abs u = 2$ and $\abs v = \abs w = 3$~\cite[Theorem 2.6]{Priddy}. The Steenrod squares of the generators are $\Sq(u) = u+v+w+u^2$,
$\Sq(v) = v+u^2+uw + v^2$, and $\Sq(w) = w+u^2 + uv + v^2$.\footnote{Mitchell--Priddy do not explicitly give the Steenrod squares, but they can be worked out from their proof. They are also given in~\cite[Proposition 5.1]{debray_invertible_2021}.}
One can compute that as an $\cA(1)$-module, $\widetilde H^*(BA_4;\Z/2)$ is a direct sum of countably many
free $\cA(1)$-modules whose lowest-degree elements are in degrees $6$, $8$, $12$, $12$, \dots, along with exactly
one indecomposable summand $S$; we define $N_2\coloneqq \Sigma^{-2}S$.\footnote{$N_2$ is called $\Sigma^{-2}P_2$
in~\cite[Definition 4.4]{Bru14} and $R_5$ in~\cite{BC18, debray_invertible_2021}.}

We draw a picture of $N_2$ in
\cref{N2_figure}, left. Yu~\cite[Theorem 3.1]{Yu95} showed that there is an $\E$-module isomorphism
\begin{subequations}
\begin{equation}
	\Ext(N_2)\cong \E\set{\beta, \beta', b_i\mid i>0}/\mathcal R_2,
\end{equation}
where $\beta\in\Ext^{0,0}$, $\beta'\in\Ext^{3,10}$, $b_i\in\Ext^{0,4i-3}$, and the submodule $\mathcal R_2$ of relations is
\begin{equation}
\begin{aligned}
    \mathcal R_2 = (&h_0\beta,\ h_1^2\beta,\ v\beta,\
    w\beta - h_1\beta',\ h_0\beta',\ v\beta',\
    h_0b_1 - h_1\beta,\\
    &h_0^3 b_2,\
    h_1b_i,\ vb_i - h_0^3b_{i+1},\
    wb_i - h_0^4b_{i+2} \text{ for all } i).
\end{aligned}
\end{equation}
\end{subequations}
We draw the Adams chart for $\Ext(N_2)$ in \cref{N2_figure}, right. Because $H^*(BA_4;\Z/2)$ is stably isomorphic to $\Sigma^2 N_2$, Margolis' theorem for $\ko$-modules (\cref{margolis}) implies that there is a $\ko$-module $M_2$ with $H_\ko^*(M_2)\cong N_2$ and that (the $2$-completion of) $\ko\wedge BA_4$ is $\ko$-module equivalent to a wedge sum of $\Sigma^2 M_2$ and a sum of shifts of $H\Z/2$. Thus $\Ext(N_2)$ is the $E_2$-page of the Adams spectral sequence computing $\pi_*(M_2)_2^\wedge$, which is a summand of $\widetilde{\ko}_{*+2}(BA_4)_2^\wedge$.

Similarly to what we saw in \cref{N1_exm} for $N_1$, all differentials in this spectral sequence vanish because they are $\E$-linear, and the $\E$-action determines all extensions, so there is a $\ko_*$-module isomorphism
\begin{subequations}
\begin{equation}
    \pi_*(M_2)\cong \ko_*\set{\overline \beta, \overline \beta{}', \overline b_i\mid i\ge 1}/\overline{\mathcal R}_2,
    \end{equation}
    where $\abs{\overline \beta} = 0$, $\abs{\overline\beta'} = 7$, $\abs{\overline b_i} = 4i-3$, and the submodule $\overline{\mathcal R}_2$ of relations is
    \begin{equation}
    \begin{aligned}
    \overline{\mathcal R}_2 = (&
        2\overline\beta,\ \eta^2\overline\beta,\ a\overline\beta,\ \beta\overline\beta - \eta\overline\beta{}',\
        2\overline\beta{}',\
        a\overline\beta{}',\
        2\overline b_1 - \eta\overline\beta,\\
        &8\overline b_2,\ \eta\overline b_i,\
        a\overline b_i - 8\overline b_{i+1},\
        \beta\overline b_i - 16\overline b_{i+2}
        \text{ for all } i).
    \end{aligned}
\end{equation}
\end{subequations}
Inverting $\beta$, one sees that $M$ is a $\ko$-module of EA-type; $\pi_*(M)$ has a long summand of order $2^{4j+2}$ in degree $8j+1$, a long summand of order $2^{4j+3}$ in degree $8j+5$, and short summands in degrees $7,0\bmod 8$. There are no Whitney summands.

As discussed above, $\ko\wedge BA_4\simeq \Sigma^2M_2\vee F$, where $F$ is a sum of shifts of $H\Z/2$, so $\ko\wedge BA_4$ is also of EA-type. It has long summands of order $2^{4j+2}$ in degree $8j+3$ and order $2^{4j+3}$ in degree $8j+7$ and short summands in degrees $1,2\bmod 8$, except not in degree $1$. In addition, the $H\Z/2$ summands contribute Whitney summands to $\widetilde{\ko}_*(BA_4)$. See Bruner--Greenlees~\cite[\S 7.7.E]{BG10} for prior work on $\ko_*(BA_4)$. \qedhere
\begin{figure}[h!]
\begin{subfigure}[c]{0.29\textwidth}
\includegraphics{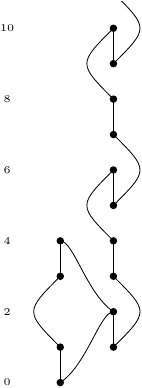}
\end{subfigure}
\begin{subfigure}[c]{0.7\textwidth}
\includegraphics{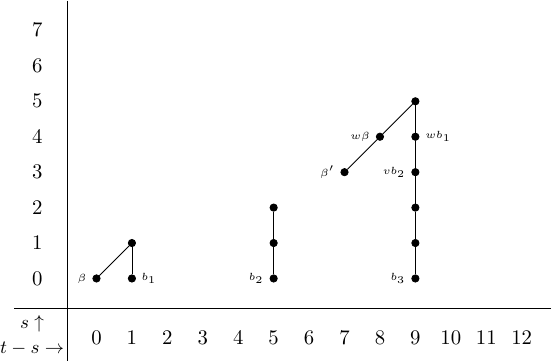}
\end{subfigure}
\caption{Left: the $\cA(1)$-module $N_2$, the unique non-free summand of $\Sigma^{-2} \widetilde H^*(BA_4;\Z/2)$.
Right: $\Ext(N_2)$, the $E_2$-page of the Adams spectral
sequence for a summand of $\ko\wedge \Sigma^{-2}BA_4$. All differentials and extension questions are trivial. We discuss $N_2$ and its Adams spectral sequence in \cref{N2_exm}.}
\label{N2_figure}
\end{figure}

\end{exm}
\begin{exm}
\label{N3_exm}
Using the Künneth theorem and the description of $H^*(BA_4;\Z/2)$ from \cref{N2_exm}, one can show that as an $\cA(1)$-module, $\widetilde H^*(BA_4\wedge B\Z/2;\Z/2)$ is a direct sum of a free $\cA(1)$-module of countably infinite rank and exactly one non-free, indecomposable summand $S'$; we let $N_3\coloneqq \Sigma^{-3}S'$. We draw a picture of $N_3$ in \cref{N3_figure}, left. Margolis' theorem for $\ko$-modules (\cref{margolis}) thus tells us that there is a connective $\ko$-module $M_3$ with $H_\ko^*(M_3)\cong N_3$ such that $\ko\wedge BA_4\wedge B\Z/2$ splits as a $\ko$-module as a sum of $\Sigma^3 M_3$ and a sum of shifts of $H\Z/2$.

Yu~\cite[Theorem 3.1]{Yu95} exhibits an $\E$-module isomorphism
\begin{subequations}
\begin{equation}
    \Ext(N_3)\cong \E\set{\gamma, c_i\mid i\ge 0}/\mathcal R_3,
\end{equation}
with $\gamma\in\Ext^{2,8}$, $c_i\in\Ext^{0,4i}$, and the submodule $\mathcal R_3$ of relations is
\begin{equation}
    \mathcal R_3 = (
        h_0\gamma,\ h_1^2\gamma - wc_0,\ v\gamma,\
        h_0c_0,\ h_0^2 c_1,\ h_1c_i,\
        vc_i - h_0^3 c_{i+1},\
        wc_i - h_0^4c_{i+2}\text{ for all }i).
\end{equation}
\end{subequations}
We draw a picture of this $\E$-module in \cref{N3_figure}, right; it is the $E_2$-page of the Baker--Lazarev Adams spectral sequence computing $\pi_*(M_3)_2^\wedge$.

In a pattern that is probably increasingly clear, all differentials vanish because they commute with the $\E$-action, and there are no hidden extensions, so we obtain a $\ko_*$-module isomorphism
\begin{subequations}
\begin{equation}
    \pi_*(M_3)\cong \ko_*\set{
        \overline\gamma, \overline c_i\mid i\ge 0}/\overline{\mathcal R}_3,
\end{equation}
with $\abs{\overline\gamma} = 6$, $\abs{\overline c_i} = 4i$, and the submodule $\overline{\mathcal R}_3$ equal to
\begin{equation}
    \overline{\mathcal R}_3 = (
        2\overline\gamma,\
        \eta^2\overline\gamma - \beta\overline c_0,\
        a\overline\gamma,\ 2\overline c_0,\
        4\overline c_1,\
        \eta\overline c_i,\
        a\overline c_i - 8\overline c_{i+1},\
        \beta\overline c_i - 16\overline c_{i+2}
        \text{ for all }i
   ),
\end{equation}
\end{subequations}
Inverting $\beta$, we see that $M_3$ is of EA-type; its homotopy groups have no Whitney summands and have short summands in degrees $6,7\bmod 8$. There is a long summand of the form $\Z/2^{4j+1}$ in degree $8j$ and another of the form $\Z/2^{4j+2}$ in degree $8j+4$ for all $j\ge 0$. Thus $\ko\wedge BA_4\wedge B\Z/2$ is also of EA-type, and the short and long summands in it homotopy can be worked out similarly to what we did for $\ko\wedge BA_4$ in \cref{N2_exm}; there are also Whitney summands. \qedhere

\begin{figure}[h!]
\begin{subfigure}[c]{0.29\textwidth}
\includegraphics{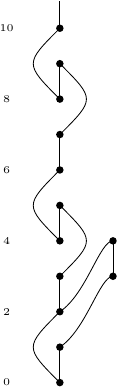}
\end{subfigure}
\begin{subfigure}[c]{0.7\textwidth}
\includegraphics{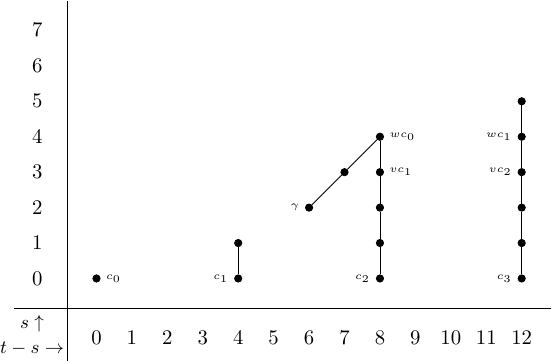}
\end{subfigure}
\caption{Left: the $\cA(1)$-module $N_3$, the unique non-free summand of $\Sigma^{-3}\widetilde H^*(BA_4\wedge B\Z/2;\Z/2)$. Right: $\Ext(N_3)$, the $E_2$-page of the Baker--Lazarev Adams spectral sequence for a $\ko$-module summand $M_3$ of $\Sigma^{-3}\ko\wedge BA_4\wedge B\Z/2$. All differentials and extension questions are trivial. See \cref{N3_exm} for more information on $N_3$, $M$, and this Adams spectral sequence.}
\label{N3_figure}
\end{figure}
\end{exm}
\begin{exm}
\label{N0_exm}
Let $N_0\coloneqq \Sigma^{-1} H^*((B\Z/2)^{1-\sigma};\Z/2)$. We draw this module in \cref{N0_figure}, left. $\Ext_{\cA(1)}(N_0)$ was computed by Gitler--Mahowald--Milgram~\cite[\S 2]{GMM68} to be the $\E$-module
\begin{subequations}
\begin{equation}
	\Ext(N_0)\cong \E\set{\delta, d_i\mid i\ge 0}/\mathcal R_4,
\end{equation}
where $\delta\in\Ext^{1,2}$, and $d_i\in\Ext^{0,4i-1}$. The submodule $\mathcal R_4$ of relations is
\begin{equation}
    \mathcal R_4 = (
        h_0d_0,\ h_0\delta,\ h_1^2\delta - vd_0,\ v\delta,\
        h_1d_i,\ vd_i - h_0^3d_{i+1},\
        wd_i - h_0^4 d_{i+2}\text{ for all } i
    ).
\end{equation}
\end{subequations}
We draw this in \cref{N0_figure}, right. All
differentials and
extension problems are trivial, as in previous examples; therefore we learn that as
$\ko_*$-modules,
\begin{subequations}
\begin{equation}
	\ko_*((B\Z/2)^{1-\sigma})\cong \ko_*\set{\overline\delta, \overline d_i\mid i\ge 0}/
    \overline{\mathcal R}_4,
\end{equation}
where $\abs{\overline \delta} = 2$ and $\abs{\overline d_i} = 4i$. (The shift in degree is because $H^*((B\Z/2)^{1-\sigma};\Z/2)\cong\Sigma N_0$.) The submodule $\overline{\mathcal R}_4$ is
\begin{equation}
    \overline{\mathcal R}_4 =
    (2\overline d_0,\ 2\overline\delta,\
     \eta^2\overline\delta - a\overline d_0,\
     a\overline\delta,\
     \eta\overline d_i,\
     a\overline d_i - 8\overline d_{i+1},\
     \beta\overline d_i - 16 \overline d_{i+2}\text{ for all } i).
\end{equation}
\end{subequations}
 These relations imply $\ko\wedge (B\Z/2)^{1-\sigma}$ is
of EA-type: for $k\ge 0$, $\beta^k\overline\delta{}'$ and $\eta\beta^k\overline\delta{}'$ generate short summands in degrees $8k+2$, resp.\ $8k+3$, and $\overline d_i$ generates a long summand isomorphic to $\Z/2^{4+4j}$ ($i = 2j+1$) or $\Z/2^{1+4j}$ ($i = 2j$) in degree $4i$. There are no Whitney summands.

This Adams spectral sequence computation is far from new: Giambalvo~\cite{Gia73} and Kirby--Taylor~\cite{KT90pinp} study the Adams spectral sequence for $\MTSpin\wedge (B\Z/2)^{1-\sigma}$, and Campbell~\cite[Example 6.6]{Cam17} studies it for $\ko\wedge (B\Z/2)^{1-\sigma}$.
\begin{figure}[h!]
\begin{subfigure}[c]{0.29\textwidth}
\includegraphics{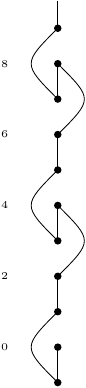}
\end{subfigure}
\begin{subfigure}[c]{0.7\textwidth}
\includegraphics{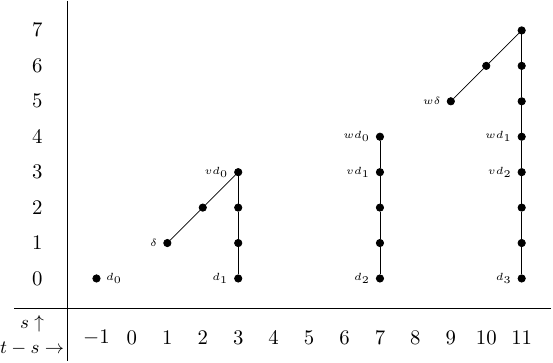}
\end{subfigure}
\caption{Left: the $\cA(1)$-module $N_0\coloneqq \Sigma^{-1} H^*((B\Z/2)^{1-\sigma};\Z/2)$. Right: $\Ext(N_0)$, the $E_2$-page of the Adams spectral sequence for $\Sigma^{-1} \ko\wedge (B\Z/2)^{1-\sigma}$. All differentials and extension questions are trivial. See \cref{N0_exm} for more information.}
\label{N0_figure}
\end{figure}
\end{exm}
\begin{lem}\label{EA-coh}
Let $X$ be a space or spectrum of finite type such that as $\cA(1)$-modules, $H^*(X;\Z/2)\cong \Sigma^n N_i\oplus F$ where
$0\le i\le 3$ and $F$ is a free $\cA(1)$-module. Then $\ko\wedge X$ is of EA-type.
\end{lem}
\begin{proof}
First assume $F = 0$. The arguments we gave in \cref{N1_exm,,N2_exm,,N3_exm,,N0_exm} showing those $\ko$-modules
are finite type used nothing more about them than the $\cA(1)$-module structure on their mod $2$ cohomology, so the
proofs generalize to $X$. If $F\not\cong 0$, \cref{margolis} implies $\ko\wedge X\simeq \overline M\vee E$,
where $E$ is a sum of shifts of $H\Z/2$ and $H_\ko^*(\overline M)\cong \Sigma^n N_i$ for some $i$,
so the lemma follows by reducing to the previous case.
\end{proof}
\begin{rem}
Not all $\ko$-modules of EA-type look like the ones in these examples. Let $C$ denote the cofiber of the map $B\Pin_2^-\hookrightarrow B\Spin_3$ sending a rank-$2$ \pinm vector bundle $V$ to $V\oplus\mathrm{Det}(V)$. Bayen--Bruner~\cite[Corollary 3]{BB96} show that $\ko\wedge C$ is a $\ko$-module of EA-type, but its cohomology looks nothing like the $N_i$, and its Adams spectral sequence has both differentials and hidden extensions (\textit{ibid.}, Theorem 4).
\end{rem}
\begin{lem}[{Yu~\cite[Lemma 2.5]{Yu95}}]
\label{yu_stable_iso}
There are stable isomorphisms of $\cA(1)$-modules (see \cref{stab_iso}) $N_0\otimes N_i\simeq N_i$, $N_1\otimes N_1\simeq N_2$, $N_1\otimes N_2\simeq N_3$, and $N_1\otimes N_3\simeq\Sigma^4 N_0$.
\end{lem}
Thus $N_i \otimes N_j$ is stably equivalent to $N_{i+j\bmod 4}$ up to suspensions.

\begin{rem}
\label{smith_periodicity_remark}
\Cref{yu_stable_iso} establishes a fourfold periodicity in $N_1^{\otimes n}$, and therefore a fourfold periodicity in the $\ko$-homology and spin bordism of $(B\Z/2)^{\wedge n}$ modulo Whitney summands. There is another fourfold periodicity over $\MTSpin$ and $\ko$, namely that of iterated Smith homomorphisms for the same vector bundle~\cite[Examples 6.14 and 7.8]{debray_smith_2024}; we saw this periodicity for the vector bundle $\sigma$ in \cref{1st_Z2_Smith_exm}.
These two periodicities are related: the diagonal map $B\Z/2\to (B\Z/2)^{\wedge n}$ induces a map of Thom spectra $\Delta\colon \ko\wedge (B\Z/2)^{(\ell-k)(\sigma-1)}\to\ko\wedge \ME_{\ell,k}$. The map $\Delta$ intertwines the Smith map on one side with the spiral map $\phi_{\ell,k}$ on the other.
\end{rem}

\begin{figure}[h!]
\includegraphics{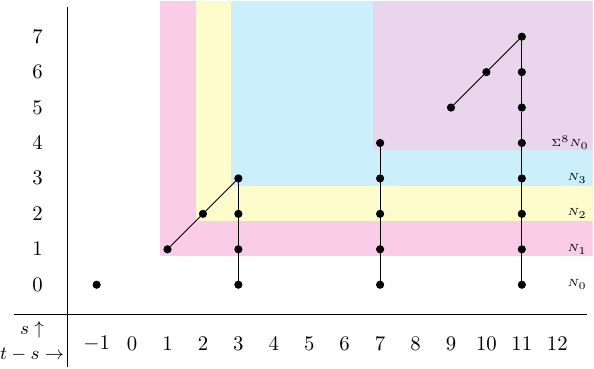}
\caption{The result of superimposing the right-hand sides of \cref{N1_figure,,N2_figure,,N3_figure,,N0_figure} on the same axes, illustrating the close relationship between the Ext charts of the corresponding four $\cA(1)$-modules. See \cref{adams_cover}.}
\label{Ni_figure}
\end{figure}

\begin{rem}
\label{adams_cover}
In \cref{Ni_figure}, we overlay the Ext charts in \cref{N1_figure,,N2_figure,,N3_figure,,N0_figure}. This illustrates a phenomenon shown in~\cite[Theorem 3.1]{Yu95}: that, up to a shift in topological degree, $\Ext(N_i)$ is the submodule of $\Ext(N_{i-1\bmod 4})$ consisting of elements in positive filtration. Of course, this is not a coincidence: in~\cite{ABP_splittings}, two of us joint with Pacheco-Tallaj will produce maps of $\ko$-modules realizing these submodule inclusions, whose cofibers are sums of shifts of $H\Z/2$; thus they are ``generalized Adams covers.'' These ideas are due to Bruner~\cite{Bru99}, who did this in the case $i = 2$ and presumably knew it for all $i$.
\end{rem}

\begin{exm}
Let $V$ be an elementary abelian $2$-group. For spaces $X$ and $Y$, there is a natural splitting
\begin{equation}
\Sigma^\infty(X\times Y)\simeq \Sigma^\infty X\vee \Sigma^\infty Y\vee \Sigma^\infty(X\wedge Y);
\end{equation}
applying this iteratively to the $B\Z/2$ pieces of $BV$, one deduces a splitting\footnote{The analogous splitting with $B\Z/2$ replaced with $\mathbb T$ is the \term{James splitting} used in the computation of groups of weak topological phases of free fermions. See~\cite[Theorem 11.8]{freed_twisted_2013} and \cite[\S 1.6]{ADKPSS24}.}
\begin{equation}\label{not_james_splitting}
    \Sigma^\infty BV\simeq \bigvee_{k=1}^n\bigvee_{i=1}^{\binom nk} \Sigma^\infty((B\Z/2)^{\wedge k}).
\end{equation}
We showed $\ko\wedge B\Z/2$ is of EA-type in \cref{N1_exm}; then \cref{EA-coh,yu_stable_iso} allow us to inductively prove that $\ko\wedge (B\Z/2)^{\wedge k}$ is also of EA-type. Thus we conclude that $\ko\wedge BV$ splits as a sum of $\ko$-modules of EA-type. This is the reason we call these modules \textit{elementary abelian}-type.

Mitchell--Priddy~\cite[Theorem A]{MP83} produce a stable splitting of $\Sigma^\infty BV$ finer than the one in~\eqref{not_james_splitting},
but
the splitting of $\ko\wedge BV$ induced by their work is coarser than the one obtained by using the methods in this section and tracking the Whitney summands. It would be interesting to compare these two splittings, which would generalize work of Bayen~\cite[Proposition 3.4.4, \S 3.6.3]{Bay94} applicable to the case $n = 2$; see also~\cite[Theorem 4.4]{Bru99}.
\end{exm}

\begin{prop}
\label{M_is_EA}
Let $i\coloneqq \ell\bmod 4\in\{0,1,2,3\}$ and $j\coloneqq\lfloor \ell/4\rfloor$.
If at least one of $\ell$ or $k$ is positive, $\ko\wedge\ME_{\ell,k}$ is of EA-type and there is a stable isomorphism
\begin{equation}\label{ME_coh_isom}
	H^*(\ME_{\ell,k};\Z/2) \simeq
        \Sigma^{4j+k} N_i.
\end{equation}
\end{prop}
\begin{proof}
By \cref{EA-coh}, the second part of this proposition implies the first part. The Künneth theorem tells us
that since $\ME_{\ell,k}$ was defined as a smash product of $\ell$ copies of $(B\Z/2)^{\sigma-1}$ and $k$ copies of
$(B\Z/2)^{1-\sigma}$, $H^*(\ME_{\ell,k};\Z/2)$ is isomorphic to a tensor product of $\ell$ copies of $N_1$ and $k$
copies of $N_0$, so it suffices to identify such a tensor product modulo free summands, which can be done by iteratively applying \cref{yu_stable_iso}.
Explicitly, we have $\ell=4j+i$ for $j\geq 0$ and $0\leq i<4$. Then
\begin{equation}
    H^*(\ME_{\ell,k};\Z/2) \simeq N_1^{\otimes \ell} \otimes (\Sigma N_0)^{\otimes k}
    \simeq \Sigma^k N_1^{\otimes (4j+i)} \otimes N_0^{\otimes k}.
\end{equation}
Since $N_1^{\otimes(4j)}\simeq \Sigma^{4j}N_0$ and $N_0\otimes N_t\simeq  N_t$ for all $t\in\Z/4$, we obtain
\begin{equation}
   H^*(\ME_{\ell,k};\Z/2)\simeq \Sigma^k (N_1)^{\otimes(4j)} \otimes N_i \otimes N_0^{\otimes k}
    \simeq \Sigma^k (\Sigma^{4j}N_i)\otimes N_0^{\otimes k} \simeq \Sigma^{4j+k} N_i.
    \qedhere
\end{equation}
\end{proof}
Recall the function $\ldeg(\ell,k)$ from \cref{ldeg_defn}. \Cref{melk_calc} implies that the lowest-degree nonzero element of $H^*(\ME_{\ell,k};\Z/2)$ not contained in a free $\cA(1)$-module summand is in degree $\ldeg(\ell, k)$. We will see below that the same is true for non-Whitney summands in $\ko_*(\ME_{\ell,k})$.
\begin{prop}\label{melk_calc}
Keep the notation from \cref{M_is_EA}, and for $n\in\Z$, let $\widetilde m\coloneqq \lfloor (n-4j-k)/8\rfloor$. Then for all $n$, %
there are isomorphisms
\begin{equation*}
	\ko_n(\ME_{\ell,k})/(\text{\rm Whitney summands}) \cong\begin{cases}
        \Z/2, & n\equiv 1+\ell+k-2i \bmod 8,\ n\ge \ldeg(\ell, k) \\
        \Z/2, & n\equiv 2+\ell+k-2i \bmod 8,\ n\ge \ldeg(\ell, k) \\
        \Z/2^{4\widetilde m + 4-i}, & n \equiv 3 + \ell+k-2i\bmod 8,\ n\ge \ldeg(\ell, k) \\
        \Z/2^{4\widetilde m + 5-i}, & n \equiv 7+\ell+k-2i \bmod 8,\ n\ge \ldeg(\ell, k) \\       
        0, &\text{\rm otherwise}.
	\end{cases}
\end{equation*}
The $\ko_*$-action on these groups is trivial on Whitney summands; the Bott class acts injectively on non-Whitney summands, and the class $\eta\in\ko_1$ acts injectively in all degrees where it is not ruled out by the orders of these groups.
\end{prop}

\begin{proof}
By Margolis' theorem for $\ko$-modules (\cref{margolis}), free $\cA(1)$-module summands in cohomology split off as $H\Z/2$ summands of a $\ko$-module, which in this case are Whitney summands. Therefore we may ignore those summands and just focus on $N_0$, $N_1$, $N_2$, or $N_3$ by \cref{M_is_EA}. Run the Adams spectral sequence for these modules; as worked out in \cref{N1_exm,,N2_exm,,N3_exm,,N0_exm}, in all four cases the spectral sequence collapses without extension problems, leading to the $\ko$-homology groups in the proposition statement.
\end{proof}
\subsection{Computing \texorpdfstring{$\mathit{f2i}_{\ell,k}$}{f2i lk} and \texorpdfstring{$\mathrm{sp}^\phi_{\ell,k}(\ko)$}{sp phi lk ko}}
\label{ss:ko_phi_f2i}

To discuss the free-to-interacting map, we need to apply Anderson duality. This modifies the discussion of short,
long, and Whitney summands slightly. Recall the ``exponential fiber sequence'' from \cref{anddual} (\cite[(5.21)]{freed_reflection_2021}):
\begin{equation}\label{exponential}
	H\C\longrightarrow I_{\C^\times} \overset{\delta}{\longrightarrow} \Sigma I_\Z.
\end{equation}
\begin{lem}\label{duals_EA}
If $M$ is a $\ko$-module of EA-type, the map $\delta_M\colon I_{\C^\times}M\to \Sigma I_\Z M$ is a $\ko$-module
equivalence.
\end{lem}
\begin{proof}
If we can show $H\C_*(M) = \pi_*(M)\otimes\C$ vanishes, then we can plug that into~\eqref{exponential} to deduce that
$\delta_M$ is an isomorphism on homotopy groups and hence an equivalence. Since $\pi_*(M)$ is finitely generated in
each degree, it suffices to show that each homotopy group is torsion. The $H\Z/2$ summands contribute $\Z/2$
summands to homotopy, which are torsion, and the rest of $\pi_*(M)$ injects into $\pi_*(M)[\beta^{-1}]$, which by
definition of EA-type is torsion, so $\pi_*(M)$ is torsion and we conclude.
\end{proof}
The universal property of $I_{\C^\times}$, namely that $\pi_{-n}(I_{\C^\times} X)\cong \Hom(\pi_n(X), \C^\times)$,
implies that a direct-sum decomposition of $\pi_n(X)$ induces a dual direct-sum decomposition of
$(I_{\C^\times}X)^n$. Thus if $X$ is a $\ko$-module of EA-type, we also get a dual direct-sum decomposition of
$(I_\Z X)^{n+1}$ by \cref{duals_EA}. We refer to the duals of long, short, and Whitney summands as long, short, and
Whitney summands respectively, as it will never be ambiguous whether we are working with an EA-type $\ko$-module or
its Anderson dual.

Now we can state our first main result, which is the computation of the $\ko$-module version of the free-to-interacting map as discussed in \cref{IZ_ABS_through_ko}.

\begin{thm}
\label{compute_F2I}
On homotopy groups, the free-to-interacting map 
\[\mathit{f2i}_{\ell, k}\colon \Sigma^{\ell-k-2}\KO\to \Sigma^2 I_\Z(\ko\wedge\ME_{\ell,k})\]
has the following explicit description.

Let $\widetilde m\coloneqq \lfloor(d-\ell+i-k+1)/8\rfloor$. %

\begin{subequations}
\begin{description}
	\item[Long summands, part 1] For $d \equiv\ell+k-2i+2\bmod 8$ such that $d\ge\ldeg(\ell,k)-2$, $\pi_{-d}(\mathit{f2i}_{\ell, k})$ is a surjective map
    \begin{equation}
    \Z\twoheadrightarrow
	\Z/2^{4+4\widetilde m-i}
    \end{equation}
    onto the unique long summand in this degree. %
	\item[Long summands, part 2]
    For $d \equiv\ell+k-2i+6\bmod 8$ such that $d\ge\ldeg(\ell,k)-2$, $\pi_{-d}(\mathit{f2i}_{\ell, k})$ is a surjective map \begin{equation}\Z\twoheadrightarrow
	\Z/2^{5+4\widetilde m-i}\end{equation} onto the unique long summand in this degree.
	\item[Short summands] For $d\equiv \ell+k-2i\bmod 8$ or $d\equiv \ell+k-2i+1\bmod 8$ such that $d\ge\ldeg(\ell,k)-2$, $\pi_{-d}(\mathit{f2i}_{\ell, k})$ is
	an isomorphism \begin{equation}
    \Z/2\xrightarrow{\cong}\Z/2\end{equation} onto the unique short summand in this degree. %
\end{description}
\end{subequations}
In all other degrees the domain of the free-to-interacting map is zero; thus the image of the free-to-interacting
map does not intersect the Whitney summands nontrivially.
\end{thm}

We now prove \cref{compute_F2I} in a series of lemmas (\cref{compute_11,,compute_1k,,compute_all}). Our argument proceeds by induction, first on $k$ and then on $\ell$.

\begin{lem}[{Freed--Hopkins~\cite[\S 10]{freed_reflection_2021}}]
\label{compute_11}
\Cref{compute_F2I} is true for $k = 1$ and $\ell = 0$, and for $k = 0$ and $\ell = 1$.
\end{lem}
\begin{proof}
By \cref{rem:same_map}, our twisted ABS maps for the cases $k = 1$, $\ell = 0$, resp.\ $k = 0$, $\ell = 1$ coincide with Freed--Hopkins' twisted ABS in their cases $s = -1$, resp.\ $s
= 1$; therefore the same is true for the respective free-to-interacting maps. Thus it suffices to observe that the description we gave in our theorem statement coincides with their
calculation in these cases.
\end{proof}
Recall from \cref{spiral_1_map_E_version} the definition of $\spint_{\ell,k}(\ko)$: the same as $\spint_{\ell,k}$, but with $\ko$ in place of $\MTSpin$.
\begin{thm}
\label{int_Bott_spiral_ko}
Modulo Whitney summands, the map $\spint_{\ell,k}(\ko)\colon I_\Z(\ko\wedge\ME_{\ell,k-1})\to\Sigma
I_\Z(\ko\wedge\ME_{\ell,k})$ is a $\pi_*$-isomorphism. 
\end{thm}

\begin{figure}[h!]
\includegraphics{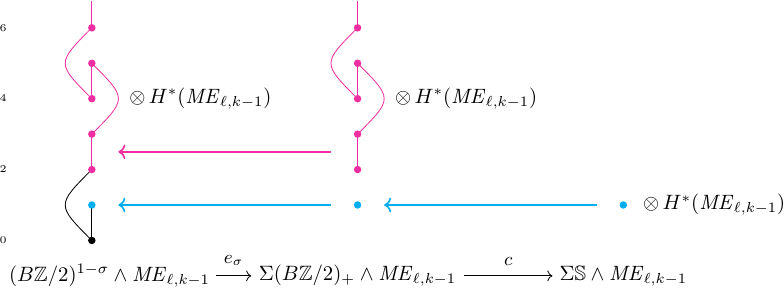}
\caption{The effect of the 
spiral map $\phi_{\ell,k}$ on mod $2$ cohomology.
This map factors as the zero section map $e_\sigma$ composed with the crush map $c$, all smashed with the identity on $\ME_{\ell,k-1}$.
We use
this map on cohomology in the proof of \cref{compute_1k}. 
}
\label{spiral_on_cohomology}
\end{figure}

\begin{lem}\label{phi_int_inj}
The map $\phi_{\ell,k}\colon\ME_{\ell,k}\to\Sigma\ME_{\ell,k-1}$ we defined in \cref{spiral_1_defn} induces an injective map on mod $2$ cohomology.
\end{lem}
\begin{proof}
The map $\phi_{\ell,k}$ in \cref{spiral_1_defn} is equivalently given by smashing
$\ME_{\ell,k-1}$ with a map $(B\Z/2)^{1-\sigma}\to \Sigma\Sph$ given by the composition of the zero section map
$e_\sigma\colon (B\Z/2)^{1-\sigma}\to \Sigma (B\Z/2)_+$ with the crush map $c\colon \Sigma
(B\Z/2)_+\to\Sigma\mathbb S$. The pullback $c^*\colon H^*(\Sigma\Sph;\Z/2)\to H^*(\Sigma(B\Z/2)_+;\Z/2)$ is
injective, since it is the inclusion of the degree-$1$ part. Since we are over a field, it is still injective after tensoring
with $\ME_{\ell,k-1}$. For the second part of $\phi_{\ell,k}$, namely the zero section map, we
make use of the fiber sequence (see, e.g., \cite[(7.1)]{debray_smith_2024})
\begin{equation}
	\Sph\longrightarrow (B\Z/2)^{1-\sigma} \xrightarrow{e_\sigma} \Sigma (B\Z/2)_+.
\end{equation}
Using the induced long exact sequence in mod $2$ cohomology\footnote{After applying the Thom isomorphism to $H^*((B\Z/2)^{1-\sigma};\Z/2)$, this long exact sequence can be identified with the usual Gysin sequence for the tautological line bundle over $B\Z/2$. See~\cite[Remark 5.11]{debray_smith_2024}.}
and the fact that $H^*(\Sph;\Z/2)\cong\Z/2$
concentrated in degree $0$, we see that $e_\sigma$ is injective on mod $2$ cohomology. Thus
after tensoring with $H^*(\ME_{\ell,k-1};\Z/2)$, the map is still injective. %
\end{proof}
See \cref{spiral_on_cohomology} for a proof of \cref{phi_int_inj}.
\begin{proof}[Proof of \cref{int_Bott_spiral_ko}]
By the universal property of $I_\Z$, this is equivalent to
asking that $\id_\ko\wedge \phi_{\ell, k}\colon \ko\wedge\ME_{\ell,k}\to\Sigma
\ko\wedge\ME_{\ell,k-1}$ is a $\pi_*$-isomorphism modulo Whitney summands, and this is what we prove.

In \cref{M_is_EA} we showed that $H^*(\ME_{\ell,k};\Z/2)\cong F\oplus \Sigma^{4j+k}N_{i}$, where $j =
\lfloor \ell/4\rfloor$, $i=\ell\bmod 4$, and $F$ is a free summand. Thus the non-free summands in the cohomologies of $\ME_{\ell,k}$
and $\Sigma \ME_{\ell,k-1}$ are graded isomorphic. By \cref{phi_int_inj}, $\phi_{\ell,k}^*\colon
H^*(\Sigma\ME_{\ell,k-1};\Z/2) \to H^*(\ME_{\ell,k};\Z/2) $ %
is injective. Thus the $\Sigma^{4j+k}N_{i}$ summand %
in the domain maps isomorphically onto its image. Because $\Sigma^{4j+k}N_i$ contains no nonzero free submodule, the only way for this to happen is when the codomain is $\Sigma^{4j+k}N_{i}$ plus a free $\cA(1)$-module is for it to map isomorphically to
a $\Sigma^{4j+k}N_{i}$ direct summand. Taking Ext groups, this implies that the map of Adams spectral sequences induced by $\id_\ko\wedge\phi_{\ell,k}$ is an isomorphism except possibly for the pieces coming from the free summands. By Margolis' theorem, those free summands do not have nontrivial
differentials or extension problems, and so pass to Whitney summands in $\ko$-homology. Thus,
modulo Whitney summands, $\id_\ko\wedge\phi_{\ell,k}$ is a $\pi_*$-isomorphism.
\end{proof}
\begin{lem}
\label{spiral_F2I_commute}
The diagram
\begin{equation}
\label{the_Bott_spiral_square}
\begin{tikzcd}%
 & {\Sigma^{2}I_\Z(\ko\wedge \ME_{\ell,k})} \ar[dd,"\mathrm{sp}^\phi_{\ell,k+1}(\ko)"] \\
{\Sigma^{\ell-k-2}\KO} \ar[ur,"{\mathit{f2i}_{\ell,k}}"] \ar[dr,"{\mathit{f2i}_{\ell,k+1}}"'] & \\
& {\Sigma^{3}I_\Z(\ko\wedge \ME_{\ell,k+1})}
\end{tikzcd}
\end{equation}
commutes.
\end{lem}
\begin{proof}
If one expands out the definitions of $\mathit{f2i}_{\ell,k}$ and $\mathrm{sp}_{\ell,k}^\phi(\ko)$, given in \cref{IZ_ABS_through_ko,spiral_maps_of_IFTs}, \eqref{the_Bott_spiral_square} factors into a smash product of a number of simpler diagrams of $\ko$-modules, each of which trivially commutes.
\end{proof}

\begin{lem}\label{compute_1k}
\Cref{compute_F2I} is true for $\ell = 1$ and $k$ arbitrary.
\end{lem}
\begin{proof}
Induct on $k$; the case $k = 1$ is \cref{compute_11}. In \cref{spiral_F2I_commute}, we proved that the the spiral map $\mathrm{sp}^\phi$ commutes with the free-to-interacting
map. Therefore it suffices to show that $\spint_{\ell,k}(\ko)\colon I_\Z(\ko\wedge\ME_{\ell,{k-1}})\to \Sigma
I_\Z(\ko\wedge\ME_{\ell,k})$ is a $\pi_*$-isomorphism modulo Whitney summands, which we proved in \cref{int_Bott_spiral_ko}. %
\end{proof}
\begin{rem}
Another way to show that $\spint_{\ell,k}(\ko)$ is an isomorphism modulo Whitney summands is to identify the fiber $F$ of $\id_\ko\wedge\phi_{\ell,k}$. It is possible to compute that $F$ is, as a $\ko$-module, equivalent to a sum of shifts
of $H\Z/2$, and so in particular the Bott class acts trivially on $F$. This implies $F\wedge_\ko \KO\simeq
F[\beta^{-1}]$ is trivial (see, e.g.,~\eqref{bott_on_modules}), so after base-changing to $\KO$, the Bott spiral map is an isomorphism. Since
base-changing to $\KO$ is injective on non-Whitney summands of homotopy groups for $\ko$-modules of EA-type, including both the domain and the codomain of $\id_\ko\wedge\phi_{\ell,k}$, that map is injective on homotopy groups modulo Whitney summands, which is enough to recover the consequences for the free-to-interacting map that prove
\cref{compute_1k}. We will use a similar strategy to prove \cref{compute_all}.
\end{rem}

\begin{lem}\label{compute_all}
\Cref{compute_F2I} is true for all values of $k$ and $\ell$.
\end{lem}
We will prove this with another inductive argument, this time inducting on $\ell$. To do this, we need a map relating $\ko\wedge\ME_{\ell,k}$ and $\ko\wedge\ME_{\ell-1,k}$. We will first construct and study this map, then return to \cref{compute_all}.

\begin{figure}[h!]
\includegraphics{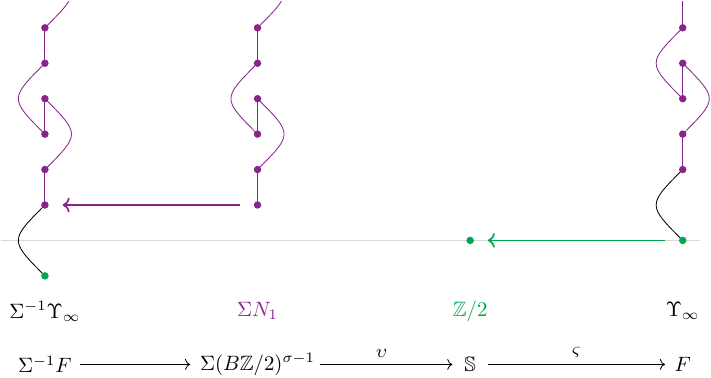}
\caption{A picture of the fiber sequence including the maps $\varsigma$ and $\upsilon$ from~\cite[\S 2]{Bru99}, introduced in the text in \cref{F_appears}. We also label the $\cA(1)$-modules appearing as the mod $2$ cohomology of the spectra in this fiber sequence, including the infinite seagull $\Upsilon_\infty$ from \cref{seagull_defn}.}
\label{upsilon_fiber}
\end{figure}

\begin{prop}
\label{no_1_cell}
The Thom spectrum $(B\Z/2)^{1-\sigma}$ admits the structure of a CW-spectrum with one $k$-cell for each $k\ge 0$, and such that the attaching map of the $2$-cell to the $1$-cell is trivial.
\end{prop}
\begin{proof}[Proof sketch]
This can be directly checked by showing that the attaching map of the $2$-cell to the $1$-cell in $(\RP^n)^{1-\sigma}$ is trivial for all $n\ge 2$. If $\lambda(n)$ is the order of $\sigma$ in $\widetilde{\KO}{}^0(\RP^n)$, then $(\RP^n)^{1-\sigma}\simeq (\RP^n)^{(\lambda(n)-1)(\sigma-1)}$, so one can finish the proof by checking that the attaching map from the $2$-cell to the $1$-cell is trivial in the Thom spectrum of $(\lambda(n)-1)(\sigma-1)\to\RP^n$, and Adams~\cite[Theorem 7.4]{Ada76} calculates $\lambda(n)$ for all $n$.
\end{proof}
\begin{cor}\label{seagull_cor}
The $k$-cells of the above CW structure on $(B\Z/2)^{1-\sigma}$ for $k\ne 1$, with the same attaching maps, define a CW spectrum $F$ with an ``inclusion map'' $F\to (B\Z/2)^{1-\sigma}$ whose cofiber is the $1$-cell $\Sigma\Sph$.\footnote{The spectrum $\mathit{TJF}_\infty$ studied in~\cite{lin_topological_2024, bauer_topological_2025, lin_genuine_2025, tominaga_computations_2026} appears to be a (nonconnective version of a) height-$2$ analog of $F$.}
\end{cor}
In essence, we want to ``forget to include'' the $1$-cell of $(B\Z/2)^{1-\sigma}$ to define $F$. This works if and only if no $2$-cell is nontrivially attached to the $1$-cell, which \cref{no_1_cell} verifies.
The following lemma appears in Bruner~\cite[\S 2]{Bru99}, who writes that it is probably originally due to Davis or Mahowald, and that he learned it from Stolz.
\begin{lem}\label{F_appears}
With $F$ as in \cref{seagull_cor}, the fiber of the map $\varsigma\colon \Sph\to F$ defined by including the $0$-cell is a nontrivial map $\upsilon\colon \Sigma (B\Z/2)^{\sigma-1}\to \Sph$.
\end{lem}
We draw this fiber sequence in \cref{upsilon_fiber}.
\begin{defn}\label{seagull_defn}
Let $\Upsilon_\infty$ denote the \term{infinite seagull}, the $\cA(1)$-module defined to be the submodule of $\Sigma^{-3}\widetilde H^*(\RP^\infty;\Z/2)$ of elements in degrees $d\ge 0$, $d\ne 1$.\footnote{\label{seagull_names} This $\cA(1)$-module is called $P/F_{-1,1}$ in~\cite{LDMA80}, $R$ in~\cite[\S A.9]{BG10} and~\cite[Definition 4.1]{Bru14}, $P$ in~\cite[Figure 5.7]{Cam17}, and $M_\infty$ in~\cite[Example 4.4.2]{BC18}.} Though a priori this is only a $\Z/2$-vector space, it is closed under $\Sq^1$ and $\Sq^2$, so it is in fact an $\cA(1)$-module.
\end{defn}
The name ``infinite seagull'' and the seagull-like notation $\Upsilon_\infty$ are due to Adamyk~\cite[Definition 2.5]{Ada23}. See \cref{upsilon_fiber}, right, for a depiction of $\Upsilon_\infty$.
\begin{lem}[{Bruner~\cite[Notation 2.1]{Bru99}}]\label{seagull_fiber}
There is an isomorphism $H^*(F;\Z/2)\cong \Upsilon_\infty$.\footnote{Bruner~\cite[Notation 2.1]{Bru99} writes that in the notation of~\cite{LDMA80}, $H^*(F;\Z/2)\cong \Sigma F_{-1,1}$. This appears to be a typo: $\Sigma F_{-1,1}$ should be replaced with $P/F_{-1,1}$, as in Footnote~\ref{seagull_names}.}\textsuperscript{\rm ,}\footnote{Since $\Upsilon_\infty$ is not an unstable $\cA(1)$-module, it cannot be the $\cA(1)$-module structure on the cohomology of any space. But if it were, that space would be a complete seagull space.}
\end{lem}
\begin{defn}\label{lam_sig_prime}
Recall the map $\lambda_\sigma\colon (B\Z/2)^{\sigma-1}\to \Sigma^{-1}\KO$ from \cref{lambda_V_defn} (specifically~\eqref{shift_lambda}). Because $(B\Z/2)^{\sigma-1}$ is $(-2)$-connected, $\lambda_\sigma$ factors through the $(-2)$-connected cover $\Sigma^{-1}\ko\to\Sigma^{-1}\KO$ as a map $\lambda_\sigma'\colon (B\Z/2)^{\sigma-1}\to \Sigma^{-1}\ko$. Smash $\lambda_\sigma'$ with $\id_\ko$ to define a $\ko$-module spectrum map $\lambda_\sigma''\colon \ko\wedge (B\Z/2)^{\sigma-1}\to \Sigma^{-1}\ko$.
\end{defn}
\begin{prop}\label{now_isnt_that_odd}
As $\ko$-module maps, the map $\lambda_\sigma''$ in \cref{lam_sig_prime} is equivalent to an odd multiple of $\id_\ko\wedge \Sigma^{-1}\upsilon$ after $2$-completing.
\end{prop}
\begin{proof}
Using the 
isomorphism from \cref{N1_exm} and the equivalence $\mathrm{Map}_R(R\wedge A, B)\simeq\mathrm{Map}_\Sph(A, B)$,
\begin{equation}\label{Z_worth_of_maps}
    \pi_0(\mathrm{Map}_\ko(\ko\wedge (B\Z/2)^{\sigma-1}, \Sigma^{-1}\ko)_2^\wedge \cong \widetilde{\ko}{}^0(B\Z/2).
\end{equation}
Since $\ko$ is the connective cover of $\KO$, $\widetilde{\ko}{}^0(B\Z/2)\cong\widetilde{\KO}{}^0(B\Z/2)$, which is isomorphic to $\Z_2$~\cite{And64}. We will show that both $\lambda_\sigma''$ and $\id_\ko\wedge\Sigma^{-1}\upsilon$ define odd elements of this $\Z_2$ after $2$-completion. Odd numbers can be inverted in the $2$-adics, so each map is an odd multiple of the other.
Given a map $f\colon \ko\wedge (B\Z/2)^{\sigma-1}\to \Sigma^{-1}\ko$, consider its effect on $\pi_0$, where it induces a map $\pi_0f\colon \Z/2\to\Z/2$. If $f\simeq 2g$, then $\pi_0f = 2\pi_0g$, so it must be the zero map on $\Z/2$. We will show that for $f = \lambda_\sigma''$ and $f = \id_\ko\wedge \Sigma^{-1}\upsilon$, $\pi_0f\ne 0$, so that both of these maps are odd.

For $\lambda_\sigma''$, precompose with the Atiyah--Bott--Shapiro map $\MTSpin\to\ko$ and postcompose with $\Sigma^{-1}\ko\to\Sigma^{-1}\KO$, as in \cref{1_lam_defn}, to obtain a map $\MTSpin\wedge (B\Z/2)^{\sigma-1}\to\Sigma^{-1}\KO$ that is precisely Freed--Hopkins' twisted ABS map~\cite[\S 10]{freed_reflection_2021} (see \cref{positive_s_ABS,1_lam_defn}). Freed--Hopkins also show (\textit{ibid.}, Corollary 9.85) that this map is an isomorphism $\Z/2\to\Z/2$ in degree $0$,\footnote{Freed--Hopkins only report the effect of the Anderson dual of their map, but since the homotopy groups of $\MTSpin\wedge (B\Z/2)^{\sigma-1}$ are torsion~\cite{ABP69}, this implies their original map is also an isomorphism in degree $0$, using the universal property of the Anderson dual.} so the middle piece $\lambda_\sigma''$ cannot be the zero map on $\pi_0$.

The cofiber of $\id_\ko\wedge\Sigma^{-1}\upsilon$ is $\ko\wedge \Sigma^{-1}F$. Mahowald~\cite[Lemma 7.2]{Mah82} shows that, $2$-locally,\footnote{This is an equivalence of spectra, but Hill~\cite[Lemma 3.3]{Hil07} showed that the two sides are \emph{not} equivalent as $\ko$-modules, where we give $H\Z$ the $\ko$-algebra structure arising from the truncation map $\ko\to H\Z$ of ring spectra. (This lemma is present in the arXiv version of~\cite{Hil07} but not the published version.)
Instead, we hypothesize that $\ko\wedge F$ is $\ko$-module equivalent to the connective cover of the $L$-theory spectrum of $\R$, which is also known to split as $H\Z\vee \Sigma^4 H\Z\vee\Sigma^8 H\Z\vee\dotsb$ as spectra but not $\ko$-modules. See~\cite{TW79, HLN20}.}
\begin{equation}
    \ko\wedge F\simeq \bigvee_{n\ge 0} \Sigma^{4n} H\Z.
\end{equation}
Plug this into the long exact sequence of homotopy groups induced by the fiber sequence $\ko\wedge (B\Z/2)^{1-\sigma}\to\Sigma^{-1}\ko\to\ko\wedge F$; exactness forces $\pi_0(\id_\ko\wedge\Sigma^{-1}\upsilon)$ to be surjective, hence an isomorphism $\Z/2\to\Z/2$.
\end{proof}
\begin{prop}\label{not_bott_commute}
Let
\begin{equation}\label{upsdefn}
    \upsilon_{\ell,k}\coloneqq \id_{\ko\wedge\ME_{\ell-1,k}}\wedge\Sigma^{-1}\upsilon\colon\ko\wedge\ME_{\ell,k}\to\Sigma\ko\wedge\ME_{\ell-1,k}.
\end{equation}
The diagram%
\begin{equation}\label{opposite_day_Bott_spiral}
\begin{tikzcd}
	& {\Sigma^2 I_\Z(\ko\wedge\ME_{\ell,k})} \\
	{\Sigma^{\ell-k-2}\KO} \\
	& {\Sigma I_\Z(\ko\wedge\ME_{\ell-1,k})}
	\arrow["{\mathit{f2i}_{\ell,k}}", from=2-1, to=1-2]
	\arrow["{\mathit{f2i}_{\ell-1,k}}"'{pos=0.5}, from=2-1, to=3-2]
	\arrow["{\Sigma^2 I_\Z\upsilon_{\ell,k}}"', from=3-2, to=1-2]
\end{tikzcd}\end{equation}
commutes after $2$-completion up to homotopy and multiplication by an odd integer (which is a unit in the $2$-adic integers).
\end{prop}
The diagram~\eqref{opposite_day_Bott_spiral} will play a key role in the proof of \cref{compute_all} completely analogous to the role of \eqref{the_Bott_spiral_square} in the proof of \cref{compute_1k}.
\begin{proof}
The proof is completely analogous to the proof of \cref{spiral_F2I_commute}, but replacing ``data inducing commutativity'' with ``data inducing commutativity after $2$-completion, up to homotopy and multiplication by an odd integer.'' Also, keep the factors of $\ko$ instead of getting rid of them. At the end of this simplification process, we have learned that if the diagram
\begin{equation}
\begin{tikzcd}
	{\ko\wedge (B\Z/2)^{\sigma-1}} & {\Sigma^{-1}\ko} \\
	& {\Sigma^{-1}\ko}
	\arrow["{\lambda_\sigma''}", from=1-1, to=1-2]
	\arrow["{\id_\ko\wedge\Sigma^{-1}\upsilon}"', from=1-1, to=2-2]
	\arrow["\id", from=1-2, to=2-2]
\end{tikzcd}\end{equation}
commutes after $2$-completion, up to homotopy and multiplication by an odd integer, then we have finished the proof. And the former is exactly what we proved in \cref{now_isnt_that_odd}.
\end{proof}
\begin{lem}\label{cof_almost_trivial}
For $\ell,k\ge 1$, $H^*(F\wedge\ME_{\ell-1,k};\Z/2)$ is a free $\cA(1)$-module.
\end{lem}
We will use a technique called \term{Margolis homology} to prove \cref{cof_almost_trivial}, so we pause here to review this tool.
\begin{defn}\label{margdefns}
Let $Q_0\coloneqq\Sq^1$ and $Q_1\coloneqq \Sq^1\Sq^2 + \Sq^2\Sq^1$. Then for $i = 0,1$, the Adem relations imply $Q_i^2 = 0$, so one may regard any $\cA(1)$-module $N$ as a cochain complex with differential $Q_i$. The homology of this complex is called the \term{Margolis homology} $H^*(M;Q_i)$~\cite{AM71, Mar83}. We say a bounded-below $\cA(1)$-module $M$ is \term{$Q_i$-local} if $H^*(M;Q_i)\ne 0$ and $H^*(M;Q_{1-i}) = 0$.
\end{defn}
We need a few key facts about Margolis homology.
\begin{thm}[{Wall, Adams--Margolis~\cite[Theorem 4.2]{AM71}}]
\label{margfree}
Let $f\colon M\to N$ be a map of bounded-below $\cA(1)$-modules that is an isomorphism on $Q_0$- and $Q_1$-Margolis homology. Then there are free $\cA(1)$-modules $F_1$ and $F_2$ and an isomorphism $M\oplus F_1\overset\cong\to N\oplus F_2$ extending $f$. In particular, if $M$ is a bounded-below $\cA(1)$-module such that $H^*(M;Q_0) = 0$ and $H^*(M; Q_1) = 0$, then $M$ is free.
\end{thm}
Adams--Margolis attribute this special case of their more general result to unpublished work of Wall.
\begin{prop}[{Yu~\cite[Chapter 2]{Yu95}}]
\label{NiQlocal}
For $i = 0,1,2,3$, $N_i$ is $Q_1$-local.
\end{prop}
\begin{prop}[{Künneth formula for Margolis homology~\cite[Chapter 19, Proposition 17(b)]{Mar83}}]
\label{kunneth_margolis}
Let $M$ and $N$ be $\cA(1)$-modules and $i = 0,1$. Then there are natural isomorphisms $H^*(M; Q_i)\otimes H^*(N; Q_i)\xrightarrow{\cong} H^*(M\otimes N;Q_i)$.
\end{prop}
\begin{proof}[Proof of \cref{cof_almost_trivial}]
By \cref{N0_exm,seagull_fiber}, $H^*(F\wedge (B\Z/2)^{1-\sigma};\Z/2)\cong \Sigma\Upsilon_\infty\otimes N_0$. Since $N_0$ is bounded below and $Q_1$-local (\cref{NiQlocal}) and $\Upsilon_\infty$ is $Q_0$-local~\cite[Proposition 4.2]{Bru14}, then by the Künneth formula (\cref{kunneth_margolis}), the $Q_0$- and $Q_1$-Margolis homology of $\Sigma\Upsilon_\infty\otimes N_0$ both vanish, so $\Sigma\Upsilon_\infty\otimes N_0$ is a free $\cA(1)$-module by \cref{margfree}.%

To finish, use the (usual) Künneth formula  on $F\wedge \ME_{\ell-1,k}\simeq (F\wedge\ME_{\ell-1,k-1})\wedge (B\Z/2)^{1-\sigma}$ to show that $H^*(F\wedge\ME_{\ell-1,k};\Z/2)\cong H^*(\ME_{\ell-1,k-1};\Z/2)\otimes (\Sigma\Upsilon_\infty\otimes N_0)$. Since the second term in the tensor product is free, the entire tensor product is a free module.
\end{proof}

\begin{proof}[Proof of \cref{compute_all}]
We will show that the map $I_\Z\upsilon_{\ell,k}$ in~\eqref{upsdefn} is a $\pi_*$-isomorphism modulo Whitney summands after $2$-completion. With this in hand, the same strategy from the proof of \cref{compute_1k}, but this time using the commutativity result from \cref{not_bott_commute}, implies the conclusion of the lemma, albeit only after $2$-completion and multiplication by an odd integer. Since the homotopy groups of $I_\Z(\ko\wedge\ME_{\ell,k})$ are $2$-torsion, this suffices to prove the lemma.

Now back to $I_\Z\upsilon_{\ell,k}$. The domain and codomain of this map are the Anderson duals of $\Sigma^{-1}\ko\wedge\ME_{\ell-1,k}$, resp.\ $\ko\wedge\ME_{\ell,k}$, which have torsion homotopy groups by \cref{melk_calc}. Therefore by \cref{duals_EA} it suffices to prove $I_{\C^\times}\chi$ is a $\pi_*$-isomorphism modulo Whitney summands after $2$-completion, and by the universal property of $I_{\C^\times}$, it suffices to prove the same for $\upsilon_{\ell,k}$.

Since the domain and codomain of $\upsilon_{\ell,k}$, namely $\ko\wedge\ME_{\ell,k}$, resp.\ $\Sigma^{-1}\ko\wedge\ME_{\ell-1,k}$, are $\ko$-modules of EA-type, the vertical maps in the diagram
\begin{equation}\begin{tikzcd}[column sep=0.9in]
	{\ko\wedge\ME_{\ell,k}} & {\Sigma^{-1}\ko\wedge\ME_{\ell-1,k}} \\
	{\KO\wedge\ME_{\ell,k}} & {\Sigma^{-1}\KO\wedge\ME_{\ell-1,k}}
	\arrow["\upsilon_{\ell,k}", from=1-1, to=1-2]
	\arrow["{\beta^{-1}}"', from=1-1, to=2-1]
	\arrow["{\beta^{-1}}", from=1-2, to=2-2]
	\arrow["{\widetilde\upsilon_{\ell,k}\coloneqq \upsilon_{\ell,k}[\beta^{-1}]}", from=2-1, to=2-2]
\end{tikzcd}\end{equation}
are injective modulo Whitney summands on homotopy groups, so it suffices to prove the map $\widetilde\upsilon_{\ell,k}\colon \KO\wedge\ME_{\ell,k}\to \Sigma^{-1}\KO\wedge\ME_{\ell-1,k}$, i.e.\ the map induced from $\upsilon_{\ell,k}$ after base changing to $\KO$, is an equivalence after $2$-completion. The cofiber of $\widetilde\upsilon_{\ell,k}$ is $\mathrm{cofib}(\upsilon_{\ell,k})[\beta^{-1}]$; since $\upsilon_{\ell,k}$ is obtained by smashing $\Sigma^{-1}\upsilon$ with $\id_{\ko\wedge\ME_{\ell-1,k}}$, \cref{F_appears} implies the cofiber of $\upsilon_{\ell,k}$ is $\ko\wedge F\wedge \ME_{\ell-1,k}$. By \cref{cof_almost_trivial}, $H^*(F\wedge\ME_{\ell-1,k};\Z/2)$ is a free $\cA(1)$-module, so Margolis' theorem for $\ko$-modules (\cref{margolis}) implies that after $2$-completion $\ko\wedge F\wedge\ME_{\ell-1,k}$ is a sum of shifts of $H\Z/2$---in particular, the Bott class $\beta$ acts by $0$ on this $\ko$-module. Thus, $\mathrm{cofib}(\upsilon_{\ell,k})[\beta^{-1}]\simeq 0$, but this was $\mathrm{cofib}(\widetilde\upsilon_{\ell,k})$, so $\widetilde\upsilon_{\ell,k}$ is an equivalence after $2$-completion, and as noted above this suffices.
\end{proof}
\subsection{Computation of \texorpdfstring{$\mathrm{sp}^\psi_{\ell,k}(\ko)$}{sp psi lk ko}}
\label{ss:ko_psi}
\label{computing_spiral_2_map}

We defined the maps $\mathrm{sp}^\psi_{\ell,k}$ in two cases: $\ell>0$ and $k>0$ (\cref{spiral_maps_of_spectra}) and $\ell\ge 3$ and $k = 0$ (\cref{Q8_psi}).  \textbf{Throughout this subsection, assume both $\ell$ and $k$ are positive,} so we are in the first case. We will return to the second case in \cref{other_real_AZ_computation}.

Recall that in \cref{spiral_1_map_E_version}, we defined $\mathrm{sp}_{\ell,k}^\psi(\ko)$ to be the analog of $\mathrm{sp}_{\ell,k}^\psi$ from \cref{spiral_maps_of_IFTs}, but with $\ko$ in place of $\MTSpin$.
Our goal in this subsection is to prove the following theorem, where we keep the following notation from \cref{melk_calc}: $i\coloneqq\ell\bmod 4$ (thought of as an integer between $0$ and $3$, not as an element of $\Z/4$), $j\coloneqq \lfloor \ell/4\rfloor$, and $\widetilde n \coloneqq n - 4j - k$. In addition, let $m\ge 0$, and let $c(m)$ be equal to $1$ if $m$ is odd and $0$ if $m$ is even. %

\begin{thm}
\label{psicalc}
The map $\mathrm{sp}^\psi_{\ell,k}(\ko)\colon \ko\wedge \ME_{\ell,k}\to\Sigma \ko\wedge \ME_{\ell+1,k}$ sends long summands to long summands. Specifically, restricted to the long summand $\Z/2^{2m+c(m) + i+4}\subset\pi_{4j+k+4m-i+3}(\ko\wedge\ME_{\ell,k})$, $\mathrm{sp}^\psi_{\ell,k}(\ko)$ surjects onto the long summand $\Z/2^{2m+c(m) + i+3}\subset\pi_{4j+k+4m-i+2}(\ko\wedge\ME_{\ell+1,k})$.
\end{thm}
The orders of these long summands were computed in \cref{melk_calc}; the key takeaway is that $\mathrm{sp}_{\ell,k}^\psi(\ko)$ is always surjective on these long summands, with kernel isomorphic to $\Z/2$.

Recall that $\mathrm{sp}^\psi_{\ell,k}(\ko)$ (\cref{spiral_maps_of_IFTs}) is the composition of an Euler map $e_\sigma$ (\cref{smith_map_of_spectra_specific}) and a map induced from the diagonal $\Delta\colon (B\Z/2)_+\to \ME_{1,1}$ (\cref{diag_map_defn}). We will compute the effect of $e_\sigma$ on the long summand in~\S\ref{subsubsec_Euler} and of $\Delta$ on the long summand in~\S\ref{subsubsec_diag}. We summarize these two calculations in \cref{pinp_euler_comp,diag_on_ko}, respectively, and these jointly imply \cref{psicalc}.

\subsubsection{The Euler map}
\label{subsubsec_Euler}
Let $r_{\ell,k}\colon\ME_{\ell,k-1}\wedge (B\Z/2)_+\to \ME_{\ell,k-1}\wedge B\Z/2$ denote the smash product of $\id_{\ME_{\ell.k-1}}$ and the map $r\colon (B\Z/2)_+\to B\Z/2$ taking the cofiber of the inclusion of the basepoint. Then, after applying $\Sigma^\infty$ (in particular, for homology and cohomology), $r$ is projection onto a direct summand, and $\iota\colon B\Z/2\to (B\Z/2)_+$ is a section.

In this subsubsection, we make the following computation.
\begin{prop}
\label{pinp_euler_comp}
The composition $\widetilde{e}_\sigma\coloneqq r_{\ell,k}\circ e_\sigma\colon \ko\wedge \ME_{\ell,k}\to \ko\wedge \Sigma \ME_{\ell,k-1}\wedge B\Z/2$ sends long summands to long summands; specifically, restricted to the long summand 
\begin{subequations}\begin{equation}\Z/2^{2m+c(m) + i+4}\subset\pi_{\ell+k+4m-2i+3}(\ko\wedge\ME_{\ell,k}),\end{equation} $\widetilde{e}_\sigma$ surjects onto the long summand 
\begin{equation}\Z/2^{2m+c(m) + i+3}\subset\pi_{\ell+k+4m-2i+2}(\ko\wedge\ME_{\ell,k-1}\wedge (B\Z/2)).\end{equation}
\end{subequations}
\end{prop}
By \cref{melk_calc}, $\pi_{4j+k+4m-i+2}(\ko\wedge\ME_{\ell,k-1}\wedge \Sigma (B\Z/2)_+)$ contains two long summands isomorphic to $\Z/2^{2m+c(m) + i+3}$ and $\Z/2^{2m+c(m) + i+4}$, coming respectively from $\Sigma B\Z/2$ and from the basepoint. The diagonal map, which we study in \S\ref{subsubsec_diag}, will destroy the long summand coming from the basepoint, so for the rest of this subsubsection, we will throw out the basepoint, as the long summand we want to surject onto is in the $B\Z/2$ summand.

\begin{lem}
\label{e1_fib}
The fiber of $\widetilde{e}_\sigma\colon (B\Z/2)^{1-\sigma}\to \Sigma B\Z/2$ is $\Sigma^{-1}\RP^2$. Thus the fiber of $\widetilde{e}_\sigma\wedge\id_{\ME_{\ell,k-1}}$ is $\Sigma^{-1}\RP^2\wedge \ME_{\ell,k-1}$.
\end{lem}
\begin{proof}
Compose $\widetilde{e}_\sigma$ with the 
homotopy equivalence
$e_\sigma\colon \Sigma B\Z/2\xrightarrow{\simeq} \Sigma^2 (B\Z/2)^{\sigma-1}$; additivity of Euler classes implies the composition is $e_{2\sigma}\colon (B\Z/2)^{1-\sigma}\to \Sigma^2 (B\Z/2)^{\sigma-1}$. Kirby--Taylor~\cite[Lemma 7]{KT90pinp} identify the fiber of $e_{2\sigma}$ as $\Sigma^{-1}\RP^2$; see also~\cite[Example 7.37]{debray_smith_2024}.
\end{proof}
We thus would like to understand $\ko\wedge\Sigma^{-1}\RP^2\wedge \ME_{\ell,k-1}$. %

\begin{figure}[h!]
\includegraphics{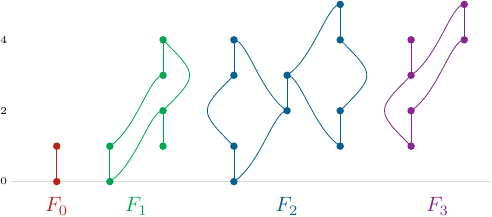}
  \caption{The $\cA(1)$-modules $F_i$ from \cref{defn_Fi}. Figure adapted from Pearson~\cite[(8-14)]{Pea14}.}
  \label{modules_Fi}
\end{figure}

\begin{defn}
\label{defn_Fi}
We introduce four more $\cA(1)$-modules, which we draw in \cref{modules_Fi}.
\begin{itemize}
    \item $F_0\coloneqq \Sigma^{-1}\widetilde H^*(\RP^2;\Z/2)$, the unique nonsplit extension of $\Z/2$ by $\Sigma\Z/2$.
    \item $F_3\coloneqq \Sigma \cA(1)/(\Sq^2\Sq^1\Sq^2)$.
    \item $F_2$ is the unique nonsplit extension of $J$ by $\Sigma J$.
    \item $F_1\coloneqq \Sigma^5 F_3^\vee$.
\end{itemize}
\end{defn}
Here $(\bl)^\vee$ denotes the $\Z/2$-vector space dual; see Baker~\cite[\S 3]{Bak20} for a discussion of the $\cA(1)$-module structure on the dual of an $\cA(1)$-module. In practice, if one is familiar with the $\Sq^1$ and $\Sq^2$ diagrams of $\cA(1)$-modules, the dual of a finitely generated $\cA(1)$-module is obtained by turning the diagram upside down; the dual of a class in degree $t$ lies in degree $-t$.
\begin{defn}
\label{defn_R2}
Let $R_2$ denote the kernel of the unique nonzero $\cA(1)$-module map $\Sigma^{-1}\cA(1)\to\Sigma^{-1}\Z/2$, let $\Omega\colon\Mod_{\cA(1)}\to\Mod_{\cA(1)}$ be $\Sigma R_2\otimes\bl$, and let $\Omega^n$ denote the result of applying $\Omega$ $n$ times.
\end{defn}
\begin{lem}[{Powell~\cite[\S 6]{powell_2015_truncated}}]
\label{loops_F0}
For $i = 0,1,2,3$, $\Omega^iF_0\simeq\Sigma^{2i}F_i$.
\end{lem}
\begin{lem}
\label{F0_tensor_Fi}
For $i\in\Z/4$, there is a stable isomorphism $N_i\otimes F_0\simeq F_i$.
\end{lem}
\begin{proof}
Bruner~\cite[Theorem 4.3]{Bru14} shows that for $i$ as above and any $Q_1$-local (see \cref{margdefns}) $\cA(1)$-module $M$, such as $F_0$~\cite[\S 6]{powell_2015_truncated},
\begin{equation}
    N_i\otimes M\simeq \Sigma^{-2i}\Omega^i M.
    \qedhere
\end{equation}
\end{proof}
\begin{cor}
\label{Fextcor}
Let $i\coloneqq \ell\bmod 4$ and $j\coloneqq \lfloor \ell/4\rfloor$. If at least one of $\ell$ or $k$ is positive, there is a stable isomorphism of $\cA(1)$-modules
\begin{equation}
    H^*(\Sigma^{-1}\RP^2\wedge \ME_{\ell,k};\Z/2) \simeq \Sigma^{4j+k}F_i.
\end{equation}
\end{cor}
\begin{proof}
Combine \cref{M_is_EA,F0_tensor_Fi}.
\end{proof}
Thus, for $i = 0,3$, the lowest-degree non-free $\cA(1)$-module summand of $H^*(\Sigma^{-1}\RP^2\wedge\ME_{\ell,k};\Z/2)$ begins in degree $\ldeg(\ell, k) + 1$; for $i = 1,2$, it begins in degree $\ldeg(\ell, k)$ (here $\ldeg$ is the function defined in \cref{ldeg_defn}).

\begin{figure}
\begin{subfigure}[c]{0.7\textwidth}
\includegraphics{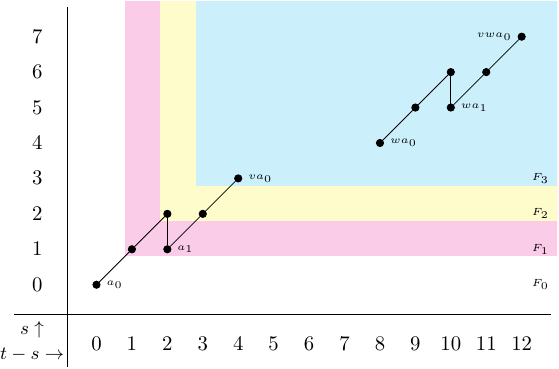}
\end{subfigure}
\caption{$\Ext_{\cA(1)}(F_i)$, as calculated in \cref{ExtFi}. The entire chart is $\Ext(F_0)$; for $\Ext(F_i)$, move everything $i$ degrees down and to the left, then delete all classes not in the first quadrant.}
\label{F_ext_figure}
\end{figure}

\begin{prop}\hfill
\label{ExtFi}
\begin{enumerate}
    \item\label{ExtF1} There is an $\E$-module isomorphism
    \begin{subequations}
    \begin{equation}
     \Ext_{\cA(1)}(F_0) \cong\E\set{a_0, a_1}/(h_0a_0,\ h_1^2a_0 + h_0a_1,\ va_0 + h_1^2a_1,\ va_1),
    \end{equation}
    with $a_0\in\Ext^{0,0}$ and $a_1\in\Ext^{1,3}$.
    \item\label{restExt} For $i = 0,1,2,3$, for $s,t\ge 0$ there are $\E$-module isomorphisms
    \begin{equation}
        \Ext_{\cA(1)}^{s,t}(F_i) \overset\cong\longrightarrow \Ext_{\cA(1)}^{s+i, t+2i}(F_0).
    \end{equation}
\end{subequations}
\end{enumerate}
\end{prop}
We draw the corresponding Ext charts in \cref{F_ext_figure}.
\begin{proof}
Part~\eqref{ExtF1} is in~\cite[Figure 6]{Pea14} or~\cite[\S 5.3.1]{debray_invertible_2021}. Part~\eqref{restExt} follows from \cref{loops_F0} together with Adams--Priddy's calculation~\cite[Proof of Lemma 3.8]{AP76} of $\Ext(\Omega M)$ in terms of $\Ext(M)$. A priori, Adams--Priddy's calculation ignores the contributions to Ext from free summands of an $\cA(1)$-module, but by construction, there are no free summands in $F_i$.
\end{proof}
Various pieces of \cref{ExtFi} are already in the literature: for example, \cite{Wil73a, BB96, Pea14, WWZ20, debray_invertible_2021} partially describe $\Ext(F_i)$ for $i = 0$, $2$, and $3$.

Thus $\Ext_{\cA(1)}(F_i)$ is a sort of ``highly connected cover'' of $\Ext_{\cA(1)}(F_0)$, much like the computations of Adams--Priddy~\cite[\S 3]{AP76} and Yu~\cite[Theorem 3.1]{Yu95} for stably invertible, resp.\ $Q_1$-locally stably invertible $\cA(1)$-modules (see \cref{Q1stab}).
\begin{prop}
\label{e1fibcalc}
Let $i,j$ be as in \cref{melk_calc} and $\ldeg$ be as in \cref{ldeg_defn}. Let $a(i) = 1$ if $i = 0$ or $3$ and $a(i) = 0$ otherwise.
There is an isomorphism
\begin{equation*}
    \ko_n(\Sigma^{-1}\RP^2\wedge \ME_{\ell,k})/(\text{\rm Whitney summands}) \cong\begin{cases}
        \Z/2, & \widetilde n+i\equiv 0,1,3,4 \bmod 8,\, n\ge\ldeg(\ell, k) + a(i)\\
        \Z/2, & i = 2,\ \widetilde n\equiv 0\bmod 8, n\ge\ldeg(\ell,k)\\
        \Z/4, &i\ne 2,\ \widetilde n + i\equiv 
            2\bmod 8,\, n\ge\ldeg(\ell, k) + a(i)\\
        0, &\text{\rm otherwise}.
    \end{cases}
\end{equation*}
The Bott class acts injectively on non-Whitney summands and trivially on Whitney summands; the class $\eta\in\ko_1$ acts injectively in all degrees on non-Whitney summands where it is not ruled out by the orders of these groups, except when $i = 2$ and $\widetilde n = 0$.
\end{prop}
\begin{proof}
Use the Adams spectral sequence as usual.
\Cref{Fextcor} gives us the $\cA(1)$-module structure on $H^*(\Sigma^{-1}\RP^2\wedge\ME_{\ell,k};\Z/2)$ modulo free summands, and \cref{ExtFi} calculates the Ext of this module (again, modulo Ext of free summands). Margolis' theorem (\cref{margolis}) implies free summands lift to Whitney summands, so we ignore them. The Adams spectral sequence then collapses for degree reasons, implying the $\ko$-homology groups in the proposition statement. The actions of $\beta$, resp.\ $\eta$ follow from the actions of $w$, resp.\ $h_1$ on the $E_\infty$-page.
\end{proof}
\begin{proof}[Proof of \cref{pinp_euler_comp}]
We will consider the long exact sequence in $\ko$-homology associated to the fiber sequence $\Sigma^{-1}\RP^2\wedge\ME_{\ell,k-1}\to \ME_{\ell,k}\xrightarrow{e_\sigma} \Sigma \ME_{\ell,k-1}\wedge (B\Z/2)_+$ constructed in \cref{e1_fib}. This is a long exact sequence of $\ko$-modules, in that it commutes with the action of classes in $\ko_*$, such as $\eta\in\ko_1$ and the Bott class $\beta\in\ko_8$. Thus in particular for any $n$ we have the following commutative diagram whose rows are exact.
\begin{equation}
\label{e1_LES_general}
\begin{tikzcd}
	\dotsb & {\ko_{n+8}(\ME_{\ell,k})} & {\ko_{n+7}(\ME_{\ell,k-1}\wedge (B\Z/2))} & {\ko_{n+7}(\Sigma^{-1}\RP^2\wedge\ME_{\ell,k-1})} & \dotsb \\
	\dotsb & {\ko_n(\ME_{\ell,k})} & {\ko_{n-1}(\ME_{\ell,k-1}\wedge (B\Z/2))} & {\ko_{n-1}(\Sigma^{-1}\RP^2\wedge\ME_{\ell,k-1})} & \dotsb
	\arrow[from=1-1, to=1-2]
	\arrow["{\widetilde{e}_\sigma}", from=1-2, to=1-3]
	\arrow["\partial", from=1-3, to=1-4]
	\arrow[from=1-4, to=1-5]
	\arrow[from=2-1, to=2-2]
	\arrow["\beta", from=2-2, to=1-2]
	\arrow["{\widetilde{e}_\sigma}", from=2-2, to=2-3]
	\arrow["\beta", from=2-3, to=1-3]
	\arrow["\partial", from=2-3, to=2-4]
	\arrow["\beta", from=2-4, to=1-4]
	\arrow[from=2-4, to=2-5]
\end{tikzcd}\end{equation}
There are two cases where this diagram contains long summands: $\widetilde n +i\equiv 3\bmod 8$ and $\widetilde n+i \equiv 7\bmod 8$. In both cases we will plug in the $\ko$-homology groups that we calculated in \cref{melk_calc,e1fibcalc} and let exactness finish the proof for us, but the two cases will differ a little. In the rest of the proof, $W$ will denote a Whitney summand, which is a finite-dimensional $\Z/2$-vector space which $\eta$ and $\beta$ act trivially on. In particular, we do not assert all $W$s appearing in the same diagram are isomorphic.

First assume $\widetilde n+i\equiv 3\bmod 8$. \Cref{melk_calc,e1fibcalc} imply that~\eqref{e1_LES_general} takes the following form:
\begin{equation}\label{diag_nmod3}
\begin{tikzcd}
	{\Z/2^{4\widetilde m + 8 + i}\oplus W} & {\Z/2^{4\widetilde m + 7 + i}\oplus W} & {\Z/2\oplus W} \\
	{\Z/2^{4\widetilde m + 4 + i}\oplus W} & {\Z/2^{4\widetilde m + 3 + i}\oplus W} & {\Z/2\oplus W}
	\arrow["{\widetilde{e}_\sigma}", from=1-1, to=1-2]
	\arrow["\partial", from=1-2, to=1-3]
	\arrow["{\begin{bsmallmatrix}16\, 0\\0\,\, 0\end{bsmallmatrix}}"', from=2-1, to=1-1]
	\arrow["{\widetilde{e}_\sigma}", from=2-1, to=2-2]
	\arrow["{\begin{bsmallmatrix}16\, 0\\0\,\, 0\end{bsmallmatrix}}"', from=2-2, to=1-2]
	\arrow["\partial", from=2-2, to=2-3]
	\arrow["{\begin{bsmallmatrix}1\, 0\\0\, 0\end{bsmallmatrix}}"', from=2-3, to=1-3]
\end{tikzcd}\end{equation}
Let $x\in\ko_{n-1}(\ME_{\ell,n-1}\wedge (B\Z/2))$ be a generator of the $\Z/2^{4\widetilde m + 3+i}$ summand. We would like to show that $\partial(x) = 0$, so that by exactness, $\widetilde{e}_\sigma$ surjects onto the summand generated by $x$; the degrees of the summands in $\ko_n(\ME_{\ell,k})$ imply that the long summand of $\ko_n(\ME_{\ell,k})$ surjects onto the long summand of $\ko_{n-1}(\ME_{\ell,n-1}\wedge (B\Z/2))$, which would finish the proof of the proposition in this case.

With respect to the splitting $\ko_{n-1}(\Sigma^{-1}\RP^2\wedge\ME_{\ell,k-1})\cong\Z/2\oplus W$ chosen in~\eqref{diag_nmod3}, let $\partial(x) = (a, b)$. By \cref{e1fibcalc}, $\eta$ zeroes out the Whitney summand and carries the $\Z/2$ summand isomorphically onto the non-Whitney summand of $\ko_n(\Sigma^{-1}\RP^2\wedge\ME_{\ell,k-1})$. Thus $\eta(\partial(x))$ is nonzero if and only if $a\ne 0$; since $\eta(\partial(x)) = \partial(\eta(x))$ and $\eta(x) = 0$ by \cref{melk_calc}, we get $a = 0$. Thus $\partial(x)$ is contained in the Whitney summand, so
\begin{equation}
    \partial(\beta(x)) = \beta(\partial(x)) = 0.
\end{equation}
By exactness, there is a $z\in\ko_{n+8}(\ME_{\ell,k})$ such that $\widetilde{e}_\sigma(z) = \beta(x)$. Since $\beta(x)$ is four times a generator of $\Z/2^{4\widetilde m+7+i}$, $2\beta(x)\ne 0$ and therefore $2z\ne 0$, so $z$ is not contained in a Whitney summand. Moreover, $z = 16w$ for some $w$: if it were not, then the order-$16$ elements of $\Z/2^{4\widetilde m+8+i}$ would be in the kernel of $\widetilde{e}_\sigma$, which is the image of $\ko_{n+8}(\Sigma^{-1}\RP^2\wedge\ME_{\ell,k-1})\to\ko_{n+8}(\ME_{\ell,k}$, but by \cref{e1fibcalc}, $\ko_{n+8}(\Sigma^{-1}\RP^2\wedge\ME_{\ell,k-1})$ contains no elements of order $16$. Now, since $z = 16w$, $z$ is in the image of $\beta\colon\ko_n(\ME_{\ell,k})\to\ko_{n+8}(\ME_{\ell,k})$. Let $\widetilde z$ be a preimage of $z$; then $\widetilde{e}_\sigma(\widetilde z) = x$, so $\partial(x) = 0$ by exactness and we are done.

The case $\widetilde n+i\equiv 7\bmod 8$ is nearly the same. The orders of the long summands are slightly different in ways that do not affect the proof; the only meaningful difference is that, when $\widetilde n+i\equiv 7\bmod 8$, $\ko_n(\Sigma^{-1}\RP^2\wedge\ME_{\ell,k-1}) \cong 0$ and $\ko_{n-1}(\Sigma^{-1}\RP^2\wedge\ME_{\ell,k-1})$ vanishes apart from Whitney summands, so the argument with $\eta$ is vacuous.
\end{proof}

\subsubsection{The diagonal}
\label{subsubsec_diag}
Our main result in this subsubsection is \cref{diag_on_ko}, that $\Delta\colon B\Z/2\to \ME_{1,1}$ (\cref{diag_map_defn}) induces an isomorphism on non-Whitney summands after smashing with $\ko\wedge \ME_{\ell,k}$. Together with \cref{pinp_euler_comp}, this will prove     \cref{psicalc}, finishing the calculation of the effect of $\mathrm{sp}^\psi_{\ell,k}$ on long summands.
\begin{lem}
\label{valar_margolis}
$\id\wedge\Delta\colon\ME_{\ell,k}\wedge B\Z/2\to \ME_{\ell+1,k+1}$ is an isomorphism on $Q_0$- and $Q_1$-Margolis homology.
\end{lem}
\begin{proof}
The Künneth formula for Margolis homology (\cref{kunneth_margolis}) immediately reduces us to the case $\ell = k = 0$, i.e.\ showing that $\Delta\colon \Sigma^\infty B\Z/2\to \ME_{1,1}$ is an isomorphism on Margolis homology.

Write $H^*(B\Z/2;\Z/2)\cong\Z/2[x]$ and $H^*(\ME_{1,1};\Z/2)\cong U\cdot H^*(B\Z/2\times B\Z/2;\Z/2)\cong U\cdot \Z/2[a, b]$, where $\abs x = \abs a = \abs b = 1$, where $a$, resp.\ $b$ are dual to the left, resp.\ right, factors in $B\Z/2\times B\Z/2$ and $U$ is the Thom class.
Using our identifications of the mod $2$ cohomology of $B\Z/2$ and $(B\Z/2)^{\pm(1-\sigma)}$ as shifts of the $\cA(1)$-modules $N_0$ and $N_1$ (\cref{N1_exm,N0_exm}) together with Yu's calculation~\cite[Chapter 2]{Yu95} of the $Q_0$- and $Q_1$-homology of $N_0$ and $N_1$ (see \cref{NiQlocal}), we get
\begin{itemize}
    \item $\widetilde H^*(B\Z/2; Q_0) = 0$ and $H^*((B\Z/2)^{\pm(\sigma-1)}; Q_0) = 0$.
    \item $\widetilde H^*(B\Z/2; Q_1) = \Z/2\cdot x^2$ and $H^*((B\Z/2)^{\pm(\sigma-1)}; Q_1) = \Z/2\cdot Ux$.
\end{itemize}
Thus, by the Künneth formula for Margolis homology (\cref{kunneth_margolis}), $H^*(\ME_{1,1};Q_0) = 0$ and $H^*(\ME_{1,1}; Q_1) = \Z/2\cdot Uab$. That is, for $i = 0,1$, the $Q_i$-Margolis homologies of $\Sigma^\infty B\Z/2$ and $\ME_{1,1}$ are abstractly isomorphic, and vacuously $\Delta^*$ is an isomorphism on $Q_0$-homology. For $Q_1$-homology, since $\Delta$ pulls both $a$ and $b$ back to $x$, $\Delta^*(Uab) = x^2$, so $\Delta^*$ is also an isomorphism on $Q_1$-homology.
\end{proof}
\begin{cor}
\label{diag_on_ko}
$\id\wedge\Delta\colon \ko\wedge\ME_{\ell,k}\wedge B\Z/2\to \ko\wedge\ME_{\ell+1,k+1}$ is a $\pi_*$-isomorphism modulo Whitney summands.
\end{cor}
\begin{proof}
\Cref{Hko_of_ko,valar_margolis} imply that $\id_\ko\wedge\Delta$ induces an isomorphism of both $Q_0$- and $Q_1$-homology on $H_\ko^*$, so by \cref{margfree}, it is an $\cA(1)$-module isomorphism modulo free summands. By Margolis' theorem (\cref{margolis}), these free summands lift to Whitney summands, and the non-free summand(s) are isomorphic, inducing an isomorphism in the Baker--Lazarev Adams spectral sequence for the complements of the Whitney summands. Thus we obtain an isomorphism of $2$-completed homotopy groups modulo Whitney summands, but since $\ME_{\ell,k}$ is trivial at odd primes if $(k,\ell)\ne (0, 0)$, this suffices.
\end{proof}
We discuss a consequence for dpin bordism specifically. Since the worldsheet in Type I superstring theory has a dpin structure~\cite[\S 6]{kaidi_topological_2020}, this may be of independent interest.
\begin{prop}
\label{dpin_decomp}
Let $X$ be a space. Then for $k<8$, there is an isomorphism
\begin{equation}
\label{dpin_decomp2}
\begin{aligned}
    \Omega_k^{\mathrm{DPin}}(X) \overset\cong\longrightarrow\,
        &\,\Omega_k^\Spin(B\Z/2\wedge X) \oplus
        H_k(X;\Z/2) \oplus
        H_{k-2}(X;\Z/2) \oplus H_{k-4}(X;\Z/2)\\
        &\mathbin\oplus H_{k-4}(X;\Z/2) \oplus H_{k-6}(X;\Z/2) \oplus
        H_{k-6}(X;\Z/2).
\end{aligned}
\end{equation}
Moreover, under the interpretation of dpin structures as spin structures on the orientation double cover~\cite[\S 6.1]{kaidi_topological_2020}, the $\Omega_k^\Spin(B\Z/2\wedge X)$ summand is precisely the summand represented by oriented manifolds.
\end{prop}
\Cref{dpin_decomp} provides a conceptual explanation for the appearance of some of the generators and bordism invariants for low-degree dpin bordism in~\cite[Appendices E and F]{kaidi_topological_2020}.
\begin{proof}
By \cref{dpinequiv}, dpin bordism is exactly the spin bordism of $\ME_{1,1}$, which in degrees $7$ and below coincides with the $\ko$-homology of $\ME_{1,1}$. Thus~\eqref{dpin_decomp2} follows from \cref{diag_on_ko} together with the claim that the degrees of the Whitney summands in $\mathit{MT}\DPin$ in degrees $7$ and below are $0$, $2$, $4$, $4$, $6$, and $6$. This second claim is proven in~\cite[(F.11)]{kaidi_topological_2020} (see also (\textit{ibid.}, Figure 6)). The rest of the proof, identifying the unique non-Whitney summand as the oriented dpin manifolds, follows from the description of the map $\Spin\times\Z/2\to\DPin$ in~\cite[\S 6.1]{kaidi_topological_2020}.
\end{proof}
One can extend \cref{dpin_decomp} into higher degrees by computing the degrees of more Whitney summands in $\mathit{MT}\DPin$. %

\section{Computing over \texorpdfstring{$\MTSpin$}{MTSpin}}
\label{s:spin_computations}
In this section, we finish the proofs of our main computational theorems by lifting our results from $\ko$ to $\MTSpin$. In \S\ref{spin_computation}, we give a complete description of the maps $F2I_{\ell,k}$, $\mathrm{sp}_{\ell,k}^\phi$, and $\mathrm{sp}_{\ell,k}^\psi$ in all dimensions, modulo Whitney summands, in \cref{only_first_summand_real,int_Bott_spiral_1,MSpin_psi}.\footnote{As in the previous section, we assume $\ell,k>0$ for $\mathrm{sp}^\psi_{\ell,k}$.} This corresponds in physics to starting the Bott spiral in one of the Altland--Zirnbauer classes D, BDI, or DIII, as well as any of the discrete real Altland--Zirnbauer classes.

In \S\ref{other_real_AZ_computation}, we consider Bott spirals beginning with a continuous Altland--Zirnbauer symmetry type, then applying $\mathrm{sp}^\phi$ and $\mathrm{sp}^\psi$ as above. These ``mixed discrete-continuous'' symmetry types are described by fermionic tensor products of a continuous Altland--Zirnbauer symmetry group $G_s$ (see \cref{disc_cont_defn}) with $E_{\ell,k}$. In \cref{gen_AZ_f2i,,remaining_real_int_Bott_thm,,MSpinHs_psi}, we compute the free-to-interacting maps, the $\mathrm{sp}^\phi$ maps, and the $\mathrm{sp}^\psi$ maps for these cases, and our results suggest these mixed discrete-continuous spiral maps could model the Bott spiral starting in the Altland--Zirnbauer classes AI, AII, C, CI, and CII. However, it is not possible in these cases to write down a full period of the spiral (see \cref{mixed_DC_spiral}), providing evidence that our fully discrete model of the Bott spiral is the correct one. We also use these results to finish the computation of $\mathrm{sp}_{\ell,0}^\psi$ in \cref{special_case_psi_ko}.

    Our results rely on \cref{thm:spinh_ABP,thm:pin_tilde_c_ABP,thm:pinh_ABP}, generalizations of the Anderson--Brown--Peterson splitting~\cite{ABP67} of $\MTSpin$ to the twisted spin bordism spectra $\mathit{MT}H(K_f(s))$, where $K_f(s)$ is the continuous fermionic group defined in \cref{disc_cont_defn} (see \cref{tenfold_table}),
    together with their interactions with Freed--Hopkins' free-to-interacting maps~\cite[\S 9.2.2]{freed_reflection_2021}. For $\MTSpin^h$ ($K_f(4) = \SU_2$), this is a theorem of Buchanan--McKean~\cite{BM23} and Mills~\cite{Mil24}; in the remaining four cases, this is new. Two of us, joint with Pacheco-Tallaj, will prove these remaining cases in a separate paper~\cite{ABP_splittings}.

In \S\ref{complex_computation}, we perform the complex analogs of these computations, working first over $\ku$, then over $\MTSpin^c$. Generally the statements are easier, and the proofs very similar, compared to their real analogs. Our main results are \cref{cplx_F2I_image_thm,,ku_int_Bott,,complex_psi}, which compute the free-to-interacting and spiral maps in this setting, modeling the Bott spiral beginning in classes A or AIII. See \cref{fig:cplx_Bott_spiral_maps} for the Bott spiral beginning in class AIII, built by assembling these computations.

Lastly, in \S\ref{spinQ8}, we compute $\Spin\times_{\set{\pm 1}}Q_8$ bordism, extending work of Pedrotti~\cite[Theorem 8.0.8]{Ped17} in low dimensions. This is the last remaining computation of the twisted spin bordism groups for our discrete Altland--Zirnbauer symmetry groups.

\subsection{Lifting to \texorpdfstring{$\MTSpin$}{MTSpin} and spiraling from \texorpdfstring{$\MTSpin$ and $\MTPin^{\pm}$}{MTSpin and MTPinpm}}
\label{spin_computation}

In this subsection we lift from theorems for $\ko$-modules, where the story is simpler, to theorems for $\MTSpin$-modules, which are the actual spectra classifying bordism and invertible field theories of interest. The lift essentially amounts to Anderson--Brown--Peterson's splitting of $\MTSpin$ and the description of the Atiyah--Bott--Shapiro orientation as the projection onto the first summand of this splitting.

Our main theorems in this section are \cref{only_first_summand_real,int_Bott_spiral_1,MSpin_psi}. Essentially they say that the description of the free-to-interacting map is the same as in \cref{compute_F2I} and that $\spint_{\ell,k}$ and $\mathrm{sp}^\psi_{\ell,k}$ behave similarly to their analogues over $\ko$.
However, to precisely state these theorems we must first introduce some definitions, so we do that, then state and prove our main theorems.

Our primary tool for lifting $\ko$-theoretic statements to $\MTSpin$-level statements is the foundational work of Anderson--Brown--Peterson:
\begin{thm}[Anderson--Brown--Peterson~\cite{ABP67}]
\label{ABP_thm}
There are numbers $a_i,b_j,c_k\in\N$ and a $2$-local homotopy equivalence
\begin{equation}
    \MTSpin\overset\simeq\longrightarrow
        \bigvee_{i\ge 1} \textcolor{Rhodamine}{\Sigma^{a_i}\ko} \vee
        \bigvee_{j\ge 1} \textcolor{Orange}{\Sigma^{b_j} \tau_{\ge 2}\ko} \vee
        \bigvee_{k\ge 1} \textcolor{Cyan}{\Sigma^{c_k} H\Z/2},
\end{equation}
where $\tau_{\ge 2}\ko$ denotes the $1$-connective cover of $\ko$. The degrees of the first few shifts are
\begin{equation}\label{spin_shifts}
    \begin{aligned}
        \textcolor{Rhodamine}{(a_i)} &= (0, 8, 16, 16, 24, 24, 24, 24, \dotsc)\\
        \textcolor{Orange}{(b_j)} &= (8, 16, 16, 24, 24, 24, 24, \dotsc)\\
        \textcolor{Cyan}{(c_k)} &= (20, 22, 24, 26, 26, \dotsc).
    \end{aligned}
\end{equation}
Moreover, the projection $\mathrm{proj}_{\textcolor{Rhodamine}{a_1}}\colon \MTSpin\to \textcolor{Rhodamine}{\ko}$ onto the first factor is equivalent to the (connective cover of the) Atiyah--Bott--Shapiro map.
\end{thm}
For any spectrum $X$, this induces a wedge-sum decomposition of $\MTSpin\wedge X$ which we call the \term{ABP decomposition}:
\begin{equation}\label{ABP_smash_X}
    \MTSpin\wedge X\overset\simeq\longrightarrow
        \bigvee_{i\ge 1} \textcolor{Rhodamine}{\Sigma^{a_i}\ko\wedge X} \vee
        \bigvee_{j\ge 1} \textcolor{Orange}{\Sigma^{b_j} \tau_{\ge 2}\ko\wedge X} \vee
        \bigvee_{k\ge 1} \textcolor{Cyan}{\Sigma^{c_k} H\Z/2\wedge X}.
\end{equation}
\begin{rem}\label{product_placement}
One can verify using the defining property of $I_\Z$ that the Anderson dual of a wedge sum is the product of the Anderson duals of the summands. Therefore~\eqref{ABP_smash_X} induces a direct product decomposition of $I_\Z(\MTSpin\wedge X)$.
\end{rem}
\begin{cor}
\label{spinbord_torsion}
Let $X$ be a space or spectrum such that $\pi_*(\ko\wedge X)$ and $\widetilde H_*(X;\Z)$ are both torsion. Then $\pi_*(\MTSpin\wedge X)$ is also torsion.
\end{cor}
\begin{proof}
Using the splitting~\eqref{ABP_smash_X}, all we need to verify is that $\pi_*(\textcolor{Orange}{\tau_{\ge 2}\ko\wedge X})$ and $\pi_*(\textcolor{Cyan}{H\Z/2\wedge X})$ are torsion. The latter is isomorphic to the mod $2$ homology of $X$, which is by definition torsion, so we focus on the former.

Taking the Postnikov cover induces a fiber exact sequence $\tau_{\ge 2}\ko\wedge X\to \ko\wedge X\to \tau_{\le 1}\ko \wedge X$, which induces a long exact sequence on homotopy groups. By assumption, $\pi_*(\ko\wedge X)$ is torsion, so if we also know that $\pi_*(\tau_{\le 1}\ko\wedge X)$ is torsion, we would be done. Since $\tau_{\le 1}\ko$ has only two nonzero homotopy groups, $\pi_0\cong\Z$ and $\pi_1\cong\Z/2$, the Atiyah--Hirzebruch spectral sequence can be used to show that, since $\widetilde H_*(X;\Z)$ is torsion, so is $\pi_*(\tau_{\le 1}\ko\wedge X)$.
\end{proof}
This in particular applies to $X = \ME_{\ell,k}$---\cref{melk_calc} shows its $\ko$-homology is torsion, and for its integral homology, use the Thom isomorphism to identify $H_*(\ME_{\ell,k};\Z)$ with certain twisted reduced homology groups of $(B\Z/2)^{\ell+k}$, which have been computed by Lastovecki~\cite{lastovecki_cohomology_2005}.

The definition of a Whitney summand (i.e.\ the pieces we will ignore) is a little delicate in this setting, ultimately because \cref{ABP_thm} is not an $\MTSpin$-module equivalence. We would like to define Whitney summands by comparing with $\ko$-theory, but to do that we would need a map of ring spectra $\ko\to\MTSpin$, which Stolz~\cite[\S 7]{Sto94} showed does not exist. Fortunately, Stolz (\textit{ibid.}, \S 1) introduced a weaker criterion that is good enough for us.

Recall that the isomorphism $\Omega_0^\O\xrightarrow{\cong} \Z/2$ counting the number of points mod $2$ lifts to an $E_\infty$-ring map $p\colon \MO\to H\Z/2$. Mahowald~\cite[2.6]{Mah79} constructed a section $s_\O\colon H\Z/2\to \MO$ of $p$ as homotopy-commutative ring spectra, which Hopkins observed refines to an $E_2$-ring map (see~\cite[\S 1]{HW20}).\footnote{See~\cite{Pri78, CMT81, MRS01, Blu10, MNN15, BW18, ACB19, Kit20, HW20, Lev22, Mao23} for additional proofs and generalizations of Mahowald's result.}
\begin{defn}
A section $s\colon \ko\to\MTSpin$ of the Atiyah--Bott--Shapiro map (as a map of spectra) is a \term{weak algebra section} if
\begin{enumerate}
    \item the induced map $s_*\colon H_*(\ko;\Z/2)\to H_*(\MTSpin;\Z/2)$ is an algebra homomorphism; and
    \item the following diagram commutes up to homotopy:
\begin{equation}\begin{tikzcd}
	\ko & \MTSpin \\
	{H\Z/2} & \MO.
	\arrow["s", from=1-1, to=1-2]
	\arrow["{\tau_0}", from=1-1, to=2-1]
	\arrow["\phi", from=1-2, to=2-2]
	\arrow["{s_\O}", from=2-1, to=2-2]
\end{tikzcd}\end{equation}
    Here $\phi$ is induced by the forgetful map $B\Spin\to B\O$ and $\tau_0$ is the Postnikov $0$-truncation map followed by mod $2$ reduction.
\end{enumerate}
\end{defn}
\begin{prop}[{Stolz~\cite[\S 6]{Sto94}}]
Weak algebra sections of the Atiyah--Bott--Shapiro orientation exist, and all of them induce equal maps on mod $2$ homology.
\end{prop}
Without loss of generality, choose such a map $s$.
\begin{defn}[{Stolz~\cite[\S 1]{Sto94}}]
\label{htpKO}
A \term{homology $\ko$-module} is a spectrum $Y$ together with a map $\mu_Y\colon \ko\wedge Y\to Y$ of spectra such that the induced map on homology
\begin{equation}
    (\mu_Y)_*\colon H_*(\ko;\Z/2)\otimes_{\Z/2} H_*(Y;\Z/2)\longrightarrow H_*(Y;\Z/2)
\end{equation}
defines the structure of an $H_*(\ko;\Z/2)$-module on $H_*(Y;\Z/2)$.
\end{defn}
\begin{rem}\hfill
\label{making_htpy_ko}
\begin{enumerate}
    \item\label{ko_to_htpy} Any $\ko$-module, including $\ko$, $\tau_{\ge 2}\ko$, and $H\Z/2$, is a homology $\ko$-module in a canonical way.
    \item Any $\MTSpin$-module $Y$ with action map $\mu^s_Y\colon\MTSpin\wedge Y\to Y$ has a canonical homology $\ko$-module structure with $\mu_Y\coloneqq \mu_Y^s\circ (s\wedge\id_Y)\colon \ko\wedge Y\to Y$ (\textit{ibid.}, (1.2)). Moreover, if the $\MTSpin$-module structure arises from a $\ko$-module structure through the ABS map, this homology $\ko$-module structure agrees with the one from~\eqref{ko_to_htpy}.
    \item The Anderson--Brown--Peterson splitting in \cref{ABP_thm} is a splitting of homology $\ko$-modules, as is implicit in (\textit{ibid.}, \S 6).
    \qedhere
\end{enumerate}
\end{rem}
\begin{defn}
\label{bosonic_MSpin}
Let $M$ be an $\MTSpin$-module. A \term{Whitney summand} of $M$ is an $H\Z/2$-summand of $M$ as homology $\ko$-modules.
\end{defn}
If $M$ acquired its $\MTSpin$-module structure from a $\ko$-module structure of EA type via the ABS map, \cref{def:summands,bosonic_MSpin} coincide.
\begin{rem}
\label{blue_bosonic}
In~\eqref{ABP_smash_X}, each $\textcolor{Cyan}{H\Z/2\wedge X}$ piece splits as a sum of shifts of $H\Z/2$ indexed by a basis of $H_*(X;\Z/2)$. Each of these $H\Z/2$ summands is by definition a Whitney summand.
\end{rem}

Whitney summands of $M$ are dual to Whitney summands of $I_\Z M$, as one can verify using the compatibility of Anderson duality with wedge sums.
\begin{thm}\label{only_first_summand_real}\hfill
\begin{enumerate}
    \item 
    Let $(\mathrm{Dir}_{\ell,k})_*$ be the group of IFTs whose partition functions vanish on all spin-$(\ell,k)$ manifolds whose images under the projection $\mathrm{proj}_{\textcolor{Rhodamine}{a_1}}\colon \Omega_*^\Spin(\ME_{\ell,k})\to\textcolor{Rhodamine}{\ko_*(\ME_{\ell,k})}$ vanish.
    The image of the free-to-interacting map $\Sigma^{\ell-k-2}\KO\to \Sigma^2 I_\Z(\MTSpin\wedge\ME_{\ell,k})$ on homotopy groups is a subgroup of $(\mathrm{Dir}_{\ell,k})_*$.
    \item $(\mathrm{Dir}_{\ell,k})_*\cong \pi_{-*}(\Sigma^2 I_\Z(\textcolor{Rhodamine}{\ko}\wedge\ME_{\ell,k}))$, and the image of the free-to-interacting map inside these groups is exactly as given in \cref{compute_F2I}.
\end{enumerate}
\end{thm}
This is a corollary of \cref{IZ_ABS_through_ko}, where we saw that $F2I_{\ell,k}$ factors as the composition of the map $\mathit{f2i}_{\ell,k}\colon \Sigma^{\ell-k-2}\KO\to \Sigma^2 I_\Z(\ko\wedge\ME_{\ell,k})$ followed by the dual ABS map $\Sigma^2 I_\Z(\ko\wedge\ME_{\ell,k})\to \Sigma^2 (I_\Z\MTSpin\wedge\ME_{\ell,k})$. By \cref{ABP_thm,product_placement}, this is the inclusion of a direct summand on $\pi_n$ for any $n$.
\begin{thm}
\label{int_Bott_spiral_1}
The spiral map $\spint_{\ell,k}\colon I_\Z(\MTSpin\wedge\ME_{\ell,k-1})\to \Sigma I_\Z(\MTSpin\wedge\ME_{\ell,k})$ is an isomorphism modulo Whitney summands. %
\end{thm}
To prove this, we need a little more information on $\tau_{\ge 2}\ko$.
\begin{lem}\label{dissect_joker}\hfill
\begin{enumerate}
    \item\label{joker_Sph} There is a finite, connective spectrum $\mathbf J$ and a $\ko$-module equivalence $\tau_{\ge 2}\ko\simeq \ko\wedge \Sigma^2 \mathbf J$.
    \item\label{joker_J} Let $J\coloneqq\cA/(\Sq^3)$. There is an $\cA(1)$-module equivalence $H_\ko^*(\tau_{\ge 2}\ko)\cong \Sigma^2 J$.
\end{enumerate}
\end{lem}
Part~\eqref{joker_Sph} was apparently a folklore theorem; one reference is Baker~\cite[Corollary 4.2]{Bak18}.\footnote{The spectrum $\mathbf J$ was constructed by Hopkins~\cite{Hop84}; what is folklore is the connection to $\tau_{\ge 2}\ko$. In fact Hopkins shows there are exactly two such homotopy classes of spectra, distinguished by whether $\Sq^4$ acts nontrivially on their mod $2$ cohomology, and they are Spanier--Whitehead dual to each other.} Part~\eqref{joker_J} then follows by combining~\cite[Corollary 4.2]{Bak18} and~\cite[Example 5.4]{Bak20}.

\begin{proof}[Proof of \cref{int_Bott_spiral_1}]
In \cref{spiral_1_map_E_version}, we defined the spiral map $\spint_{\ell,k}$ by defining a map $\ME_{\ell,k}\to \Sigma \ME_{\ell,k-1}$, smashing with $\MTSpin$, and applying Anderson duality. Therefore, as we discussed above, we can compute the effect of the Bott spiral map by checking what it does on each of the three kinds of summands in the ABP decomposition of $I_\Z(\MTSpin\wedge\text{--})$.
\begin{enumerate}
    \item For the $\textcolor{Rhodamine}{\Sigma^{a_i}\ko}$ summands, we get the map $\spint_{\ell,k}(\ko)$, which we proved is an isomorphism modulo Whitney summands in \cref{int_Bott_spiral_ko}.
    \item Next the $\textcolor{Orange}{\Sigma^{b_j}\tau_{\ge 2}\ko}$ summands. By \cref{dissect_joker}, $\ko\wedge \mathbf{J}\simeq \tau_{\ge 2}\ko$ as $\ko$-modules. Therefore on these summands the interacting Bott spiral map has the form
    \begin{equation}
        \mathrm{sp}_{\ell,k}^\phi(\textcolor{Orange}{\Sigma^{b_j}\mathbf J})\colon \mathrm{Map}(\textcolor{Orange}{\Sigma^{b_j} \mathbf{J}}, I_\Z(\textcolor{Orange}{\ko}\wedge\ME_{\ell,k+1})) \to
        \mathrm{Map}(\textcolor{Orange}{\Sigma^{b_j} \mathbf{J}}, I_\Z(\Sigma \textcolor{Orange}{\ko}\wedge\ME_{\ell,k})).
    \end{equation}
    That is, we can think of it as a natural transformation of cohomology theories, evaluated on the spectrum $\textcolor{Orange}{\Sigma^{b_j}\mathbf{J}}$. We can compute the effect of this map by considering the induced map from the Atiyah--Hirzebruch spectral sequence computing the $I_\Z(\textcolor{Orange}{\ko}\wedge\ME_{\ell,k+1})$-cohomology of $\textcolor{Orange}{\Sigma^{b_j}\mathbf J}$ to the $I_\Z(\Sigma \textcolor{Orange}{\ko}\wedge\ME_{\ell,k})$-cohomology of the same spectrum. On $E_2^{\bullet, q}$, this map is exactly the map induced by applying $\pi_{-q}$ to $\spint_{\ell,k}(\ko)$, then applying $H^*(\textcolor{Orange}{\Sigma^{b_j}\mathbf{J}};\bl)$. By \cref{int_Bott_spiral_ko}, the map of $E_2$-pages is an isomorphism modulo those $\Z/2$ summands coming from Whitney summands. Since those Whitney summands are Eilenberg--Mac Lane summands of the spectra we applied the Atiyah--Hirzebruch spectral sequence to, they split off of the spectral sequence. Specifically, they (1) do not have nontrivial differentials or extensions to or from other summands and (2) they converge to Whitney summands of $\mathrm{Map}(\textcolor{Orange}{\Sigma^{b_j} \mathbf{J}}, I_\Z(\textcolor{Orange}{\ko}\wedge\ME_{\ell,k+1}))$ (resp.\ $\mathrm{Map}(\textcolor{Orange}{\Sigma^{b_j} \mathbf{J}}, I_\Z(\Sigma \textcolor{Orange}{\ko}\wedge\ME_{\ell,k}))$). Therefore we can focus on the unique, non-Whitney summand of both; by the above discussion, the map of Atiyah--Hirzebruch spectral sequences on this summand is an isomorphism on $E_2$, hence is an isomorphism on homotopy groups.
    \item The $\textcolor{Cyan}{\Sigma^{c_k}H\Z/2}$ summands are Whitney summands (see \cref{blue_bosonic}), so we ignore them. \qedhere
\end{enumerate}
\end{proof}
To discuss the behavior of $\mathrm{sp}_{\ell,k}^\psi$, we need to understand the structure of $\MTSpin\wedge\ME_{\ell,k}$ a little better.
\begin{defn}[{\cite{AP76, Yu95}}]
\label{Q1stab}
A bounded-below, finite-type $\cA(1)$-module $M$ is \term{stably invertible} if there is some other $\cA(1)$-module $N$ such that $M\otimes N$ is stably isomorphic (see \cref{stab_iso}) to $\Z/2$. $M$ is \term{$Q_1$-locally stably invertible} if $M$ is $Q_1$-local (\cref{margdefns}) and there is another $Q_1$-local $\cA(1)$-module $N$ such that $M\otimes N\simeq N_0$.
\end{defn}
Thus if $M$ and $N$ are stably invertible, resp.\ $Q_1$-locally stably invertible, $M\otimes N$ is stably invertible, resp.\ $Q_1$-locally stably invertible.
\begin{lem}[{\cite[Theorem 5.2]{Bru14}}]
\label{tensor_thru_Q1loc}
Let $M$ and $N$ be $\cA(1)$-modules, and assume that $M$ is stably invertible and $N$ is $Q_1$-locally stably invertible. Then $M\otimes N$ is $Q_1$-locally stably invertible, and is stably isomorphic to $(M\otimes N_0)\otimes N$.
\end{lem}
\begin{prop}[{Yu~\cite[Theorem 2.1]{Yu95}}]
\label{Q1_Pic}
$N_i$ is $Q_1$-locally stably invertible for all $i\in\Z/4$. Conversely, let $M$ be a $Q_1$-locally stably invertible $\cA(1)$-module. Then there are $\alpha\in\Z$ and $i'\in\Z/4$ such that $M\simeq\Sigma^\alpha N_{i'}$.
\end{prop}
\begin{cor}
\label{stab_inv_EA}
Let $M$ be a finite type $\ko$-module such that $H_\ko^*(M)$ is stably invertible or $Q_1$-locally stably invertible and $(\ell,k)\ne (0, 0)$. Then $M\wedge\ME_{\ell,k}$ is of EA-type.
\end{cor}
\begin{proof}
By \cref{M_is_EA,Q1_Pic}, $H^*(\ME_{\ell,k};\Z/2)$ is a $Q_1$-locally stably invertible $\cA(1)$-module; since $H_\ko^*(M)$ is either stably invertible or $Q_1$-locally stably invertible, $H_\ko^*(M\wedge\ME_{\ell,k})$ is $Q_1$-locally stably invertible. By \cref{Q1_Pic},
$H_\ko^*(M\wedge \ME_{\ell,k})\cong\Sigma^\alpha N_{i'}$ for some $\alpha\in\Z$ and $i\in\Z/4$. The result then follows from \cref{EA-coh}.
\end{proof}
\begin{lem}[{Adams--Priddy~\cite[\S 3]{AP76}}]
\label{joker_stabinv}
The $\cA(1)$-module $J \cong \Sigma^{-2} H_\ko^*(\tau_{\ge 2}\ko)$ defined in \cref{dissect_joker} is stably invertible.
\end{lem}
Thus every non-Whitney summand in the ABP decomposition~\eqref{ABP_smash_X} of $\MTSpin\wedge\ME_{\ell,k}$ is of EA-type, whence we may refer to long, short, and Whitney summands of each, and likewise for the Anderson dual. 
\begin{thm}
\label{MSpin_psi}
For $k,\ell>0$, the spiral map $\mathrm{sp}_{\ell,k}^\psi\colon I_\Z(\MTSpin\wedge\ME_{\ell+1,k}) \to I_\Z(\MTSpin \wedge\Sigma\ME_{\ell,k})$ sends long summands to long summands; on each long summand it is an inclusion $\Z/2^N\hookrightarrow \Z/2^{2N}$ with cokernel $\Z/2$.
\end{thm}
Hence we can think of this map as ``multiplication by $2$'' on long summands.
\begin{proof}
Just like for the proof of \cref{int_Bott_spiral_1}, because $\mathrm{sp}_{\ell,k}^\psi$ was defined in \cref{spiral_maps_of_spectra} by applying $I_\Z\MTSpin$ to a map $\ME_{\ell,k}\to\Sigma\ME_{\ell+1,k}$, we may smash that map with each non-Whitney summand in the ABP decomposition to give the proof. For the $\textcolor{Rhodamine}{\Sigma^{a_i}\ko}$ summands, we are done by \cref{psicalc}. For the $\textcolor{Orange}{\Sigma^{b_j}\tau_{\ge 2}\ko}$ summands, we need to check that \cref{pinp_euler_comp,diag_on_ko} both still hold after smashing with $\tau_{\ge 2}\ko$. For \cref{diag_on_ko}, this is completely straightforward: the proof applies \textit{mutatis mutandis}. For \cref{pinp_euler_comp}, the proof can be adapted in essentially the same way, but one needs to know that the fiber of
\begin{equation}
    \tau_{\ge 2}\ko\wedge\Sigma^{-1}\RP^2\wedge \ME_{\ell,k}\longrightarrow \tau_{\ge 2}\ko\wedge \Sigma^{-1}\RP^2\wedge\ME_{\ell,k-1}\wedge \Sigma B\Z/2
\end{equation}
has $H_\ko^*$ stably equivalent to $\Sigma^\alpha F_{i'}$ for certain $\alpha$ and $i'$. This follows from \cref{F0_tensor_Fi,Q1_Pic}, and the fact that $H_\ko^*(\tau_{\ge 2}\ko)$ is stably invertible, as we observed after the proof of \cref{stab_inv_EA}. The remaining details are the same as for $\ko$.
\end{proof}
\begin{rem}[Put a pin in it]
\label{BDI_DIII}
This completes the story for class D; in this remark we explain how these theorems also apply to Bott spirals starting in classes BDI or DIII. In these three cases, the discrete and continuous fermionic groups coincide (see \cref{disc_cont_defn,tenfold_table}), so we may continue with the discrete groups, corresponding to the tangential structures \pinm and \pinp, respectively. Moreover, Freed--Hopkins' free-to-interacting maps~\cite[\S 9.2.2]{freed_reflection_2021} for classes BDI and DIII coincide with our $(\ell,k) = (1, 0)$ resp.\ $(0, 1)$ free-to-interacting maps by construction (see \cref{rem:same_map}), so there is no difference between the free-to-interacting map, $\mathrm{sp}^\phi$, or $\mathrm{sp}^\psi$ for $\MTSpin\wedge\ME_{\ell,k}$, $\MTPin^-\wedge\ME_{\ell-1,k}$, and $\MTPin^+\wedge\ME_{\ell, k-1}$.
\end{rem}

\subsection{Spiraling from \texorpdfstring{$\MTSpin^h$, $\MTPin^{h\pm}$, and $\MTPin^{\tilde c\pm}$}{MTSpinh, MTPin hpm, and MTPin tilde c}}
\label{other_real_AZ_computation}
The ideas we used in \S\ref{spin_computation} to calculate the free-to-interacting and Bott spiral maps when the starting symmetry type is one of spin, \pinp, or \pinm apply with only slight modifications to \spinh, pin\textsuperscript{$\tilde c\pm$}, and pin\textsuperscript{$h\pm$}, which correspond to the \emph{continuous} Altland--Zirnbauer classes C, AI, AII, CI, and CII, respectively, from \cref{disc_cont_defn,tenfold_table}.\footnote{In the \emph{discrete} case, except for class C, these groups are $E_{\ell,k}$ groups, and the free-to-interacting maps match, so we may proceed like in \cref{BDI_DIII}. For discrete class C, corresponding to $\Spin\times_{\set{\pm 1}}Q_8$-structure, \cref{class_C_prime} implies that if $(\ell,k)\ne(0,0)$, $\mathit{MT}(Spin\times_{\set{\pm 1}}Q_8)\wedge\ME_{\ell,k}\simeq\MTSpin\wedge\ME_{\ell+3,k-1}$ (see \cref{using_Q8}), taking care of the computation of the bordism groups and the maps $\mathrm{sp}^\phi$ and $\mathrm{sp}^\psi$. There is then the question of the free-to-interacting map, which we solve in \cref{disc_F2I_compute}.} Thus, in this section, we are considering Altland--Zirnbauer types which are ``mixed discrete-continuous:'' we start with a continuous fermionic group, and then in the process of running the Bott spiral take the fermionic tensor product with the discrete symmetry groups $E_{\ell,k}$. 

We study these mixed discrete-continuous spiral maps in order to better understand how to mathematically model the Bott spiral. We were surprised to discover that, even though the behavior of the free-to-interacting map and both spiral maps is qualitatively very similar to the purely discrete maps we studied in the previous subsection, assembling them into a mixed continuous-discrete Bott spiral does not work very well. We discuss this further in \cref{mixed_DC_spiral}.

In all cases, the key ingredient in our Bott spiral calculations is an Anderson--Brown--Peterson-style splitting generalizing \cref{ABP_thm}: \cref{thm:spinh_ABP} for $\MTSpin^h$, \cref{thm:pin_tilde_c_ABP} for $\MTPin^{\tilde c\pm}$, and \cref{thm:pinh_ABP} for $\MTPin^{h\pm}$. The \spinh splitting is a recent result of Buchanan--McKean~\cite{BM23} and Mills~\cite{Mil24}; the others are new to our knowledge. All of these splittings are strictly finer than what one obtains by combining~\eqref{ABP_smash_X}, Freed--Hopkins' shearing equivalences~\cite[(10.20)]{freed_reflection_2021}, and the stable splittings of $(B\O_2)^{\pm(V-2)}$ and $(B\O_3)^{\pm(V-3)}$ constructed by Mitchell--Priddy~\cite{MP83, Priddy} and Kashiwabara--Zare~\cite[Theorem 1]{kashiwabara_splitting_2023}.

Once we have these ABP-style splittings, the main theorems in this section are \cref{gen_AZ_f2i} on the image of the F2I map, \cref{remaining_real_int_Bott_thm} computing the spiral map $\spint_{\ell,k}(\mathit{MTH})$  modulo Whitney summands (where $H$ is one of $\Spin^h$, $\Pin^{\tilde c\pm}$, or $\Pin^{h\pm}$), and \cref{MSpinHs_psi} computing the spiral map $\mathrm{sp}^\psi_{\ell,k}(\mathit{MTH})$ on long summands. We also finish the computation of $\mathrm{sp}^\psi_{\ell,0}$, for $\ell\ge 3$, in \cref{special_case_psi_ko}.
The proofs of these theorems are similar to those of \cref{only_first_summand_real,int_Bott_spiral_1}. Together with the work of the previous sections, these theorems account for all eight of the continuous fermionic groups over the real numbers.

\begin{thm}[{Buchanan--McKean~\cite[Theorem 7.2]{BM23}, Mills~\cite[Theorem 1.1]{Mil24}}]
\label{thm:spinh_ABP}
Let $\widetilde{\ko}$ denote the fiber of the algebra map $\ko\to H\Z/2$ and $\ksp\coloneqq\Sigma^{-4}(\tau_{\ge 4}\ko)$, so that $\ksp$ is the connective cover of the quaternionic $K$-theory spectrum $\mathit{KSp}$. Then there are numbers $d_i,e_j,f_k\in\N$ and a $2$-local equivalence of homology $\ko$-modules
\begin{equation}
\label{spinh_ABP}
    \MTSpin^h\overset\simeq\longrightarrow  
        \bigvee_{i\ge 1} \textcolor{Rhodamine}{\Sigma^{d_i}\ksp} \vee
        \bigvee_{j\ge 1} \textcolor{Orange}{\Sigma^{e_j}\widetilde{\ko}} \vee
        \bigvee_{k\ge 1} \textcolor{Cyan}{\Sigma^{f_k} H\Z/2},
\end{equation}
where $d_1 = 1$.
Moreover, the composition
\begin{equation}
    \MTSpin^h\xrightarrow{\mathrm{proj}_{\textcolor{Rhodamine}{d_1}}} \ksp\xrightarrow{\tau_{\ge 0}} \Sigma^4\KO
\end{equation}
is equivalent as $\MTSpin$-module maps to Freed--Hopkins' twisted ABS map~\cite[\S 9.2.2]{freed_reflection_2021}.
\end{thm}
The homology $\ko$-module structures on the spectra appearing in \cref{thm:spinh_ABP} are as defined in \cref{making_htpy_ko}. The fact that~\eqref{spinh_ABP} is a homology $\ko$-module splitting is not discussed in \textit{loc.\ cit.}, but follows from Mills' proof~\cite[\S 3]{Mil24}, which uses a theorem of Stolz~\cite{Sto94} which produces splittings of homology $\ko$-modules. %
The comparison between $\mathrm{proj}_{\textcolor{Rhodamine}{d_1}}$ and the twisted Atiyah--Bott--Shapiro map is implicit in~\cite[\S 5.5]{BM23} via the definition of their $\mathit{KSp}$-Pontrjagin classes.

\begin{lem}\hfill
\label{dissect_Q_R2}
\begin{enumerate}
    \item\label{ksp_tensor_factors} There is a finite, connective spectrum $Y_1$ and a $\ko$-module equivalence $\ksp\simeq \ko\wedge Y_1$.
    \item\label{ksp_Q} Let $\uQ\coloneqq\cA(1)/(\Sq^1, \Sq^2\Sq^3)$. Then $\uQ$ is stably invertible and there is an $\cA(1)$-module isomorphism $H_\ko^*(\ksp)\cong\uQ$.
    \item\label{F_tensor_factors} There is a finite, connective spectrum $Y_2$ and an equivalence of $\ko$-modules $\widetilde{\ko}\simeq\ko\wedge Y_2$.
    \item\label{F_R2} Let $R_2$ be the $\cA(1)$-module defined in \cref{defn_R2}. Then $R_2$ is stably invertible and there is a $\cA(1)$-module isomorphism $H_\ko^*(\widetilde\ko)\simeq R_2$.
\end{enumerate}
\end{lem}
Factoring a $\ko$-module $M$ as $\ko\wedge X$ for a spectrum $X$ is far from automatic: for example, it is impossible for $M = H\Z$~\cite[\S 5]{Bak20}.
\begin{proof}
Item~\eqref{ksp_tensor_factors} is proven in~\cite[Theorem 1.5]{MM76}, where it is attributed to Anderson~\cite{And64a}, and Stolz~\cite[\S 8]{Sto94} proves~\eqref{F_tensor_factors}. Except for stable invertibility, which was proven for these $\cA(1)$-modules by Adams--Priddy~\cite[\S 3]{AP76}, items~\eqref{ksp_Q} and~\eqref{F_R2} are implicit in~\cite[\S 5]{Bak20}.
\end{proof}
The remaining four ABP-style splittings we need in this subsection have not been previously shown to our knowledge.

\begin{thm}[\cite{ABP_splittings}]
\label{thm:pin_tilde_c_ABP}
There are numbers $g_i^\pm,h_j^\pm,m_k^\pm\in\N$ and $2$-local equivalences of homology $\ko$-modules
\begin{equation}\label{eqn:pin_tilde_c_ABP}
    \MTPin^{\tilde c\pm} \overset\simeq\longrightarrow
        \bigvee_{i\ge 1} \textcolor{Rhodamine}{\Sigma^{g_i^\pm}\ko} \vee
        \bigvee_{j\ge 1} \textcolor{Orange}{\Sigma^{h_j^\pm}\tau_{\ge 2}\ko} \vee
        \bigvee_{k\ge 1} \textcolor{Cyan}{\Sigma^{m_k^\pm} H\Z/2},
\end{equation}
in which $\textcolor{Rhodamine}{g_1^+} = 2$ and $\textcolor{Orange}{h_1^-} = -2$. The splittings~\eqref{eqn:pin_tilde_c_ABP} may be chosen so that the compositions
\begin{subequations}\label{pinc_pm_ABP_splitting}
    \begin{align}
        \MTPin^{\tilde c+} &\xrightarrow{\mathrm{proj}_{\textcolor{Rhodamine}{g_1^+}}} \Sigma^2\ko \xrightarrow{\tau_{\ge 0}} \Sigma^2 \KO\\
        \MTPin^{\tilde c-} & \xrightarrow{\mathrm{proj}_{\textcolor{Orange}{h_1^-}}} \Sigma^{-2}\tau_{\ge 2}\ko \xrightarrow{\tau_{\ge 0}} \Sigma^{-2} \KO
    \end{align}
\end{subequations}
are $\MTSpin$-module equivalent to Freed--Hopkins' twisted ABS maps for $s = \mp 2$~\cite[\S 9.2.2]{freed_reflection_2021}.
\end{thm}
These splittings are finer than the ones induced by smashing $\MTSpin$ with the splittings of $(B\O_2)^{\pm(V-2)}$ constructed by Mitchell--Priddy~\cite{MP83, Priddy} and Kashiwabara--Zare~\cite[Theorem 1]{kashiwabara_splitting_2023}.
\begin{thm}[\cite{ABP_splittings}]\label{thm:pinh_ABP}
Let $M_2$ and $M_3$ be the $\ko$-module spectra we defined in \cref{N2_exm,N3_exm}. Then there are numbers $p_i^\pm,q_j^\pm, r_k^\pm$, with $p_1^+ = 0$ and $q_1^-=2$, and splittings of homology $\ko$-modules
\begin{subequations}
\label{eqn:pinh_ABP}
    \begin{align}
        \MTPin^{h+} &\overset\simeq\longrightarrow
            \bigvee_{i\ge 1} \textcolor{Orange}{\Sigma^{p_i^+} M_3}\vee
            \bigvee_{j\ge 1} \textcolor{Rhodamine}{\Sigma^{q_j^+}\ko\wedge (B\Z/2)^{\sigma-1}}\vee
            \bigvee_{k\ge 1} \textcolor{Cyan}{\Sigma^{r_k^+}H\Z/2}\\
        \MTPin^{h-} &\overset\simeq\longrightarrow
            \bigvee_{i\ge 1} \textcolor{Orange}{\Sigma^{p_i^-} M_2}\vee
            \bigvee_{j\ge 1} \textcolor{Rhodamine}{\Sigma^{q_j^-}\ko\wedge (B\Z/2)^{1-\sigma}}\vee
            \bigvee_{k\ge 1} \textcolor{Cyan}{\Sigma^{r_k^-}H\Z/2}.
    \end{align}
\end{subequations}
\end{thm}
These splittings are also finer than the ones induced by smashing $\MTSpin$ with the splittings of $(B\O_3)^{\pm(V-3)}$ constructed by Mitchell--Priddy~\cite[Theorem C]{MP83} and Kashiwabara--Zare~\cite[Theorem 1]{kashiwabara_splitting_2023}.

We will also identify how Freed--Hopkins' Atiyah--Bott--Shapiro maps $\MTPin^{h\pm}\to\Sigma^{\mp 3}\KO$ behave with respect to the decompositions in \cref{thm:pinh_ABP}. First, recall from \cref{ABS_through_ko} that there are maps $\mathit{abs}_{\mp 1}\colon \ko\wedge (B\Z/2)^{\pm(1-\sigma)}\to\Sigma^{\pm 1}\KO$ such that the following diagram commutes.
\begin{equation}
\begin{tikzcd}[column sep = 4em]
	{\MTPin^{\pm}} & {\ko\wedge (B\Z/2)^{\pm(1-\sigma)}} & {\Sigma^{\pm 1}\KO.}
	\arrow["{\text{--}\wedge_{\MTSpin}\ko}", from=1-1, to=1-2]
	\arrow["{\mathit{ABS}_{\mp 1}~(\ref{1_lam_defn},\,\ref{-1_lam_defn})}"', curve={height=18pt}, from=1-1, to=1-3]
	\arrow["{\mathit{abs}_{\mp 1}}", from=1-2, to=1-3]
\end{tikzcd}
\end{equation}
\begin{thm}[\cite{ABP_splittings}]
\label{thm:pinh_ABP_2}
\hfill
\begin{enumerate}
    \item There are maps of $\ko$-modules $a_3\colon\Sigma M_3\to M_2$ and $a_2\colon \Sigma M_2\to \ko\wedge (B\Z/2)^{\sigma-1}$, which are Adams covers in the sense of~\cite[\S 4]{Bru99}, and which are isomorphisms on short summands and injective, with cokernel $\Z/2$, on long summands.
    \item The splittings in~\eqref{eqn:pinh_ABP} may be chosen so that the compositions
    \begin{subequations}
    \begin{align}
        \label{pinhp_twABS}
        \MTPin^{h+} &\xrightarrow{\mathrm{proj}_{\textcolor{Orange}{p_1^+}}} \textcolor{Orange}{M_3} \xrightarrow{a_3} \Sigma^{-1}M_2 \xrightarrow{a_2} \Sigma^{-2}\ko\wedge (B\Z/2)^{\sigma-1} \xrightarrow{\mathit{abs}_{+1}} \Sigma^{-3}\KO\\
        \MTPin^{h-} &\xrightarrow{{\mathrm{proj}_{\textcolor{Rhodamine}{q_1^-}}}} \textcolor{Rhodamine}{\Sigma^2 \ko\wedge (B\Z/2)^{1-\sigma}} \xrightarrow{\mathit{abs}_{-1}} \Sigma^3\KO
    \end{align}
    \end{subequations}
    are $\MTSpin$-module equivalent to Freed--Hopkins' twisted ABS maps for $s = \mp 3$~\cite[\S 9.2.2]{freed_reflection_2021}.
\end{enumerate}
\end{thm}
\begin{rem}[No obvious shortcut in classes CI and CII]
By \cite[(3.11)]{DK24}, $\MTPin^{h+}\simeq\MTSpin^h\wedge\ME_{0,1}$ and $\MTPin^{h-}\simeq\MTSpin^h\wedge\ME_{1,0}$. This suggests that we could understand the Bott spirals starting with pin\textsuperscript{$h\pm$} in terms of the spiral starting with \spinh, parallel to \cref{BDI_DIII}, but it is not obvious that the free-to-interacting maps match (see \cite[Figure 5, cases $s = \pm 3$ and $s = 4$]{freed_reflection_2021}), and indeed we had to give a more complicated description in \cref{thm:pinh_ABP_2}.
\end{rem}
\begin{defn}
\label{cts_AZ_f2i}
For $s = \pm 2$, $\pm 3$, and $4$, let $K_f(s)$ be the continuous fermionic group that we associated to the parameter $s$ in \cref{tenfold_table} and $\mathit{ABS}_s\colon\mathit{MTH}(K_f(s))\to\Sigma^{-s}\KO$ be Freed--Hopkins' twisted ABS map~\cite[(9.43)]{freed_reflection_2021}. Let $\mathit{ABS}_{\ell,k}$ be the twisted ABS map defined in \cref{defn:gen_ABS}.
Then define $\mathit{ABS}_{\ell,k,s}$ to be the composition
\begin{equation}\label{lks_ABS}
\begin{tikzcd}[column sep=1em]
	{\mathit{ABS}_{\ell,k,s}\colon \textcolor{BrickRed}{\mathit{MTH}(K_f(s))}\wedge \textcolor{MidnightBlue}{\ME_{\ell,k}} } & {\textcolor{BrickRed}{\mathit{MTH}(K_f(s))}\wedge_{\MTSpin} \textcolor{MidnightBlue}{\MTSpin\wedge \ME_{\ell,k}}} \\
	& { \textcolor{BrickRed}{\Sigma^{-s}\KO} \wedge_{\MTSpin} \textcolor{MidnightBlue}{\Sigma^{k-\ell}\KO}} \\
	& {\Sigma^{\textcolor{MidnightBlue}{k-\ell}-\textcolor{BrickRed}{s}}\textcolor{Fuchsia}{\KO}}
	\arrow["\simeq", from=1-1, to=1-2]
	\arrow["{\textcolor{BrickRed}{\mathit{ABS}_s} \wedge \textcolor{MidnightBlue}{\mathit{ABS}_{\ell,k}}}", from=1-2, to=2-2]
	\arrow["\textcolor{Fuchsia}{\mu}", from=2-2, to=3-2]
\end{tikzcd}
\end{equation}
where $\mu$ denotes multiplication. Taking the Anderson dual of~\eqref{lks_ABS}, we obtain a free-to-interacting map
\begin{equation}
\label{eqn:mixed_F2I}
    \mathit{F2I}_{\ell,k,s} \colon \Sigma^{\ell-k+s-2}\KO\to \Sigma^2 I_\Z(\mathit{MTH}(K_f(s))\wedge\ME_{\ell,k}).
\end{equation}
\end{defn}

Recall the modules $N_i$, $i\in\{0,1,2,3\}$, from \S\ref{ss:EA_type}. For $i\ge 4$, define $N_i\coloneqq\Sigma^4 N_{i-4}$.
\begin{lem}
\label{oops_all_EA}\hfill
\begin{enumerate}
    \item\label{tensor_MElk} Let $\ell,k\ge 0$ such that $(\ell,k)\ne(0,0)$.
Let $i \coloneqq\ell\bmod 4$, thought of in the set $\{0,1,2,3\}$ and $j\coloneqq\lfloor \ell/4\rfloor$. Then there are stable equivalences of $\cA(1)$-modules
\begin{subequations}\label{QJ_ME}
    \begin{align}
        \label{ksp_lk_coh}
        H_\ko^*(\ksp\wedge\ME_{\ell,k}) &\cong \uQ\otimes H_\ko^*(\ME_{\ell,k}) \simeq\Sigma^{4j+k-1} N_{i+3}\\
        \label{joker_lk_coh}
        H_\ko^*(\Sigma^{-2}\tau_{\ge 2}\ko\wedge\ME_{\ell,k}) &\cong J\otimes H_\ko^*(\ME_{\ell,k}) \simeq \Sigma^{4j+k}N_{i+2}.
    \end{align}
\end{subequations}
\item\label{lots_EA} If $M$ is one of $\ko$, $\tau_{\ge 2}\ko$, $\ksp$, $\widetilde{\ko}$, $\ko\wedge (B\Z/2)^{\pm (1-\sigma)}$, $M_2$ (see \cref{N2_exm}), or $M_3$ (see \cref{N3_exm}), or a suspension of one of these, then $M\wedge\ME_{\ell,k}$ is of EA type.
\end{enumerate}
\end{lem}
\begin{proof}
The first isomorphisms in~\eqref{QJ_ME} are from the Künneth formula; for the stable equivalences, since $\uQ$ and $J$ are stably invertible (see \cref{dissect_Q_R2,joker_stabinv}), \cref{yu_stable_iso} immediately reduces these equivalences to the assertions
\begin{equation}
\label{N0_QJ}
    \uQ\otimes N_0\cong \Sigma^{-1}N_3\qquad\qquad\text{and}\qquad\qquad J\otimes N_0\cong N_2.
\end{equation}
Since $\uQ$ and $J$ are stably invertible, then by \cref{tensor_thru_Q1loc}, $\uQ\otimes N_0$ and $J\otimes N_0$ are $Q_1$-locally stably invertible. By \cref{Q1_Pic}, this implies these tensor products are stably equivalent to a shift of some $N_{i'}$, so we can prove~\eqref{N0_QJ} by just calculating these tensor products in low degrees until the shift and $i'$ are uniquely determined.

On to part~\eqref{lots_EA}. Since $H_\ko^*(\tau_{\ge 2}\ko)\cong\Sigma^2 J$ and $H_\ko^*(\ksp)\cong\uQ$, part~\eqref{tensor_MElk} implies part~\eqref{lots_EA} for these two choices of $M$.  We already showed the case of $\ko$ just before \cref{MSpin_psi}. By \cref{dissect_Q_R2}, $H_\ko^*(\widetilde{\ko})$ is stably invertible, so we can use \cref{stab_inv_EA}. Finally, as we saw in \cref{N1_exm,,N2_exm,,N3_exm,,N0_exm}, $H_\ko^*$ of $\ko\wedge (B\Z/2)^{\pm(\sigma-1)}$, $M_2$, and $M_3$ are isomorphic to the $\cA(1)$-modules $N_i$ for different $i$; by \cref{Q1_Pic}, these $\cA(1)$-modules are $Q_1$-locally stably invertible, so \cref{stab_inv_EA} implies that, as we wanted, smashing $\ko\wedge (B\Z/2)^{\pm(\sigma-1)}$, $M_2$, and $M_3$ with $\ME_{\ell,k}$ yields a $\ko$-module of EA-type.
\end{proof}
As a consequence of \cref{oops_all_EA} and our generalized ABP splittings in \cref{thm:spinh_ABP,thm:pin_tilde_c_ABP,thm:pinh_ABP}, we have long, short, and Whitney summands in the homotopy groups of $I_\Z(\mathit{MTH}(K_f(s))\wedge\ME_{\ell,k})$. This will be helpful in our descriptions of $\mathit{F2I}_{\ell,k,s}$ and $\mathrm{sp}_{\ell,k}^\psi(\mathit{MTH}(K_f(s))$ in \cref{gen_AZ_f2i,MSpinHs_psi} respectively.
\begin{defn}
\label{variant_m}
Given $\ell,k,n\in\Z$, let $j\coloneqq \lfloor\ell/4\rfloor$ and $\widetilde n\coloneqq n - 4j - k$. For $s\in\{-3,-2,0,2,3\}$, define $\widetilde m(s)\coloneqq \lfloor(\widetilde n+s)/8\rfloor$. Also, let $\widetilde m(4)\coloneqq\widetilde m(0)$.
\end{defn}
The implicit dependence of $\widetilde m(s)$ on $\widetilde n$ is an abuse of notation, but in the places $\widetilde m(s)$ will appear, $\widetilde n$ will be clear from context. Recall the notation $\ldeg(\ell,k)$ from \cref{ldeg_defn}.
\begin{prop}[Locations of long summands]
\label{C_AI_etc_longs}
Let $\ell$, $k$, $i$, $j$, $n$, and $\widetilde n$ be as in \cref{melk_calc}, and $\widetilde m(s)$ be as in \cref{variant_m}.
\begin{enumerate}
    \setcounter{enumi}{1}
    \item %
    $\Sigma^{-2}\tau_{\ge 2}\ko\wedge\ME_{\ell,k}$ and $\MTPin^{\tilde c-}\wedge\ME_{\ell,k}$ have the following long summands.
    \begin{itemize}
        \item A $\Z/2^{4\widetilde m(2) + 2-i}$ in degree $n$ if $\widetilde n+i\equiv 1\bmod 8$ and $n\ge\ldeg(\ell+2, k)$.%
        \item A $\Z/2^{4\widetilde m(2) + 3-i}$ in degree $n$ if $\widetilde n+i\equiv 5\bmod 8$ and $n\ge\ldeg(\ell+2, k)$.%
    \end{itemize}
    \setcounter{enumi}{-3}
    \item %
    $\Sigma^2\ko\wedge\ME_{\ell,k}$ and $\MTPin^{\tilde c+}\wedge\ME_{\ell,k}$ have the following long summands.
    \begin{itemize}
        \item A $\Z/2^{4\widetilde m(-2) + 4-i}$ in degree $n$ if $\widetilde n+i\equiv 5\bmod 8$ and $n\ge\ldeg(\ell, k)+2$.
        \item A $\Z/2^{4\widetilde m(-2) + 5-i}$ in degree $n$ if $\widetilde n+i\equiv 1\bmod 8$ and $n\ge\ldeg(\ell, k)+2$.
    \end{itemize}
    \setcounter{enumi}{2}
    \item %
     $M_3\wedge\ME_{\ell,k}$ and $\MTPin^{h+}\wedge\ME_{\ell,k}$ have the following long summands.
    \begin{itemize}
        \item A $\Z/2^{4\widetilde m(3) + 1-i}$ in degree $n$ if $\widetilde n+i\equiv 0\bmod 8$ and $n\ge\ldeg(\ell+3, k)$.
        \item A $\Z/2^{4\widetilde m(3) + 2-i}$ in degree $n$ if $\widetilde n+i\equiv 4\bmod 8$ and $n\ge\ldeg(\ell+3, k)$.
    \end{itemize}
    \setcounter{enumi}{-4}
    \item %
     $\Sigma^2\ko\wedge (B\Z/2)^{1-\sigma}\wedge \ME_{\ell,k}$ and $\MTPin^{h-}\wedge\ME_{\ell,k}$ have the following long summands.
    \begin{itemize}
        \item A $\Z/2^{4\widetilde m(-3) + 4-i}$ in degree $n$ if $\widetilde n+i\equiv 6\bmod 8$ and $n\ge\ldeg(\ell, k)+3$.
        \item A $\Z/2^{4\widetilde m(-3) + 5-i}$ in degree $n$ if $\widetilde n+i\equiv 2\bmod 8$ and $n\ge\ldeg(\ell, k)+3$.
    \end{itemize}
    \setcounter{enumi}{3}
    \item %
     $\ksp\wedge \ME_{\ell,k}$ and $\MTSpin^{h}\wedge\ME_{\ell,k}$ have the following long summands.
    \begin{itemize}
        \item A $\Z/2^{4\widetilde m(4) + 5-i}$ in degree $n$ if $\widetilde n+i\equiv 7\bmod 8$ and $n\ge\ldeg(\ell+3, k)-1$.
        \item A $\Z/2^{4\widetilde m(4) + 2-i}$ in degree $n$ if $\widetilde n+i\equiv 3\bmod 8$ and $n\ge\ldeg(\ell+3, k)-1$.
    \end{itemize}
\end{enumerate}
\end{prop}
\begin{proof}
By \cref{thm:spinh_ABP,thm:pin_tilde_c_ABP,thm:pinh_ABP}, in each of these five cases, the first spectrum $\mathcal M_1$ that appears is a homology $\ko$-module summand of the second (e.g.\ $\mathcal M_1 = \Sigma^{-2}\tau_{\ge 2}\ko$ is a summand of $\MTPin^{\tilde c-}$, $\mathcal M_1 = \ksp$ is a summand of $\MTSpin^h$, etc.), so it suffices to prove the proposition for the first spectrum in each of the five cases. By \cref{oops_all_EA}, part~\eqref{lots_EA}, in each of these cases, $\mathcal M_1\wedge\ME_{\ell,k}$ is of EA-type, so we can prove the proposition in the same manner as \cref{melk_calc} as soon as we know $H_\ko^*(\mathcal M_1\wedge\ME_{\ell,k})$. For $\mathcal M_1 = \Sigma^{-2}\tau_{\ge 2}\ko$ and $\ksp$, we did this in \cref{oops_all_EA}, part~\eqref{tensor_MElk}; for $\mathcal M_1 = \Sigma^2\ko$, use that $H_\ko^*(\Sigma^t\ko)\cong \Sigma^t\Z/2$ by definition; and for $\mathcal M_1 = \Sigma^2\ko\wedge (B\Z/2)^{1-\sigma}$ and $M_3$, use \cref{N3_exm,N0_exm,yu_stable_iso}.
\end{proof}
As a corollary, we get another characterization of the long summands that we are interested in.
\begin{prop}
\label{only_one_long_summand}
In $\pi_n(\MTSpin\wedge\ME_{\ell,k})$, hence also in $\mho^n_{\ell,k}$, exactly one of the following is true.
\begin{enumerate}
    \item $n\equiv \ell+k-2i-1\bmod 4$ and $n\ge\ldeg(\ell,k)$. In this case, $\pi_n(\MTSpin\wedge\ME_{\ell,k})$ is the direct sum of the long summand $L$ from the $\textcolor{Rhodamine}{\Sigma^0\ko}\wedge\ME_{\ell,k}$ piece of the ABP decomposition~\eqref{ABP_smash_X} and cyclic groups of orders at most that of $L$, and strictly smaller order than $L$ if $\abs L>2$.
    \item $n\not\equiv\ell+k-2i-1\bmod 4$ or $n<\ldeg(\ell,k)$, and $\pi_n(\MTSpin\wedge\ME_{\ell,k})$ is simple $2$-torsion.
\end{enumerate}
\end{prop}
\begin{proof}
\Cref{ABP_thm} gives a direct-sum decomposition of the spectrum $\MTSpin\wedge\ME_{\ell,k}$, and \cref{melk_calc,C_AI_etc_longs} compute the homotopy groups of the summands modulo simple $2$-torsion, implying the result.
\end{proof}
Using the twisted ABP splittings in \cref{thm:spinh_ABP,thm:pin_tilde_c_ABP,thm:pinh_ABP}, the reader can provide analogs of \cref{only_one_long_summand} in which $\MTSpin$ is replaced with $\mathit{MTH}(K_f(s))$ for the other continuous Altland--Zirnbauer groups $K_f(s)$.
\begin{thm}
\label{gen_AZ_f2i}
Let $s = \pm 2$, $\pm 3$, or $4$, and $\ell,k\ge 0$ but not both equal to $0$. The image of the free-to-interacting map $\mathit{F2I}_{\ell,k,s}$ of \cref{cts_AZ_f2i} consists of the short and long summands of the summand of $\Sigma I_\Z(\mathit{MTH}(K_f(s))\wedge\ME_{\ell,k}$ dual to the unique lowest-degree summands in \cref{thm:spinh_ABP,thm:pin_tilde_c_ABP,thm:pinh_ABP}, i.e.\ $\Sigma^{-2}\tau_{\ge 2}\ko$, $\Sigma^2\ko$, $M_3$, $\Sigma^2\ko\wedge (B\Z/2)^{1-\sigma}$, and $\ksp$
for $s = 2$, $-2$, $3$, $-3$, and $4$ respectively. Specifically, in spatial dimension $d\equiv k-\ell-s+2\bmod 4$, $\mathit{F2I}_{\ell,k,s}$ maps $\Z\cong\KO^{d+\ell-k+s-2}$ surjectively onto this long summand in $I_\Z^{d+2}(\mathit{MTH}(K_f(s))\wedge\ME_{\ell,k})$. The isomorphism type of this summand is the same as the corresponding long summand in \cref{C_AI_etc_longs}, where $n = d+1$.
\end{thm}
The lower bounds on $\widetilde n$ in \cref{C_AI_etc_longs} appear implicitly in \cref{gen_AZ_f2i} as follows: if $d$, $\ell$, and $k$ are such that $\widetilde n$ does not reach this lower bound, the image of the free-to-interacting map in that degree is $0$.
\begin{proof}
For $s = 2$, $-2$, and $4$, \cref{thm:spinh_ABP,thm:pin_tilde_c_ABP} identify Freed--Hopkins' twisted ABP maps with projection onto the summand $\Sigma^2\ko$, $\Sigma^{-2}\tau_{\ge 2}\ko$, resp.\ $\ksp$, followed by a connective cover map to the requisite shift of $\KO$, and so an analogous dual statement is true for these free-to-interacting maps. Then we smash with the free-to-interacting map for $\ME_{\ell,k}$. Therefore in the degrees that these free-to-interacting maps $\mathit{F2I}_{\ell,k,s}$ are nontrivial, they are equivalent to shifts of $I_\Z$ of the connective cover $\ko\to\KO$ (the Anderson dual of the connective cover is an isomorphism in nonnegative cohomological degrees) followed by shifts of the map $\mathit{f2i}_{\ell,k}$ from \cref{IZ_ABS_through_ko}, and we computed the latter in \cref{compute_F2I}.

The case $s = -3$ is scarcely more difficult: by \cref{thm:pinh_ABP,thm:pinh_ABP_2}, we can similarly identify $\mathit{F2I}_{\ell,k}$ as $I_\Z(\ko\to\KO)$ followed by the smash product of $I_\Z\mathit{abs}_{-1}$ and $\mathit{f2i}_{\ell,k}$. Ultimately by \cref{rem:same_map}, this smash product is equivalent to $\mathit{f2i}_{\ell,k+1}$, so we have once again reduced to \cref{compute_F2I}.

Finally $s = 3$. By \cref{thm:pinh_ABP_2}, this free-to-interacting map factors as $I_\Z\mathit{abs}_{1}$ followed by the Anderson duals of $a_2$ and $a_3$, then the inclusion of the direct summand claimed in the current theorem statement. If not for $a_2$ and $a_3$, the proof would be identical to that of $s = -3$, except that we would get $\mathit{f2i}_{\ell+1,k}$ instead of $\mathit{f2i}_{\ell,k+1}$. But by \cref{thm:pinh_ABP_2}, since the maps $a_2$ and $a_3$ are injective on long summands, their Anderson duals are surjective on long summands, so their presence in the decomposition~\eqref{pinhp_twABS} of $\mathit{F2I}_{\ell,k,3}$ does not change the proof or the final outcome.
\end{proof}

\begin{thm}
\label{remaining_real_int_Bott_thm}
With $s$ and $K_f(s)$ as in \cref{cts_AZ_f2i}, the
spiral map
\begin{equation}
    \spint_{\ell,k}(\mathit{MTH}(K_f(s)))\colon I_\Z(\mathit{MTH}(K_f(s))\wedge\ME_{\ell,k})\to \Sigma I_\Z(\mathit{MTH}(K_f(s))\wedge\ME_{\ell,k+1})
\end{equation}
is an isomorphism on homotopy, modulo Whitney summands.
\end{thm}
\begin{proof}
The proof is essentially the same as that of \cref{int_Bott_spiral_1}: consider the map of Atiyah--Hirzebruch spectral sequences induced by $\spint_{\ell,k}$ on each summand of the Anderson--Brown--Peterson decomposition of $\mathit{MTH}(K_f(s))$ proven in \cref{thm:spinh_ABP,thm:pin_tilde_c_ABP,thm:pinh_ABP} (for $s = 4$, $\pm 2$, $\pm 3$, respectively), observe that the $E_2$-page is an isomorphism modulo Whitney summands, and therefore the same is true for the $E_\infty$-page.

To make this argument, though, we need to show that each summand appearing in these ABP decompositions factors as $\ko\wedge X$, so that we can use the Atiyah--Hirzebruch spectral sequence.
\begin{enumerate}
    \item For $\ko$ this is tautological. $H\Z/2$ is always a Whitney summand, so we ignore it.
    \item For $\tau_{\ge 2}\ko$, we did this in \cref{dissect_joker}.
    \item $\ko\wedge (B\Z/2)^{\pm(\sigma-1)}$ is already in this form.
    \item We factored $\ksp$ and $\widetilde{\ko}$ in this way in \cref{dissect_Q_R2}.
    \item We do not realize $M_2$ and $M_3$ in this way, but \cref{N2_exm} proves $\ko\wedge (BA_4)\simeq \Sigma^2 M_2\vee B$, where $B$ consists of Whitney summands. Thus we apply the first paragraph to show that $\spint_{\ell,k}$ applied to $\ko\wedge BA_4$ is an isomorphism modulo Whitney summands, so the same is true for $M_2$. \Cref{N3_exm} shows the same for $M_3$, with $BA_4\wedge B\Z/2$ in place of $BA_4$.
    \qedhere
\end{enumerate}
\end{proof}
\begin{lem}
\label{ko_mods_psi_piece}
\hfill
\begin{enumerate}
    \item\label{smash_EA} Let $M$ be a $\ko$-module and $\ell,k$ be such that $M\wedge\ME_{\ell,k}$ is of EA-type. Then $\mathrm{sp}_{\ell,k}^\psi\colon I_\Z(M\wedge\ME_{\ell,k})\to \Sigma I_\Z(M\wedge\ME_{\ell,k+1})$ is an isomorphism on homotopy, modulo Whitney summands.
    \item\label{whichtens__} The above holds when $M\in\set{\ko, \tau_{\ge 2}\ko, \ksp, \widetilde{\ko}, \ko\wedge(B\Z/2)^{\pm(\sigma-1)}, M_2, M_3}$ and $\ell,k>0$. 
\end{enumerate}
\end{lem}
\begin{proof}
The proof of part~\eqref{smash_EA} is completely analogous to that of \cref{MSpin_psi}. Thus part~\eqref{whichtens__} follows as soon as we know that, for these $M$, $M\wedge \ME_{\ell,k}$ is of EA-type, which we proved in \cref{oops_all_EA}.
\end{proof}
\begin{thm}
\label{special_case_psi_ko}
For all $\ell\ge 3$, $\mathrm{sp}_{\ell,0}^\psi\colon I_\Z(\MTSpin\wedge\ME_{\ell,0})\to \Sigma I_\Z(\MTSpin\wedge\ME_{\ell,1})$ is an isomorphism on homotopy, modulo Whitney summands.
\end{thm}
\begin{proof}
By \cref{MSpin_psi}, the corollary is true if $k>0$, so for the rest of the proof we assume $k = 0$, meaning $\ell \ge 3$ and we are in the situation of \cref{Q8_psi}. Looking at the definition of $\overline\psi_{\ell,0}$ there, as the identity map on $\mathit{MTH}(Q_8)$ smashed with $\psi_{\ell-3,1}$, up to $\MTSpin$-module equivalence, we see that it suffices to show that $\mathrm{sp}_{\ell-3,1}^\psi(\mathit{MTH}(Q_8))$ (\cref{spiral_1_map_E_version}) is a $\pi_*$-isomorphism modulo Whitney summands. This uses the more generic definition of $\mathrm{sp}^\psi$ (\cref{spiral_maps_of_spectra}).

In \cref{spin_Q8_bordism}, we will show that, as a homology $\ko$-module, $\mathit{MTH}(Q_8)$ splits as a sum of shifts of $\ksp$, $\widetilde{\ko}$, and $H\Z/2$. (That proof will not use any facts about the maps $\mathrm{sp}^\psi_{\ell,k}$.) The $H\Z/2$ summands are Whitney summands, so we ignore them. Thus to prove the corollary, it suffices to show that $\mathrm{sp}_{\ell-3,1}^\psi(M)$ is a $\pi_*$-isomorphism modulo Whitney summands for $M \in\set{\ksp,\widetilde{\ko}}$, which we did in \cref{ko_mods_psi_piece}, part~\eqref{whichtens__}.
\end{proof}
\begin{defn}\label{MQ_defn}
For conciseness, let
\begin{equation}
\mathit{MQ}_{s,\ell,k}\coloneqq \mathit{MTH}(K_f(s))\wedge_{\MTSpin} \mathit{MTH}(Q_8)\wedge\ME_{\ell,k}.
\end{equation}
We give $\mathit{MQ}_{s,\ell,k}$ the homology $\ko$-module structure induced from its $\MTSpin$-module structure (see \cref{making_htpy_ko}).
\end{defn}
\begin{lem}
\label{MQ_break_down}
For $s \in\set{\pm 2, \pm 3, 4}$ and $(\ell,k)\ne(0,0)$, $\mathit{MQ}_{s,\ell,k}$ is a sum of shifts of $H\Z/2$ and $\ko$-modules of EA-type. 
\end{lem}
This proof uses a few facts about $\mathit{MTH}(Q_8)$ that we prove later, in \cref{spinQ8}. Those facts do not depend on anything we do in this or the next subsection.
\begin{proof}
In \cref{shearQ8}, we construct a real vector bundle $V\to B(\Z/2\times \Z/2)$ and show there is an $\MTSpin$-module equivalence $\mathit{MTH}(Q_8)\simeq \MTSpin\wedge (B\Z/2\times B\Z/2)^{V-3}$. Therefore there are $\MTSpin$-module equivalences
\begin{equation}\label{complicated_decomp}
    \begin{aligned}
        \mathit{MTH}(K_f(s)) &{}\wedge_{\MTSpin} \mathit{MTH}(Q_8)\\
        &\simeq \mathit{MTH}(K_f(s))\wedge_{\MTSpin}\MTSpin\wedge(B\Z/2\times B\Z/2)^{V-3}\\
        &\simeq \mathit{MTH}(K_f(s))\wedge (B\Z/2\times B\Z/2)^{V-3}.
    \end{aligned}
\end{equation}
Let $\mathcal S\coloneqq \set{\ko, \tau_{\ge 2}\ko, \ksp, \widetilde{\ko}, \ko\wedge (B\Z/2)^{\pm(\sigma-1}), N_2, N_3}$.
By \cref{thm:spinh_ABP,thm:pin_tilde_c_ABP,thm:pinh_ABP}, $\mathit{MTH}(K_f(s))$ is, as a homology $\ko$-module, equivalent to a sum of shifts of $\ko$-modules equivalent to some $M\in\mathcal S$ or to $H\Z/2$. For any spectrum $X$, $H\Z/2\wedge X$ is a sum of shifts of $H\Z/2$. Thus for any $\Sigma^t H\Z/2$ summand of $\mathit{MTH}(K_f(s))$, the summands coming from $\Sigma^t H\Z/2\wedge (B\Z/2\times B\Z/2)^{V-3}$ in~\eqref{complicated_decomp}
are equivalent to shifts of $H\Z/2$ as promised in the lemma statement. Therefore it suffices to consider summands that are shifts of the $\ko$-modules in $\mathcal S$---and since the lemma statement is insensitive to such shifts, we may assume the summands are in $\mathcal S$ without a shift.

For any $M\in\mathcal S$, the homology $\ko$-module structure on $M$ is induced from an actual $\ko$-module structure (see \cref{making_htpy_ko}). Thus there are equivalences of homology $\ko$-modules
\begin{equation}
  \begin{aligned}
    M\wedge (B\Z/2\times B\Z/2)^{V-3} &\simeq M\wedge_\ko \ko\wedge (B\Z/2\times B\Z/2)^{V-3}\\
    &\simeq M\wedge_\ko \paren*[\Bigg]{\ksp\vee \bigvee_{i\ge 0} \Sigma^{n_\ell}H\Z/2}
  \end{aligned}
\end{equation}
for some integers $n_\ell\ge 0$, with the latter equivalence by \cref{ko_Q8}. As in the previous paragraph, we may without loss of generality lose the $\Sigma^{n_\ell}H\Z/2$ summands. Therefore it suffices to show that $M\wedge_\ko \ksp\wedge \ME_{\ell,k}$ is a sum of shifts of $H\Z/2$ and a $\ko$-module of EA-type.

By the Künneth theorem for $H_\ko^*$, there is an $\cA(1)$-module isomorphism
\begin{equation}
    H_\ko^*(M\wedge_\ko\ksp) \cong H_\ko^*(M) \otimes H_\ko^*(\ksp).
\end{equation}
We claim that for all $M\in\mathcal S$, $H_\ko^*(M)$ is either stably invertible or $Q_1$-locally stably invertible, and likewise that $H_\ko^*(\ksp)$ is stably invertible. This amounts to checking, case by case, that we have already seen this: for $\ko$, stable invertibility of $H_\ko^*(\ko)\cong\Z/2$ is true by definition. For $\tau_{\ge 2}\ko$, see \cref{joker_stabinv}. For $\ksp$ and $\widetilde\ko$, see \cref{dissect_Q_R2}. For $\ko\wedge (B\Z/2)^{\pm(\sigma-1)}$, $N_2$, and $N_3$, see \cref{Q1_Pic}.
Thus by \cref{Q1stab,tensor_thru_Q1loc}, $H_\ko^*(M) \otimes H_\ko^*(\ksp)$ is also either stably invertible or $Q_1$-locally stably invertible, so by \cref{stab_inv_EA}, $M\wedge_\ko\ksp\wedge\ME_{\ell,k}$ is of EA-type, finishing the proof.
\end{proof}
\begin{thm}
\label{MSpinHs_psi}
With $s$ and $K_f(s)$ as in \cref{cts_AZ_f2i} and $\ell\ge 1$,
the spiral map
\begin{equation}
    \mathrm{sp}_{\ell,k}^\psi(\mathit{MTH}(K_f(s)))\colon I_\Z\mathit{MTH}(K_f(s))\wedge\ME_{\ell,k} \to I_\Z\mathit{MTH}(K_f(s)) \wedge\Sigma\ME_{\ell-1,k}
\end{equation}
sends long summands to long summands; on each long summand it is an inclusion $\Z/2^N\hookrightarrow \Z/2^{N+1}$ with cokernel $\Z/2$.
\end{thm}
\begin{proof}
Thanks to the generalized ABP splittings in \cref{thm:spinh_ABP,thm:pin_tilde_c_ABP,thm:pinh_ABP}, $\mathit{MTH}(K_f(s))$ is homology $\ko$-module equivalent to a sum of Whitney summands and shifts of $\ko$-modules $M$, where $M$ is as in \cref{ko_mods_psi_piece}. Therefore if $k>0$, \cref{ko_mods_psi_piece}, part~\eqref{whichtens__} finishes the proof.

This leaves the case $k = 0$, $\ell\ge 3$. In this case we defined $\mathrm{sp}^\psi_{\ell,0}$ using $\overline\psi_{\ell,k}$ in \cref{Q8_psi}, and the definition of $\mathrm{sp}^\psi_{\ell,0}(\mathit{MTH}(K_f(s)))$ in \cref{spiral_1_map_E_version} is the Anderson dual of the map
\begin{equation}
     \id_{\mathit{MTH}(K_f(s))}\wedge\overline\psi_{\ell,0}\colon \mathit{MQ}_{s,\ell,0} \longrightarrow
    \Sigma \mathit{MQ}_{s,\ell+1,1}
\end{equation}
(where $\mathit{MQ}_{s,\ell,k}$ is as defined in \cref{MQ_defn}).
By \cref{MQ_break_down}, $\mathit{MQ}_{s,\ell,0}$ is a sum of $\ko$-modules of EA-type and shifts of $H\Z/2$, so we can invoke part~\eqref{smash_EA} of \cref{ko_mods_psi_piece} and finish.
\end{proof}

\subsection{The complex cases: spiraling from 
\texorpdfstring{$\MTSpin^c$ and $\MTPin^c$}{MTSpinc and MTPinc}
}
\label{complex_computation}
The story over $\C$, i.e.\ using $\KU$ and $\MTSpin^c$, is similar but easier. In this section we compute the F2I and interacting Bott spiral maps in the complex case, again modulo Whitney summands. Our main results are \cref{cplx_F2I_image_thm,,ku_int_Bott,,complex_psi}, in which we compute the free-to-interacting map, $\mathrm{sp}_{\ell,k}^\phi$, and $\mathrm{sp}_{\ell,k}^\psi$ respectively in Altland--Zirnbauer class A. Analogously to \cref{BDI_DIII}, the identification of $\MTPin^c\simeq\MTSpin^c\wedge\ME_1$ in a manner compatible with the free-to-interacting map allows us to extend these results to class AIII.

The most striking change in our computation is that our two indices $\ell$ and $k$ behave identically, as discussed in \cref{cpx_collapse}. 
Recall the definition of $\ME_m$ from \cref{MEm_defn}.
\subsubsection{Working over \texorpdfstring{$\ku$}{ku}} 
\label{sss:ku}
We begin with $\ku$-module computations.

There is an isomorphism
\begin{equation}
    \ku_*\cong \Z[\beta_\C],\qquad\qquad \abs{\beta_\C} = 2,
\end{equation}
where $\beta_\C$ is the complex Bott class. As for $\beta$ in $\KO$-theory, $\beta_\C$ implements Bott periodicity: there is a natural isomorphism $\KU_*(X)\cong \ku_*(X)[\beta_\C^{-1}]$, and for any $\ku$-module $M$, $\KU\wedge_\ku M\simeq M[\beta_\C^{-1}]$. 

There is a $\ku$-module structure on $H\Z/2$ defined analogously to the $\ko$-module structure.
\begin{lem}[{Wilson~\cite[Proposition 1.2]{Wil73}}]
\label{cpx_K_RPinfty}
The reduced $\KU$-cohomology of $\RP^\infty$ is
\begin{equation}
    \widetilde{\KU}{}^i(\RP^\infty) = \begin{cases}
        ~~0, & i \text{ even} \\
        ~~\mu_{2^\infty}, & i \text{ odd}.
    \end{cases} \qedhere
\end{equation}
\end{lem}
\begin{defn}
We define $\ku$-modules of EA-type exactly as in \cref{EA_type}, but with $\ku$ and $\KU$ in place of $\ko$, resp.\ $\KO$. 
We define Whitney summands and long summands in the same way as \cref{def:summands}, mutatis mutandis.
\end{defn}

Because $\widetilde{\KU}{}^*(\RP^\infty)$ is simpler than $\widetilde{\KO}{}^*(\RP^\infty)$, the analog of \cref{def:summands} is simpler. There is no analog of short summands in the complex setting. %

Define $H_\ku^*$ analogously to $H_\ko^*$ from \cref{Hko_defn}: for a $\ku$-module $M$, $H_\ku^*(M)\coloneqq\pi_{-*}(\mathrm{Map}(M, H\Z/2))$. When $M = H\Z/2$, this is a graded $\Z/2$-algebra with respect to composition.

Recall the stable cohomology operations $Q_0\coloneqq\Sq^1$ and $Q_1\coloneqq\Sq^1\Sq^2 + \Sq^2\Sq^1$ from \cref{margdefns}, and let $\cE(1)\coloneqq\ang{Q_0, Q_1}\subset\cA$. \begin{lem}[{Baker~\cite[Theorem 5.1]{Bak20}}]
There is an isomorphism of $\Z$-graded $\Z/2$-algebras $H_\ku^*(H\Z/2)\cong\cE(1)$.
\end{lem}
Thus for any $\ku$-module $M$, $H_\ku^*(M)$ is naturally an $\cE(1)$-module. The analog of \cref{Hko_of_ko} holds for $\ku$-modules of the form $\ku\wedge X$.

We also have the analogue of \cref{BL} for $\ku$: a Baker--Lazarev Adams spectral sequence for a $\ku$-module $M$ with analogous convergence hypotheses:
\begin{subequations}
\begin{equation}
    E_2^{s,t} = \Ext_{\cE(1)}^{s,t}(H_\ku^*(M), \Z/2) \Longrightarrow \pi_*(M)_2^\wedge.
\end{equation}
And, just as in~\eqref{usuAdams}, if $M = \ku\wedge X$, this spectral sequence simplifies to one more commonly constructed using the change-of-rings theorem:
\begin{equation}\label{ku_X_Adams}
    E_2^{s,t} = \Ext_{\cE(1)}^{s,t}(H^*(X;\Z/2), \Z/2) \Longrightarrow \ku_*(X)_2^\wedge.
\end{equation}
\end{subequations}
The algebra $\Ext_{\cE(1)}(\Z/2)$ is the Adams $E_2$-page for $\ku$, and it acts on the Adams $E_2$-page for any $\ku$-module. Differentials commute with this action.
\begin{prop}[{Liulevicius~\cite[\S 2]{Liu62}}]
There is an isomorphism of $\Z^2$-graded $\Z/2$-algebras
\begin{equation}\label{usu_ku_Adams}
    \Ext_{\cE(1)}(\Z/2) \overset\cong\longrightarrow
    \Z/2[h_0, v_1],
\end{equation}
with $h_0\in\Ext^{1,1}$ and $v_1\in\Ext^{1,3}$.
\end{prop}
See also~\cite[Example 4.5.6]{BC18}.

The class $h_0$ in~\eqref{usu_ku_Adams} lifts to multiplication by $2$, and $v_1$ lifts to multiplication by $\beta_\C$. The analog of Margolis' theorem (\cref{margolis}) in this setting is true; see Bruner--Greenlees~\cite[\S 2.1]{BG03}.

\begin{figure}[h!]
\begin{subfigure}[c]{0.29\textwidth}
\includegraphics{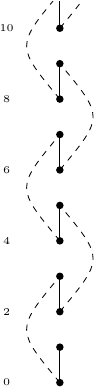}
\end{subfigure}
\begin{subfigure}[c]{0.7\textwidth}
\includegraphics{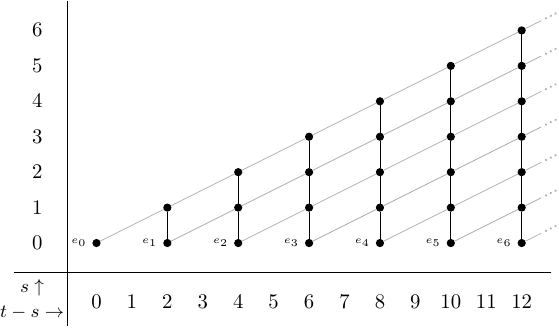}
\end{subfigure}
\caption{Left: the $\cE(1)$-module structure on $N_1$; straight lines are $Q_0$-actions and dashed curved lines are $Q_1$-actions. Right: $\Ext_{\cE(1)}(N_1)$, the $E_2$-page of the Adams spectral sequence for $\ku\wedge (B\Z/2)^{\sigma-1}$. Gray diagonal lines denote $v_1$-actions. All differentials and extension questions are trivial. We discuss this Adams spectral sequence in \cref{N1_exm_E1}. This figure adapted from~\cite[Figure 6, right]{debray_invertible_2021}.}
\label{N1_figure_E1}
\end{figure}

\begin{exm}\label{N1_exm_E1}
The prototypical $\ku$-module of EA-type is $\ku\wedge\RP^\infty$. Just like in \cref{N1_exm}, we will work with $\ku\wedge\ME_1\simeq \Sigma^{-1}(\ku\wedge\RP^\infty)$. To run the Adams spectral sequence, we need $\Ext_{\cE(1)}(N_1)$, which was computed by~\cite[Proposition 4.48]{debray_invertible_2021}\footnote{This source computes $\Ext_{\cE(1)}(\Sigma N_0)$, but $\Sigma N_0\cong N_1$ as $\cE(1)$-modules, which is yet another avatar of the fact that $\Pin^+\times_{\set{\pm 1}}\U_1$ and $\Pin^-\times_{\set{\pm 1}}\U_1$ define the same symmetry type: $\Pin^c$.}\textsuperscript{,}\footnote{See also Dhankhar--Field--Nigam--Quigley--Yang~\cite[Figure 2]{dhankhar_effective_2025} for an independent calculation of the $\Z/2[h_0]$-module structure on $\Ext_{\cE(1)}(N_1)$.} to be
\begin{equation}
    \Ext_{\cE(1)}(N_1)\cong \Z/2[h_0, v_1]\set{
        e_0, e_1, \dotsc
    }/(h_0e_0,\ v_1e_i - h_0e_{i+1})
\end{equation}
with $e_i\in\Ext^{0,2i}$. We draw this in \cref{N1_figure_E1}, right. All differentials vanish for degree reasons, and all extension questions are solved by the $h_0$- and $v_1$-actions, so we learn that
\begin{equation}
    \widetilde{\ku}_*(\ME_1)\cong\Z\set{\overline e_i\colon i\ge 0}/(2\overline e_0,\ \beta_\C \overline e_i - 2\overline e_{i+1})
\end{equation}
with $\abs{\overline e_i} = 2i$; see \cref{N1_figure_E1}. Therefore $\widetilde{\ku}_{2j+1}(\ME_1) = 0$ and $\widetilde{\ku}_{jk}(\ME_1)\cong\Z/2^{j+1}$ generated by $\overline e_j$; the latter is a long summand. Inverting $\beta_\C$, one concludes $\ku\wedge\ME_1$ is EA-type.

Just as for $\ko$, there are no Whitney summands, and we can obtain the same results for $\RP^\infty$ by shifting degrees by $1$. The groups $\ku_*(\RP^\infty)$ were first computed by Hashimoto~\cite[Theorem 3.1]{Has83}.
\end{exm}

\begin{prop}\label{ku_EA_N1}
For $m\ge 1$, $\ku\wedge\ME_m$ is of EA-type, and there is a stable isomorphism of $\cE(1)$-modules $H^*(\ME_m;\Z/2)\simeq \Sigma^{m-1} N_1$.
\end{prop}
\begin{proof}[Proof sketch]
The proof is identical to that of \cref{M_is_EA}, except using the stable equivalence $N_1\otimes N_1\simeq \Sigma N_1$ of $\cE(1)$-modules due to Ossa~\cite[\S 3]{Oss89}. We also need the $\ku$-module analog of \cref{EA-coh}, that if $H^*(X;\Z/2)\simeq \Sigma^n N_1$ as $\cE(1)$-modules then $\ku\wedge X$ is of EA-type, but this can be proven in a similar way using the analysis of \cref{N1_exm_E1}, which used nothing about $\ME_1$ save for the $\cE(1)$-module structure on its cohomology.
\end{proof}

\begin{rem}
Comparing $N_1^{\otimes n}$ as an $\cE(1)$-module versus as an $\cA(1)$-module, one discovers that the $\cA(1)$-modules $N_0$, $N_1$, $N_2$, and $N_3$ are all stably equivalent as $\cE(1)$-modules (in the sense of \cref{stab_iso}) up to a shift.
This can be verified directly (c.f.~\cite[Chapter 2]{Yu95} or~\cite[\S 4.4.4, \S 5.2.4]{debray_invertible_2021}).
\end{rem}
\begin{prop}\label{this}
There is an isomorphism
\begin{equation}
    \ku_n(\ME_m)/(\text{\rm Whitney summands}) \cong
    \begin{cases}
        \Z/2^{j+1}, & n = m + 2j-1,\ j\ge 0\\
        0, &\text{\rm otherwise.}
    \end{cases}
\end{equation}
\end{prop}
\begin{proof}
Just as in the proof of \cref{melk_calc}, use Margolis' theorem to reduce to studying the Ext of the unique non-free $\cE(1)$-module summand of $H^*(\ME_m;\Z/2)$. By \cref{ku_EA_N1}, this amounts to just studying $N_1$ up to a shift, which we did in \cref{N1_exm_E1}.
\end{proof}
\subsubsection{Lifting to \texorpdfstring{$\MTSpin^c$}{MTSpinc}}
\label{ss:comp_spinc}
Before we describe the image of the free-to-interacting map, we recall Anderson--Brown--Peterson's splitting of $\MTSpin^c$. It is a little simpler than the $\MTSpin$ version from \cref{ABP_thm}. %
\begin{thm}[{Anderson--Brown--Peterson~\cite{ABP67}, Stong~\cite[Chapter XI]{Sto68}}]
\label{ABP_cpx_thm}
There are numbers $a_i,b_j\in\N$ and a $2$-local homotopy equivalence
\begin{equation}
    \MTSpin^c \overset\simeq\longrightarrow
        \bigvee_{i\ge 1} \textcolor{Rhodamine}{\Sigma^{s_i}\ku} \vee
        \bigvee_{j\ge 1} \textcolor{Cyan}{\Sigma^{t_j}H\Z/2}.
\end{equation}
The degrees of the first few shifts are
\begin{equation}
\begin{aligned}\label{spinc_shifts}
    \textcolor{Rhodamine}{(s_i)} &= (0, 4, 8, 8, 12, 12, 12, 16, 16, 16, 16, 16, \dotsc)\\
    \textcolor{Cyan}{(t_j)} &= (10, 14, 18, 18, 18, 20, 22, 22, 22, 22, 22, \dotsc)
\end{aligned}
\end{equation}
The projection $\mathrm{proj}_{\textcolor{Rhodamine}{s_1}}\colon\MTSpin^c\to \textcolor{Rhodamine}{\ku}$ onto the first summand is equivalent to the connective cover of the Atiyah--Bott--Shapiro map.
\end{thm}
Anderson--Brown--Peterson write that their proof of the splitting of $\MTSpin$ (\cref{ABP_thm}) can be generalized to the complex case; Stong is the first to write it down. See Abdallah--Salch~\cite{AS24} for discussion of the degrees $\textcolor{Cyan}{(t_j)}$ in \cref{ABP_cpx_thm}.
\begin{rem}
The real Atiyah--Bott--Shapiro map is $7$-connected, which means that in physics applications to IFTs of (spatial) dimension $6$ and below, one can essentially ignore the difference between $\MTSpin$ and $\ko$. However, because $\textcolor{Rhodamine}{s_2} = 4$ in the complex ABP splitting, the complex Atiyah--Bott--Shapiro map is only $3$-connected, and therefore using $\ku$ instead of $\MTSpin^c$ misses IFTs beginning in spatial dimension $2$. This includes, for example, the example studied 
 by Wang--Senthil in~\cite[\S III.E]{WS14}.
\end{rem}
We define a Whitney summand of an $\MTSpin^c$-module to be a Whitney summand of the underlying $\MTSpin$-module.\footnote{One could try to generalize Stolz' analysis in~\cite{Sto94} to maps $\ku\to\MTSpin^c$ splitting the Atiyah--Bott--Shapiro map, and therefore obtain an principle narrower definition of a Whitney summand and therefore a sharper result on the Bott spiral. We felt this would digress too far from the present discussion.}

\begin{thm}\label{cplx_F2I_image_thm}
On homotopy groups, the image of the free-to-interacting map $\Sigma^{m-2}\KU\to \Sigma^2 I_\Z(\MTSpin^c\wedge\ME_m)$ is contained in the subgroup  $(\Sigma^2 I_\Z(\textcolor{Rhodamine}{\ku}\wedge\ME_m))^*\subset (\Sigma^2 I_\Z(\MTSpin^c\wedge\ME_m))^*$ corresponding to the first factor in the Atiyah--Bott--Shapiro decomposition. Restricted to that subgroup of the codomain, the free-to-interacting map is
a surjection $\Z\to\Z/2^{j+1}$ 
onto the unique long summand in degree $m + 2j-4$ for each $j\ge 0$,
and the zero map in degrees $n\equiv m+1\bmod 2$;
the Whitney summands are not in the image.
\end{thm}

\begin{proof}[Proof sketch]
The proof is quite similar to the proofs of \cref{compute_F2I,only_first_summand_real}. The only significant difference is a simplification: when studying the image of the F2I map $\Sigma^{m-2}\KU\to \Sigma^2 I_\Z(\ku\wedge\ME_m)$, instead of a two-step induction on $k$, then $\ell$, we only need to induct on $m$. Therefore this proof includes analogs of \cref{compute_11,compute_1k}, but we do not need to adapt \cref{compute_all} or the arguments supporting it.
\end{proof}
Recall that $m = \ell+k$.
\begin{thm}\label{ku_int_Bott}
The 
spiral map $\spint_{\ell,k}\colon I_\Z(\MTSpin^c\wedge\ME_{m+1})\to \Sigma I_\Z(\MTSpin^c\wedge\ME_m)$ is an isomorphism modulo Whitney summands.
\end{thm}
The argument proceeds completely analogously to those for \cref{int_Bott_spiral_ko,int_Bott_spiral_1}.
\begin{thm}\label{complex_psi}
The spiral map $\mathrm{sp}_{\ell,k}^\psi\colon I_\Z(\MTSpin^c\wedge\ME_{m+1})\to \Sigma I_\Z(\MTSpin^c\wedge\ME_{m})$ sends long summands to long summands; on each long summand it is an inclusion $\Z/2^N\hookrightarrow \Z/2^{N+1}$, with cokernel $\Z/2$.
\end{thm}
Since we can treat the indices $\ell$ and $k$ identically when working over $\MTSpin^c$, the extra complication in the definition of $\mathrm{sp}^\psi_{\ell,0}$ in the real case (see \cref{Q8_psi}) does not appear here.
\begin{proof}
As usual, first work over $\ku$, then use \cref{ABP_cpx_thm} to lift to $\MTSpin^c$. Since all non-Whitney summands in the ABP decomposition of $\MTSpin^c$ are shifts of $\ku$, the second step is immediate, and we focus on the first step.

The calculation over $\ku$ is also analogous to the proof of \cref{psicalc}. The part with the diagonal is exactly the same; the part with $e_\sigma$ is similar, except that we want to work over $\cE(1)$, rather than $\cA(1)$. For $i = 0,1,2,3$, there is a stable isomorphism of $\cE(1)$-modules $F_i\simeq\Sigma^i F_0$, so we focus on $F_0$. \cite[(4.49)]{debray_invertible_2021} shows $\Ext_{\cE(1)}(F_0)\cong\Z/2[v_1]$ with $v_1\in\Ext^{1,3}$ as a module over $\Ext_{\cE(1)}(\Z/2)$.\footnote{As just a bigraded abelian group, this Ext computation was done earlier by Davis--Mahowald~\cite[\S 2]{DM81}.} Thus, analogously to \cref{e1fibcalc},
\begin{equation}
    \ku_n(\Sigma^{-1}\RP^2\wedge\ME_m)/(\text{Whitney summands})\cong\begin{cases}
        \Z/2, & n = 2j+m,\ j\ge 0\\
        0, &\text{otherwise,}
    \end{cases}
\end{equation}
and the Bott class acts injectively on the non-Whitney summands. Thus one can imitate the rest of the proof of \cref{pinp_euler_comp} and finish.
\end{proof}
We will occasionally need to know the groups $\mho_{\Spin^c}^*(\ME_2)$---corresponding to class $\mathrm{A}'$ in the language of \S\ref{primed_AZ}---in low degrees, including Whitney summands. This is the \spinc analog of the dpin computations in~\cite[Appendices E and F]{kaidi_topological_2020}.
\begin{prop}
\label{dpin_c}
There are isomorphisms
\begin{equation}
    \begin{alignedat}{2}
        \mho_{\Spin^c}^0(\ME_2) &\cong \Z/2 \qquad\qquad\qquad
            & \mho_{\Spin^c}^4(\ME_2) &\cong (\Z/2)^{\oplus 4}\\
        \mho_{\Spin^c}^1(\ME_2) &\cong \Z/2
            & \mho_{\Spin^c}^5(\ME_2) &\cong \Z/8\oplus\Z/2\\    
        \mho_{\Spin^c}^2(\ME_2) &\cong (\Z/2)^{\oplus 2}
            & \mho_{\Spin^c}^6(\ME_2) &\cong (\Z/2)^{\oplus 6}.\\
        \mho_{\Spin^c}^3(\ME_2) &\cong \Z/4
    \end{alignedat}
\end{equation}
In the groups given above, a $\Z/2$ summand is a Whitney summand if and only if it has even degree.
\end{prop}
\begin{proof}
As usual, the universal property~\eqref{IZproperty} of $I_\Z$ reduces the question to the computation of $\Omega_*^{\Spin^c}(\ME_2)$ in degrees $6$ and below, together with the assertion that $\Omega_7^{\Spin^c}(\ME_2)$ is torsion. The proof of the latter fact is closely analogous to the proof over $\MTSpin$, discussed in and after \cref{spinbord_torsion}. For the bordism groups, use the \spinc ABP decomposition (\cref{ABP_cpx_thm}) to identify $\Omega_k^{\Spin^c}(\ME_2)\cong\ku_k(\ME_2)\oplus\ku_{k-4}(\ME_2)$ for $k\le 7$.

Since $\ME_2$ is trivial at odd primes, it suffices to work $2$-locally, so we can use the Adams spectral sequence over $\cE(1)$, given in~\eqref{ku_X_Adams}. The first step is to compute $H^*(\ME_2;\Z/2)$ as an $\cE(1)$-module in low degrees, this time up to isomorphism, not stable equivalence. From~\cite[(F.11)]{kaidi_topological_2020} we learn that as $\cA(1)$-modules,
\begin{subequations}
\begin{equation}\label{A1_dpinc}
    H^*(\ME_2;\Z/2)\cong F\oplus \Sigma N_1 \oplus P,
\end{equation}
where $P$ is concentrated in degrees $7$ and above, so we may ignore it, and $F$ is a free $\cA(1)$-module with a basis of homogeneous elements in degrees $0$, $2$, $4$, $4$, $6$, and $6$. Since $\cE(1)$ is a subalgebra of $\cA(1)$, this determines the $\cE(1)$-module structure: because there is an $\cE(1)$-module isomorphism $\cA(1)\cong\cE(1)\oplus\Sigma^2\cE(1)$~\cite[\S 4.7.2]{Bay94}, we learn that as $\cE(1)$-modules,
\begin{equation}
    H^*(\ME_2;\Z/2) \cong F^c\oplus \Sigma N_1 \oplus P.
\end{equation}
\end{subequations}
Here $P$ is as in~\eqref{A1_dpinc} and $F^c$ is a free $\cE(1)$-module with a basis of homogeneous elements in degrees $0$, $2$, $2$, $4$, $4$, $4$, $6$, $6$, $6$, $6$, $8$, and $8$. By Margolis' theorem, the summands in $F^c$ lift to Whitney summands in $\ku_*(\ME_2)$, giving the Whitney summands in the proposition statement. The rest of the proof is identical to the Adams computation in \cref{N1_exm_E1}, but with the degrees shifted up by one.
\end{proof}

\subsection{\texorpdfstring{$\Spin\times_{\set{\pm 1}}Q_8$}{spinQ8} bordism}
\label{spinQ8}
In this subsection, we compute $\Spin\times_{\set{\pm 1}}Q_8$ bordism (contingent as usual on the specific degrees of the suspensions in the Anderson--Brown--Peterson splitting from \cref{ABP_thm}) and the corresponding free-to-interacting map. This is the exceptional case from \cref{discrete_ne_cts}: the fermionic group $Q_8$ is not isomorphic to $E_{\ell,k}$ for any $(\ell,k)$, so the bordism groups we need are not computed in \cref{melk_calc}. Fortunately, the computation is not any more difficult.
\begin{lem}[{Mitchell--Priddy~\cite[\S 2]{Priddy}}]
\label{xtn_class_Q8}
The central extension $1\to\Z/2\to Q_8\to \Z/2\times\Z/2\to 1$ is classified by $x^2+xy+y^2\in H^2(B(\Z/2\times \Z/2); \Z/2)$, where $x$, resp.\ $y$ are dual to the first, resp.\ second $\Z/2$ summands.
\end{lem}
Since the grading $\theta\colon Q_8\to\Z/2$ is zero, $H(Q_8)\cong \Spin\times_{\set{\pm 1}}Q_8$~\cite[Example 9]{stehouwer_interacting_2022} and \cref{xtn_class_Q8} gives us the following result.
\begin{cor}
\label{what_is_Q8_cor}
A $\Spin\times_{\set{\pm 1}}Q_8$ structure is equivalent to a $(B\Z/2\times B\Z/2, 0, x^2+xy+y^2)$-twisted spin structure.
\end{cor}
\begin{lem}\label{shearQ8}
For $i = 1,2$, let $\sigma_i\to B\Z/2\times B\Z/2$ be the line bundle that is nontrivial on the $i^{\mathrm{th}}$ $\Z/2$ summand and trivial on the other summand. Let $V\coloneqq \sigma_1 \oplus\sigma_2\oplus (\sigma_1\otimes\sigma_2)$;
then there is an $\MTSpin$-module equivalence
\begin{equation}
    \mathit{MT}(\Spin\times_{\set{\pm 1}}Q_8) \overset\simeq\longrightarrow \MTSpin\wedge (B\Z/2\times B\Z/2)^{V-3}.
\end{equation}
\end{lem}
\begin{proof}
By the Whitney sum formula, $(w_1(V), w_2(V)) = (0, x^2+xy+y^2)$, matching the twist in \cref{what_is_Q8_cor}. Thus \cref{shearing} finishes the proof.
\end{proof}
Moreover, interpreting the representation $V$ as a homomorphism $\rho\colon \Z/2\times\Z/2\inj\SO_3$, we see that upon taking spin covers, $\rho$ is the map $Q_8\inj\SU_2$ from the discrete to the continuous class C fermionic group that we discussed in \cref{disc_cont_defn}.

As usual, we start with the corresponding twisted $\ko$-homology.
\begin{lem}
\label{split_ko_BSO}
There is a $\ko$-module splitting
\begin{equation}
    \label{ko_spinh_ABP}
    \ko\wedge (B\SO_3)^{V-3} \simeq\ksp\vee\dotsb.
\end{equation}
such that the projection $p_0\colon \ko\wedge (B\SO_3)^{V-3}\to\ksp$ is $3$-connected.
\end{lem}
\begin{proof}
Buchanan--McKean~\cite{BM23} and Mills~\cite{Mil24} (\cref{thm:spinh_ABP}) split $\MTSpin^h\simeq\MTSpin\wedge (B\SO_3)^{V-3}$; using the original Anderson--Brown--Peterson splitting (\cref{ABP_thm}) $\MTSpin\simeq\ko\vee\dotsb$ and smashing with $(B\SO_3)^{V-3}$, Buchanan--McKean--Mills' splitting induces a splitting of the form~\eqref{ko_spinh_ABP} by taking the unique summands of each splitting which are not $1$-connected. Inspecting the remaining summands, the projection onto the $\ksp$ factor is a $3$-connected map.
\end{proof}
\begin{prop}
\label{ko_Q8}\hfill
\begin{enumerate}
    \item There is a $\ko$-module equivalence
    \begin{equation}
    \label{eqncase8}
        \ko\wedge (B\Z/2\times B\Z/2)^{V-3} \simeq \ksp\vee \bigvee_{i\ge 0} \Sigma^{n_\ell}H\Z/2.
    \end{equation}
    If $n = 4k+m$, $0\le m<4$, the number of $\Sigma^n H\Z/2$ summands in~\eqref{eqncase8} is $0$ if $m$ is even, $k+2$ if $m = 1$, and $k+1$ if $m = 3$.
    \item\label{Q8_to_SU2} With respect to the decomposition~\eqref{ko_spinh_ABP}, the map
    \begin{equation}\label{rho_ko_defn}
        \rho\colon \ko\wedge (B\Z/2\times B\Z/2)^{V-3}\to\ko\wedge (B\SO_3)^{V-3}
    \end{equation}
    induced by the fermionic group homomorphism $Q_8\inj\SU_2$ is the identity on the $\ksp$ summands.
\end{enumerate}
\end{prop}
Before we prove this proposition, we make a few Margolis homology calculations that we will need.
\begin{lem}
\label{wrong_marg_hom}
Let $V' \coloneqq \sigma_1 \oplus \sigma_2\oplus 3(\sigma_1\otimes\sigma_2)$ and $U\in H^0((B\Z/2\times B\Z/2)^{V'-5};\Z/2)$ denote the mod $2$ Thom class. Then $H^*((B\Z/2\times B\Z/2)^{V'-5}; Q_0)\cong\Z/2$, generated by $U$, and $H^*((B\Z/2\times B\Z/2)^{V'-5}; Q_1)\cong \Sigma^2\Z/2$, generated by $U(a^2 + ab + b^2)$.
\end{lem}
\begin{proof}
By the Whitney sum formula, $w_1(V') = 0$ and $w_2(V') = ab$. Thus $Q_0(U) = \Sq^1(U) = 0$, so the Thom isomorphism $H^*(B\Z/2\times B\Z/2;\Z/2)\to H^*((B\Z/2\times B\Z/2)^{V'-5};\Z/2)$ commutes with the $Q_0$-action. Thus $H^*((B\Z/2\times B\Z/2)^{V'-5}; Q_0)\cong H^*(B\Z/2\times B\Z/2; Q_0)\cong\Z/2$ concentrated in degree $0$ by the Künneth formula (\cref{kunneth_margolis}). Giambalvo~\cite[Proof
of Lemma 3.4]{Gia76} calculates the $Q_1$-homology (in his notation, this is $H(\Phi(Z_2[\alpha, \beta]), Q_1)$), showing that it is one-dimensional and $U(a^2 + ab + b^2)$ generates. %
\end{proof}
\begin{cor}
\label{rightmarg}
\Cref{wrong_marg_hom} is true mutatis mutandis with $V-3$ in place of $V'-5$.
\end{cor}
\begin{proof}
If $U$, resp.\ $U'$ are the Thom classes for $(B\Z/2\times B\Z/2)^{V-3}$, resp.\ $(B\Z/2\times B\Z/2)^{V'-5}$, respectively, then by the Wu formula, the Thom isomorphisms to the cohomology of $B\Z/2\times B\Z/2$ identify $Q_i(U)$ and $Q_i(U')$ for $i = 0,1$. Therefore these isomorphisms induce an isomorphism of the $Q_0$- and $Q_1$-Margolis homologies of $(B\Z/2\times B\Z/2)^{V-3}$ and $(B\Z/2\times B\Z/2)^{V' - 5}$.\footnote{This strategy generalizes to any pair of twists of $\ko$-homology over a space $X$ which induce equivalent twists of $\ku$, because the $Q_0$- and $Q_1$-homology only depend on the underlying $\cE(1)$-module structure of an $\cA(1)$-module.}
\end{proof}
\begin{proof}[Proof of \cref{ko_Q8}]
Localized away from $2$, this is trivial (as $B\Z/2\times B\Z/2\simeq \pt$ at odd primes), so in the rest of the proof, we localize at $2$. We will show that the map $\rho$ in~\eqref{rho_ko_defn} induces an isomorphism on $Q_0$- and $Q_1$-Margolis homology, once we restrict to the $\ksp$ summand of $\ko\wedge (B\SO_3)^{V-3}$; since $\ksp$ has no free summands, this and the stable Whitehead theorem imply that $\ko\wedge (B\Z/2\times B\Z/2)^{V-3}$ is $\ko$-module equivalent to a sum of $\ksp$ and some $\Sigma^k H\Z/2$ summands. The specific degrees of the $\Sigma^k H\Z/2$ summands can be computed using a Poincaré polynomial computation similar to those in~\cite[\S\S D.3--D.6]{freed_reflection_2021}.

After applying \cref{Hko_of_ko} and the Thom isomorphism, the effect of $\rho$ on $H_\ko^*$ is
the same as $(B\rho^*)\colon H^*(B\SO_3;\Z/2)\to H^*(BV;\Z/2)$. Recall from \cref{split_ko_BSO} that there is a $3$-connected map $p_0\colon \ko\wedge (B\SO_3)^{V-3}\to \ksp$ which is the projection onto a direct $\ko$-module summand. %
Let $\widetilde\rho\coloneqq p_0\circ\rho$; we will show
$\widetilde\rho$ is an isomorphism on $Q_0$- and $Q_1$-Margolis homology. %

Baker~\cite[\S 5]{Bak20} shows $H_\ko^*(\ksp)\cong\uQ$, where $\uQ\coloneqq\cA(1)/(\Sq^1, \Sq^2\Sq^3)$. Adams--Priddy~\cite[\S 3]{AP76} showed that $H_\ko^*(\ksp; Q_0)\cong\Z/2$ and $H_\ko^*(\ksp;Q_1)\cong\Sigma^2\Z/2$. Since
$p_0$ is $3$-connected, the generators pull back to the unique nonzero classes of $H_\ko^*(\ko\wedge (B\SO_3)^{V-3})\cong H^*((B\SO_3)^{V-3};\Z/2)$ (\cref{Hko_of_ko}) in
degrees $0$ and $2$, which are $U$ and $Uw_2$, respectively. Then $\rho$ pulls these back to $U$, resp.
\begin{equation}
        U(w_2(\sigma_1\oplus\sigma_2\oplus (\sigma_1\otimes\sigma_2))) = U(a^2+ab+b^2),
\end{equation}
so by \cref{rightmarg}, $\widetilde\rho$ induces an isomorphism on $Q_0$- and $Q_1$-Margolis homology as promised, and the proof is
complete.
\end{proof}
\begin{lem}
\label{joker_ksp}
There is a $\ko$-module equivalence $\tau_{\ge 2}\ko\wedge_\ko\ksp\simeq \Sigma^2 H\Z/2\vee \Sigma^4\widetilde{\ko}$.
\end{lem}
\begin{proof}
Baker~\cite[\S 5]{Bak20} shows $H_\ko^*(\tau_{\ge 2}\ko)\simeq \Sigma^2 J$, $H_\ko^*(\ksp)\cong\uQ$, and $H_\ko^*(\widetilde{\ko})\cong R_2$, where $J$, $\uQ$, and $R_2$ are the $\cA(1)$-modules we defined in \cref{defn_R2,dissect_joker,dissect_Q_R2}. Mills~\cite[Lemma 3.4(2)]{Mil24} proves $J\otimes \uQ\cong \cA(1)\oplus \Sigma^2 R_2$; then the $2$-localization of the lemma statement follows from Stolz' splitting theorem~\cite[Theorem 4.1]{Sto94}.

At odd primes, the lemma is less exciting: there are $\ko[1/2]$-module equivalences $\tau_{\ge 2}\ko [1/2]\simeq\Sigma^4\ko[1/2]$, $\ksp[1/2]\simeq\ko[1/2]$, and $\widetilde\ko[1/2]\simeq\ko[1/2]$, which can be proven from the definitions of these $\ko$-modules and the fact that at localized at an odd prime $p$, the generator of $\pi_4\ko$ is a unit in the ring $\pi_*\ko_{(p)}$. After these identifications the result is straightforward.
\end{proof}
We use the following theorem, which splits the spectrum $\mathit{MT}(\Spin\times_{\set{\pm 1}}Q_8)$ and computes its homotopy groups, in the proof of \cref{special_case_psi_ko}, calculating the maps $\mathrm{sp}^\psi_{\ell,0}$ on long summands. This is not circular: our proof of \cref{spin_Q8_bordism} does not use the maps $\mathrm{sp}_{\ell,k}^\psi$ at all, nor anything that depends on them.
\begin{thm}
\label{spin_Q8_bordism}
There are numbers $u_i$, $v_j$, and $w_k$ and a homology $\ko$-module equivalence
\begin{equation}
    \mathit{MTH}(Q_8)\simeq \mathit{MT}(\Spin\times_{\set{\pm 1}} Q_8) \overset\simeq\longrightarrow \bigvee_{i\ge 1} \Sigma^{u_i}\ksp\vee \bigvee_{j\ge 1} \Sigma^{v_j}\widetilde{\ko}\vee\bigvee_{k\ge 1} \Sigma^{w_k}H\Z/2.
\end{equation}
Explicitly, on homotopy groups, recall the numbers $\textcolor{Rhodamine}{a_i}$, $\textcolor{Orange}{b_j}$, and $\textcolor{Cyan}{c_k}$ from \cref{ABP_thm}, and the numbers $n_\ell$ from \cref{ko_Q8}.
$\Omega_*^{\Spin\times_{\set{\pm 1}}Q_8}$ is the direct sum of the following abelian groups.
\begin{enumerate}
    \item For each $i\ge 1$:
    \begin{enumerate}
        \item a copy of $\pi_{*+\textcolor{Rhodamine}{a_i}}(\ksp)$, and
        \item for each $\ell\ge 1$, a $\Z/2$ in degree $\textcolor{Rhodamine}{a_i}+n_\ell$.
    \end{enumerate}
    \item For each $j\ge 1$:
    \begin{enumerate}
        \item a copy of $\pi_{*+\textcolor{Orange}{b_j}+4}(\widetilde{\ko})$, a $\Z/2$ in degree $\textcolor{Orange}{b_j} + 2$, and
        \item for each $\ell\ge 1$ and $m\in \{2,\dotsc,6\}$, a $\Z/2$ in degree $\textcolor{Orange}{b_j}+n_\ell+m$.
    \end{enumerate}
    \item For each $k\ge 1$:
    \begin{enumerate}
        \item for each $m\in\{0,2,3\}$, a $\Z/2$ in degree $\textcolor{Cyan}{c_k} + m$, and
        \item for each $\ell\ge 1$ and $m\in\{0,1,2,4,5,6\}$, a $\Z/2$ in degree $\textcolor{Cyan}{c_k} + n_\ell + m$, and a $\Z/2\oplus\Z/2$ in degree $\textcolor{Cyan}{c_k} + n_\ell + 3$.
    \end{enumerate}
\end{enumerate}
\end{thm}
In degrees $4$ and below, these bordism groups were computed by Pedrotti~\cite[Theorem 8.0.8]{Ped17}; the remaining computations are new. From the definitions of $\widetilde\ko$ and $\ksp$ in \cref{thm:spinh_ABP}, we see that $\pi_*(\widetilde\ko)\cong\pi_*(\ko)$ (though there is no map of $\ko$-modules implementing this isomorphism), and $\pi_n(\ksp)\cong\pi_{n+4}(\ko)$ as long as $n\ge 0$. In negative degrees $\pi_*(\ksp)$ vanishes.
\begin{proof}
Plug~\eqref{eqncase8} into the Anderson--Brown--Peterson splitting \eqref{ABP_smash_X}, then compute $\tau_{\ge 2}\ko\wedge_\ko \ksp$, $\tau_{\ge 2}\ko\wedge_\ko H\Z/2$, $H\Z/2\wedge_\ko \ksp$, and $H\Z/2\wedge_\ko H\Z/2$. The first of these is \cref{joker_ksp}; for the rest, use the identification $\pi_*(H\Z/2\wedge_\ko M)\cong H_\ko^*(M)$ for any $\ko$-module $M$, which can be proven using the Künneth formula for $H_\ko^*$ and Margolis' theorem (\cref{margolis}).
\end{proof}
Explicitly, we get the following bordism groups in low degrees.
\begin{equation}\label{explicit_Q8}
\begin{alignedat}{2}
    \Omega_0^{\Spin\times_{\set{\pm 1}}Q_8} &\cong\Z \qquad\qquad & \Omega_7^{\Spin\times_{\set{\pm 1}}Q_8} &\cong (\Z/2)^{\oplus 3}\\
    \Omega_1^{\Spin\times_{\set{\pm 1}}Q_8} &\cong (\Z/2)^{\oplus 2} \qquad\qquad & \Omega_8^{\Spin\times_{\set{\pm 1}}Q_8} &\cong \Z^2\\
    \Omega_2^{\Spin\times_{\set{\pm 1}}Q_8} &\cong 0 \qquad\qquad & \Omega_9^{\Spin\times_{\set{\pm 1}}Q_8} &\cong (\Z/2)^{\oplus 8}\\
    \Omega_3^{\Spin\times_{\set{\pm 1}}Q_8} &\cong\Z/2 \qquad\qquad & \Omega_{10}^{\Spin\times_{\set{\pm 1}}Q_8} &\cong \Z/2\\
    \Omega_4^{\Spin\times_{\set{\pm 1}}Q_8} &\cong\Z \qquad\qquad & \Omega_{11}^{\Spin\times_{\set{\pm 1}}Q_8} &\cong (\Z/2)^{\oplus 8}\\
    \Omega_5^{\Spin\times_{\set{\pm 1}}Q_8} &\cong (\Z/2)^{\oplus 4} \qquad\qquad & \Omega_{12}^{\Spin\times_{\set{\pm 1}}Q_8} &\cong \Z^3\oplus\Z/2.\\
    \Omega_6^{\Spin\times_{\set{\pm 1}}Q_8} &\cong\Z/2
\end{alignedat}
\end{equation}
\begin{rem}
\label{signature_rem}
Teichner~\cite[\S 1]{teichner_signature_1993} showed that the signature $\sigma\colon\Omega_4^{\Spin\times_{\set{\pm 1}}Q_8}\to\Z$ has image $4\Z$; Pedrotti~\cite[Theorem 8.0.8]{Ped17} showed $\tfrac 14\sigma$ is an isomorphism onto $\Z$. One can choose the generator to be the quotient $H$ of a smooth Enriques surface by a free antiholomorphic involution, as Teichner~\cite[\S 1]{teichner_signature_1993} showed this is a closed $\Spin\times_{\set{\pm 1}}Q_8$ $4$-manifold with signature $4$. Hitchin~\cite[\S 4]{hitchin_compact_1974} showed that Enriques surfaces with these involutions exist, and gave an example.

By \cref{ko_Q8}, part~\eqref{Q8_to_SU2}, $H$ also generates a $\Z$ summand of $\Omega_4^{\Spin^h}\cong\Z^2$, though Hu~\cite[\S A]{hu_invariants_2023} has already shown this summand is generated by $\mathbb{HP}^1$. \Cref{ko_Q8}, together with Buchanan--McKean's work~\cite{BM23} on \spinh bordism, can also be used to show that $\Spin\times_{\set{\pm 1}}Q_8$ bordism is detected by Buchanan--McKean's $\mathit{KSp}$-characteristic classes~\cite[\S 5.5]{BM23}, Anderson--Brown--Peterson's $\KO$-Pontrjagin classes~\cite[\S 4]{anderson_SU_1967}, and characteristic classes in mod $2$ cohomology (``Whitney classes'' in the language of Conner--Floyd~\cite[\S 17]{ConnerFloyd}; see \cref{def:summands}).
\end{rem}
Recall from \cref{Q8_F2I} that we defined a free-to-interacting map $F2I_{s=4}^\delta\colon \Sigma^{2}\KO\to \Sigma^2 I_\Z(\mathit{MT}(\Spin\times_{\set{\pm 1}}Q_8))$, essentially by composing the usual \spinh free-to-interacting map with $\rho\colon Q_8\inj\SU_2$.
\begin{prop}\label{disc_F2I_compute}
$I_\Z\rho\colon I_\Z\MTSpin^h\to I_\Z\mathit{MT}(\Spin\times_{\set{\pm 1}}Q_8)$ restricts to an isomorphism from the image of the continuous class C free-to-interacting map to the image of the discrete class C free-to-interacting map.
\end{prop}
Of course, the cokernels are quite different. For example, the above proposition implies the entirety of $\mho_{\Spin\times_{\set{\pm 1}Q_8}}^1\cong (\Z/2)^{\oplus 2}$ is in the cokernel of the discrete F2I map, whereas by~\cite[Corollary 9.99]{freed_reflection_2021}, the cokernel of the continuous F2I map vanishes in degrees $2$ and below.
\begin{proof}
By \cref{IZ_ABS_through_ko}, we may replace $I_\Z(\mathit{MT}(\Spin\times_{\set{\pm 1}}Q_8))$ with $I_\Z(\ko\wedge (B\Z/2\times B\Z/2)^{V-3})$ and $I_\Z\MTSpin^h$ with $I_\Z(\ko\wedge (B\SO_3)^{V-3})$.
After this, the proposition follows from a combination of the fact that the continuous free-to-interacting map has image contained in the $I_\Z\ksp$ summand of $I_\Z\MTSpin^h$, which follows from \cref{thm:spinh_ABP}, and the fact that $I_\Z\rho$ preserves this $I_\Z\ksp$ summand and kills all other summands of $I_\Z\ko\wedge (B\SO_3)^{V-3}$, which is \cref{ko_Q8}, part~\eqref{Q8_to_SU2}.
\end{proof}
Freed--Hopkins~\cite[Corollary 9.99]{freed_reflection_2021} calculate the image of the continuous free-to-interacting map: $\Z$ in degrees $d\equiv 3\bmod 4$, $d\ge -1$, and $\Z/2$ in degrees $d\equiv 5,6\bmod 8$, $d\ge 0$.

\bibliographystyle{alpha}
\bibliography{CK_Zotero}

\end{document}